\documentclass[aps,prx,reprint,superscriptaddress,longbibliography,nofootinbib]{revtex4-2}
\usepackage[T1]{fontenc}
\usepackage{lmodern}
\usepackage{microtype}
\usepackage[english]{babel}
\usepackage{enumitem}
\usepackage{algorithm}
\usepackage{algpseudocode}
\usepackage{comment}
\setlist[enumerate,itemize]{itemsep=0.4ex,topsep=0.8ex,leftmargin=1.5em}
\usepackage{amsmath,amssymb,amsthm,mathtools,bm,booktabs,graphicx}
\usepackage[dvipsnames]{xcolor}
\usepackage{tikz}
\usetikzlibrary{arrows.meta,positioning}
\tikzset{
 irrepbox/.style={draw=black!55,rounded corners=2pt,line width=.55pt,
   align=center,inner xsep=5pt,inner ysep=6pt,font=\small,
   text width=143mm},
 irrepcommon/.style={irrepbox,fill=blue!4},
 irrepbasis/.style={irrepbox,fill=teal!6},
 irrepnoise/.style={irrepbox,fill=orange!7},
 irrepleft/.style={irrepbox,text width=65mm,fill=blue!4,font=\footnotesize},
 irrepright/.style={irrepbox,text width=65mm,fill=violet!5,font=\footnotesize},
 irrepout/.style={irrepbox,fill=black!3},
 irreparrow/.style={-{Stealth[length=2mm]},draw=black!65,line width=.65pt}
}

\usepackage[colorlinks=true,allcolors=blue,hypertexnames=false]{hyperref}
\makeatletter
\renewcommand{\l@section}[2]{\ifnum\c@tocdepth>0\relax
    \l@@sections{}{section}{#1}{#2}\fi}
\renewcommand{\l@subsection}[2]{\ifnum\c@tocdepth>1\relax
    \l@@sections{section}{subsection}{#1}{#2}\fi}
\renewcommand{\l@subsubsection}[2]{\ifnum\c@tocdepth>2\relax
    \l@@sections{subsection}{subsubsection}{#1}{#2}\fi}
\renewcommand{\appendix@toc}{\setcounter{tocdepth}{2}}
\makeatother
\newtheorem{theorem}{Theorem}[section]
\newtheorem{lemma}[theorem]{Lemma}
\newtheorem{proposition}[theorem]{Proposition}
\newtheorem{corollary}[theorem]{Corollary}
\theoremstyle{definition}
\newtheorem{definition}[theorem]{Definition}
\newtheorem{remark}[theorem]{Remark}
\DeclareMathOperator{\Tr}{Tr}
\DeclareMathOperator{\Sym}{Sym}
\DeclareMathOperator{\Var}{Var}
\DeclareMathOperator{\TV}{TV}
\DeclareMathOperator{\Bin}{Bin}
\DeclareMathOperator{\supp}{supp}
\newcommand{\E}{\mathbb E}
\newcommand{\Prob}{\mathbb P}
\newcommand{\fall}[2]{(#1)^{\underline{#2}}}
\newcommand{\rise}[2]{(#1)^{\overline{#2}}}
\newcommand{\ket}[1]{\lvert#1\rangle}
\newcommand{\bra}[1]{\langle#1\rvert}
\newcommand{\cL}{\mathcal L}
\newcommand{\cR}{\mathcal R}
\newcommand{\poly}{\operatorname{poly}}
\newcommand{\eps}{\varepsilon}
\algrenewcommand{\algorithmicrequire}{\textbf{Input:}}
\algrenewcommand{\algorithmicensure}{\textbf{Output:}}
\allowdisplaybreaks
\numberwithin{equation}{section}
\begin{document}
\title{Polynomial-time classical simulation of boson and fermion sampling under thermal loss}
\author{Armando Angrisani}
\affiliation{Department of Mathematical Sciences, University of Copenhagen, Copenhagen, Denmark}
\affiliation{Institute of Physics, Ecole Polytechnique Fédérale de Lausanne (EPFL), Lausanne, Switzerland}
\affiliation{Centre for Quantum Science and Engineering, Ecole Polytechnique F\'{e}d\'{e}rale de Lausanne (EPFL),   Lausanne, Switzerland}
\author{Hela Mhiri}
\affiliation{Laboratoire d’Informatique de Paris 6 (LIP6), CNRS, Sorbonne Université, Paris, France}
\affiliation{Institute of Physics, Ecole Polytechnique Fédérale de Lausanne (EPFL), Lausanne, Switzerland}

\date{\today}

\begin{abstract}
Sampling experiments are a leading approach to demonstrating quantum
computational advantage, yet their robustness to experimental noise
is not completely understood. We develop a common framework for
classical simulation of boson and fermion sampling under thermal
loss, including pure loss. For Fock-state boson sampling, Gaussian boson
sampling with fixed uniform squeezing, and fermion sampling with magic
inputs, we obtain polynomial-time sampling at inverse-polynomial total
variation error,
for any fixed nonzero overall loss rate, and with high probability over Haar-random interferometers. A milder flatness assumption
on the ideal output distribution suffices for quasipolynomial-time
simulation without Haar randomness. With constant thermal loss per
layer, the same flatness assumption gives polynomial-time simulation at
logarithmic depth. 
Our bounds for Fock-state boson sampling cover both dilute and saturated regimes and
become stronger near full input filling: when only a bounded number
of input modes are empty, the noisy output for Haar-random
interferometers converges, as the number of modes grows, to a
distribution independent of the interferometer.
\end{abstract}

\maketitle

\section{Introduction}
\label{sec:introduction}

Sampling experiments can achieve quantum computational advantage
using devices with substantially less control than a universal quantum
computer~\cite{zhong2020quantum,madsen2022quantum, bao2023very}.
A prominent example is Fock-state boson sampling, where indistinguishable photons propagate through a passive
interferometer and are measured in the occupation basis, producing
distributions whose efficient classical simulation would contradict
widely studied complexity-theoretic
conjectures~\cite{aaronson2011computational,bouland2026complexity,bouland2025exponential,bouland2022noise}.
Related proposals include Gaussian boson
sampling~\cite{hamilton2017gaussian} and fermion sampling with
 magic inputs~\cite{oszmaniec2022fermion}.
Proposed applications of these architectures include generative
modelling~\cite{kurkin2026universality,kolarovszki2026generative,bako2025fermionic}
and the photonic simulation of molecular vibronic
spectra in quantum chemistry~\cite{huh2015boson}.

The prospects for these experiments depend on how physical noise
changes their classical simulation complexity. We consider thermal
loss, in which each mode exchanges particles with a thermal bath.
A vacuum bath gives pure particle loss.
Conditional hardness can survive sufficiently weak
loss~\cite{aaronson2016bosonsampling,oszmaniec2022fermion,go2025sufficient},
particularly in regimes where the transmissivity approaches one as the
system grows. At fixed transmissivity below one, loss instead
removes a constant fraction of the input particles on average.
This regime is therefore central to understanding the robustness
of sampling advantage to particle loss.

In this work, we develop a unified framework for classical simulation
of particle sampling experiments under local thermal loss.
For Fock-state boson sampling, Gaussian boson sampling with fixed uniform squeezing and fermion sampling with magic inputs  we obtain polynomial-time sampling to
inverse-polynomial accuracy in total variation distance at any fixed nonzero
overall loss rate, with high probability over
Haar-random interferometers. 

Our approach uses a low-degree approximation of the noisy output
distribution. Previous applications of low-degree methods addressed Fock-state boson sampling for specific noise models and
parameter regimes. Kalai and
Kindler~\cite{kalai2014gaussian} and Oh, Jiang, and
Fefferman~\cite{oh2023classical} developed such approximations
under an idealized Gaussian perturbation model.
More recently, Go and Jeong~\cite{go2026polynomial} obtained
polynomial-time simulation at any fixed nonzero pure-loss rate
for Haar-random interferometers in the dilute regime.
Related approximations were also established for qubit circuits under local depolarizing noise\ \cite{aharonov2022polynomial, schuster2024polynomial}.

Complementary approaches establish classical simulation of boson sampling under
sufficient loss conditions that depend on particle number, target
accuracy, or interferometer
structure~\cite{oszmaniec2018classical,garcia2019simulating,brod2020classical}.
For constant-depth interferometers, sufficiently strong loss can
reduce interference to small connected components, yielding efficient
simulation above a depth-dependent noise
threshold~\cite{oh2025classical}.
Phase-space methods provide further simulation criteria involving
photon loss, detector noise, and bath
temperature~\cite{rahimi2016sufficient,qi2020regimes,bressanini2024gaussian}.

We express the output probabilities using polynomial families of the
occupation operators that are orthogonal with respect to an appropriate
reference distribution. These
polynomials correspond to the diagonal parts of the operator space’s irreducible components under the
adjoint action of passive
transformations~\cite{wilkens2024benchmarking,arienzo2024bosonic,mhiri2026boson}.
Thermal loss suppresses contributions exponentially in their degree,
extending the noise-induced truncation
principle~\cite{gao2018efficient,aharonov2022polynomial, fontana2023classical, schuster2024polynomial, mele2024noise, gonzalez2024pauli, angrisani2025simulating, martinez2025efficient, angrisani2026dissipation, facelli2026open}
to boson and fermion sampling, with polynomial degree playing the role
of Pauli weight. 
We evaluate the retained low-degree polynomials either deterministically
or through unbiased estimators whose variance is controlled on average
over random interferometers. 
We then turn this approximation into an efficient classical algorithm using rejection sampling.

We also provide guarantees for a broader class of
interferometers, without assuming Haar randomness. Under a
flatness condition on the ideal output distribution, analogous
to the anticoncentration property used in
Refs.~\cite{aharonov2022polynomial,schuster2024polynomial},
we obtain quasipolynomial-time classical simulation.
For logarithmic-depth interferometers, thermal loss
at a fixed nonzero rate after each layer improves this guarantee
to polynomial time. Combining this result with the recent anticoncentration
bound of Mathur and Kerenidis~\cite{mathur2026logarithmic}
establishes polynomial-time classical simulation, with high probability,
for their logarithmic-depth fermion sampling ensemble under thermal
loss.

\section{Framework}
\label{sec:main-framework}

{We consider $m$ bosonic or fermionic modes undergoing passive
linear-optical evolution, followed by occupation-number measurements.
We label configurations of $n$ particles by occupation vectors
$x=(x_1,\ldots,x_m)$, where $x_i$ counts the particles in mode $i$
and $|x|:=\sum_i x_i=n$. The allowed occupations are
\begin{equation}
 \Omega_{n,m}=
 \begin{cases}
  \{x\in\mathbb N_0^m:|x|=n\},&\text{bosons},\\
  \{x\in\{0,1\}^m:|x|=n\},&\text{fermions}.
 \end{cases}
 \label{eq:main-occupation-space}
\end{equation}
The Fock states $\{\ket{x}:x\in\Omega_{n,m}\}$ form an orthonormal
basis of the $n$-particle space $\mathcal H_n$. The full Fock space
is the direct sum of these sectors, over $n\ge0$ for bosons and
$0\le n\le m$ for fermions.}
The interferometer is specified by a matrix $U\in\mathrm{U}(m)$,
and we write $\mathcal U$ for its action on Fock space.
Our sampling guarantees apply to three input families:

\begin{itemize}
    \item \emph{Fock-state boson sampling.}
    The input contains one photon in each of $n\le m$ modes,
    \begin{equation}
        \ket{\Psi_{\rm in}}
        =\ket{1}^{\otimes n}\otimes\ket{0}^{\otimes(m-n)}.
        \label{eq:main-fock-input}
    \end{equation}

    \item \emph{Gaussian boson sampling.}
    Every mode is initialized in a squeezed vacuum with the same
    squeezing parameter $r>0$, {fixed independently of system size},
    \begin{equation}
        \ket{\Psi_{\rm in}}
        =\ket{\mathrm{sq}(r)}^{\otimes m}.
        \label{eq:main-gbs-input}
    \end{equation}
    The mean input occupation per mode is
    $\bar n=\sinh^2 r$.

    \item \emph{Fermion sampling.}
    The input is a product of four-mode magic states,
    \begin{equation}
        \begin{gathered}
            \ket{\Psi_{\rm in}}
            =\left(\frac{\ket{1100}+\ket{0011}}{\sqrt{2}}\right)^{\otimes B},\\
            m=4B,\qquad n=2B.
        \end{gathered}
        \label{eq:main-fermion-input}
    \end{equation}
    These non-Gaussian inputs distinguish the model from
    efficiently simulable Gaussian fermionic
    evolution~\cite{bravyi2004lagrangian,oszmaniec2022fermion}.
\end{itemize}

We model noise by coupling every system mode to an independent
thermal environment mode with the same mean occupation $\beta\geq 0$\ \cite{weedbrook2012gaussian}.
For transmissivity $\eta$, the coupling transforms annihilation
operators according to
\begin{equation}
    \hat a_i\longmapsto
    \sqrt{\eta}\,\hat a_i+\sqrt{1-\eta}\,\hat b_i,
    \label{eq:main-thermal-coupling}
\end{equation}
where $\hat b_i$ acts on the environment. Tracing out the environment
defines the single-mode channel $\mathcal N_{\eta,\beta}$.
We assume $0<\eta<1$, with $0 \leq \beta \leq m^{\mathcal{O}(1)}$ for bosons and
$0\le\beta\le1$ for fermions. The case $\beta=0$ is pure loss.
After averaging over the thermal bath, the number operators transform as
\begin{equation}
    \hat n_i \longmapsto \eta\hat n_i+(1-\eta)\beta I,
    \qquad \hat n_i=\hat a_i^\dagger\hat a_i.
    \label{eq:main-thermal-mean}
\end{equation}

Let $\rho_U$ denote the ideal output state after dephasing in the
occupation basis. Occupation dephasing commutes with thermal loss,
so the measured noisy distribution is
\begin{equation}
    P_U(x)
    =
    \bra{x}\mathcal N_{\eta,\beta}^{\otimes m}(\rho_U)\ket{x}.
    \label{eq:main-target}
\end{equation}

Our approximation uses a diagonal reference state $\tau$.
For the fixed-particle models, $\tau=u_n$ is uniform over all
occupation configurations with total particle number $n$:
\begin{equation}
    u_n=\frac{1}{d_n}\sum_{x \in \Omega_{n,m}}\ket{x}\bra{x},
    \label{eq:main-uniform-reference}
\end{equation}
where $d_n= |\Omega_{n,m}|$ is the size of the occupation space defined in Eq.\ \eqref{eq:main-occupation-space}.
For Gaussian boson sampling, $\tau$ is a product thermal state
with mean occupation
$a=\bar n+(1+\bar n)/m$ per mode. 

The noisy reference $\mathcal N_{\eta,\beta}^{\otimes m}(\tau)$
provides the candidate occupation configurations for our rejection sampling algorithm.

We quantify the \emph{flatness} of the ideal output distribution using the chi-squared divergence
\begin{equation}
    \chi^2(\rho_U\|\tau)
    =
    \sum_x\frac{[p_U(x)-\tau(x)]^2}{\tau(x)},
    \label{eq:main-flatness}
\end{equation}
where $p_U(x)=\bra{x}\rho_U\ket{x}$ and
$\tau(x)=\bra{x}\tau\ket{x}$. The sum is over the support of $\tau$.
A polynomial bound on this quantity limits the concentration of
the ideal output relative to its reference.

We expand diagonal observables in the Fock basis into polynomials of the local number
operators. 
Writing their elements as
$X_j$, where $\deg(X_j)$ denotes the polynomial degree of $X_j$, we choose families such that $X_0=I$ and
\begin{equation}
    \operatorname{Tr}(\tau X_iX_j)=\delta_{ij}h_j,
    \qquad h_j>0.
    \label{eq:main-polynomial-orthogonality}
\end{equation}
{At each degree $k$, the basis spans the occupation polynomials of
degree at most $k$ that are orthogonal to all lower degrees. }
In particular, we employ Young polynomials for fermion sampling
\cite{filmus2014orthogonal}, Hahn polynomials for Fock-state
boson sampling~\cite{khare2009rates}, and Meixner polynomials for
Gaussian boson sampling~\cite{griffiths2011multivariate}. 
{Intuitively, these components capture the $k$-particle correlation
information accessible through occupation measurements, isolated from
lower-degree contributions. This correspondence is highlighted in Section \ref{sec:irreps}.}

\section{Results}
\label{sec:main-res}
Unless otherwise stated, we assume the overall transmissivity $\eta$ of the interferometer be a constant bounded away from one.
\begin{theorem}[Polynomial-time simulation for Haar-random interferometers]
\label{thm:main}
For each of the three sampling models of Sec.~\ref{sec:main-framework} and every
$\varepsilon,\delta \in(0,1)$, there is a classical randomized
algorithm whose output law $Q_U$ satisfies
\begin{equation}
    \Pr_{U\sim\mathrm{Haar}}
    \left[
        \operatorname{TV}(Q_U,P_U)\le\varepsilon
    \right]
    \ge1-\delta.
    \label{eq:main-haar-guarantee}
\end{equation}
{Its runtime is polynomial in $m$, $\varepsilon^{-1}$,
and $\delta^{-1}$.} 
\end{theorem}

The proof is given by Theorem~\ref{thm:final-sketch}, using the
model-specific bounds in Table~\ref{tab:haar-guarantees} and the
finite-precision analysis in Appendix~\ref{sec:precision}.

The theorem requires no dilute-output assumption {and hence covers experimentally relevant regimes where the number of modes scales linearly with the input particle number and collisions dominate. These regimes are also covered by complexity-theoretic hardness results in the noiseless setting \cite{bouland2026complexity}.
Moreover, it} does not require the loss rate need to exceed a
specific threshold: every fixed $\eta<1$ is covered. 

We also obtain a simulation guarantee under a milder flatness assumption, without requiring Haar randomness.

\begin{theorem}[Quasipolynomial-time simulation under flatness]
\label{thm:main-flatness}
For each of the three sampling models, every
interferometer $U$ satisfying
$\chi^2(\rho_U\|\tau)\le m^{\mathcal{O}(1)}$, and every
$\varepsilon\in(0,1)$, there is a classical randomized algorithm
whose output law $Q_U$ satisfies
\begin{equation}
    \operatorname{TV}(Q_U,P_U)\le\varepsilon.
    \label{eq:main-flatness-guarantee}
\end{equation}
Its runtime is bounded by
    $(m/\varepsilon)^{\mathcal{O}(\log(m/\varepsilon))}$.
\end{theorem}

The proof combines Theorem~\ref{thm:master} with the model-specific
approximation bounds in Appendices~\ref{sec:young-approx},
\ref{sec:hahn-approx}, and~\ref{sec:gbs-approx}, and the deterministic
sampling guarantee in Corollary~\ref{thm:final-enumeration}.
Eq.~\eqref{eq:flatness-runtime} gives the stated runtime.

{When loss accumulates over circuit layers, the same
flatness condition yields a stronger runtime guarantee.}
Homogeneous thermal loss commutes with passive interferometers.
If each of $L$ layers is followed by thermal loss with
transmissivity $\eta_0$ and the same bath occupation $\beta$,
the entire noisy circuit is equivalent to the ideal interferometer
followed by a single thermal-loss channel with total
transmissivity $\eta=\eta_0^L$.

\begin{corollary}[Polynomial-time simulation at logarithmic depth]
\label{cor:main-logarithmic-depth}
{Consider an interferometer composed of
$L=\Omega(\log m)$ passive layers, each followed by thermal loss
with fixed transmissivity $\eta_0\in(0,1)$, so that
$\eta=\eta_0^L$.}
Suppose that the ideal output satisfies
$\chi^2(\rho_U\|\tau)\le m^{\mathcal{O}(1)}$.
For every $\varepsilon\in(0,1)$, there is a classical randomized
algorithm whose output law $Q_U$ satisfies
\begin{equation}
    \operatorname{TV}(Q_U,P_U)\le\varepsilon
    \label{eq:main-logarithmic-depth-guarantee}
\end{equation}
in time polynomial in $m$ and $\varepsilon^{-1}$.
\end{corollary}

The proof is given in Appendix~\ref{sec:precision}.
In particular, Mathur and
Kerenidis~\cite{mathur2026logarithmic} recently showed that
logarithmic-depth random fermionic circuits satisfy the
flatness condition with high probability. Consequently,
these circuits admit polynomial-time classical simulation
under homogeneous thermal loss at any fixed nonzero loss
rate per layer.

For Fock-state boson sampling, our approximation becomes particularly strong near full input
filling. In this saturated regime, the noisy output concentrates
around an interferometer-independent distribution, and the
degree-zero approximation alone becomes sufficient.

\begin{proposition}[Concentration near full input filling]
\label{prop:main-saturated-concentration}
Consider Fock-state boson sampling with $m-n=\mathcal O(1)$.
For every integer $D\ge0$, the degree-$D$ approximation
$\widetilde\rho_{D,U}$ defined below in
Eq.~\eqref{eq:main-truncated-state} satisfies
\begin{equation}
    \mathbb E_{U\sim\mathrm{Haar}}
        \left\|
            \mathcal N_{\eta,\beta}^{\otimes m}(\rho_U)
            -\widetilde\rho_{D,U}
        \right\|_1^2
    =\mathcal O\!\left(\frac{e^{-\Omega(D)}}{m}\right).
    \label{eq:main-saturated-degree-bound}
\end{equation}
Consequently, for every $\varepsilon,\delta\in(0,1)$,
a degree
\begin{equation}
    D=\mathcal O\!\left(
        \log\!\left(1+\frac{1}{m\delta\varepsilon^2}\right)
    \right)
    \label{eq:main-saturated-cutoff}
\end{equation}
suffices to achieve
\begin{equation}
    \Pr_{U\sim\mathrm{Haar}}
    \left[
        \left\|
            \mathcal N_{\eta,\beta}^{\otimes m}(\rho_U)
            -\widetilde\rho_{D,U}
        \right\|_1
        \le\varepsilon
    \right]
    \ge1-\delta.
    \label{eq:main-saturated-guarantee}
\end{equation}
\end{proposition}

{The proof is given in Appendix~\ref{sec:hahn-approx}.}

In particular, degree zero suffices once
$m\delta\varepsilon^2$ is sufficiently large.
Thus approaching full input filling strengthens the approximation:
at fixed accuracy and confidence, sufficiently large systems are
well approximated by the interferometer-independent distribution
\begin{equation}
\mathcal N_{\eta,\beta}^{\otimes m}(u_n),
    \label{eq:main-saturated-reference}
\end{equation}
obtained by applying thermal loss to the uniform occupation state.
{This phenomenon arises from the combined effect of weak low-degree
correlations and the suppression of higher degrees by loss.
Near full input filling, retaining a fixed number of particles
and tracing out the rest leaves a reduced state that approaches
the maximally mixed state as the system grows. From a
group-theoretic perspective, its projections onto nontrivial
irreducible components become small, while the scalar component
represents the maximally mixed part. The information distinguishing
the full state from its maximally mixed reference is therefore
carried primarily by higher-degree correlations. Thermal loss
suppresses these contributions exponentially with their degree.
The irrep-norm and noise-suppression formulas in
Appendix~\ref{sec:hahn-model} quantify these two effects.
}

\section{Outlook}
\label{sec:outlook}

Our results provide evidence that superpolynomial quantum advantage with Haar-random interferometers requires the loss rate to vanish as the system grows, even in experimentally relevant regimes where the mean detected particle number is proportional to the number of modes. Detecting a constant fraction of the input particles is therefore insufficient to certify such an advantage. 

Our more general guarantees under a flatness assumption reflect a tension
familiar from random circuit
sampling~\cite{aaronson2011computational,aharonov2022polynomial,schuster2024polynomial}:
anticoncentration supports hardness arguments for ideal sampling, yet
also helps control classical approximation errors once noise suppresses
high-degree correlations.
Moreover, this dependence on flatness leaves unresolved the effect of noise on
peaked distributions, such as those recently proposed for demonstrating
verifiable quantum
advantage~\cite{aaronson2024verifiable, zhang2025complexity, gheorghiu2026verifiable}. Families with sufficiently
pronounced peaks violate polynomial flatness relative to the uniform
distribution, placing them beyond the reach of our error bound.
It remains open whether these constructions
admit low-degree approximation under noise, perhaps choosing a different reference distribution which exploits the structure of their peaks.

Even in the regimes covered by our guarantees, a polynomial runtime
does not necessarily imply that finite-size experiments can be simulated
in practice. Our bounds grow exponentially with
the inverse loss rate, leaving open the possibility of substantial
polynomial quantum advantage, growing as loss weakens. Determining the optimal classical
sampling cost, and in particular obtaining matching lower bounds,
could help identify parameter regimes that remain out of reach
for practical classical simulation.

The literature on Pauli propagation~\cite{rudolph2023classical, rudolph2025pauli,shrikhande2025rapid,beguvsic2023fast,beguvsic2024real}
and Majorana propagation~\cite{miller2025simulation, d2025majorana, facelli2026fast}
suggests that low-degree expansions can be useful well beyond the regimes
in which their accuracy has been proved. For particle sampling, trained
bosonic and fermionic generative models provide concrete test
cases~\cite{kurkin2026universality,kolarovszki2026generative,bako2025fermionic}:
although their optimized interferometers are not Haar-random, their noisy
outputs may still be accurately captured by expansions of modest degree.
The polynomial bases and noise-dependent truncation bounds developed here
could also inform the design of propagation methods for
bosonic systems, complementing recent proposals based on displacement
operators~\cite{upreti2025quantum} and coherent
states~\cite{guseynov2026coherent}.

\section{Methods}

{We first show how thermal loss and flatness justify a low-degree polynomial approximation, and how rejection sampling turns this signed approximation into a sampler with controlled error. We evaluate the retained contributions through noiseless overlaps on smaller particle spaces, recovering the noisy approximation by combining them with known noise-dependent weights, as sketched in Figure \ref{fig:common-irrep-roadmap}. These overlaps admit exact deterministic evaluation or unbiased randomized estimation, with Haar-average variance bounds yielding polynomial-time sampling for the latter.}
Alternatively, we can backpropagate the retained low-degree occupation polynomials through the circuit layers and evaluate their expectation values on the input state.
\subsection{Conditions for low-degree approximation}\label{sec:main-approximation-conditions}

Let $\tau$ be a reference distribution and $\mu =\mathcal N_{\eta,\beta}^{\otimes m}(\tau)$ be its evolution under noise. For each model, we choose the reference and the polynomial
families as in Sec.~\ref{sec:main-framework}.
The low-degree approximation relies on four properties:

\begin{itemize}
\item \emph{Orthogonality before noise.}
The occupation polynomials $X_j$, with $X_0=I$, form a
complete basis for the dephased noiseless output and satisfy the orthogonality relation introduced
in Eq.~\eqref{eq:main-polynomial-orthogonality}:
 \begin{equation}
    \operatorname{Tr}(\tau X_iX_j)=\delta_{ij}h_j,
    \qquad h_j>0.
    \label{eq:main-approximation-orthogonality}
\end{equation}

    \item \emph{Orthogonality after noise.}
    Orthogonality is preserved by the noise channel: 
    \begin{equation}
        \operatorname{Tr}\!\left[
            \mathcal N_{\eta,\beta}^{\otimes m}(\tau X_i)\mathcal N_{\eta,\beta}^{\otimes m}(\tau X_j)\mu^{-1}
        \right]
        =\delta_{ij}\Gamma_j h_j,
        \label{eq:main-noisy-orthogonality}
    \end{equation}
    where $\Gamma_j > 0$.

    \item \emph{Suppression with degree.}
    For some $0<q<1$,
    \begin{equation}
        \Gamma_j\le q^{\deg(X_j)},
        \qquad j\ge1,
        \label{eq:main-noise-suppression}
    \end{equation}

    \item \emph{Flatness.}
    The ideal diagonal output $\rho_U$ is supported on $\tau$ and
    satisfies
    \begin{equation}
    \chi^2(\rho_U\|\tau)
    =\sum_{j\ge1}
      \frac{|\operatorname{Tr}(\rho_UX_j)|^2}{h_j}
    \le C,
    \label{eq:main-approximation-flatness}
\end{equation}
where the identity follows by Parseval.
    When $\tau$ is the uniform distribution and $C = \mathcal{O}(1)$,
    this assumption corresponds to the anticoncentration property used in Refs.\ \cite{aharonov2022polynomial, schuster2024polynomial}.
\end{itemize}

The approximation retains degrees through $D$ before applying
the physical channel:
\begin{equation}
    \widetilde\rho_{D,U}
    =
    \mathcal N_{\eta,\beta}^{\otimes m}\!\left(
        \tau\sum_{j:\deg(X_j)\le D}
        \frac{\operatorname{Tr}(\rho_UX_j)}{h_j}X_j
    \right).
    \label{eq:main-truncated-state}
\end{equation}
In general, this is a \emph{signed} approximation: its entries sum
to one but may be negative. Nevertheless, its distance
from the target distribution is controlled by
Theorem~\ref{thm:master}, which gives
\begin{equation}
\begin{aligned}
    \left\|
        \mathcal N(\rho_U)-\widetilde\rho_{D,U}
    \right\|_1^2
    &\le
    \sum_{j:\deg(X_j)>D}\Gamma_j
        \frac{|\operatorname{Tr}(\rho_UX_j)|^2}{h_j}\\
    &\le q^{D+1}\chi^2(\rho_U\|\tau).
\end{aligned}
    \label{eq:main-approximation-bound}
\end{equation}
Under the flatness bound, a sufficient cutoff for
trace-norm error at most $\varepsilon$ is
\begin{equation}
    D=\mathcal O\!\left(
        1+\frac{\log[(1+C)/\varepsilon^2]}{\log(1/q)}
    \right).
    \label{eq:main-degree-cutoff}
\end{equation}

Under the conventions of
Sec.~\ref{sec:main-framework}, $q$ is bounded away from one.
Assuming $C = m^{\mathcal{O}(1)}$, inverse-polynomial accuracy
requires only $D=\mathcal O(\log m)$.
For the layered circuits in
Corollary~\ref{cor:main-logarithmic-depth}, the model-specific
bounds give $q\le K\eta_0^L$, with $K$ independent of system
size. Logarithmic depth then reduces the required cutoff to
$D=\mathcal O(1)$.

\subsection{From signed approximation to sampling}

We now turn the low-degree approximation in
Eq.~\eqref{eq:main-truncated-state} into an approximate sampler.
Previous work has used sequential sampling based on marginal
evaluations~\cite{bremner2017achieving,aharonov2022polynomial,oh2023classical,go2026polynomial}.
Although we expect this approach to apply here, it would require
computing the marginals of our approximation.
We therefore use rejection sampling for its conceptual simplicity:
given an efficiently sampled reference distribution
$\mu=\mathcal N_{\eta,\beta}^{\otimes m}(\tau)$, it requires
only evaluations of the approximation at individual occupation
configurations.

Let the relative density be
\begin{equation}
    f_{D,U}(x)
    =\frac{\bra{x}\widetilde\rho_{D,U}\ket{x}}{\mu(x)}.
    \label{eq:main-signed-density}
\end{equation}
Using the flatness bound $\chi^2(\rho_U\|\tau)\le C$, we choose a clipping threshold
$M=(1+C)/\varepsilon$, which limits the clipping error to
$\mathcal{O}(\varepsilon)$. The procedure is as follows:
\begin{enumerate}
    \item Draw an occupation $x$ with probability $\mu(x)$.
    \item Accept and return $x$ with probability
    \begin{equation}
        \min\left\{1,\frac{\max\{0,f_{D,U}(x)\}}{M}\right\}.
        \label{eq:main-rejection-acceptance}
    \end{equation}
    Otherwise, reject $x$ and repeat from step~1.
\end{enumerate}
To bound the runtime, we impose a
maximum number of proposals and return a fixed occupation if
none is accepted. Lemma~\ref{lem:rejection-sampling} bounds
the total variation error by the approximation error together
with the clipping and stopping errors.

The relative density $f_{D,U}$ can be evaluated from expectation
values of low-degree occupation polynomials. For the input families
considered here, polynomials of degree at most $k$ can be
backpropagated through the interferometer and their expectations
on the input states computed in time $m^{\mathcal O(k)}$ (Lemma~\ref{lem:efficient-occupation-moments}).
For Haar-random interferometers, we obtain faster sampling
using randomized evaluations with cost
$\operatorname{poly}(m)\,2^{\mathcal O(k)}$ per estimate.
The following subsections develop these procedures and establish
the Haar-average moment bounds that control their sampling cost.

\subsection{Adjoint degrees and Polynomial occupation bases}\label{sec:irreps}

We now connect the polynomial degree bases used above to the
irreducible components of passive dynamics. On the Fock space
$\mathcal H_n$ defined in Sec.~\ref{sec:main-framework}, passive
transformations act by conjugation on $\operatorname{End}(\mathcal H_n)$.
This adjoint representation
decomposes particle-number preserving operators into multiplicity free irreducible components $\mathcal V_{n,k}$, each
isolating $k$-particle correlation information beyond the lower
degrees~\cite{wilkens2024benchmarking,arienzo2024bosonic,mhiri2026boson},
as established in Proposition~\ref{prop:common-adjoint-decomposition}.
Occupation measurement probes their diagonal parts.
Since the irrep projectors commute with occupation dephasing,
the diagonal operator space decomposes as
\begin{align}
 \mathcal D_{n,m}:=&\operatorname{span}
 \{\ket{x}\bra{x}:x\in\Omega_{n,m}\}=\bigoplus_{k=0}^{J_n}\operatorname{Diag}(\mathcal V_{n,k}),
 \label{eq:methods-diagonal-degrees}
\end{align}
where $\operatorname{Diag}(\mathcal V_{n,k})
:=\mathcal V_{n,k}\cap \mathcal D_{n,m}$,
with $J_n=n$ for bosons and $J_n=\min(n,m-n)$ for fermions.

Each space $\operatorname{Diag}(\mathcal V_{n,k})$ consists of occupation
polynomials of degree at most $k$ orthogonal to all lower degrees.
Thus, for the fixed-particle models, the polynomials $X_j$ with $\mathrm{deg}(X_j)=k$ used in
Sec.~\ref{sec:main-approximation-conditions} form a basis of this space:
Young polynomials for fermions and Hahn polynomials for bosons.
The polynomial cutoff therefore selects the diagonal parts of the
corresponding adjoint degrees. Proposition~\ref{prop:common-diagonal-basis}
establishes this identification; Sec.~\ref{sec:exact-degree-projectors}
constructs the projectors.
For Gaussian boson sampling, radial polynomials additionally provide bases for the multiplicity spaces arising from equivalent adjoint irreps across particle-number sectors. The approximation degree is then $d=k+\ell$, where
$k$ is the angular adjoint degree and $\ell$ the radial degree, as
described in Sec.~\ref{sec:gbs-representation}.

\subsection{Whole-degree overlaps and deterministic evaluation}

For an occupation outcome $x\in\Omega_{n,m}$, the ideal probability
$p_U(x)=\bra xU_n\rho_{\rm in}U_n^\dagger\ket x$ decomposes as
\begin{equation}
 \begin{aligned}
 p_U(x)&=\sum_{k=0}^{J_n}p_{k,U}(x),\\
 p_{k,U}(x)
 &=\operatorname{Tr}\!\left[
 \mathsf P_{n,k}(\ket x\bra x)\,
 \mathsf P_{n,k}(U_n\rho_{\rm in}U_n^\dagger)\right],
 \end{aligned}
 \label{eq:methods-probability-decomposition}
\end{equation}
where $U_n$ denotes the action of a passive interferometer on $n$ particles and $\mathsf P_{n,k}$ denotes the projector onto irreducible components $\mathcal V_{n,k}$.

We introduce the $k$-particle reduction as
\begin{equation}
    \rho_{\rm in}^{(k)}
 :=\operatorname{Tr}_{k+1,\ldots,n}(\rho_{\rm in}),
\end{equation}
where the partial trace removes $n-k$ indistinguishable particles.

Using the above definition, each overlap $ p_{k,U}(x)$ can be evaluated on the smaller $k$-particle space.
The reduction $\rho_{\rm in}^{(k)}$ retains correlation
information through degree $k$, and applying $\Pi_k$ removes
the lower-degree contributions.
Thus $T_k:=\Pi_k(\rho_{\rm in}^{(k)})$ describes the same correlations as
$\mathsf P_{n,k}(\rho_{\rm in})$, using only $k$ particles.
Writing $\rho_x^{(k)}$ for the normalized reduction of
$\ket x\bra x$, we have
\begin{equation}
 p_{k,U}(x)
 \propto
 \operatorname{Tr}\!\left[
 \rho_x^{(k)}U_kT_kU_k^\dagger\right],
 \label{eq:methods-degree-overlap}
\end{equation}
with a known proportionality factor depending on $m,n,k$, as detailed in Corollary \ref{cor:common-output-degrees}.
Since $\rho_x^{(k)}$ is diagonal, this expression reads the
diagonal degree-$k$ contribution of the evolved reduced operator.

For exact evaluation, we tabulate the diagonal entries of $U_k\rho_{\rm in}^{(k)}U_k^\dagger$, apply $\Pi_k$, and contract the resulting entries with the occupation weights of $\rho_x^{(k)}$.  For the specified input families, preprocessing through degree $D$ costs $m^{\mathcal O(D)}$ arithmetic operations.
Each subsequent degree-$k$ overlap costs $\mathcal O(kd_k)$, where $d_k=\dim\mathcal H_k$, and the tables are reused across
outcomes. The evaluation is therefore polynomial in system size at fixed degree.

\subsection{Randomized overlap evaluation}
\label{sec:main-methods-randomized}

To avoid the exact evaluation described in the previous section, the randomized method replaces the large occupation tables used
for exact evaluation by compact random operators whose overlap
reproduces the same quantity on average. To do so, we introduce the following independent random operators satisfying
\begin{equation}
 \mathbb E_h\widehat\rho_{\rm in}^{(k)}(h)
 =\rho_{\rm in}^{(k)},
 \qquad
 \mathbb E_g\widehat\rho_x^{(k)}(g)
 =\rho_x^{(k)},
 \label{eq:methods-random-operators}
\end{equation}
where $h$ and $g$ denote the random choices defining the input and output operator estimates, respectively.

Applying the degree projector gives
\begin{equation}
     \widehat T_k(h)
 :=\Pi_k\bigl(\widehat\rho_{\rm in}^{(k)}(h)\bigr),
 \qquad
 \mathbb E_h\widehat T_k=T_k.
\end{equation}

The full projector is applied to each realization, so
$\widehat T_k$ has exact degree $k$. Consequently,
\begin{equation}
 \begin{aligned}
 Z_k(U,x;h,g)
 &:=\operatorname{Tr}\!\left[
 \widehat\rho_x^{(k)}(g)
 U_k\widehat T_k(h)U_k^\dagger\right],\\
 \mathbb E_{h,g}Z_k(U,x;h,g)
 &=\operatorname{Tr}\!\left[
 \rho_x^{(k)}U_kT_kU_k^\dagger\right].
 \end{aligned}
 \label{eq:methods-unbiased-overlap}
\end{equation}
Multiplying by the known normalization estimates the degree
contribution in Eq.~\eqref{eq:methods-degree-overlap}.
Thus randomization estimates the whole diagonal degree-$k$
overlap, without estimating its occupation entries separately. These estimates are unbiased for every fixed interferometer with a controlled variance under random interferometers.

The models use different random representations. For fermions,
random exterior matrices represent the input reduction, while the
output reduction remains exact. Fock-state boson sampling uses
independent colorings for both factors. Gaussian boson sampling keeps its explicit
angular input operator $T_k$ exact and randomizes only the output
test, with deterministic radial weights. The bosonic constructions
adapt color coding from Ref.~\cite{go2026polynomial}: selected particle
configurations are retained and rescaled by their selection
probability, with weighted output slots accounting for repeated
occupations. 

Determinant formulas compute each random overlap directly, without
constructing the full $k$-particle matrices. Averaging $R$ independent
realizations costs
$R\,\operatorname{poly}(m,|x|,k)2^{\mathcal O(k)}$.
The constructions are given in Sections~\ref{sec:young-sketch},
\ref{sec:hahn-sketch}, and~\ref{sec:gbs-sketch}.

\medskip

\emph{Note added.}
During the completion of this work, Go and Jeong~\cite{go2026polynomial}
established polynomial-time classical simulation of Fock-state
boson sampling under pure loss in the dilute regime for Haar-random interferometers, using a different low-degree approximation. Before their work appeared, we had
already derived the low-degree approximations presented here
and the resulting quasipolynomial-time simulation at fixed overall
transmissivity, as well as polynomial-time simulation for
logarithmic-depth circuits with fixed transmissivity per layer. Their work inspired our randomized
evaluation procedures, which improve the runtime to polynomial
for Haar-random interferometers even at fixed overall
transmissivity. In particular, our randomized evaluation for boson sampling uses their color-coding
method.

After completing our manuscript, we became aware of the independent and concurrent work of
Lim~\cite{lim2026boson}, which uses closely related proof techniques
to establish polynomial-time classical simulation of Fock-state
and Gaussian boson sampling under pure loss at fixed
overall transmissivity, with approximation error controlled
on average over Haar-random interferometers.

\begin{acknowledgments}
The authors thank Ulysse Chabaud and Zoë Holmes for useful discussions.
AA acknowledges support from the European Union’s Horizon Europe research and innovation programme under the Marie Skłodowska-Curie grant agreement No. 101282902 (PROQSY) and from the Sandoz Family Foundation--Monique de Meuron program for Academic Promotion. HM acknowledges support from the grant ANR-22-PETQ-0007.
\end{acknowledgments}

\section*{AI usage statement}
During the preparation of this work, the authors used ChatGPT
(OpenAI) from January 2026 to October 2026, specifically GPT-5.2, 5.4, 5.5, 5.6 Sol, and 6 Astra. ChatGPT was used to help identify the relevant
polynomial basis and associated literature, prove technical results, and draft parts of the manuscript.

The authors devised the overall proof strategy and worked through
the technical derivations to bring out the main ideas and make
the proofs easier to follow. They identified the argument shared
by the three particle-sampling models and presented it as a general low-degree
approximation result under explicit assumptions and a representation theoretic interpretation of degrees, so that the
remaining proofs could focus on how each model satisfies those
assumptions and adapts the representation-theoretic constructions. The authors verified all derivations and take full
responsibility for the content of the manuscript.
\bibliographystyle{unsrt}
\bibliography{quantum, ref}

\clearpage
\onecolumngrid
\appendix
\renewcommand{\tocname}{Appendix contents}
\setcounter{tocdepth}{-1}
\tableofcontents
\clearpage

\begingroup
\raggedbottom
\section*{Guide to the proofs}
\phantomsection
\label{sec:proof-guide}

We approximate the noisy output distribution by a low-degree polynomial
and use this approximation to construct a rejection sampler.
The proof proceeds through the following stages.
\begin{itemize}[itemsep=2pt,topsep=4pt,parsep=0pt,partopsep=0pt]
\item \textbf{Representation-theoretic framework.}
Section~\ref{sec:particle-representation} develops a common framework for decomposing output probabilities into orthogonal contributions
of increasing degree, corresponding to irreducible components under
passive transformations. 
This representation-theoretic identification
enables a systematic construction of a polynomial basis adapted to
the measured occupations and yields identities for evaluating its
coefficients from reduced states. In particular, degree-$k$
contributions to an $n$-particle output probability can be computed
through overlaps on the smaller $k$-particle space. To establish
these results, we introduce adjoint degrees and intertwiners,
connecting different particle sectors, and identify their diagonal
degree components with orthogonal occupation polynomials, as summarized in Figure\ \ref{fig:common-irrep-roadmap}. This provides the common foundation for
the polynomial expansions used to analyse noise and the overlap
formulas used in the subsequent evaluation algorithms.
\item \textbf{Low-degree approximation.}
We establish a general approximation theorem
(Theorem~\ref{thm:master}), which identifies assumptions under which
noise suppresses the high-degree contributions of orthogonal
polynomials to the output distribution. One of these is
a flatness assumption, which limits
how concentrated the ideal output distribution can be relative to
a reference distribution. It plays the same role as the second-moment
conditions used in earlier work on noisy qubit-circuit
simulation~\cite{aharonov2022polynomial,schuster2024polynomial}.
When the flatness bound grows at most polynomially with the system
size, a logarithmic degree cutoff suffices for inverse-polynomial
accuracy. We verify the theorem's assumptions for fermion sampling,
Fock-state boson sampling, and Gaussian boson sampling, using Young,
Hahn, and Meixner polynomials, respectively
(Subsections~\ref{sec:young-approx}, \ref{sec:hahn-approx}, and \ref{sec:gbs-approx}).
\item \textbf{Operator overlap formulas.}
Using the identities established in Section~\ref{sec:particle-representation},
we express the low-degree approximations in terms of operator overlaps in
Subsections~\ref{sec:young-evaluation}, \ref{sec:hahn-evaluation}, and
\ref{sec:gbs-evaluation}. These overlaps are computed by the deterministic
algorithms and estimated by the randomized algorithms.

\item \textbf{Rejection sampling with deterministic evaluation.}
We compute the overlap coefficients and use the resulting density in
Algorithm~\ref{alg:rejection-sampling}. Uniform density bounds or
second-moment bounds control the number of reference proposals.
Under polynomial flatness, this gives quasipolynomial time at
inverse-polynomial accuracy.

\item \textbf{Faster rejection sampling with randomized evaluations.}
We estimate the retained density at a proposed occupation and increase
the number of evaluations only when the acceptance decision remains
undetermined. The bosonic estimators build on the color-coding
approach of Go and Jeong~\cite[Appendix B]{go2026polynomial}.
A second-moment bound controls the acceptance threshold, while the
integrated estimator variance controls the additional evaluations.
For Haar-random interferometers, the resulting sampler has polynomial
runtime.
\end{itemize}

Section~\ref{sec:precision} combines these
constructions with the approximation bounds, evaluation counts, and
finite-precision analysis. 

\begingroup
\setlength{\intextsep}{6pt}
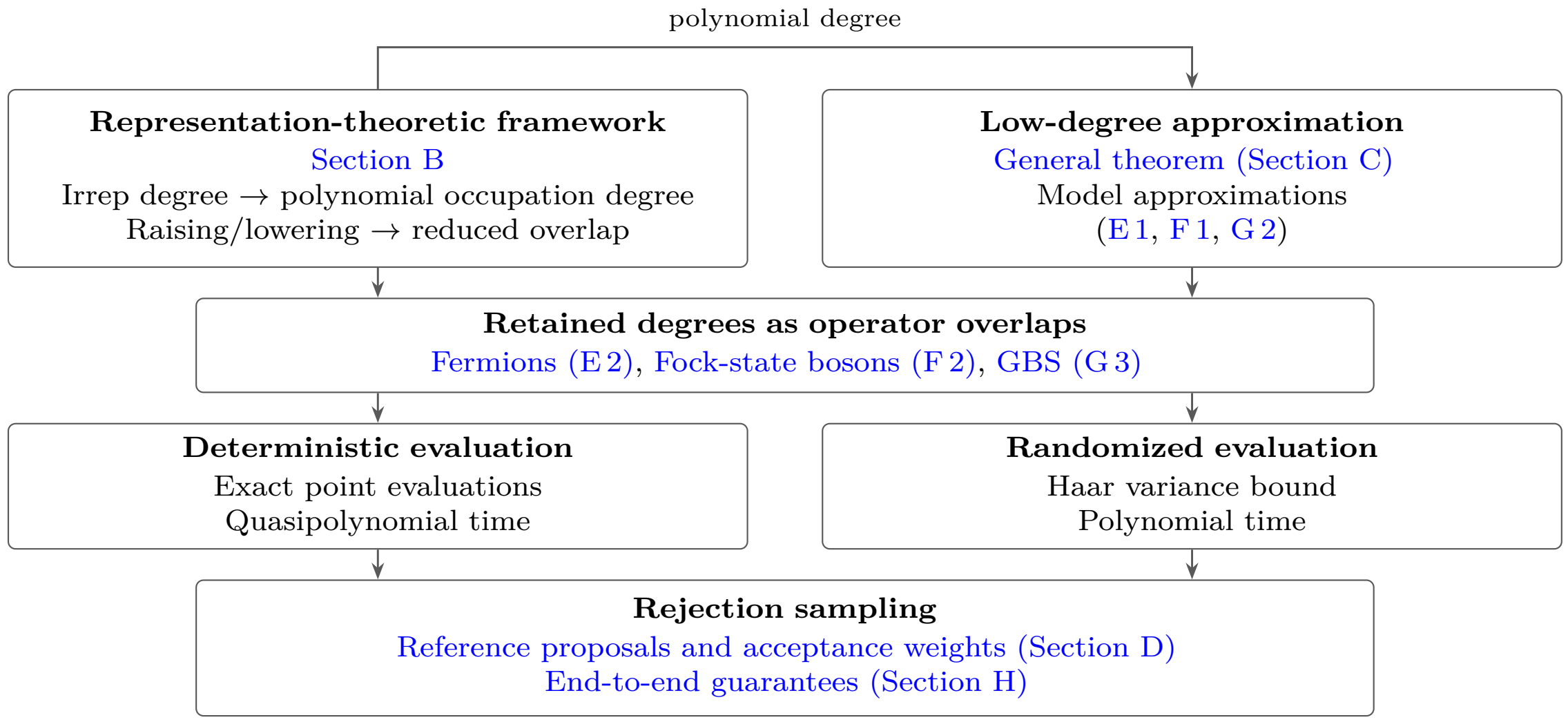
\begin{figure}[H]
\centering
\begin{tikzpicture}[>=Stealth,
 box/.style={draw=black!65,rounded corners=2pt,align=center,
 inner xsep=4pt,inner ysep=3pt,font=\footnotesize,
 anchor=north,execute at begin node={\hyphenpenalty=10000}},
 flow/.style={-{Stealth[length=1.6mm]},draw=black!65,line width=.55pt}]
\node[box,text width=68mm,minimum height=17mm]
 (representation) at (-39mm,0)
 {\textbf{Representation-theoretic framework}\\[1pt]
 \hyperref[sec:particle-representation]{Section~\ref*{sec:particle-representation}}\\
 Irrep degree $\to$ polynomial occupation degree\\
 Raising/lowering $\to$ reduced overlap};
\node[box,text width=68mm,minimum height=17mm]
 (approx) at (39mm,0)
 {\textbf{Low-degree approximation}\\[1pt]
 \hyperref[sec:framework]{General theorem (Section~\ref*{sec:framework})}\\
 Model approximations\\
 (\ref{sec:young-approx}, \ref{sec:hahn-approx}, \ref{sec:gbs-approx})};
\node[box,text width=110mm,minimum height=9mm]
 (overlap) at (0,-20mm)
 {\textbf{Retained degrees as operator overlaps}\\[1pt]
 \hyperref[sec:young-evaluation]{Fermions (\ref*{sec:young-evaluation})},
 \hyperref[sec:hahn-evaluation]{Fock-state bosons (\ref*{sec:hahn-evaluation})},
 \hyperref[sec:gbs-evaluation]{GBS (\ref*{sec:gbs-evaluation})}};
\node[box,text width=68mm,minimum height=12mm]
 (exact) at (-39mm,-32mm)
 {\textbf{Deterministic evaluation}\\[1pt]
 Exact point evaluations\\
 Quasipolynomial time};
\node[box,text width=68mm,minimum height=12mm]
 (sketch) at (39mm,-32mm)
 {\textbf{Randomized evaluation}\\[1pt]
 Haar variance bound\\
 Polynomial time};
\node[box,text width=110mm,minimum height=13mm]
 (sample) at (0,-47mm)
 {\textbf{Rejection sampling}\\[1pt]
 \hyperref[sec:sampling]{Reference proposals and acceptance weights
 (Section~\ref*{sec:sampling})}\\
 \hyperref[sec:precision]{End-to-end guarantees
 (Section~\ref*{sec:precision})}};
\draw[flow] (representation.north) -- ++(0,4mm)
 -- node[above=1pt,font=\scriptsize] {polynomial degree}
 ([yshift=4mm]approx.north) -- (approx.north);
\draw[flow] (representation.south) -- (overlap.north -| representation.south);
\draw[flow] (approx.south) -- (overlap.north -| approx.south);
\draw[flow] (overlap.south -| exact.north) -- (exact.north);
\draw[flow] (overlap.south -| sketch.north) -- (sketch.north);
\draw[flow] (exact.south) -- (sample.north -| exact.south);
\draw[flow] (sketch.south) -- (sample.north -| sketch.south);

\end{tikzpicture}
\caption{Proof dependencies. Representation theory defines polynomial degrees and supplies the
reduced-overlap identities, while low-degree approximation determines
which contributions to retain. The runtime labels refer to
inverse-polynomial accuracy at fixed nonzero loss and fixed squeezing
in GBS. The deterministic bound assumes polynomial flatness, while
the randomized bound additionally uses Haar randomness.}
\label{fig:dependency-roadmap}
\end{figure}
\endgroup
\clearpage
\endgroup

\clearpage
\section{Preliminaries}
\label{sec:notation}

We collect the general notation and group-theoretic facts used throughout
the paper. 

\subsection{Notation}
\label{sec:basic-notation}

We fix the scalar, Hilbert-space, and operator conventions needed for
both the representation-theoretic construction and the sampling bounds.

We write \(\mathbb R\) and \(\mathbb C\) for the real and complex numbers,
\(\mathbb N_0=\{0,1,2,\ldots\}\), and
\([m]=\{1,\ldots,m\}\).  For \(k\in\mathbb N_0\), falling and rising
factorials mean
\begin{equation}
 \fall{x}{k}=\prod_{j=0}^{k-1}(x-j),\qquad
 \rise{x}{k}=\prod_{j=0}^{k-1}(x+j),\qquad
 \fall{x}{0}=\rise{x}{0}=1.
\end{equation}
For a polynomial or power series \(f(z)=\sum_{j\ge0}f_jz^j\),
\([z^k]f(z)=f_k\) denotes coefficient extraction.  The same notation is
used for coefficients in several variables.

Vectors belong to a complex Hilbert space \(\mathcal H\), with inner
product \(\langle u,v\rangle=\bra u v \rangle\), conjugate-linear in its
first argument.  The adjoint \(A^\dagger\) is characterized by
\(\langle u,Av\rangle=\langle A^\dagger u,v\rangle\).
We write \(\operatorname{End}(\mathcal H)\) for the space of linear maps
from \(\mathcal H\) to itself. Such maps are called endomorphisms.
For any linear map \(L:\mathcal V\to\mathcal W\), its kernel is
\begin{equation}
 \ker L=\{v\in\mathcal V:L(v)=0\}.
\end{equation}
When \(L\) acts on operators, its kernel is accordingly a space of
operators.  The identity operator is \(I\), and \(\Tr(A)\) denotes the
trace of \(A\).  The inner product on an operator space is
\(\langle A,B\rangle=\Tr(A^\dagger B)\). Its associated norm is the
Schatten \(2\)-norm,
\begin{equation}
 \|A\|_2=\sqrt{\Tr(A^\dagger A)}.
\end{equation}
A density operator is a positive operator of trace one.  An observable
is a self-adjoint operator.  A quantum channel is a completely positive,
trace-preserving linear map on operators. Its action on a state is
written \(\Phi(\rho)\).

\subsection{Schur orthogonality}
\label{sec:technical-preliminaries}

\begin{lemma}[Schur orthogonality, adapted from Proposition~5.2 of Chapter~3 in Ref.~\cite{murnaghan2013representations}]\label{lem:schur-orthogonality}
Let \(V\) be a continuous irreducible unitary representation of a compact
group \(G\) on a complex Hilbert space of dimension \(d\).
For any vectors \(a,b\) in that space, normalized Haar measure satisfies
\begin{equation}
 \E_{g\in G}\bigl|\langle a,V(g)b\rangle\bigr|^2
 =\frac{\|a\|^2\|b\|^2}{d}.
 \label{eq:schur-second-moment}
\end{equation}
\end{lemma}

\begin{proof}
The irreducible twirling identity~\cite[Chapter~3, Proposition~5.2]{murnaghan2013representations},
applied to the rank-one operator $\ket b\bra b$, gives
\begin{equation}
 \E_{g\in G}V(g)\ket b\bra b V(g)^\dagger
 =\frac{\Tr(\ket b\bra b)}{d}I
 =\frac{\|b\|^2}{d}I.
 \label{eq:schur-rank-one-twirl}
\end{equation}
Taking the matrix element in $a$ gives
\eqref{eq:schur-second-moment}.
\end{proof}

In an operator space, the same lemma uses the inner product
\(\Tr(A^\dagger B)\) and the norm \(\|\cdot\|_2\).
For example, conjugation \(B\mapsto U(g)BU(g)^\dagger\) is a unitary
representation on operators.  If \(\mathcal V\) is an invariant
subspace on which this representation is irreducible and
\(A,B\in\mathcal V\), then
\begin{equation}
\E_g\left|\Tr\!\left(A^\dagger U(g)BU(g)^\dagger\right)\right|^2
 =\frac{\|A\|_2^2\|B\|_2^2}{\dim\mathcal V}.
 \label{eq:schur-operator-second-moment}
\end{equation}

\clearpage
\section{Representation theory for non-interacting particle-number-preserving dynamics}
\label{sec:particle-representation}

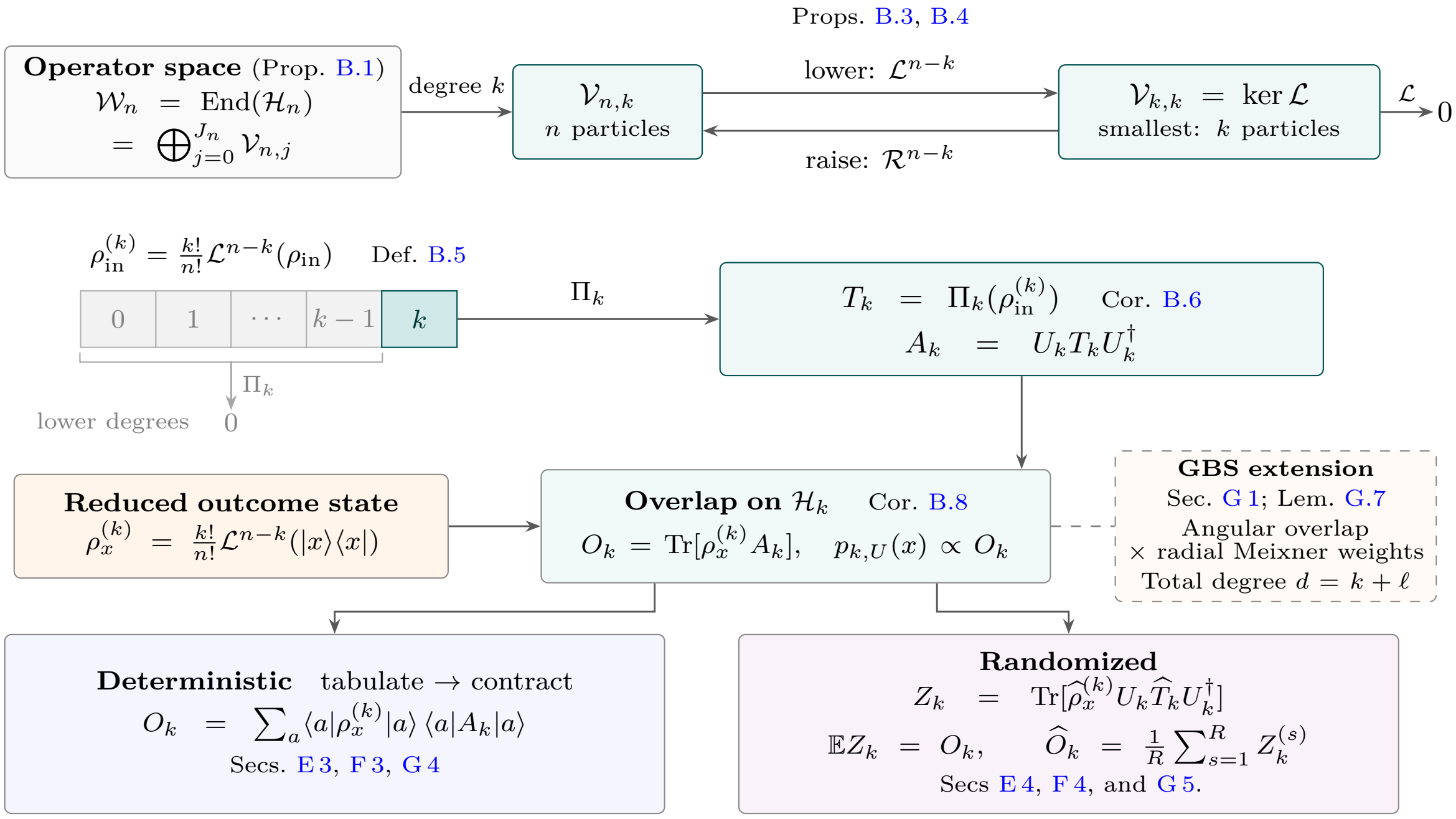
\begin{figure}[!htbp]
\centering
\begin{tikzpicture}[
 carrier/.style={draw=teal!65!black,fill=teal!6,rounded corners=2pt,
   align=center,inner xsep=3pt,inner ysep=2pt,font=\small,
   minimum height=10mm},
 box/.style={draw=black!45,rounded corners=2pt,align=center,
   inner xsep=3pt,inner ysep=2pt,font=\footnotesize},
 chip/.style={draw=black!30,fill=black!5,text=black!50,
   minimum width=8mm,minimum height=6mm,inner sep=0pt,font=\footnotesize},
 note/.style={font=\footnotesize,align=center,inner sep=1pt},
 flow/.style={-{Stealth[length=1.6mm]},draw=black!65,line width=.55pt},
 keep/.style={draw=teal!65!black,fill=teal!18,text=teal!40!black}]
\node[box,text width=40mm,minimum height=14mm,fill=black!2]
 (decomposition) at (-53mm,0)
 {\textbf{Operator space}
 {\scriptsize (Prop.~\ref{prop:common-adjoint-decomposition})}\\[1pt]
 $\mathcal W_n=\operatorname{End}(\mathcal H_n)$\\[1pt]
 $=\bigoplus_{j=0}^{J_n}\mathcal V_{n,j}$};
\node[carrier,text width=18mm] (large) at (-10mm,0)
 {$\mathcal V_{n,k}$\\[-1pt]{\scriptsize $n$ particles}};
\draw[flow] (decomposition.east) --
 node[above=1pt,font=\scriptsize] {degree $k$} (large.west);
\node[carrier,text width=32mm] (small) at (55mm,0)
 {$\mathcal V_{k,k}=\ker\cL$\\[-1pt]{\scriptsize smallest: $k$ particles}};
\node[note,font=\scriptsize] at (19mm,10mm)
 {Props.~\ref{prop:properties_RL}, \ref{prop:ident}};
\draw[flow] ([yshift=2mm]large.east) --
 node[above=1pt,font=\footnotesize] {lower: $\cL^{n-k}$}
 ([yshift=2mm]small.west);
\draw[flow] ([yshift=-2mm]small.west) --
 node[below=1pt,font=\footnotesize] {raise: $\cR^{n-k}$}
 ([yshift=-2mm]large.east);
\node[inner sep=1pt] (zero) at (79mm,0) {$0$};
\draw[flow] (small.east) -- node[above,font=\scriptsize] {$\cL$} (zero.west);

\node[note] at (-45mm,-15mm)
 {$\rho_{\rm in}^{(k)}=\tfrac{k!}{n!}\cL^{n-k}(\rho_{\rm in})$
 \quad {\scriptsize Def.~\ref{def:common-particle-reductions}}};
\node[chip] (d0) at (-62mm,-22mm) {$0$};
\node[chip] (d1) at (-54mm,-22mm) {$1$};
\node[chip] (dots) at (-46mm,-22mm) {$\cdots$};
\node[chip] (dprev) at (-38mm,-22mm) {$k-1$};
\node[chip,keep] (dk) at (-30mm,-22mm) {$k$};
\node[carrier,text width=62mm,minimum height=12mm]
 (harmonic) at (34mm,-22mm)
 {$T_k=\Pi_k(\rho_{\rm in}^{(k)})$
 \quad {\scriptsize Cor.~\ref{cor:common-projectors}}\\[2pt]
 $A_k=U_kT_kU_k^\dagger$};
\draw[flow] (dk.east) -- node[above=1pt,font=\footnotesize] {$\Pi_k$}
 (harmonic.west);
\draw[draw=black!35] ([yshift=-.5mm]d0.south west) -- ++(0,-1mm)
 -| ([yshift=-.5mm]dprev.south east);
\node[text=black!50,font=\footnotesize,inner sep=1pt]
 (discarded) at (-50mm,-33mm) {$0$};
\draw[flow,draw=black!35] (-50mm,-26.5mm) --
 node[right,font=\scriptsize,text=black!50] {$\Pi_k$} (discarded.north);
\node[note,anchor=east,font=\scriptsize,text=black!50]
 at (-54mm,-33mm) {lower degrees};

\node[box,text width=44mm,minimum height=11mm,fill=orange!8]
 (test) at (-50mm,-44mm)
 {\textbf{Reduced outcome state}\\[2pt]
 $\rho_x^{(k)}=\tfrac{k!}{n!}\cL^{n-k}(\ket x\bra x)$};
\node[box,text width=52mm,minimum height=12mm,fill=teal!5]
 (overlap) at (10mm,-44mm)
 {\textbf{Overlap on $\mathcal H_k$}
 \quad {\scriptsize Cor.~\ref{cor:common-output-degrees}}\\[2pt]
 $O_k=\Tr[\rho_x^{(k)}A_k],\quad p_{k,U}(x)\propto O_k$};
\draw[flow] (test.east) -- (overlap.west);
\draw[flow] (harmonic.south) -- (overlap.north -| harmonic.south);
\node[box,dashed,text width=32mm,minimum height=16mm,fill=orange!4,
 font=\scriptsize] (gbs) at (61mm,-44mm)
 {\textbf{GBS extension}\\[1pt]
 Sec.~\ref{sec:gbs-representation}; Lem.~\ref{lem:gbs-meixner-representation}\\[1pt]
 Angular overlap\\[-1pt]
 $\times$ radial Meixner weights\\[1pt]
 Total degree $d=k+\ell$};
\draw[dashed,draw=black!50,line width=.5pt] (overlap.east) -- (gbs.west);

\node[box,text width=68mm,minimum height=19mm,fill=blue!4]
 (exact) at (-39mm,-65mm)
 {\textbf{Deterministic}\quad tabulate $\to$ contract\\[3pt]
 $O_k=\sum_a\bra a\rho_x^{(k)}\ket a\,\bra a A_k\ket a$\\[3pt]
 {\scriptsize Secs.~\ref{sec:young-enumeration}, \ref{sec:hahn-enumeration},
 \ref{sec:gbs-enumeration}}};
\node[box,text width=68mm,minimum height=19mm,fill=violet!5]
 (random) at (39mm,-65mm)
 {\textbf{Randomized}\\[1pt]
 $Z_k=\Tr[\widehat\rho_x^{(k)}U_k\widehat T_kU_k^\dagger]$\\[2pt]
 $\E Z_k=O_k,\qquad
 \widehat O_k=\tfrac1R\sum_{s=1}^{R}Z_k^{(s)}$\\[2pt]
 {\scriptsize Secs~\ref{sec:young-sketch},
\ref{sec:hahn-sketch}, and~\ref{sec:gbs-sketch}.}};
\draw[flow] ([xshift=-15mm]overlap.south) -- ++(0,-3mm) -| (exact.north);
\draw[flow] ([xshift=15mm]overlap.south) -- ++(0,-3mm) -| (random.north);

\end{tikzpicture}
\caption{An adjoint degree is followed from the operator-space decomposition
to its smallest particle sector. Reductions to $k$ particles retain
degrees through $k$; $\Pi_k$ removes lower degrees before evolution and
diagonal readout. Probability normalization is suppressed. Both branches
compute the same overlap. Randomization uses independent unbiased operator
factors and averages independent samples; integrated Haar variance controls
repetitions. For GBS, $T_k$ comes from the pair-state construction, the outcome
reduction uses $n=|x|$, and radial Meixner weights resolve $\mathcal M_k$.}
\label{fig:common-irrep-roadmap}
\end{figure}

 We first introduce the particle spaces and
decompose their operator spaces into irreducible representations under
passive mode transformations. Raising and lowering intertwiners connect
equivalent components across particle sectors and yield explicit irrep
projectors. We then identify the diagonal, or zero-weight, components
with spaces of orthogonal occupation polynomials, linking the operator
decomposition to functions of measured occupations. This construction,
summarized in Figure~\ref{fig:common-irrep-roadmap}, provides the common
basis for the degree truncations and evaluation formulas used in the
subsequent sampling sections.

\subsection{Particle spaces, passive dynamics, and adjoint irreps}
\label{sec:particle-framework}
\label{sec:adjoint-degrees}
\label{sec:young-adjoint}
\label{sec:young-haar-contractions}

The Hilbert space of a single particle in $m\ge2$ modes is
$\mathbb C^m$, with orthonormal basis $e_1,\ldots,e_m$. For $n\ge1$,
the permutation group $S_n$ acts on $(\mathbb C^m)^{\otimes n}$ by
\begin{equation}
 T_\pi(v_1\otimes\cdots\otimes v_n)
 =v_{\pi^{-1}(1)}\otimes\cdots\otimes v_{\pi^{-1}(n)},
 \qquad \pi\in S_n.
\end{equation}
The symmetrizing and antisymmetrizing projectors are
\begin{equation}
 P_n^+=\frac1{n!}\sum_{\pi\in S_n}T_\pi,
 \qquad
 P_n^-=\frac1{n!}\sum_{\pi\in S_n}\operatorname{sgn}(\pi)T_\pi,
\end{equation}
where $\operatorname{sgn}(\pi)=+1$ for even permutations and $-1$
for odd permutations.

The bosonic and fermionic $n$-particle Hilbert spaces are the images
of these projectors:
\begin{equation}
 \begin{aligned}
 \mathcal H_n^+
 &:=P_n^+\bigl((\mathbb C^m)^{\otimes n}\bigr)
   =\Sym^n(\mathbb C^m),\\
 \mathcal H_n^-
 &:=P_n^-\bigl((\mathbb C^m)^{\otimes n}\bigr)
   =\bigwedge^n\mathbb C^m.
 \end{aligned}
\end{equation}
They are spanned by the projected tensor-product basis vectors
$P_n^\pm(e_{i_1}\otimes\cdots\otimes e_{i_n})$, with
$i_1,\ldots,i_n\in[m]$. Reordering the tensor factors leaves the
symmetric vector unchanged and changes the antisymmetric vector only
by a sign. Repeated mode indices give zero in the latter case.

We therefore label the resulting basis vectors by their mode
occupations $x=(x_1,\ldots,x_m)$, where $x_i$ counts how often $e_i$
occurs in the tensor product. Writing $|x|=\sum_i x_i$ and
$x!=\prod_i x_i!$, the allowed labels are
\begin{equation}
 \Omega_{n,m}^{+}=\{x\in\mathbb N_0^m:|x|=n\},
 \qquad
 \Omega_{n,m}^{-}=\{x\in\{0,1\}^m:|x|=n\}
\end{equation}
for bosons and fermions, respectively. For $x\in\Omega_{n,m}^{\pm}$,
define the normalized Fock vector
\begin{equation}
 \ket x_\pm=\sqrt{\frac{n!}{x!}}\,
 P_n^\pm(e_1^{\otimes x_1}\otimes\cdots\otimes e_m^{\otimes x_m}).
\end{equation}
The fixed mode ordering specifies the fermionic sign convention.
The vectors $\{\ket x_\pm:x\in\Omega_{n,m}^{\pm}\}$ form an
orthonormal basis of $\mathcal H_n^\pm$.
Thus $\Sym^n(\mathbb C^m)$ denotes the symmetric subspace, whose
vectors are invariant under every $T_\pi$, while
$\bigwedge^n\mathbb C^m$ denotes the antisymmetric subspace, whose
vectors acquire the factor $\operatorname{sgn}(\pi)$. Counting the
allowed occupations gives
\begin{equation}
 \dim\mathcal H_n^+=\binom{m+n-1}{n},
 \qquad
 \dim\mathcal H_n^-=\binom mn.
\end{equation}
The bosonic sectors exist for all $n\ge0$, whereas the fermionic
sectors vanish for $n>m$. In both cases,
$\mathcal H_0^\pm=\mathbb C\ket0$ is the vacuum sector.
Allowing particle number to vary gives the full Fock spaces
\[
 \mathcal F_+=\bigoplus_{n=0}^{\infty}\mathcal H_n^+,
 \qquad
 \mathcal F_-=\bigoplus_{n=0}^{m}\mathcal H_n^-.
\]
In what follows,  we omit the
statistics label whenever the distinction is clear or irrelevant and write $d_n=\dim\mathcal H_n$.

Since $U^{\otimes r}$ commutes with particle permutations, it preserves
$\mathcal H_r$ and defines a unitary representation of $\mathrm U(m)$
on this space:
\begin{equation}
 U\longmapsto U_r=\left.U^{\otimes r}\right|_{\mathcal H_r}.
 \label{eq:particle-mode-action}
\end{equation}
The direct-sum representation $U\mapsto\mathcal U=\bigoplus_r U_r$
acts on the full Fock space $\mathcal F$, with $r\ge0$ for bosons
and $0\le r\le m$ for fermions. It describes passive transformations,
which apply the same mode unitary to each particle and preserve
particle number. On the operator space
$\mathcal W_r=\operatorname{End}(\mathcal H_r)$, conjugation induces
the adjoint representation $U\mapsto\operatorname{Ad}_r(U)$,
\begin{equation}\label{eq:adjoint_action}
    \operatorname{Ad}_r(U)(A)=U_rAU_r^\dagger.
\end{equation}

While, the representation of $\mathrm U(m)$ on $\mathcal H_r$ in
Eq.~\eqref{eq:particle-mode-action} is irreducible, its adjoint
representation on $\mathcal W_r$, defined in
Eq.~\eqref{eq:adjoint_action}, admits a multiplicity-free decomposition
into irreducible subspaces, which we identify and construct explicitly
below.

\begin{proposition}[Operator space irrep decomposition]
\label{prop:common-adjoint-decomposition}
The operator space $\mathcal{W}_r$ over $r$ particles admits the multiplicity-free decomposition
\begin{equation}
 \mathcal W_r=\bigoplus_{k=0}^{J_r}\mathcal V_{r,k},
 \qquad
 J_r=\begin{cases}
 r,&\text{bosons},\\
 \min(r,m-r),&\text{fermions}.
 \end{cases}
 \label{eq:common-adjoint-decomposition}
\end{equation}
 Moreover,  the irrep label $k$ is
independent of the particle number $r$: copies with the same $k$ in
different sectors are equivalent. Their dimensions are given by 
\begin{equation}
 v_k:=\dim\mathcal V_{r,k}=d_k^2-d_{k-1}^2,
 \qquad 0\le k\le J_r,
 \label{eq:common-irrep-dimensions}
\end{equation}
where we recall that $d_k=\rm dim \mathcal{H}_k$.
\end{proposition}
\begin{proof}

Both fermionic and bosonic irrep decomposition of a fixed particle number operator space are given in \cite{wilkens2024benchmarking}. For completeness, we give here an overview of the proof sketch. 

Fix $m\ge2$ and put $V=\mathbb C^m$. Every $U\in\mathrm U(m)$
can be written as $U=zS$, with $S\in\mathrm{SU}(m)$ and $|z|=1$.
Since $z$ acts as the scalar $z^r$ on $\mathcal H_r$, it cancels
under conjugation. It therefore suffices to decompose the
restriction of the adjoint representation to $\mathrm{SU}(m)$.

We label the irreducible representations of $SU(m)$ by Young diagrams
$\lambda=(\lambda_1,\ldots,\lambda_m)$ with
$\lambda_1\ge\cdots\ge\lambda_m\ge0$ where each $\lambda_i$ denotes the number of boxes in the $i^{\rm th}$ row. 
For $\mathrm{SU}(m)$, diagrams differing by full columns of height
$m$ represent equivalent irreps. Removing all such columns gives
a unique diagram with at most $m-1$ nonzero rows. All representation
isomorphisms in the diagram calculations below are understood
for $\mathrm{SU}(m)$.

 For fermions, the operator space $\mathcal W_r$ over $r$ particles is given by
\begin{equation}
    \mathcal W_r \simeq (\bigwedge^rV) \otimes (\bigwedge^rV)^*  \simeq(\det V)^{-1}\otimes
 (\bigwedge^rV\otimes\bigwedge^{m-r}V),
\end{equation}
where $\simeq$ denotes equality up to a change of basis and the last equality follows from the exterior power duality.

The exterior power representation $\bigwedge^rV$ corresponds to the column
$(1^r)$. The exterior Pieri
rule~\cite[Section~6.1, Eq.~(6.9), p.~79]{fulton1991representation}
adds $m-r$ boxes to this column, at most one added box per row.
If $k$ existing rows receive a second box, the remaining
$m-r-k$ boxes form new rows. The resulting diagrams are precisely
\begin{equation}
 \lambda_k^{\mathrm F}=(2^k,1^{m-2k}),
 \qquad 0\le k\le\min(r,m-r),
\end{equation}
each with multiplicity one. For $k=0$, the full column $(1^m)$
is equivalent to the empty diagram and gives the trivial irrep.
For $k\ge1$, the diagrams already have fewer than $m$ rows and
are mutually distinct. This proves the multiplicity-free
fermionic decomposition.

For bosons, a similar decomposition is stated in
\cite[Appendix~B.1.1, Lemma~1]{wilkens2024benchmarking}. For $r \ge 0$ bosons   in $m$ modes, the bosonic operator space is given by
\begin{equation}\label{eq:op_bosons}
    \mathcal W_r \simeq  \rm Sym^r(V) \otimes (\rm Sym^r(V))^*.
\end{equation}
 The symmetric power representation corresponds to the single row $(r)$,
while its dual, as an $\mathrm{SU}(m)$ representation,
corresponds to the rectangle $(r^{m-1})$. Applying Littlewood–Richardson rules\ \cite{stembridge2002concise}, the tensor product in Eq.\eqref{eq:op_bosons} further decomposes into irreps described by Young diagrams of the form 

\begin{equation}
 \lambda_k^{\mathrm B}=(2k,k^{m-2}),
 \qquad 0\le k\le r.
\end{equation}
These reduced diagrams are mutually distinct, with $k=0$
corresponding to the empty diagram. This proves the
multiplicity-free bosonic decomposition.

Let us denote by $J_r$, 
\begin{align}
    J_r=\begin{cases}
 r,&\text{bosons},\\
 \min(r,m-r),&\text{fermions}.
 \end{cases}
\end{align}
In both cases, the reduced diagram depends on $k$ and $m$,
but not on $r$. Thus copies indexed by $k$ in sectors $r$ and
$r'$ are equivalent whenever $k\le\min(J_r,J_{r'})$.
Because the center acts trivially on every operator sector,
these equivalences also hold for the full $\mathrm U(m)$ action.

The dimensions of these irreps follow from the hook-content formula\ \cite[Exercise~6.4, p.~78]{fulton1991representation} for a Young diagram denoted by $\lambda$,
\[
 \dim \lambda
 =\prod_{(i,j)\in\lambda}\frac{m+j-i}{h_{ij}},
\]
where $h_{ij}$ is one plus the number of boxes directly to the
right of or below $(i,j)$. Evaluating it on the diagrams above gives
\begin{align}
 \dim\mathcal V_{r,k}
 &=\frac{m-2k+1}{m+1}\binom{m+1}{k}^{\!2}
   =\binom mk^{\!2}-\binom m{k-1}^{\!2},
 &&\text{fermions},\\
 \dim\mathcal V_{r,k}
 &=\frac{m+2k-1}{m-1}\binom{m+k-2}{k}^{\!2}
   =\binom{m+k-1}{k}^{\!2}
    -\binom{m+k-2}{k-1}^{\!2},
 &&\text{bosons}.
\end{align}
Thus, with $d_{-1}=0$, both cases satisfy
\[
 v_k:=\dim\mathcal V_{r,k}=d_k^2-d_{k-1}^2.
\]

\end{proof}

We denote the orthogonal projector onto $\mathcal V_{r,k}$ by
$\mathsf P_{r,k}$.  We next recall how to construct these projectors through particle reductions and intertwiners.

\subsection{Particle reductions, intertwiners, and irrep projectors}
\label{sec:exact-degree-projectors}
\label{sec:young-projectors}
\label{sec:particle-reductions}

Particle reductions connect copies of the same adjoint irrep in
different particle sectors. We use this connection to extract one
exact degree from a reduced state and then lift it to the original
sector. All identities below hold for each fixed interferometer.

\begin{definition}[Particle raising and lowering maps]
\label{def:common-particle-maps}
Let $\hat a_i,\hat a_i^\dagger$ be the canonical annihilation and creation
operators. Put $\varsigma=+1$ for bosons and $\varsigma=-1$ for
fermions, so that
$\hat a_i \hat a_j^\dagger-\varsigma \hat a_j^\dagger \hat a_i=\delta_{ij}I$.
Define
\begin{equation}
 \cL(A)=\sum_{i=1}^m \hat a_iA\hat a_i^\dagger:\mathcal W_r\to\mathcal W_{r-1},
 \qquad
 \cR(A)=\sum_{i=1}^m \hat a_i^\dagger A\hat a_i:\mathcal W_r\to\mathcal W_{r+1}.
 \label{eq:common-lowering-raising}
\end{equation}
We use $\mathcal W_{-1}=\{0\}$ and, for fermions,
$\mathcal W_{m+1}=\{0\}$. Thus $\cL=0$ at the vacuum and
$\cR=0$ on the fully occupied fermionic sector.
An operator $B\in\mathcal W_k$ is called \emph{harmonic} if
$\cL(B)=0$.
\end{definition}

\begin{proposition}[Properties of the raising and lowering maps]
\label{prop:properties_RL}
The maps of Definition~\ref{def:common-particle-maps} are
Hilbert--Schmidt adjoints, $\cR=\cL^\dagger$, and intertwine passive
conjugation on adjacent particle sectors:
\begin{equation}
 \cL\operatorname{Ad}_r(U)=\operatorname{Ad}_{r-1}(U)\cL,
 \qquad
 \cR\operatorname{Ad}_r(U)=\operatorname{Ad}_{r+1}(U)\cR.
 \label{eq:common-particle-map-covariance}
\end{equation}
\end{proposition}

\begin{proof}
For $A\in\mathcal W_r$ and $B\in\mathcal W_{r+1}$, cyclicity of
trace gives
\[
 \langle\cR(A),B\rangle_{\rm HS}
 =\sum_i\Tr\!\left[A^\dagger\hat a_iB\hat a_i^\dagger\right]
 =\langle A,\cL(B)\rangle_{\rm HS},
\]
which proves adjointness. Passive transformations mix creation
operators according to
$\mathcal U\hat a_i^\dagger\mathcal U^\dagger
=\sum_jU_{ji}\hat a_j^\dagger$. Consequently,
\[
 \begin{aligned}
 \operatorname{Ad}_{r+1}(U)\cR(A)
 &=\sum_{i,j,\ell}U_{ji}\overline{U_{\ell i}}\,
   \hat a_j^\dagger\operatorname{Ad}_r(U)(A)\hat a_\ell\\
 &=\cR\operatorname{Ad}_r(U)(A),
 \end{aligned}
\]
where unitarity gives
$\sum_iU_{ji}\overline{U_{\ell i}}=\delta_{j\ell}$.
The same calculation with annihilation operators proves the
lowering identity. These arguments apply to both statistics;
see also Ref.~\cite[Appendix~B.2--B.3]{mhiri2026boson} for the
bosonic case.
\end{proof}

\begin{proposition}[Identification of irreps through raising and lowering maps]
\label{prop:ident}
For $k\ge0$ (with $k\le\lfloor m/2\rfloor$ for fermions),
the smallest carrier of degree $k$ is the harmonic subspace
\begin{equation}
 \ker(\cL|_{\mathcal W_k})=\mathcal V_{k,k},
 \label{eq:ker_ident}
\end{equation}
and every equivalent copy is obtained by raising:
\begin{equation}
 \mathcal V_{r,k}=\cR^{r-k}(\mathcal V_{k,k}),
 \qquad k\le J_r.
 \label{eq:raise_ident}
\end{equation}
For $B\in\mathcal V_{k,k}$ and $t\ge0$ such that $k\le J_{k+t}$,
\begin{equation}
 \cL^t\cR^t(B)=\alpha_{k+t,k}B,
 \qquad
 \alpha_{k+t,k}:=t!\begin{cases}
  \rise{m+2k}{t},&\text{bosons},\\
  \fall{m-2k}{t},&\text{fermions}.
 \end{cases}
 \label{eq:common-harmonic-ladder}
\end{equation}
\end{proposition}

\begin{proof}
Let $\hat N=\sum_i\hat a_i^\dagger\hat a_i$.
Expanding $\cL\cR(A)$ with the canonical commutation or
anticommutation relations, for $A\in\mathcal W_r$, gives
\begin{equation}
 [\cL,\cR](A)
 =mA+\varsigma(\hat NA+A\hat N)
 =(m+2\varsigma r)A.
 \label{eq:common-lr-commutator}
\end{equation}
Together with adjointness, this implies
\[
 \|\cR(A)\|_2^2
 =\|\cL(A)\|_2^2+(m+2\varsigma r)\|A\|_2^2.
\]
For $k\ge1$ in the stated range, the coefficient is positive at
$r=k-1$, so $\cR:\mathcal W_{k-1}\to\mathcal W_k$ is injective.
By covariance and the multiplicity-free decomposition of
Proposition~\ref{prop:common-adjoint-decomposition}, its image is
$\bigoplus_{j<k}\mathcal V_{k,j}$.
Since $\cL=\cR^\dagger$, its kernel in $\mathcal W_k$ is the
orthogonal complement of this image, namely $\mathcal V_{k,k}$.
For $k=0$, the identification follows from $\mathcal W_{-1}=\{0\}$.

Now let $B\in\mathcal V_{k,k}$, so $\cL(B)=0$.
Commuting $\cL$ through $t\ge1$ raising maps yields
\[
 \begin{aligned}
 \cL\cR^t(B)
 &=\sum_{j=0}^{t-1}\cR^j[\cL,\cR]\cR^{t-1-j}(B)\\
 &=t\bigl[m+\varsigma(2k+t-1)\bigr]\cR^{t-1}(B).
 \end{aligned}
\]
Iterating gives the product
$\prod_{s=1}^t s[m+\varsigma(2k+s-1)]$, which is precisely
$\alpha_{k+t,k}$ in Eq.~\eqref{eq:common-harmonic-ladder}.
For the allowed $t$, this scalar is positive and adjointness gives
$\|\cR^t(B)\|_2^2=\alpha_{k+t,k}\|B\|_2^2$.
Thus $\cR^t$ is injective on $\mathcal V_{k,k}$; covariance and
multiplicity-freeness identify its image with $\mathcal V_{k+t,k}$,
proving Eq.~\eqref{eq:raise_ident}.
\end{proof}

For the same range of $k$, the resulting orthogonal decomposition is
\begin{equation}
 \mathcal W_k
 =\cR(\mathcal W_{k-1})\oplus\ker(\cL|_{\mathcal W_k}).
\end{equation}
It separates the raised lower degrees from the harmonic degree-$k$
component used to construct the projectors below. To connect these
maps with physical reduced states, we normalize repeated lowering
as follows.

\begin{definition}[Particle reductions]
\label{def:common-particle-reductions}
For a state $\rho_{\rm in} \in \mathcal W_n$, denote by $\rho_{\rm in}^{(k)} \in \mathcal W_k$ the reduced state by partial tracing $n-k$ indistinguishable particles
\begin{equation}
 \rho_{\rm in}^{(k)}=\Tr_{k+1,\ldots,n}(\rho_{\rm in}),
 \qquad 0\le k\le n.
 \label{eq:particle-reduction-definition}
\end{equation}
The above partial trace can be expressed in terms of successive application of the lowering map defined in Eq.\eqref{eq:common-lowering-raising} as 
\begin{equation}
 \mathcal T_{n\to k}(A):=\Tr_{k+1,\ldots,n}[A]
 =\frac{k!}{n!}\cL^{n-k}(A),
 \label{eq:common-normalized-reduction}
\end{equation}
for arbitrary operators $A \in \mathcal W_n$.
\end{definition}

Now, that we have introduced the necessary notation and the formulation of consecutive application of the lowering map as a partial trace operation, we present in the following Corollary a closed form expression of a projector onto a degree $k$ irrep.

We recall that the projectors below are standard: the bosonic formula is the
contraction-free projection in
\cite[Appendix~F, Eq.~(F2)]{kronke2018born}, after normalization,
while the fermionic formula is the extremal projector in
\cite[Section~3, Eqs.~(3.16)--(3.19)]{tolstoy2005fortieth}, with
$J_+=\cL$, $J_-=\cR$, and $J_0|_{\mathcal W_r}=(m/2-r)I$.

However, we give here a common proof for completeness.

\begin{corollary}[Exact-degree projectors]
\label{cor:common-projectors}
For $k\ge0$ (with $k\le\lfloor m/2\rfloor$ for fermions), define
\begin{equation}
 D_{k,q}=\begin{cases}
  \fall{m+2k-2}{q},&\text{bosons},\\
  \rise{m-2k+2}{q},&\text{fermions},
 \end{cases}
 \qquad 0\le q\le k,\qquad D_{k,0}=1.
 \label{eq:common-projector-coefficients}
\end{equation}
The orthogonal projector onto $\mathcal V_{k,k}$ is
\begin{equation}
 \displaystyle
 \Pi_k=\sum_{q=0}^k\frac{(-1)^q}{q!D_{k,q}}\cR^q\cL^q.
 \label{eq:common-projector}
\end{equation}
For $k\le J_n$, its lift to $\mathcal W_n$ yields the orthogonal projector $\displaystyle
 \mathsf P_{n,k}$ onto $\mathcal V_{n,k}$
\begin{equation}
 \displaystyle
 \mathsf P_{n,k}=\frac{1}{\alpha_{n,k}}\cR^{n-k}\Pi_k\cL^{n-k},
 \label{eq:common-full-sector-projector}
\end{equation}
with $\alpha_{n,k}$ given by Eq.~\eqref{eq:common-harmonic-ladder}.
\end{corollary}
\begin{proof}
For $C\in\mathcal W_r$ and $q\ge1$, repeated use of the commutator
identity in Proposition~\ref{prop:ident} gives
\begin{equation}
    \mathcal L\mathcal R^q(C)=
\mathcal R^q\mathcal L(C)
+q\bigl[m+\varsigma(2r+q-1)\bigr]\mathcal R^{q-1}(C).
\end{equation}
The term involving $\mathcal L(C)$ vanishes when $C$ is harmonic.

Now take $A\in\mathcal W_k$ and substitute $C=\mathcal L^q(A)$. This operator lies in $\mathcal W_{k-q}$. choosing $r=k-q$, we get
\begin{align}\label{eq:recursion_RL}
\begin{aligned}
\mathcal L\mathcal R^q\mathcal L^q(A)
&=
\mathcal R^q\mathcal L^{q+1}(A)+
q\bigl[m+\varsigma(2k-q-1)\bigr]
\mathcal R^{q-1}\mathcal L^q(A).
\end{aligned}
\end{align}

Define the ansatz $\Pi_k = \sum_{q=0}^k c_q R^q\mathcal L^{q} $. Setting $c_0=1$ enforces that $\Pi_k$ acts as the identity on harmonic operators verifying $\mathcal L(B)=0$.

The projector should also satisfy $\mathcal L \Pi_k =0$, ensuring that every $k-$particle operator after the $\Pi_k$ projector is harmonic. 
Invoking Eq.\eqref{eq:recursion_RL} applied to $A \in \mathcal W_k$ produces the two sums
\begin{align}
\begin{aligned}
\mathcal L\Pi_k(A)
&=
\sum_{q=0}^{k}c_q\mathcal R^q\mathcal L^{q+1}(A)+
\sum_{q=1}^{k}c_q q
\bigl[m+\varsigma(2k-q-1)\bigr]
\mathcal R^{q-1}\mathcal L^q(A).
\end{aligned}
\end{align}
Reindex the second sum and use $\mathcal L^{k+1}(A)=0$,
\begin{equation}
    \mathcal L\Pi_k(A)=
\sum_{q=0}^{k-1}
\left[
c_q+(q+1)\bigl[m+\varsigma(2k-q-2)\bigr]c_{q+1}
\right]\mathcal R^q\mathcal L^{q+1}(A).
\end{equation}

Each remaining operator expression receives two contributions, from adjacent terms of the projector. Choosing their coefficients to cancel gives

\begin{equation}\label{eq:cq_recursion}
   c_{q+1}
=-\frac{c_q}
{(q+1)[m+\varsigma(2k-q-2)]}. 
\end{equation}

Solving for Eq.\eqref{eq:cq_recursion} proves the $\Pi_k$ projector formula in Eq.\eqref{eq:common-projector}.

Finally, these conditions give a map that fixes \(\ker\mathcal L\) and has image inside it. Since every \(\mathcal R^q\mathcal L^q\) is self-adjoint and the coefficients are real, this map is the orthogonal projector onto that kernel.

Set $t=n-k$ and let $S=\cR^t|_{\mathcal V_{k,k}}$.
Since its domain is restricted to $\mathcal V_{k,k}$, adjointness
gives $S^\dagger=\Pi_k\cL^t$. Equation~\eqref{eq:common-harmonic-ladder} gives
$S^\dagger S=\alpha_{n,k}I$, and Proposition~\ref{prop:ident}
identifies $\operatorname{im}S=\mathcal V_{n,k}$.
Consequently, $SS^\dagger/\alpha_{n,k}$ is self-adjoint, maps into
$\mathcal V_{n,k}$, and fixes every $S(B)$. Hence
\begin{equation}
 \mathsf P_{n,k}
 =\frac{1}{\alpha_{n,k}} SS^\dagger
 =\frac{1}{\alpha_{n,k}} \cR^{n-k}\Pi_k\cL^{n-k}.
\end{equation}

\end{proof}

\subsection{Occupation polynomials as bases for diagonal degree-$k$ operators}
\label{sec:zero-weight-bases}

Occupation measurements access the diagonal part of an output state.
We now identify how this measured part decomposes under the adjoint
degrees of Proposition~\ref{prop:common-adjoint-decomposition}. The
construction has three steps: identify diagonal operators with
occupation functions, identify their orthogonal polynomial degrees
with adjoint degrees, and construct each degree by raising a harmonic
operator from its smallest particle sector. This gives a common
framework for bosons and fermions before any particular polynomial
basis is chosen. Particle reductions then provide the coefficients
needed to evaluate the output probabilities.

Throughout, $\Omega_r$ denotes $\Omega_{r,m}^{+}$ for bosons and
$\Omega_{r,m}^{-}$ for fermions.

\paragraph{The measured operator space.}
The projectors $\ket x\bra x$, $x\in\Omega_r$, span the space of
operators diagonal in the occupation basis. To locate this space
within the adjoint decomposition, consider mode phase shifts
$D_\theta=\operatorname{diag}(e^{i\theta_1},\ldots,e^{i\theta_m})$.
Their action on $\mathcal H_r$ is the induced representation
\[
 D_{\theta,r}:=\left.D_\theta^{\otimes r}\right|_{\mathcal H_r},
 \qquad D_{\theta,r}\ket x=e^{i\theta\cdot x}\ket x.
\]
Thus the adjoint action is
\begin{equation}
 \operatorname{Ad}_r(D_\theta)(\ket x\bra y)
 =D_{\theta,r}\ket x\bra yD_{\theta,r}^\dagger
 =e^{i\theta\cdot(x-y)}\ket x\bra y.
 \label{eq:common-occupation-weight}
\end{equation}
The matrix unit $\ket x\bra y$ has weight $x-y$, so the operators
fixed by every mode phase shift are exactly the diagonal ones.
They form the \emph{zero-weight subspace}
\begin{equation}
 (\mathcal W_r)_0
 :=\{A\in\mathcal W_r:\operatorname{Ad}_r(D_\theta)(A)=A
                         \text{ for all }\theta\}=\operatorname{span}\{\ket x\bra x:x\in\Omega_r\}
.
\end{equation}
Zero weight describes phase invariance. It is distinct from the
trivial irrep label $k=0$, which describes scalar operators.

The projectors $\mathsf P_{r,k}$ onto degree $k$ irreducible components given in \ref{cor:common-projectors} commute with the phase-shift
action. For $A\in(\mathcal W_r)_0$, this gives
\begin{equation}\operatorname{Ad}_r(D_\theta)(\mathsf P_{r,k}A)
 =\mathsf P_{r,k}\operatorname{Ad}_r(D_\theta)(A)
 =\mathsf P_{r,k}A.
\end{equation}
Each degree component of $A$ is therefore diagonal. Since
$A=\sum_k\mathsf P_{r,k}A$, the orthogonal irrep decomposition
restricts to
\begin{equation}
 \begin{gathered}
 (\mathcal W_r)_0
 =\bigoplus_{k=0}^{J_r}(\mathcal V_{r,k})_0,\quad
 (\mathcal V_{r,k})_0:=\mathcal V_{r,k}\cap(\mathcal W_r)_0.
 \end{gathered}
 \label{eq:common-diagonal-decomposition}
\end{equation}
We call elements of $(\mathcal V_{r,k})_0$ diagonal degree-$k$
operators. A general interferometer preserves $\mathcal V_{r,k}$
but need not preserve its diagonal part. Consequently, when
evaluating output probabilities, we must rotate the full operator
before taking its measured diagonal part.

\paragraph{Occupation functions and polynomial degree.}
With $\hat n_i\ket x=x_i\ket x$, a function
$f:\Omega_r\to\mathbb C$ corresponds to the diagonal operator
\begin{equation}
 A_f:=f(\hat n_1,\ldots,\hat n_m)|_{\mathcal H_r}
 =\sum_{x\in\Omega_r}f(x)\ket x\bra x.
 \label{eq:common-diagonal-function}
\end{equation}
Every such function has a polynomial representative. Indeed, write
$\binom yx=\prod_i\binom{y_i}{x_i}$. For $x,y\in\Omega_r$,
this is $\delta_{xy}$: if $y_i\ge x_i$ for every $i$, equality
of their total occupations forces $x=y$. Therefore
\begin{equation}
 \left.\prod_i\binom{\hat n_i}{x_i}\right|_{\mathcal H_r}
 =\ket x\bra x,
 \label{eq:fock-projector-polynomial}
\end{equation}
and the polynomial
\begin{equation}\label{eq:polynomial_f_rep}
    p_f(t)=\sum_{x\in\Omega_r}f(x)\binom tx
\end{equation}
has degree at most $r$ and restricts to $f$ on $\Omega_r$.
The identification also preserves inner products when diagonal
operators are equipped with the normalized trace:
\begin{equation}
 \langle f,g\rangle_{u_r}
 =\frac1{d_r}\sum_{x\in\Omega_r}\overline{f(x)}g(x)
 =\Tr(u_r A_f^\dagger A_g).
 \label{eq:common-diagonal-inner-product}
\end{equation}

Polynomial representatives are not unique. On $\Omega_r$,
$\sum_i x_i=r$, and for fermions $x_i^2=x_i$. In particular,
a homogeneous polynomial can restrict to a function of lower degree, e.g.
$\sum_i x_i$ becomes a constant. The relevant degree spaces are
therefore defined after restriction and orthogonal removal of lower
degrees:
\begin{equation}
 \begin{aligned}
 \mathcal P_{r,\le k}
 &=\{p|_{\Omega_r}:\deg p\le k\},\\
 \mathcal P_{r,k}
 &=\mathcal P_{r,\le k}\cap\mathcal P_{r,\le k-1}^{\perp},
 \end{aligned}
 \label{eq:common-orthogonal-degree-subspace}
\end{equation}
where orthogonality is in $L^2(u_r)$ and
$\mathcal P_{r,\le-1}=\{0\}$. Thus exact polynomial degree means
orthogonality to all lower-degree occupation functions.

\paragraph{Constructing orthogonal bases degree by degree.}
We now show that these polynomial degrees coincide with the adjoint
degrees in \eqref{eq:common-diagonal-decomposition}. The construction
starts with diagonal harmonic operators on the smallest sector
$\mathcal H_k$, where $\ker\cL=\mathcal V_{k,k}$, and raises them to
$\mathcal H_r$.

For a diagonal operator $B=\sum_{a\in\Omega_k}b_a\ket a\bra a \in (\mathcal{W}_k)_0$, define
\begin{equation}
 h_B(x)=\sum_{a\in\Omega_k}b_a\binom xa,
 \label{eq:common-harmonic-polynomial}
\end{equation}
corresponding to its polynomial representation according to Eq.\eqref{eq:polynomial_f_rep} with $ h_B(\hat n)|_{\mathcal H_k}= B$.  The same polynomial
can be restricted to other occupation sectors.

\begin{proposition}[Diagonal degree spaces and normalization]
\label{prop:common-diagonal-basis}
For $0\le k\le J_r$, normalized raising gives the isomorphism
\begin{equation}
 \begin{aligned}
 (\mathcal V_{k,k})_0
 &\longrightarrow(\mathcal V_{r,k})_0,\\
 B&\longmapsto
 \frac{\cR^{r-k}(B)}{(r-k)!}
 =h_B(\hat n)|_{\mathcal H_r}.
 \end{aligned}
 \label{eq:common-polynomial-lift}
\end{equation}
Equivalently, $B\mapsto h_B|_{\Omega_r}$ identifies
$(\mathcal V_{k,k})_0$ with $\mathcal P_{r,k}$, and
\begin{equation}
 \{A_f:f\in\mathcal P_{r,k}\}=(\mathcal V_{r,k})_0,
 \qquad
 z_k:=\dim\mathcal P_{r,k}
     =\dim(\mathcal V_{k,k})_0=d_k-d_{k-1}.
 \label{eq:common-diagonal-degree-dimension}
\end{equation}
For $B,C\in(\mathcal V_{k,k})_0$, the normalization is
\begin{equation}
 \langle h_B,h_C\rangle_{u_r}
 =\beta_{r,k}\Tr(B^\dagger C),
 \qquad
 \beta_{r,k}:=\frac{\alpha_{r,k}}{d_r((r-k)!)^2}.
 \label{eq:common-lift-inner-product}
\end{equation}
\end{proposition}

\begin{proof}
We first compute raising on a single occupation projector, without
assuming harmonicity. Fix $a\in\Omega_j$, put $t=r-j$, and consider
$x\in\Omega_r$ with $x\ge a$ componentwise.
For bosons, reaching $x$ requires $x_i-a_i$ additions in mode $i$.
There are $t!/\prod_i(x_i-a_i)!$ possible orders, and each order
contributes the factor $\prod_i x_i!/a_i!$. The coefficient of
$\ket x\bra x$ is therefore
\begin{equation}
 \frac{t!}{\prod_i(x_i-a_i)!}\prod_i\frac{x_i!}{a_i!}
 =t!\binom xa.
 \label{eq:common-raising-occupation-coefficient}
\end{equation}
For fermions, an admissible $x$ extends $a$ by $t$ distinct occupied
modes. Each of their $t!$ orders contributes one, because the
fermionic signs cancel between the ket and bra. Here $\binom xa$
equals one precisely when $a\subseteq x$.
In either statistics the coefficient vanishes unless $x\ge a$, so
\begin{equation}
 \cR^{r-j}(\ket a\bra a)
 =(r-j)!\sum_{x\in\Omega_r}\binom xa\ket x\bra x.
 \label{eq:common-occupation-projector-lift}
\end{equation}
By linearity and \eqref{eq:common-harmonic-polynomial},
\eqref{eq:common-occupation-projector-lift} proves the operator
identity in \eqref{eq:common-polynomial-lift} for every diagonal
$B$, whether harmonic or not.

We next restrict this identity to a single adjoint degree.
For $j\le J_r$, the full-irrep correspondence
\eqref{eq:raise_ident} makes $\cR^{r-j}$ an equivariant
isomorphism from $\mathcal V_{j,j}$ onto $\mathcal V_{r,j}$.
Equivariance sends operators fixed by all mode phase shifts to
operators with the same property. Conversely, if
$\cR^{r-j}(B)$ is fixed by every diagonal mode unitary $D_\theta$,
then
\begin{equation}
 \begin{aligned}
 \cR^{r-j}\bigl(\operatorname{Ad}_j(D_\theta)(B)-B\bigr)
 &=\operatorname{Ad}_r(D_\theta)\bigl(\cR^{r-j}(B)\bigr)
   -\cR^{r-j}(B)=0.
 \end{aligned}
 \label{eq:common-zero-weight-preimage}
\end{equation}
Injectivity on $\mathcal V_{j,j}$ implies that $B$ is itself fixed
by every mode phase shift. Since these fixed operators are exactly
the diagonal ones, \eqref{eq:common-zero-weight-preimage} proves
\begin{equation}
 \cR^{r-j}\bigl((\mathcal V_{j,j})_0\bigr)
 =(\mathcal V_{r,j})_0.
 \label{eq:common-zero-weight-lift}
\end{equation}
Thus raising identifies the entire zero-weight subspace on each
sector, not merely a subspace of it.

To relate this correspondence to polynomial degree, observe that
bosonic binomial monomials span polynomials of bounded degree:
their leading terms are $x^a/a!$, so the change from ordinary
monomials is triangular with nonzero diagonal.
For fermions, first use $x_i^q=x_i$ on binary occupations to obtain
multilinear representatives; their monomials are precisely the
restrictions of $\binom xa$ with binary $a$.
Consequently, for $k\le J_r$,
\begin{equation}
 \mathcal P_{r,\le k}
 =\operatorname{span}\left\{
   \binom xa\big|_{\Omega_r}:
   a\in\Omega_j,\ 0\le j\le k
 \right\}.
 \label{eq:common-binomial-filtration}
\end{equation}

By \eqref{eq:common-occupation-projector-lift}, each generator in
\eqref{eq:common-binomial-filtration} is represented by raising an
occupation projector from $\mathcal H_j$. That projector contains
only adjoint degrees through $j$, and raising preserves the irrep
label. Its image therefore belongs to
$\bigoplus_{\ell=0}^j(\mathcal V_{r,\ell})_0$.
Conversely, \eqref{eq:common-zero-weight-lift} represents every
element of $(\mathcal V_{r,j})_0$ by a diagonal harmonic seed on
$\mathcal H_j$, whose polynomial
\eqref{eq:common-harmonic-polynomial} has degree at most $j$.
These two inclusions give
\begin{equation}
 \{A_f:f\in\mathcal P_{r,\le k}\}
 =\bigoplus_{j=0}^k(\mathcal V_{r,j})_0.
 \label{eq:common-polynomial-filtration}
\end{equation}
The function--operator identification preserves orthogonality by
\eqref{eq:common-diagonal-inner-product}. Taking the orthogonal
complement of the degrees below $k$ in
\eqref{eq:common-polynomial-filtration} identifies
$\mathcal P_{r,k}$ with $(\mathcal V_{r,k})_0$.
Together with \eqref{eq:common-zero-weight-lift} and the nonzero
normalization in \eqref{eq:common-polynomial-lift}, this establishes
the claimed isomorphisms.

The isomorphism \eqref{eq:common-zero-weight-lift} also shows that
the dimension $z_j$ is independent of the particle sector wherever
degree $j$ occurs. Applying the diagonal decomposition
\eqref{eq:common-diagonal-decomposition} in sectors $k$ and $k-1$
gives
\begin{equation}
 d_k=\sum_{j=0}^kz_j,
 \qquad
 d_{k-1}=\sum_{j=0}^{k-1}z_j,
 \qquad
 z_k=d_k-d_{k-1}.
 \label{eq:common-diagonal-dimension-difference}
\end{equation}
For fermions, the displayed ranges are valid because $k\le J_r$
implies $k\le\lfloor m/2\rfloor$. At $k=0$, the vacuum seed space
is one-dimensional, giving
\begin{equation}
 z_0=1=d_0-d_{-1}.
 \label{eq:common-diagonal-vacuum-dimension}
\end{equation}

It remains to determine the normalization. For
$B,C\in(\mathcal V_{k,k})_0$ and $t=r-k$, the polynomial lift
\eqref{eq:common-polynomial-lift}, adjointness, and the harmonic
ladder identity \eqref{eq:common-harmonic-ladder} give
\begin{equation}
 \begin{aligned}
 \langle h_B,h_C\rangle_{u_r}
 &=\frac{1}{d_r(t!)^2}
   \langle\cR^t(B),\cR^t(C)\rangle_{\rm HS}\\
 &=\frac{1}{d_r(t!)^2}
   \langle B,\cL^t\cR^t(C)\rangle_{\rm HS}\\
 &=\frac{\alpha_{r,k}}{d_r(t!)^2}\Tr(B^\dagger C).
 \end{aligned}
 \label{eq:common-diagonal-norm-transfer}
\end{equation}
This proves \eqref{eq:common-lift-inner-product}. Substituting the
bosonic and fermionic ladder coefficients yields
\begin{equation}
 \beta_{r,k}
 =\begin{cases}
  \displaystyle\frac{\rise{m+2k}{r-k}}{d_r(r-k)!},
       &\text{bosons},\\[5pt]
  \displaystyle\frac{\binom{m-2k}{r-k}}{\binom mr},
       &\text{fermions}.
 \end{cases}
 \label{eq:common-diagonal-beta}
\end{equation}
The factor $1/d_r$ in \eqref{eq:common-diagonal-norm-transfer}
distinguishes the uniform-measure inner product on the polynomial
side from the ordinary Hilbert--Schmidt inner product on the seeds.
\end{proof}

\paragraph{Complete diagonal bases.}
For each $0\le k\le J_r$, choose a Hilbert--Schmidt orthonormal
basis $\{B_{k,\mu}\}_{\mu=1}^{z_k}$ of $(\mathcal V_{k,k})_0$.
Then
\begin{equation}
 \phi_{r,k,\mu}(x)
 =\beta_{r,k}^{-1/2}h_{B_{k,\mu}}(x),
 \qquad
 0\le k\le J_r,\quad 1\le\mu\le z_k,
 \label{eq:common-diagonal-polynomial-basis}
\end{equation}
forms an orthonormal basis of each $\mathcal P_{r,k}$.
Their union is complete in $L^2(u_r)$ by
\eqref{eq:common-diagonal-decomposition}.
Thus harmonic seeds across all $k$ provide bases of every diagonal
irrep component on sector $r$, not only its top degree.

The condition $\cL(B)=0$ is equivalent to
\begin{equation}
 \begin{cases}
 \displaystyle\sum_i(a_i+1)b_{a+e_i}=0
      &\text{for every }a\in\Omega_{k-1},
       \quad\text{bosons},\\[3pt]
 \displaystyle\sum_{i\notin T}b_{T\cup\{i\}}=0
      &\text{for every }|T|=k-1,
       \quad\text{fermions}.
 \end{cases}
 \label{eq:common-diagonal-kernel}
\end{equation}

Subsection~\ref{subsec:Young} chooses Filmus's Young
coordinates~\cite{filmus2014orthogonal}, while
Subsection~\ref{sec:hahn-ingredient-basis} uses multivariate
Hahn coordinates~\cite{khare2009rates}.
The proposition identifies their diagonal adjoint degrees and
transfers their norms between sectors. These complete bases
support the noise analysis; whole-degree reduced-overlap
evaluation requires neither constructing their raised operators
nor enumerating their elements.

\subsection{Evaluating degree-$k$ probability components in the $k$-particle space}
\label{sec:reduced_overlap}

Although the output distribution is defined on the $n$-particle
sector, its degree-$k$ component can be evaluated using operators
on $\mathcal H_k$. The preceding subsection provides the link:
Eq.~\eqref{eq:common-diagonal-degree-dimension} identifies diagonal
adjoint degrees with orthogonal polynomial degrees, and
Eq.~\eqref{eq:common-diagonal-polynomial-basis} constructs their
bases from diagonal harmonic operators on $\mathcal H_k$.
We now use particle reductions to compute the coefficients in
these bases and express the contribution of an entire degree as
a single overlap in the smaller particle sector.

The key is the projector formula
\eqref{eq:common-full-sector-projector}: the degree-$k$ part of
an $n$-particle state is obtained by raising the harmonic part of
its $k$-particle reduction. By adjointness, this raising map can
be transferred to the measurement operator, where it becomes
a particle reduction. Occupation measurement then selects the
diagonal degree-$k$ parts of these reduced operators.
The identities below hold for every
fixed interferometer and arbitrary input state.

\begin{corollary}[Degree-$k$ components of output probabilities]
\label{cor:common-output-degrees}
Let $\rho_{\rm out}=U_n\rho_{\rm in}U_n^\dagger$ and
$p_U(x)=\bra x\rho_{\rm out}\ket x$, $x\in\Omega_n$.
For $0\le k\le J_n$, define
\[
 \begin{gathered}
 T_k:=\Pi_k(\rho_{\rm in}^{(k)}),\qquad
 B_{k,U}:=\mathcal D_k(U_kT_kU_k^\dagger),\\
 \rho_x^{(k)}:=\mathcal T_{n\to k}(\ket x\bra x),
 \end{gathered}
\]
where $\mathcal D_k(A)=\sum_{a\in\Omega_k}
\bra a A\ket a\ket a\bra a$ denotes occupation dephasing.
Then $B_{k,U}\in(\mathcal V_{k,k})_0$, and
\begin{equation}
 p_U(x)=\sum_{k=0}^{J_n}p_{k,U}(x),\qquad
 p_{k,U}(x):=\bra x\mathsf P_{n,k}(\rho_{\rm out})\ket x.
 \label{eq:common-output-degree-decomposition}
\end{equation}
In the basis of Eq.~\eqref{eq:common-diagonal-polynomial-basis},
\begin{equation}
 p_{k,U}(x)=\frac1{d_n}\sum_{\mu=1}^{z_k}
   \Tr[\rho_{\rm out}A_{\phi_{n,k,\mu}}^\dagger]\,
   \phi_{n,k,\mu}(x),
 \label{eq:common-output-degree-coordinates}
\end{equation}
whose coefficients can be evaluated on $\mathcal H_k$ as
\begin{equation}
 \Tr[\rho_{\rm out}A_{\phi_{n,k,\mu}}^\dagger]
 =\binom nk\beta_{n,k}^{-1/2}
   \Tr[B_{k,\mu}^\dagger B_{k,U}].
 \label{eq:common-output-reduced-coefficients}
\end{equation}
The factor $1/d_n$ in
Eq.~\eqref{eq:common-output-degree-coordinates} converts the
expansion of the density $d_np_U$ with respect to $u_n$ into
the probability mass function.

Equivalently, the entire degree is given by the reduced overlaps
\begin{equation}
 \begin{aligned}
 p_{k,U}(x)
 &=\frac{(n!/k!)^2}{\alpha_{n,k}}
      \Tr[\rho_x^{(k)}U_kT_kU_k^\dagger]\\
 &=\frac{(n!/k!)^2}{\alpha_{n,k}}
      \Tr[\Pi_k(\rho_x^{(k)})B_{k,U}]\\
 &=\frac{n!(n-k)!}{k!\alpha_{n,k}}\,h_{B_{k,U}}(x).
 \end{aligned}
 \label{eq:common-output-degree-overlap}
\end{equation}
The constant component is $p_{0,U}(x)=1/d_n$; for $k\ge1$,
$\sum_xp_{k,U}(x)=0$, and these components can take either sign.
Truncation through degree $D$ retains
$\sum_{k=0}^{\min(D,J_n)}p_{k,U}$.
\end{corollary}
\begin{proof}
The diagonal decomposition
\eqref{eq:common-diagonal-decomposition} and its polynomial basis
give Eqs.~\eqref{eq:common-output-degree-decomposition}
and~\eqref{eq:common-output-degree-coordinates}.
By the normalized reduction
\eqref{eq:common-normalized-reduction} and the projector formula
\eqref{eq:common-full-sector-projector},
\begin{equation}
 \cL^{n-k}(\rho_{\rm in})=\frac{n!}{k!}\rho_{\rm in}^{(k)},
 \qquad
 \mathsf P_{n,k}(\rho_{\rm in})
 =\frac{n!}{k!\alpha_{n,k}}\cR^{n-k}(T_k).
 \label{eq:common-projected-reduced-input}
\end{equation}
Using covariance
\eqref{eq:common-particle-map-covariance} and moving
$\cR^{n-k}$ onto $\ket x\bra x$ by adjointness proves the first
line of Eq.~\eqref{eq:common-output-degree-overlap}.
Since $\rho_x^{(k)}$ is diagonal, only the diagonal part of
$U_kT_kU_k^\dagger$ contributes. Dephasing preserves each adjoint
degree, so this part is $B_{k,U}\in(\mathcal V_{k,k})_0$.
Self-adjointness of $\Pi_k$ then gives the second line.
Finally, covariance under phase rotations implies
$\mathcal D_n\cR^{n-k}=\cR^{n-k}\mathcal D_k$;
the polynomial lift \eqref{eq:common-polynomial-lift} gives
the third line.

For the coefficient identity, the same lift and normalized
reduction give
\[
 \Tr[\rho_{\rm out}A_{\phi_{n,k,\mu}}^\dagger]
 =\binom nk\beta_{n,k}^{-1/2}
   \Tr[B_{k,\mu}^\dagger U_k\rho_{\rm in}^{(k)}U_k^\dagger].
\]
Because $B_{k,\mu}$ is diagonal and harmonic, the last trace
selects precisely $B_{k,U}$, proving
Eq.~\eqref{eq:common-output-reduced-coefficients}.
The constant and zero-sum claims follow from the scalar irrep
and orthogonality to the identity.
\end{proof}

Equation~\eqref{eq:common-output-degree-overlap} collects the
diagonal-basis expansion into one overlap between diagonal
harmonic operators on $\mathcal H_k$. Thus, given the reduced
input data, a whole degree can be evaluated without constructing
operators on $\mathcal H_n$ or enumerating the basis coordinates.
The full $T_k$ must nevertheless be evolved before dephasing:
its off-diagonal entries can contribute to $B_{k,U}$.

The coefficient identity extends to arbitrary occupation
polynomials, including those containing several degrees.

\begin{lemma}[Occupation moments from particle reductions]
\label{lem:rdm-moments}
Let $\rho_{\rm out}=U_n\rho_{\rm in}U_n^\dagger$ and let
$a\in\Omega_k$, with $1\le k\le n$. Then
\begin{equation}
 \Tr\!\left[\rho_{\rm out}\prod_i\binom{\hat n_i}{a_i}\right]
 =\binom nk\bra a U_k\rho_{\rm in}^{(k)}U_k^\dagger\ket a.
 \label{eq:rdm-moments}
\end{equation}
More generally, for any diagonal $B$ on $\mathcal H_k$,
\[
 \Tr[\rho_{\rm out}h_B(\hat n)]
 =\binom nk\Tr[B U_k\rho_{\rm in}^{(k)}U_k^\dagger].
\]
Consequently, the reduction to $\min(D,n)$ particles determines
every occupation-polynomial expectation of degree at most $D$.
\end{lemma}
\begin{proof}
The occupation-projector lift
\eqref{eq:common-occupation-projector-lift} and normalized
reduction \eqref{eq:common-normalized-reduction} give
\begin{equation}
 \left.\prod_i\binom{\hat n_i}{a_i}\right|_{\mathcal H_n}
 =\binom nk\,\mathcal T_{n\to k}^{\dagger}(\ket a\bra a).
 \label{eq:common-reduction-intertwiner}
\end{equation}
Taking the expectation and using covariance of particle reduction
proves Eq.~\eqref{eq:rdm-moments}; linearity gives the formula
for $B$. Lower-order reductions follow by further partial
traces, and the binomial monomials span the polynomial filtration
\eqref{eq:common-binomial-filtration}.
\end{proof}

For evaluation, the harmonic input $T_k$ can be expressed directly
through reductions of order at most $k$. Substituting
$\cL^q(\rho_{\rm in}^{(k)})=\frac{k!}{(k-q)!}
\rho_{\rm in}^{(k-q)}$ into Eq.~\eqref{eq:common-projector} gives
\begin{equation}
 T_k=\sum_{q=0}^k
       \frac{(-1)^q\binom{k}{q}}{D_{k,q}}\,
       \cR^q(\rho_{\rm in}^{(k-q)}).
 \label{eq:common-reduced-projector-expansion}
\end{equation}
For the input families considered here, the required low-order
moments admit the following arithmetic bound, which also applies
to inputs with indefinite particle number.

\begin{lemma}[Computing low-degree occupation moments]
\label{lem:efficient-occupation-moments}
Let $m\ge2$, $k\ge1$, and let $X(\hat n)$ be an occupation
polynomial of degree at most $k$, given by its coefficients.
Given the mode matrix $U$, the normally ordered coefficients of
$\mathcal U^\dagger X(\hat n)\mathcal U$ and its expectation on
any input family in Sec.~\ref{sec:main-framework} can be computed
in $m^{\mathcal O(k)}$ arithmetic operations. The same bound holds
for computing all occupation moments through degree $k$.
The input scalars, including squeezed-state second moments,
are treated as arithmetic inputs.
\end{lemma}
\begin{proof}
Since $\hat n_i=\hat a_i^\dagger\hat a_i$, an occupation polynomial
of degree at most $k$ is a polynomial of degree at most $2k$ in
creation and annihilation operators. Gaussian unitary evolution
acts linearly on these operators (affinely if displacements are
included), so backpropagation does not increase this degree.
Normal ordering also does not increase degree, by the canonical
commutation or anticommutation relations.

There are at most
\begin{equation}
 \sum_{d=0}^{2k}(2m)^d=m^{\mathcal O(k)}
 \label{eq:occupation-coefficient-count}
\end{equation}
operator monomials to retain. We multiply the transformed linear
forms successively, collecting normally ordered coefficients after
each multiplication. Multiplying a normally ordered monomial by
one ladder operator produces at most two normally ordered terms.
Thus this procedure takes $m^{\mathcal O(k)}$ arithmetic operations.

For the specified inputs, monomial expectations reduce to
single-mode moments for the Fock and squeezed product states,
and four-mode matrix elements for the fermionic magic product
state. These are computable in polynomial time in $m,k$, using
Wick recurrences for the squeezed-state moments. Finally, there
are only $m^{\mathcal O(k)}$ occupation monomials through degree
$k$, so evaluating all their expectations preserves the bound.
\end{proof}

\section{Low-degree approximation under thermal loss}
\label{sec:framework}

We consider sampling experiments in which a passive bosonic or fermionic
interferometer acts on a fixed input state, followed by particle-number
measurements in the output modes. There are $m$ modes, and the
interferometer is specified by an $m\times m$ unitary matrix $U$.
We denote by $\mathcal U$  its action on the bosonic or fermionic Fock
space. We study three input families:
\begin{enumerate}
\item \textbf{Fermion sampling.} The input is a product of four-mode,
two-particle magic states,
\begin{equation}
 \ket{\Psi_{\rm in}}
 =\left(\frac{\ket{1100}+\ket{0011}}{\sqrt2}\right)^{\otimes B},\qquad
 m=4B,\qquad n=2B.
 \label{eq:framework-fermion-input}
\end{equation}
Here $B$ is the number of blocks and $n$ is the total number of particles.

\item \textbf{Fock-state boson sampling.} The input has one photon in
each of $n\le m$ modes and vacuum in the remaining modes,
\begin{equation}
 \ket{\Psi_{\rm in}}=\ket{1}^{\otimes n}\otimes\ket{0}^{\otimes(m-n)}.
 \label{eq:framework-fock-input}
\end{equation}

\item \textbf{Gaussian boson sampling.} The input is
\begin{equation}
 \ket{\Psi_{\rm in}}=\ket{\mathrm{sq}(r)}^{\otimes m},
 \label{eq:framework-gbs-input}
\end{equation}
where $\ket{\mathrm{sq}(r)}$ denotes the single-mode squeezed vacuum
state with real squeezing parameter $r$.
\end{enumerate}

We study the noisy versions of these experiments, with the same thermal
loss channel applied independently to every output mode. Pure loss is
the special case in which the environment is in the vacuum state.
To establish an approximation, we use either a flatness assumption on
the ideal output distribution, expressed by a suitable $\chi^2$
divergence, or the more structured assumption that $U$ is Haar-random.
In the latter case we derive the needed divergence and degree bounds
by averaging over the interferometer.

\subsection{Thermal loss}
\label{sec:thermal-loss}

Thermal loss couples a system mode to an environment mode, then discards
the environment. Let $\hat a$ and $\hat b$ be their annihilation operators.
For transmissivity $\eta\in[0,1]$, their coupling satisfies
\begin{equation}
 V_\eta^\dagger \hat aV_\eta
 =\sqrt\eta\,\hat a+\sqrt{1-\eta}\,\hat b.
 \label{eq:thermal-mode-coupling}
\end{equation}
For bosons, thermal loss can be modeled by a beam-splitter~\cite[Sec.~V.2]{weedbrook2012gaussian}. For fermions, it corresponds to  a number-preserving rotation of two fermionic modes.

The environment mode has mean occupation $\beta$ and thermal state
\begin{equation}
 \vartheta_\beta=
 \begin{cases}
 (1-\beta)\ket0\bra0+\beta\ket1\bra1,
       &\text{fermions},\quad 0\le\beta\le1,\\[2pt]
 \displaystyle\sum_{j=0}^{\infty}
       \frac{\beta^j}{(1+\beta)^{j+1}}\ket j\bra j,
       &\text{bosons},\quad \beta\ge0.
 \end{cases}
 \label{eq:thermal-bath}
\end{equation}
The single-mode thermal-loss channel is
\begin{equation}
 \mathcal N_{\eta,\beta}(\sigma)
 =\Tr_{\rm env}\!\left[
 V_\eta(\sigma\otimes\vartheta_\beta)V_\eta^\dagger\right].
 \label{eq:thermal-channel-definition}
\end{equation}
Coupling each of $m$ system modes to an independent environment mode
gives the channel
\begin{equation}
 \sigma\longmapsto\mathcal N_{\eta,\beta}^{\otimes m}(\sigma).
 \label{eq:thermal-product-channel}
\end{equation}
Homogeneous noise means that $\eta$ and $\beta$ are the same in every mode.
On occupation-diagonal
operators, it acts as the ordinary tensor product of the one-mode
transition maps. The environment can supply particles as well as absorb them.
When $\beta=0$, it supplies none and the channel is pure loss.

For occupation measurements, we calculate the adjoint channel $\mathcal N_{\eta,\beta}^{\dagger\otimes m}$ on
observables diagonal in the occupation basis. For the single-mode
number operator $\hat n=\hat a^\dagger \hat a$, we have
\begin{equation}
 \mathcal N_{\eta,\beta}^{\dagger}(I)=I,
 \qquad
 \mathcal N_{\eta,\beta}^{\dagger}(\hat n)
 =\eta\hat n+(1-\eta)\beta I.
 \label{eq:thermal-fermion-kernel}
\end{equation}
The first identity follows from trace preservation. To prove the second,
we expand the system--environment transformation in
\eqref{eq:thermal-mode-coupling}:
\begin{equation}
 V_\eta^\dagger\hat n V_\eta
 =\eta \hat a^\dagger \hat a+(1-\eta)\hat b^\dagger \hat b
  +\sqrt{\eta(1-\eta)}(\hat a^\dagger \hat b+\hat b^\dagger \hat a).
 \label{eq:thermal-number-expansion}
\end{equation}
Taking the expectation in the environment state removes the cross terms,
since $\Tr(\vartheta_\beta \hat b)=\Tr(\vartheta_\beta \hat b^\dagger)=0$, and replaces
$\hat b^\dagger \hat b$ by $\Tr(\vartheta_\beta \hat b^\dagger \hat b)=\beta$. This proves the
second identity in \eqref{eq:thermal-fermion-kernel} for both fermions and
bosons.

For a fermionic mode, $\hat n=\ket1\bra1$ and
$I-\hat n=\ket0\bra0$. Every diagonal observable is therefore a linear
combination of $I$ and $\hat n$, so \eqref{eq:thermal-fermion-kernel}
determines the full action on occupation observables.

For bosons, it is convenient to describe the action on polynomials of
the number operator using
\begin{equation}
 \binom{\hat n}{k}
 =\frac{\hat n(\hat n-I)\cdots(\hat n-(k-1)I)}{k!}
 =\frac{(\hat a^\dagger)^k\hat a^k}{k!},
 \qquad k\in\mathbb N_0,
 \label{eq:thermal-binomial-observable}
\end{equation}
where $\binom{\hat n}{0}=I$. For each $k$, these observables through
degree $k$ span the same space as $I,\hat n,\ldots,\hat n^k$.

\begin{lemma}[Thermal loss on factorial moments]
\label{lem:thermal-factorial-moments}
For a bosonic mode and $t=(1-\eta)\beta$, the adjoint thermal-loss channel
satisfies
\begin{equation}
 \mathcal N_{\eta,\beta}^{\dagger}\!\left(\binom{\hat n}{k}\right)
 =\sum_{j=0}^k\binom kj\eta^j t^{k-j}\binom{\hat n}{j},
 \qquad k\in\mathbb N_0.
 \label{eq:thermal-factorial-transfer}
\end{equation}
The channel also satisfies
\begin{equation}
 \mathcal N_{\eta,\beta}^\dagger(z^{\hat n})
 =\frac1{1+t(1-z)}
  \left(1-\frac{\eta(1-z)}{1+t(1-z)}\right)^{\hat n},
 \qquad 0<z<1.
 \label{eq:thermal-pgf}
\end{equation}
\end{lemma}

\begin{proof}
The thermal environment is diagonal in the occupation basis, so its
expectation of $(\hat b^\dagger)^u \hat b^v$ vanishes unless $u=v$. In the latter
case, summing its geometric occupation law gives
\begin{equation}
 \Tr\!\left[\vartheta_\beta(\hat b^\dagger)^u \hat b^v\right]
 =\delta_{uv}\sum_{\ell=u}^\infty
   \frac{\beta^\ell}{(1+\beta)^{\ell+1}}
   \frac{\ell!}{(\ell-u)!}
 =\delta_{uv}\,u!\beta^u.
 \label{eq:thermal-bath-factorial-moments}
\end{equation}
The last equality follows by differentiating the geometric series
$u$ times. We now apply the coupling
\eqref{eq:thermal-mode-coupling} to
$(\hat a^\dagger)^k\hat a^k/k!$ and expand both $k-$th powers. System and
environment operators commute. The bath expectation retains precisely
the terms with equal numbers of environment creation and annihilation
operators. Hence
\begin{equation}
\begin{aligned}
 \mathcal N_{\eta,\beta}^{\dagger}\!\left(\binom{\hat n}{k}\right)
 &=\frac1{k!}\sum_{j=0}^k\binom kj^2\eta^j(1-\eta)^{k-j}
    (k-j)!\beta^{k-j}(\hat a^\dagger)^j\hat a^j\\
 &=\sum_{j=0}^k\binom kj\eta^j[(1-\eta)\beta]^{k-j}
    \binom{\hat n}{j}.
\end{aligned}
 \label{eq:thermal-factorial-transfer-proof}
\end{equation}
Here we used $(\hat a^\dagger)^j\hat a^j=j!\binom{\hat n}{j}$ and
$\binom kj^2(k-j)!j!/k!=\binom kj$.

To prove~\eqref{eq:thermal-pgf}, we work on an occupation vector and
initially take $z$ close to one:
\begin{equation}
\begin{aligned}
 \mathcal N_{\eta,\beta}^\dagger(z^{\hat n})
 &=\sum_{\ell\ge0}(z-1)^\ell
   \sum_{j=0}^\ell\binom\ell j\eta^jt^{\ell-j}
                     \binom{\hat n}{j}\\
 &=\frac1{1-t(z-1)}
   \sum_{j\ge0}\binom{\hat n}{j}
    \left(\frac{\eta(z-1)}{1-t(z-1)}\right)^j\\
 &=\frac1{1+t(1-z)}
   \left(1-\frac{\eta(1-z)}{1+t(1-z)}\right)^{\hat n}.
\end{aligned}
 \label{eq:thermal-pgf-proof}
\end{equation}
Analytic continuation gives \eqref{eq:thermal-pgf} for $0<z<1$.
\end{proof}

We use $0<\eta<1$ in the approximation theorems.
At $\eta=0$ the output is the product environment state and is sampled
directly.

\subsection{Occupation states and reference norms}
\label{sec:occupation-reference}

An occupation string is $x=(x_1,\ldots,x_m)$, with
$x_i\in\{0,1\}$ for fermions and $x_i\in\mathbb N_0$ for bosons.
The vector $\ket x$ is the corresponding occupation basis vector.
Let $\mathcal D$ denote dephasing in this basis:
\begin{equation}
 \mathcal D(\sigma)
 =\sum_x\bra x\sigma\ket x\,\ket x\bra x.
 \label{eq:occupation-dephasing}
\end{equation}
We represent the ideal measured output and its probabilities by
\begin{equation}
\begin{aligned}
 \rho_U&=\mathcal D(\mathcal U\rho_{\rm in}\mathcal U^\dagger)
       =\sum_xp_U(x)\ket x\bra x,\quad  p_U(x)=\bra x\mathcal U\rho_{\rm in}\mathcal U^\dagger\ket x.
\end{aligned}
 \label{eq:ideal-dephased-output}
\end{equation}
Here $\rho_{\rm in}=\ket{\Psi_{\rm in}}\bra{\Psi_{\rm in}}$ is one of
the input states above. The commutation with dephasing gives
\begin{equation}
\begin{gathered}
 \mathcal D\!\left(\mathcal N_{\eta,\beta}^{\otimes m}(\sigma)\right)
 =\mathcal N_{\eta,\beta}^{\otimes m}\!\left(\mathcal D(\sigma)\right),\\
 \rho_{\rm noisy,U}=\mathcal N_{\eta,\beta}^{\otimes m}(\rho_U).
\end{gathered}
 \label{eq:dephased-noisy-output}
\end{equation}
It follows that the one-mode occupation laws determined by
\eqref{eq:thermal-fermion-kernel} and \eqref{eq:thermal-pgf} can be applied
directly to $p_U$, independently in each mode.

To state the approximation theorem, let $\rho$ be a diagonal state
whose distribution we want to approximate after a channel. We choose
a diagonal reference state $\tau$ and write
\begin{equation}
 \rho=\sum_xp(x)\ket x\bra x,\qquad
 \tau=\sum_x\tau(x)\ket x\bra x,\qquad
 \sum_xp(x)=\sum_x\tau(x)=1.
 \label{eq:diagonal-state-reference}
\end{equation}

We assume that $\tau(x)>0$ whenever $p(x)>0$, and take $\tau^{-1}$ on the
support of $\tau$. The chi-squared divergence is
\begin{equation}
 \chi^2(\rho\|\tau)
 =\Tr\!\left[(\rho-\tau)^2\tau^{-1}\right]
 =\sum_{x:\tau(x)>0}\frac{(p(x)-\tau(x))^2}{\tau(x)}
 =\sum_{x:\tau(x)>0}\frac{p(x)^2}{\tau(x)}-1.
 \label{eq:chi}
\end{equation}
It is infinite if the support condition fails or the sum diverges.

We identify real diagonal observables with functions on the support of
$\tau$. We write $L^2(\tau)$ for those with $\Tr(\tau X^2)<\infty$,
equipped with inner product $\Tr(\tau XY)$. By Cauchy--Schwarz and
$\Tr\tau=1$,
\begin{equation}
 \|\tau X\|_1\le\sqrt{\Tr(\tau X^2)}.
 \label{eq:weighted-cs}
\end{equation}
Thus convergence in $L^2(\tau)$ implies trace-norm convergence after
multiplication by $\tau$. For diagonal states $\rho_p$ and $\rho_q$
representing probability distributions $p$ and $q$,
\begin{equation}
 \TV(p,q)=\frac12\|\rho_p-\rho_q\|_1.
 \label{eq:tvd-trace-norm}
\end{equation}

\subsection{Propagating the reference through noise}
\label{sec:reference-propagation}

Let $\mathcal N$ be a channel mapping diagonal states to diagonal
states.

\begin{lemma}[Reference contraction]\label{lem:reference-contraction}
For every real diagonal $X\in L^2(\tau)$,
\begin{equation}
 \Tr\!\left[\mathcal N(\tau X)^2\mathcal N(\tau)^{-1}\right]
 \le\Tr(\tau X^2).
 \label{eq:reference-contraction}
\end{equation}
The inverse of $\mathcal N(\tau)$ is restricted to its support.
\end{lemma}
\begin{proof}
Since the operators
$\mathcal N(\tau\ket x\bra x)$ are positive and diagonal,
Cauchy--Schwarz gives
\begin{equation}
 \begin{aligned}
 \mathcal N(\tau X)^2
 &=\left(\sum_x X(x)\,\mathcal N(\tau\ket x\bra x)\right)^2\\
 &\preceq
 \left(\sum_x\mathcal N(\tau\ket x\bra x)\right)
 \left(\sum_xX(x)^2\,\mathcal N(\tau\ket x\bra x)\right)\\
 &=\mathcal N(\tau)\,\mathcal N(\tau X^2),
 \end{aligned}
 \label{eq:reference-cauchy-schwarz}
\end{equation}
where we have inserted $I=\sum_x\ket x\bra x$ in the first step.
Therefore, using cyclicity and trace preservation,
\begin{equation}
 \Tr\!\left[\mathcal N(\tau X)^2\mathcal N(\tau)^{-1}\right]
 \le \Tr\!\left[\mathcal N(\tau)\mathcal N(\tau X^2)\mathcal N(\tau)^{-1}\right] = \Tr\!\left[\mathcal N(\tau X^2)\right]
 =\Tr(\tau X^2).
 \label{eq:reference-contraction-proof}
\end{equation}
\end{proof}

\subsection{Assumptions for low-degree approximation}
\label{sec:approximation-ingredients}

Let $\mathcal N$ be a channel preserving occupation-diagonal operators,
let $\tau$ be a diagonal reference state, and let $\mu=\mathcal N(\tau)$.
These are fixed independently of the interferometer $U$.
Inverses of the reference states are taken on their supports.
We make the following four assumptions.

\begin{enumerate}
\item \textbf{Flatness.}
The ideal diagonal output $\rho_U$ is supported on $\tau$ and satisfies
\begin{equation}
 \E_U\chi^2(\rho_U\|\tau)\le C<\infty.
 \label{eq:ingredient-flatness}
\end{equation}

\item \textbf{A complete orthogonal input basis.}
The real occupation polynomials $X_0=I,X_1,\ldots$ form a complete
orthogonal basis of $L^2(\tau)$, with
\begin{equation}
 \Tr(\tau X_iX_j)=\delta_{ij}h_j,
 \qquad h_j>0.
 \label{eq:master-input-basis}
\end{equation}
Their degrees satisfy $\deg(X_0)=0$ and $\deg(X_j)\ge1$ for $j\ge1$.
There are finitely many basis elements through each degree.

\item \textbf{Orthogonality after noise.}
There are numbers $\Gamma_j\ge0$ such that, for $i,j\ge1$,
\begin{equation}
 \Tr\!\left[\mathcal N(\tau X_i)\mathcal N(\tau X_j)\mu^{-1}\right]
 =\delta_{ij}\Gamma_jh_j.
 \label{eq:ingredient-orthogonality}
\end{equation}

\item \textbf{Suppression under noise.}
For some $0<q<1$,
\begin{equation}
 \overline\Gamma_k:=\max_{j:\deg(X_j)=k}\Gamma_j\le q^k,\qquad k\ge1.
 \label{eq:ingredient-suppression}
\end{equation}
We set $\overline\Gamma_k=0$ if there are no basis elements of degree $k$.
\end{enumerate}

The degree-$k$ contribution to the ideal divergence is
\begin{equation}
 W_k(U):=\sum_{j:\deg(X_j)=k}\frac{|\Tr(\rho_UX_j)|^2}{h_j}.
 \label{eq:degree-energy}
\end{equation}

\subsection{The approximation theorem}
\label{sec:master-theorem}

\begin{theorem}[Low-degree approximation under noise]
\label{thm:master}
Under Assumptions 1--4, we define, for an integer $D\ge0$,
\begin{equation}
 \rho_{D,U}
 =\tau\sum_{j:\deg(X_j)\le D}\frac{\Tr(\rho_UX_j)}{h_j}X_j,
 \label{eq:master-input-approximation}
\end{equation}
and
\begin{equation}
 \widetilde\rho_{D,U}=\mathcal N(\rho_{D,U}).
 \label{eq:master-approximation}
\end{equation}
These are diagonal trace-one operators for almost every $U$, and
\begin{equation}
 \begin{aligned}
 \|\mathcal N(\rho_U)-\widetilde\rho_{D,U}\|_1^2
 &\le\sum_{k>D}\overline\Gamma_kW_k(U),\\
 \E_U\|\mathcal N(\rho_U)-\widetilde\rho_{D,U}\|_1^2
 &\le\sum_{k>D}\overline\Gamma_k\E_UW_k(U)\le Cq^{D+1}.
 \end{aligned}
 \label{eq:master-tail}
\end{equation}
If, in addition, some $A\ge0$ and $0<a<1$ satisfy
\begin{equation}
 \overline\Gamma_k\E_UW_k(U)\le Aa^k,\qquad k\ge1,
 \label{eq:ingredient-ensemble-degree-suppression}
\end{equation}
then
\begin{equation}
 \E_U\|\mathcal N(\rho_U)-\widetilde\rho_{D,U}\|_1^2
 \le\frac{Aa^{D+1}}{1-a}.
 \label{eq:master-degree-suppression}
\end{equation}
\end{theorem}

\begin{proof}
We fix $U$ with finite $\chi^2(\rho_U\|\tau)$. Orthogonality and
completeness give
\begin{equation}
 \rho_U=\tau\left[I+\sum_{j\ge1}
 \frac{\Tr(\rho_UX_j)}{h_j}X_j\right],
 \label{eq:master-density-expansion}
\end{equation}
where the density relative to $\tau$ converges in $L^2(\tau)$.
Parseval's identity gives
\begin{equation}
 \sum_{k\ge1}W_k(U)
 =\sum_{j\ge1}\frac{|\Tr(\rho_UX_j)|^2}{h_j}
 =\Tr\!\left[(\rho_U-\tau)^2\tau^{-1}\right]
 =\chi^2(\rho_U\|\tau).
 \label{eq:master-parseval}
\end{equation}
By Lemma~\ref{lem:reference-contraction}, applying $\mathcal N$ to
the partial sums in \eqref{eq:master-density-expansion} preserves
convergence in the output norm $B\mapsto\sqrt{\Tr(B^2\mu^{-1})}$.
Consequently, with $\Delta=\mathcal N(\rho_U)-\widetilde\rho_{D,U}$,
Assumption 3 and \eqref{eq:weighted-cs} yield
\begin{equation}
 \begin{aligned}
 \|\Delta\|_1^2
 &\le\Tr(\Delta^2\mu^{-1})\\
 &=\Tr\!\left[
 \left(\sum_{j:\deg(X_j)>D}\frac{\Tr(\rho_UX_j)}{h_j}
                  \mathcal N(\tau X_j)\right)^2\mu^{-1}\right]\\
 &=\sum_{j:\deg(X_j)>D}\Gamma_j
                   \frac{|\Tr(\rho_UX_j)|^2}{h_j}
 \le\sum_{k>D}\overline\Gamma_kW_k(U).
 \end{aligned}
 \label{eq:master-tail-proof}
\end{equation}
Tonelli's theorem, Assumptions 1 and 4, and
\eqref{eq:master-parseval} now give
\begin{equation}
 \begin{aligned}
 \E_U\|\Delta\|_1^2
 &\le\sum_{k>D}\overline\Gamma_k\E_UW_k(U)\\
 &\le q^{D+1}\E_U\sum_{k\ge1}W_k(U)
 \le Cq^{D+1}.
 \end{aligned}
 \label{eq:master-flatness-proof}
\end{equation}
Under \eqref{eq:ingredient-ensemble-degree-suppression}, the same
tail is at most $A\sum_{k>D}a^k=Aa^{D+1}/(1-a)$.
Finally, $\Tr(\tau X_j)=0$ for $j\ge1$, so
$\Tr(\rho_{D,U})=\Tr(\widetilde\rho_{D,U})=1$.
\end{proof}

In each of the three models below, $\Gamma_j$ is constant within each
degree, so $\Gamma_j=\overline\Gamma_k$ whenever $\deg(X_j)=k$.
Under this additional property, linearity extends Assumption 3 to
arbitrary positive-degree contributions: if $Z_k$ is a linear combination
of the $X_j$ with $\deg(X_j)=k$, then
\begin{equation}
 \Tr\!\left[\mathcal N(\tau Z_k)\mathcal N(\tau Z_\ell)\mu^{-1}\right]
 =\delta_{k\ell}\overline\Gamma_k\Tr(\tau Z_kZ_\ell).
 \label{eq:master-degree-contraction}
\end{equation}

\section{Rejection sampling from a low-degree approximation}
\label{sec:sampling}

The low-degree approximation from the preceding section may take
negative values. We use rejection sampling to obtain samples whose
distribution is close to the noisy output distribution.

Let $\mu$ be a reference probability distribution on a countable set
$\Omega$ of occupations, let $P$ be a probability distribution supported on $\mu$,
and let $q=\mu f$ be a signed approximation satisfying
\begin{equation}
 \E_\mu f=1,\qquad \E_\mu f^2<\infty,
 \qquad [t]_+=\max\{t,0\}.
 \label{eq:rejection-score-assumptions}
\end{equation}
Each draw from $\mu$ is a proposed occupation. Fix a truncation level
$M>0$ and a maximum number $A\ge1$ of proposals, with $A$ an integer.
If the algorithm stops without accepting a proposal, it returns a fixed
occupation $x_0\in\Omega$.

\begin{algorithm}[H]
\caption{Rejection sampling with exact evaluations}
\label{alg:rejection-sampling}
\begin{algorithmic}[1]
\Require A sampler for $\mu$ on $\Omega$, exact evaluations of $f$,
a truncation level $M>0$, a maximum number of proposals $A\ge1$,
and a fixed occupation $x_0\in\Omega$.
\Ensure An occupation in $\Omega$.
\For{$i=1,\ldots,A$}
    \State Draw $X\sim\mu$.
    \State With probability $\min\{1,\max\{f(X),0\}/M\}$, \Return $X$.
\EndFor
\State \Return $x_0$
\end{algorithmic}
\end{algorithm}

\begin{lemma}[Rejection sampling from a signed approximation]
\label{lem:rejection-sampling}
Suppose each independent proposal $X\sim\mu$ is accepted with probability $w(X)/M$,
where $0\le w\le M$. Let $w_0(x)=\min\{[f(x)]_+,M\}$ be the
positive part of $f$, truncated at $M$, and set
\begin{equation}
 \tau_M=\E_\mu(f-M)_+,\qquad
 \eta=\E_\mu|w-w_0|.
 \label{eq:rejection-errors}
\end{equation}
If $\tau_M+\eta<1$, the procedure that returns the first accepted
proposal, or $x_0$ after $A$ rejections, has output law $Q_A$ satisfying
\begin{equation}
 \begin{gathered}
 \TV(Q_A,P)\le\frac12\|q-P\|_1+\tau_M+\eta
       +\exp\!\left[-\frac{A(1-\tau_M-\eta)}{M}\right],
 \\
 \tau_M\le\frac{\E_\mu f^2}{4M}.
 \end{gathered}
 \label{eq:rejection-tv}
\end{equation}
Without the proposal limit, the expected number of proposals is at
most $M/(1-\tau_M-\eta)$.
\end{lemma}
\begin{proof}
Let $n_-=\E_\mu(-f)_+$ and $z_M=\E_\mu w$. Since $\E_\mu f=1$,
\begin{equation}
 \E_\mu w_0=1+n_--\tau_M,\qquad
 z_M\ge1+n_--\tau_M-\eta\ge1-\tau_M-\eta>0.
 \label{eq:rejection-normalization}
\end{equation}
The first accepted proposal has density $w/z_M$ relative to $\mu$,
and the number of proposals until acceptance is geometric with success
probability $z_M/M$ \cite[Sec.~1.7]{devroye2006nonuniform}. Let $p=P/\mu$.
On $\{f<0\}$, replacing $f$ by zero decreases
$|f-p|$ by $-f$. Truncating $[f]_+$ at $M$ removes total mass
$\tau_M$. Hence
\begin{equation}
 \begin{aligned}
 \frac12\E_\mu\left|\frac{w}{z_M}-p\right|
 &\le\frac12\left(\E_\mu|w-p|+|1-z_M|\right)\\
 &\le\frac12\left(\|q-P\|_1-n_-+\tau_M+\eta
                         +|n_--\tau_M|+\eta\right)\\
 &=\frac12\|q-P\|_1+(\tau_M-n_-)_++\eta
 \le\frac12\|q-P\|_1+\tau_M+\eta.
 \end{aligned}
 \label{eq:rejection-normalized-error}
\end{equation}
The first inequality uses $\E_\mu|w/z_M-w|=|1-z_M|$.
The bound on $\tau_M$ follows by averaging $(f-M)_+\le f^2/(4M)$.
The expected number of proposals is $M/z_M\le M/(1-\tau_M-\eta)$,
and the probability of $A$ consecutive rejections is
\begin{equation}
 (1-z_M/M)^A\le e^{-Az_M/M}
 \le\exp[-A(1-\tau_M-\eta)/M].
 \label{eq:rejection-attempt-tail}
\end{equation}
Returning $x_0$ after $A$ rejections changes the output law by at most
this probability. Together with \eqref{eq:rejection-normalized-error},
this proves the claim.
\end{proof}

For Algorithm~\ref{alg:rejection-sampling}, $w=w_0$ and $\eta=0$.
A uniform bound $f\le M$ also gives $\tau_M=0$.

\subsection{Rejection sampling with adaptive evaluations} \label{sec:adaptive-rejection}

To reduce the sampling cost, we give a second algorithm based on
randomized estimates, refined only while the acceptance decision
remains uncertain. Similar techniques are used in
\cite{bardenet2014towards,latuszynski2011simulating}.

Suppose $Z(x,\xi)$ is a real-valued randomized estimate of $f(x)$ satisfying
\begin{equation}
 \E_{\xi\mid x}Z(x,\xi)=f(x),\qquad
 \E_\mu f^2\le\overline C,\qquad
 \E_{X\sim\mu}\Var_{\xi\mid X}Z(X,\xi)\le\overline V.
 \label{eq:adaptive-point-assumptions}
\end{equation}
Repeated evaluations at the same occupation use independent
randomness $\xi$. For each proposed occupation $X$, draw a single
threshold $T$ uniformly from $[0,M]$, independently of these evaluations.
Compare $T$ with successive averages, reusing all values already computed.
The comparison using $n_j$ evaluations is called stage $j$, and
$Y_j$ denotes their mean. The tolerance $r_j$ determines when more
evaluations are needed: accept if $Y_j>T+r_j$, reject if $Y_j<T-r_j$,
and otherwise continue to the next comparison.

Choose $M\ge1$ and $0<\gamma<1$ so that
$z_*=1-\overline C/(4M)-\gamma>0$, and define
\begin{equation}
 \begin{gathered}
 J=\left\lceil\log_2\frac{4M}{\gamma}\right\rceil,\qquad
 a=\frac{\overline V(J+1)}{2\gamma},\\
 r_j=M2^{-j},\qquad
 n_j=\max\!\left\{1,\left\lceil\frac{a}{r_j}\right\rceil\right\},
 \qquad 0\le j\le J.
 \end{gathered}
 \label{eq:adaptive-parameters}
\end{equation}
We allow at most $A$ proposals and at most $E_{\max}$ evaluations beyond
the first in each proposal, counted across all proposals. Given
$0<\beta_{\rm p},\beta_{\rm e}<1$, set
\begin{equation}
 A=\left\lceil\frac{M}{z_*}
          \left\lceil\log_2\frac1{\beta_{\rm p}}\right\rceil\right\rceil,
 \qquad
 E_{\max}=\left\lceil
       \frac{a(1+4J)}{z_*\beta_{\rm e}}\right\rceil.
 \label{eq:adaptive-work-limits}
\end{equation}
The parameters $\beta_{\rm p}$ and $\beta_{\rm e}$ bound the probabilities
that these two limits stop the algorithm before acceptance.
The counter $e$ records these additional evaluations.

\begin{algorithm}[H]
\caption{Rejection sampling with adaptive evaluations}
\label{alg:adaptive-rejection}
\begin{algorithmic}[1]
\Require A sampler for $\mu$ and independent evaluations of $Z(X,\xi)$
\Require $M,r_j,n_j,A,E_{\max}$ from \eqref{eq:adaptive-parameters}--\eqref{eq:adaptive-work-limits}, and the fixed occupation $x_0$
\Ensure An occupation in $\Omega$
\State Set $e=0$.
\For{$i=1,\ldots,A$}
    \State Draw $X\sim\mu$ and, independently, $T$ uniformly from $[0,M]$.
    \State Set $n=0$ and $S=0$ (the number and sum of evaluations for $X$).
    \For{$j=0,\ldots,J$}
        \If{$e+n_j-\max\{n,1\}>E_{\max}$}
            \State \Return $x_0$
        \EndIf
        \State Add $n_j-\max\{n,1\}$ to $e$.
        \State Draw $n_j-n$ new evaluations of $Z(X,\xi)$ and add them to $S$.
        \State Set $n=n_j$ and $Y_j=S/n_j$.
        \If{$Y_j>T+r_j$}
            \State \Return $X$
        \ElsIf{$Y_j<T-r_j$}
            \State Reject $X$ and leave the inner loop.
        \EndIf
    \EndFor
\EndFor
\State \Return $x_0$
\end{algorithmic}
\end{algorithm}

If the comparison is still inconclusive after $n_J$ evaluations, reject $X$.

\begin{theorem}[Adaptive rejection sampling]
\label{thm:adaptive-rejection}
Under \eqref{eq:rejection-score-assumptions} and
\eqref{eq:adaptive-point-assumptions}, Algorithm~\ref{alg:adaptive-rejection}
produces an output law $Q$ satisfying
\begin{equation}
 \TV(Q,P)\le\frac12\|q-P\|_1+
       \frac{\overline C}{4M}+\gamma+\beta_{\rm p}+\beta_{\rm e}.
 \label{eq:adaptive-error}
\end{equation}
It uses at most $A$ reference proposals and $A+E_{\max}$ evaluations,
with the limits defined in \eqref{eq:adaptive-work-limits}.
\end{theorem}
\begin{proof}
Consider first the algorithm without limits on the number of proposals
or evaluations. For the analysis, draw all $n_J$ evaluations for each
proposal before drawing $T$, including those the algorithm may not use.
Each $Y_j$ is the mean of the first $n_j$ values, so $Y_j$ and $T$ are
independent conditional on $X$, and
$\E[(Y_j-f(X))^2]\le\overline V/n_j$.

The exact acceptance test is $T<f(X)$. For fixed $X$ and $Y_j$,
an incorrect decision at stage $j$ can occur only for thresholds in an
interval of length at most $(|Y_j-f(X)|-r_j)_+$.
An inconclusive final comparison requires $|Y_J-T|\le r_J$,
an interval of length at most $2r_J$.
Let $w(x)/M$ be the adaptive acceptance probability and
$w_0(x)=\min\{[f(x)]_+,M\}$. Using the same $X,T$ for the two tests gives
\begin{equation}
 \begin{aligned}
 \E_\mu|w-w_0|
 &\le\sum_{j=0}^J\E(|Y_j-f(X)|-r_j)_+ +2r_J\\
 &\le\sum_{j=0}^J\frac{\overline V}{4n_jr_j}+2r_J
 \le\gamma.
 \end{aligned}
 \label{eq:adaptive-weight-error}
\end{equation}
Here we used $(|t|-r)_+\le t^2/(4r)$ and
\eqref{eq:adaptive-parameters}. Lemma~\ref{lem:rejection-sampling}
therefore bounds the sampling error without work limits by
$\frac12\|q-P\|_1+\overline C/(4M)+\gamma$, and gives an acceptance
probability $p_{\rm acc}\ge z_*/M$ per proposal.

Reaching stage $j\ge1$ requires
$|Y_{j-1}-T|\le r_{j-1}$, which has probability at most $2r_{j-1}/M$.
Since $n_0-1\le a/M$ and $n_j-n_{j-1}\le a/r_j$, the number
$N_{\rm extra}^{(1)}$ of evaluations beyond the first in one proposal obeys
\begin{equation}
 \E N_{\rm extra}^{(1)}
 \le\frac aM+\sum_{j=1}^J\frac a{r_j}\frac{2r_{j-1}}M
 =\frac{a(1+4J)}M.
 \label{eq:adaptive-refinement-work}
\end{equation}
Each new proposal is independent of the preceding rejections.
Thus the total number $N_{\rm extra}$ of extra evaluations until acceptance
satisfies
\begin{equation}
 \E N_{\rm extra}
 =\sum_{i\ge1}(1-p_{\rm acc})^{i-1}\E N_{\rm extra}^{(1)}
 \le\frac{a(1+4J)}{z_*}.
 \label{eq:adaptive-expected-refinements}
\end{equation}
The probability of requiring more than $A$ proposals is at most
$e^{-Az_*/M}\le\beta_{\rm p}$. By Markov's inequality, the probability
of requiring more than $E_{\max}$ extra evaluations is at most
$\beta_{\rm e}$; when $a=0$, no extra evaluations are needed.
With the same random draws, the algorithm with limits returns the same
occupation as the one without limits unless the latter exceeds a limit.
Their output laws therefore differ in total variation by at most
$\beta_{\rm p}+\beta_{\rm e}$, proving \eqref{eq:adaptive-error}.
There is at most one first evaluation per proposal and at most
$E_{\max}$ additional evaluations, giving a total of at most $A+E_{\max}$.
\end{proof}

\begin{lemma}[Stability under numerical errors]
\label{lem:rejection-precision}
For either rejection-sampling algorithm, keep the limits on proposals
and evaluations and the comparison tolerances $r_j$ unchanged.
Absolute errors at most $\zeta$ in each evaluation of $f$ or sample mean
$Y_j$, and in each random threshold $T$, change the output law by total
variation at most
\begin{equation}
 \frac{8A(J+1)\zeta}{M},
 \label{eq:rejection-precision}
\end{equation}
where $J=0$ for Algorithm~\ref{alg:rejection-sampling}.
No moment assumption is needed for this bound.
\end{lemma}
\begin{proof}
Use the same random draws for the exact and approximate computations.
For Algorithm~\ref{alg:adaptive-rejection}, generate all $n_J$ evaluations
for each of the $A$ possible proposals before drawing its threshold $T$.
The computations follow the same steps until a comparison has different
outcomes. At stage $j$, this requires $T$ to lie within $2\zeta$ of either
$Y_j-r_j$ or $Y_j+r_j$, in intervals of total length at most $8\zeta$.
Each $Y_j$ is independent of $T$ conditional on $X$.
A union bound over all $A(J+1)$ possible comparisons therefore
proves \eqref{eq:rejection-precision}. For
Algorithm~\ref{alg:rejection-sampling}, the only boundary is $f(X)$,
and the same bound applies.
\end{proof}

\section{Fermion sampling}
\label{sec:young-model}
We specialize the framework of Appendix~\ref{sec:particle-representation}
to fermions and verify the assumptions in
Subsection~\ref{sec:approximation-ingredients} required for the
low-degree approximation theorem.
Young polynomials provide coordinates for the diagonal adjoint degrees
identified in Eq.~\eqref{eq:common-diagonal-degree-dimension},
while the reduced overlaps of Eq.~\eqref{eq:common-output-degree-overlap}
supply their evaluation formulas. The additional ingredients here are
the action of thermal loss, the required flatness bounds, the
magic-state reductions, and the algorithms for evaluating these overlaps.

\subsection{Young approximation}
\label{sec:young-approx}

The fermionic experiment has a fixed number of input particles. Its
input is a product of \(B\) four-mode states,
\begin{equation}
  \rho_{\rm in}=\ket{\Psi_B}\bra{\Psi_B},\qquad
 \ket{\Psi_B}=
 \left(\frac{\ket{1100}+\ket{0011}}{\sqrt2}\right)^{\otimes B},\qquad
 m=4B,\qquad n=2B.
  \label{eq:young-magic-family}
\end{equation}
The modes are ordered block by block, which fixes the signs in this
fermionic tensor product. We first prove the approximation statements
for an arbitrary state with exactly \(n\) particles in \(m\) modes,
and then evaluate their Haar averages for \eqref{eq:young-magic-family}.
We assume \(n\le m/2\) without loss of generality, since the case \(n>m/2\) follows by exchanging occupied and empty modes.
We write \(\Omega_{m,n}=\{x\in\{0,1\}^m:\sum_i x_i=n\}\) for the occupation
strings in this sector, \(u_n\) for its uniform diagonal state, and
\(\rho_{\rm ideal}\) for the dephased state after the interferometer
and before loss.

\subsubsection{Orthogonal polynomial degrees}\label{subsec:Young}

We choose Filmus's top-set basis~\cite{filmus2014orthogonal}
for the diagonal degree spaces of Subsection~\ref{sec:zero-weight-bases}. For each degree
$0\le k\le m/2$, let $\alpha$ index the sequences
$b=(b_1,\ldots,b_k)$ with $1\le b_1<\cdots<b_k\le m$ and $b_i\ge2i$.
Let $\mathcal A(b)$ consist of ordered sequences
$a=(a_1,\ldots,a_k)$ of distinct indices, disjoint from $b$, with
$a_i<b_i$ for every $i$. For $a\in\mathcal A(b)$, we define the paired polynomial
\begin{equation}
 H_{a,b}=\prod_{i=1}^k(\hat n_{a_i}-\hat n_{b_i}).
 \label{eq:young-paired-polynomial-definition}
\end{equation}
The basis polynomials are
\begin{equation}
 F_{k\alpha}(\hat n)
 =\sum_{a\in\mathcal A(b)}H_{a,b},\qquad F_{0}=I.
 \label{eq:young-basis-definition}
\end{equation}
The polynomials $F_{k\alpha}$ are multilinear, homogeneous of degree $k$,
and invariant under a common translation:
\begin{equation}
 \begin{aligned}
 F_{k\alpha}(t\hat n)&=t^kF_{k\alpha}(\hat n),\\
 F_{k\alpha}(\hat n_1+tI,\ldots,\hat n_m+tI)
 &=F_{k\alpha}(\hat n).
 \end{aligned}
 \label{eq:young-harmonic-identities}
\end{equation}
Filmus's construction directly gives one polynomial $F_{k\alpha}$
across all particle sectors. Its coefficient operator
$B_{k\alpha}:=F_{k\alpha}(\hat n)|_{\mathcal H_k}$ is harmonic:
translation invariance is the fermionic kernel condition
\eqref{eq:common-diagonal-kernel}. Hence
Eq.~\eqref{eq:common-polynomial-lift} gives
\begin{equation}
    F_{k\alpha}(\hat n)|_{\mathcal H_L}
 =\frac{\mathcal R^{L-k}(B_{k\alpha})}{(L-k)!},
 \qquad k\le L\le m-k.
\end{equation}
Thus the construction already realizes raising between equivalent
irreducible components. Raising changes the particle count $L$,
while the polynomial degree remains $k$; the sector-dependent
orthonormalization is given below.

\begin{lemma}[Young orthogonality and completeness, from Theorems~3.1, 3.2, and~4.1 of Ref.~\cite{filmus2014orthogonal}]
\label{lem:young-basis}
On the $L$-particle sector, the polynomials in
\eqref{eq:young-basis-definition} with
$0\le k\le\min(L,m-L)$ form a complete orthogonal basis, with
\begin{equation}
 \begin{gathered}
 \Tr(u_LF_{k\alpha}F_{j\gamma})
 =\delta_{kj}\delta_{\alpha\gamma}c_\alpha h_{L,k},\qquad
 h_{L,k}=\frac{2^k\fall Lk\fall{m-L}k}{\fall m{2k}},\\
 c_\alpha=\prod_{i=1}^k
 \frac{(b_i-2i+2)(b_i-2i+1)}2.
 \end{gathered}
 \label{eq:young-basis-norm}
\end{equation}
For $k>\min(L,m-L)$, $F_{k\alpha}$ vanishes on that sector.
\end{lemma}
The number of degree-$k$ polynomials is
$z_k=\binom mk-\binom m{k-1}$, as in
Eq.~\eqref{eq:common-diagonal-degree-dimension} and
Ref.~\cite[Lemma~2.2]{filmus2014orthogonal}. Thus the basis verifies
Assumption~2 of Theorem~\ref{thm:master}. The common norm transfer
\eqref{eq:common-lift-inner-product}, or the explicit Young norms
above, gives the coordinate-independent ratio
\begin{equation}
 \frac{\Tr(u_LF_{k\alpha}^2)}{\Tr(u_nF_{k\alpha}^2)}
 =\frac{h_{L,k}}{h_{n,k}}
 =\frac{\fall Lk\fall{m-L}k}{\fall nk\fall{m-n}k}.
 \label{eq:young-norm-ratio}
\end{equation}

\begin{lemma}[Paired polynomials and total occupation]
\label{lem:young-paired-reference}
For $a\in\mathcal A(b)$, we set
\begin{equation}
 R_{a,b}=[m]\setminus\{a_1,b_1,\ldots,a_k,b_k\},\qquad
 \hat N=\sum_{i=1}^m\hat n_i.
 \label{eq:young-paired-summands}
\end{equation}
For every function $g:\mathbb Z\to\mathbb C$,
\begin{equation}
 H_{a,b}\,g\!\left(\sum_{i\in R_{a,b}}\hat n_i\right)
 =H_{a,b}\,g(\hat N-k).
 \label{eq:young-paired-reference}
\end{equation}
\end{lemma}
\begin{proof}
On an occupation $x$,
\begin{equation}
 H_{a,b}(x)\ne0
 \ \Longrightarrow\ x_{a_i}+x_{b_i}=1\ (1\le i\le k)
 \ \Longrightarrow\ \sum_{i\in R_{a,b}}x_i=|x|-k.
 \label{eq:young-paired-remaining-count}
\end{equation}
Both sides of \eqref{eq:young-paired-reference} vanish otherwise.
\end{proof}

The degree energies and the approximate noisy output are
\begin{equation}
 W_k(U)=\sum_\alpha
 \frac{|\Tr(\rho_{\rm ideal}F_{k\alpha})|^2}{c_\alpha h_{n,k}},
 \qquad
 \sum_{k=1}^nW_k(U)=\chi^2(\rho_{\rm ideal}\|u_n),
 \label{eq:young-degree-energy}
\end{equation}
and
\begin{equation}
 \widetilde\rho_D
 :=\mathcal N_{\eta,\beta}^{\otimes m}\!\left(
 u_n\left[I+\sum_{k=1}^{\min(D,n)}\sum_\alpha
 \frac{\Tr(\rho_{\rm ideal}F_{k\alpha})}{c_\alpha h_{n,k}}
 F_{k\alpha}\right]\right).
 \label{eq:young-noisy-approximation}
\end{equation}

\subsubsection{Orthogonality after thermal loss}\label{sec:young-noisy-orthogonality}

We set
\begin{equation}
 a_+=(1-\eta)\beta,\qquad a_-=(1-\eta)(1-\beta).
 \label{eq:young-transition-parameters}
\end{equation}
The channel commutes with mode permutations, so its uniform reference
and total-count law are
\begin{equation}
 \begin{gathered}
 \mu_{\rm F}=\mathcal N_{\eta,\beta}^{\otimes m}(u_n)
 =\sum_{L=0}^m p_Lu_L,\\
 p_L=\Pr\!\left[\Bin(n,1-a_-)+\Bin(m-n,a_+)=L\right].
 \end{gathered}
 \label{eq:young-reference}
\end{equation}
The same count law holds for every input on the $n$-particle sector.
The two binomial variables are independent.

\begin{lemma}[Orthogonality after fermionic thermal loss]
\label{lem:young-noisy-orthogonality}
For $0<\eta<1$ and $1\le k,j\le n$, there are nonnegative numbers
$\Gamma_k^{\rm F}$, independent of $\alpha$, such that
\begin{equation}
 \begin{aligned}
 &\Tr\!\left[
 \mathcal N_{\eta,\beta}^{\otimes m}(u_nF_{k\alpha})
 \mathcal N_{\eta,\beta}^{\otimes m}(u_nF_{j\gamma})
 \mu_{\rm F}^{-1}\right]\\
 &\qquad=\delta_{kj}\delta_{\alpha\gamma}
 \Gamma_k^{\rm F}c_\alpha h_{n,k}.
 \end{aligned}
 \label{eq:young-output-orthogonality}
\end{equation}
In particular, Assumption~3 of Theorem~\ref{thm:master} holds.
\end{lemma}
\begin{proof}
Let $z=e^{-\theta}$, with $\theta>0$, so that $z^{\hat N}$ is the
unnormalized thermal operator. For a diagonal occupation polynomial
$F$, the uniform states $u_r$ on the particle-number sectors give
\begin{equation}
 \begin{aligned}
 z^{\hat N}F
 &=\sum_x z^{|x|}F(x)\ket x\bra x
 =\sum_r \binom{m}{r}z^r u_rF,\\
 \mathcal N_{\eta,\beta}^{\otimes m}(z^{\hat N}F)
 &=\sum_r \binom{m}{r}z^r\mathcal N_{\eta,\beta}^{\otimes m}(u_rF).
 \end{aligned}
 \label{eq:thermal-reference-generating}
\end{equation}
Consequently,
\begin{equation}
 \mathcal N_{\eta,\beta}^{\otimes m}(u_nF)
 =\frac1{\binom{m}{n}}[z^n]\,
 \mathcal N_{\eta,\beta}^{\otimes m}(z^{\hat N}F).
 \label{eq:thermal-coefficient-extraction}
\end{equation}

The one-mode transition probabilities in
\eqref{eq:thermal-fermion-kernel} give
\begin{equation}
 \begin{gathered}
 B_0(z)=1-a_++a_-z,\qquad B_1(z)=a_++(1-a_-)z,\\
 \mathcal N_{\eta,\beta}(z^{\hat n})
 =B_0(z)\ket0\bra0+B_1(z)\ket1\bra1.
 \end{gathered}
 \label{eq:young-local-channel}
\end{equation}
For $H_1=\ket{10}\bra{10}-\ket{01}\bra{01}$,
\begin{equation}
 \begin{aligned}
 z^{\hat n_1+\hat n_2}H_1&=zH_1,\\
 \mathcal N_{\eta,\beta}^{\otimes2}(H_1)
 &=\bigl[(1-a_-)(1-a_+)-a_-a_+\bigr]H_1
 =\eta H_1.
 \end{aligned}
 \label{eq:young-paired-channel}
\end{equation}
Tensoring over $k$ disjoint pairs gives
\begin{equation}
 \mathcal N_{\eta,\beta}^{\otimes 2k}
 \!\left(z^{\sum_{i=1}^{2k}\hat n_i}H_1^{\otimes k}\right)
 =(\eta z)^kH_1^{\otimes k}.
 \label{eq:young-paired-tensor-channel}
\end{equation}
Here $H_1^{\otimes k}$ acts on the pairs $(1,2),\ldots,(2k-1,2k)$.
Relabelling these pairs as $(a_i,b_i)$ and extending by the identity on
the remaining modes gives $H_{a,b}$. Tensoring with $z^{\hat n}$ on
those remaining modes and using \eqref{eq:young-local-channel} yields
\begin{equation}
 \begin{aligned}
 &\bra y\mathcal N_{\eta,\beta}^{\otimes m}
             (z^{\hat N}H_{a,b})\ket y\\
 &\quad=(\eta z)^kH_{a,b}(y)
      \prod_{i\in R_{a,b}} B_{y_i}(z)\\
 &\quad=(\eta z)^kB_1(z)^{L-k}B_0(z)^{m-L-k}H_{a,b}(y),
 \qquad |y|=L.
 \end{aligned}
 \label{eq:young-paired-forward}
\end{equation}
The last equality is Lemma~\ref{lem:young-paired-reference}.
It applies for $k\le L\le m-k$. Otherwise, the image vanishes.
Recalling $F_{k\alpha}=\sum_{a\in\mathcal A(b)}H_{a,b}$ from
\eqref{eq:young-basis-definition}, summing over $a\in\mathcal A(b)$ gives
\begin{equation}
 \bra y\mathcal N_{\eta,\beta}^{\otimes m}
             (z^{\hat N}F_{k\alpha})\ket y
 =\eta^kz^kB_1(z)^{L-k}B_0(z)^{m-L-k}F_{k\alpha}(y).
 \label{eq:young-forward-generating}
\end{equation}
We define
\begin{equation}
 r_k(L)=\frac1{\binom{m}{n}}[z^{n-k}]
           B_1(z)^{L-k}B_0(z)^{m-L-k},\qquad k\le L\le m-k,
 \label{eq:young-forward-coefficient}
\end{equation}
and we set $r_k(L)=0$ outside this range. Applying
\eqref{eq:thermal-coefficient-extraction} now yields
\begin{equation}
 \bra y\mathcal N_{\eta,\beta}^{\otimes m}(u_nF_{k\alpha})\ket y
 =\eta^kr_k(L)F_{k\alpha}(y),\qquad
 r_0(L)=\frac{p_L}{\binom{m}{L}},\qquad |y|=L.
 \label{eq:young-output-amplitude}
\end{equation}
These coefficients are nonnegative, and positivity implies that the
images vanish off the support of $\mu_{\rm F}$.
Using Lemma~\ref{lem:young-basis},
\begin{equation}
 \begin{aligned}
 &\Tr\!\left[
 \mathcal N_{\eta,\beta}^{\otimes m}(u_nF_{k\alpha})
 \mathcal N_{\eta,\beta}^{\otimes m}(u_nF_{j\gamma})
 \mu_{\rm F}^{-1}\right]\\
 &\quad=\eta^{k+j}\sum_{L:r_0(L)>0}
 \frac{\binom{m}{L}r_k(L)r_j(L)}{r_0(L)}
 \Tr(u_LF_{k\alpha}F_{j\gamma})\\
 &\quad=\delta_{kj}\delta_{\alpha\gamma}c_\alpha\eta^{2k}
 \sum_{L:r_0(L)>0}\frac{\binom{m}{L}h_{L,k}r_k(L)^2}{r_0(L)}.
 \end{aligned}
 \label{eq:young-forward-gram}
\end{equation}
Thus
\begin{equation}
 \Gamma_k^{\rm F}
 =\frac{\eta^{2k}}{h_{n,k}}
 \sum_{\substack{k\le L\le m-k\\r_0(L)>0}}
 \frac{\binom{m}{L}h_{L,k}r_k(L)^2}{r_0(L)},
 \label{eq:young-gamma}
\end{equation}
which proves the lemma.
\end{proof}

Equation~\eqref{eq:master-degree-contraction} extends this identity to
arbitrary exact-degree Young polynomials, with orthogonal images for
distinct degrees. At degree zero we set
$\Gamma_0^{\rm F}=1$, since
$\mathcal N_{\eta,\beta}^{\otimes m}(u_n)=\mu_{\rm F}$.

\subsubsection{Suppression under noise}

\begin{proposition}[Thermal Young suppression]
\label{prop:young-suppression}
For $0<\eta<1$,
\begin{equation}
 \Gamma_k^{\rm F}\le\kappa_{\rm F}^k,\qquad
 \kappa_{\rm F}
 =\frac{\eta^2}{\eta+2(1-\eta)^2\beta(1-\beta)}\le\eta.
 \label{eq:young-kappa}
\end{equation}
\end{proposition}
\begin{proof}
We use the paired polynomial
\begin{equation}
 H_k=\prod_{i=1}^k(\hat n_{2i-1}-\hat n_{2i}),\qquad
 \Tr(u_LH_k^2)=h_{L,k}.
 \label{eq:young-paired-polynomial}
\end{equation}
For $S=H_1^2$ and $b=\eta+2a_+a_-$, the two-mode transition
probabilities give
\begin{equation}
 \begin{aligned}
 \mathcal N_{\eta,\beta}^{\otimes2}(S)
 &=bS+2a_-(1-a_+)\ket{00}\bra{00}
       +2a_+(1-a_-)\ket{11}\bra{11}\succeq bS.
 \end{aligned}
 \label{eq:young-pair-positive-image}
\end{equation}
Since $z^{\hat n_1+\hat n_2}S=zS$, tensoring this inequality and
extracting $[z^n]/\binom{m}{n}$ as in
\eqref{eq:young-paired-forward} gives
\begin{equation}
 \mathcal N_{\eta,\beta}^{\otimes m}(u_nH_k^2)
 \succeq b^kr_k(\hat N)H_k^2.
 \label{eq:young-pair-square-forward}
\end{equation}
Here coefficient extraction preserves the inequality because each
remainder is a polynomial with positive operator coefficients.
Also $0\preceq H_k^2\preceq I$, so
\begin{equation}
 \begin{aligned}
 \mu_{\rm F}-\mathcal N_{\eta,\beta}^{\otimes m}(u_nH_k^2)
 &=\mathcal N_{\eta,\beta}^{\otimes m}(u_n(I-H_k^2))\succeq0,\\
 r_0(L)&\ge b^kr_k(L)\qquad(k\le L\le m-k).
 \end{aligned}
 \label{eq:young-reference-lower-bound}
\end{equation}
The last line evaluates the operator inequality on an occupation
with $H_k(y)^2=1$ and $|y|=L$.
Moreover, $B_0(z)+B_1(z)=1+z$ and
$\binom{m}{L}h_{L,k}=2^k\binom{m-2k}{L-k}$ imply
\begin{equation}
 \begin{aligned}
 \sum_{L=k}^{m-k}\binom{m}{L}h_{L,k}r_k(L)
 &=\frac{2^k}{\binom{m}{n}}[z^{n-k}]
       (B_0(z)+B_1(z))^{m-2k}\\
 &=\frac{2^k}{\binom{m}{n}}\binom{m-2k}{n-k}=h_{n,k}.
 \end{aligned}
 \label{eq:young-forward-coefficient-sum}
\end{equation}
Substituting into \eqref{eq:young-gamma},
\begin{equation}
 \begin{aligned}
 \Gamma_k^{\rm F}
 &\le\frac{\eta^{2k}}{b^kh_{n,k}}
       \sum_{L=k}^{m-k}\binom{m}{L}h_{L,k}r_k(L)\\
 &=\left(\frac{\eta^2}{\eta+2a_+a_-}\right)^k
 =\kappa_{\rm F}^k.
 \end{aligned}
 \label{eq:young-paired-contraction}
\end{equation}
Finally, $\eta+2a_+a_-\ge\eta$.
\end{proof}

This verifies Assumption~4 of Theorem~\ref{thm:master} with
$q=\kappa_{\rm F}$. Its deterministic bound gives
\begin{equation}
 \begin{aligned}
 &\|\mathcal N_{\eta,\beta}^{\otimes m}(\rho_{\rm ideal})
       -\widetilde\rho_D\|_1^2\\
 &\quad\le\sum_{k>D}\Gamma_k^{\rm F}W_k(U)
 \le\kappa_{\rm F}^{D+1}\chi^2(\rho_{\rm ideal}\|u_n).
 \end{aligned}
 \label{eq:young-tail}
\end{equation}
The error is zero when $D\ge n$. At $\eta=0$ the reference is the
exact output and every nonconstant degree vanishes, while at $\eta=1$
there is no noise suppression.

\subsubsection{Flatness}
\label{sec:young-flatness}

For Haar interferometers and the magic state in
\eqref{eq:young-magic-family}, the flatness assumption
 \(\E_U\chi^2(\rho_{\rm ideal}\|u_n)\le C\) follows from the collision-moment
bounds used to prove Haar anticoncentration
\cite[Theorem~1 and Appendix~E.1]{oszmaniec2022fermion}.
We use the finite-size bound of
\cite[Theorem~3 and Proposition~4]{mathur2026logarithmic}, expressed below
in our normalization.

\begin{proposition}[Haar flatness for the magic input, adapted from Theorem~3 and Proposition~4 in Ref.~\cite{mathur2026logarithmic}]
\label{prop:young-flatness}
Let \(U\) be Haar-distributed on \(U(4B)\), with \(B\ge1\), and let the
input be as in \eqref{eq:young-magic-family}. Then
\begin{equation}
 \E_U\chi^2(\rho_{\rm ideal}\|u_n)<3.
 \label{eq:young-flatness}
\end{equation}
\end{proposition}
\begin{proof}
Let $p_U(x)=\bra x\rho_{\rm ideal}\ket x$. Since
$u_n(x)=\binom mn^{-1}$ and $\sum_xp_U(x)=1$, expanding the
square in the divergence gives
\begin{equation}
 1+\chi^2(\rho_{\rm ideal}\|u_n)
 =\sum_{x\in\Omega_{m,n}}\frac{p_U(x)^2}{u_n(x)}
 =\binom mn\sum_{x\in\Omega_{m,n}}p_U(x)^2.
 \label{eq:young-collision-normalization}
\end{equation}
The right-hand side is the normalized collision probability.
For the magic input with $m=4B$ and $n=2B$,
\cite[Theorem~3 and Proposition~4]{mathur2026logarithmic} bounds its Haar
average strictly below $4$ for every $B\ge1$. Taking expectations and subtracting
one therefore yields
\begin{equation}
 \E_U\chi^2(\rho_{\rm ideal}\|u_n)
 =\binom mn\E_U\sum_{x\in\Omega_{m,n}}p_U(x)^2-1<3.
\end{equation}
\end{proof}

\begin{remark}[Flatness beyond the Haar ensemble]
The flatness assumption can also be verified for ensembles of shallower
interferometers. Mathur and Kerenidis consider independent layers of
uniform random perfect matchings, with independent Haar \(U(2)\) gates
on the matched mode pairs. For the same magic-state input, Corollary~7 of Ref.~\cite{mathur2026logarithmic}
shows that the mean normalized collision is at most twice its Haar value
once the circuit depth exceeds a suitable $O(\log m)$ threshold.
Equations~\eqref{eq:young-collision-normalization}
and~\eqref{eq:young-flatness} then give
\begin{equation}
 \begin{gathered}
 \E_U\chi^2(\rho_{\rm ideal}\|u_n)<7,\\
 \E_U\|\mathcal N_{\eta,\beta}^{\otimes m}(\rho_{\rm ideal})
              -\widetilde\rho_D\|_1^2
 \le7\kappa_{\rm F}^{D+1}.
 \end{gathered}
\end{equation}
\end{remark}

\subsubsection{Low-degree approximation}\label{sec:young-truncation}

\begin{corollary}[Low-degree approximation for fermion sampling]
\label{cor:young-approximation}
We consider any ensemble of interferometers whose ideal outputs on the
\(n\)-particle sector satisfy
\(\E_U\chi^2(\rho_{\rm ideal}\|u_n)\le C\).
For \(0<\eta<1\), \(0\le\beta\le1\), and \(0\le D<n\),
\begin{equation}
\E_U\|\mathcal N_{\eta,\beta}^{\otimes m}(\rho_{\rm ideal})
              -\widetilde\rho_D\|_1^2
 \le \sum_{k=D+1}^{n}\Gamma_k^{\rm F}\E_UW_k(U)
 \le C\kappa_{\rm F}^{D+1}.
 \label{eq:young-flatness-tail}
\end{equation}
For the magic input and a Haar interferometer, \(C=3\) suffices.
\end{corollary}
\begin{proof}
Taking expectations in \eqref{eq:young-tail} and using the assumed
flatness gives the stated bound. For the magic input under Haar
randomness, Proposition~\ref{prop:young-flatness} gives $C=3$.
\end{proof}

\subsection{Operator overlap formulas}
\label{sec:young-evaluation}

We combine the reduced-overlap identity of
Corollary~\ref{cor:common-output-degrees} with the thermal action
\eqref{eq:young-output-amplitude}. The Young basis establishes the
degree decomposition and its behavior under noise; evaluating a whole
degree uses $\Pi_k$ directly, without computing individual Young
coefficients. The fermionic particle action is
\begin{equation}
 U_k=\bigwedge\nolimits^k U,\qquad
 (U_k)_{S,T}=\det U_{S,T},\qquad |S|=|T|=k.
 \label{eq:young-compound-matrix}
\end{equation}
These overlaps determine the retained density for the magic input
\eqref{eq:young-magic-family} and will be used by both evaluation methods.

Throughout, $\mathcal L$, $\mathcal R$, and $\Pi_k$ are the
fermionic maps of Eqs.~\eqref{eq:common-lowering-raising}
and~\eqref{eq:common-projector}, with
$D_{k,q}=\rise{m-2k+2}{q}$ from
Eq.~\eqref{eq:common-projector-coefficients}.
We use the normalized reductions
\eqref{eq:common-normalized-reduction}, with
$\rho_{\rm in}^{(0)}=1$, and take $0<\eta<1$ and $0\le\beta\le1$.

\subsubsection{Reduced overlaps and normalization}
\label{sec:young-operator-ingredients}
\label{sec:young-degree-contribution}

For a binary occupation $x$ of weight $L$ and $0\le k\le L$, write
$x_S=\prod_{i\in S}x_i$ and define
\begin{equation}
 \begin{gathered}
 P_x=\sum_{i:x_i=1}\ket i\bra i,\qquad
 A_{x,k}=\bigwedge\nolimits^k P_x
        =\sum_{|S|=k}x_S\ket S\bra S
        =\binom Lk\rho_x^{(k)},\\
 \ket S=e_{s_1}\wedge\cdots\wedge e_{s_k},
 \qquad s_1<\cdots<s_k.
 \end{gathered}
 \label{eq:young-test-operator}
\end{equation}
For $k>L$, the polynomial definition gives $A_{x,k}=0$.
The fermionic norm factor in Eq.~\eqref{eq:common-diagonal-beta}
can be written as $\beta_{n,k}=\binom nk/c_{n,k}$, where
\begin{equation}
 c_{n,k}=\binom mk\frac{\binom{m-k}k}{\binom{m-n}k},
 \qquad 0\le k\le n\le m/2.
 \label{eq:young-overlap-normalization}
\end{equation}
We reserve $p_{k,U}$ for the ideal probability contribution defined
in Corollary~\ref{cor:common-output-degrees};  $f_{k,U} $ denotes
a density relative to the specified reference distribution.

\subsubsection{Magic-state reductions and parity}
\label{sec:young-magic-reductions}

\begin{lemma}[Magic-state reductions and vanishing odd degrees]
\label{lem:young-magic-reductions}
\label{lem:young-odd-degrees}
For the input \eqref{eq:young-magic-family}, let
$\ket\phi=(\ket{12}+\ket{34})/\sqrt2$ be one four-mode factor.
Its normalized reductions satisfy
\begin{equation}
 \binom20\rho_b^{(0)}=1,\qquad
 \binom21\rho_b^{(1)}=\frac{I_4}{2},\qquad
 \binom22\rho_b^{(2)}=\ket\phi\bra\phi.
 \label{eq:young-block-reductions}
\end{equation}
The global reductions are
\begin{equation}
 \binom{2B}r\rho_{\rm in}^{(r)}
 =\bigoplus_{\substack{j_1+\cdots+j_B=r\\0\le j_b\le2}}
   \bigotimes_{b=1}^B
   \left(\binom2{j_b}\rho_b^{(j_b)}\right).
 \label{eq:young-global-reduction}
\end{equation}
Moreover, for every interferometer $U$,
\begin{equation}
 W_k(U)=0\qquad\text{for odd }k.
 \label{eq:young-odd-degrees}
\end{equation}
\end{lemma}
\begin{proof}
The one-particle matrix elements in a block are
\begin{equation}
 2\bra i\rho_b^{(1)}\ket j
 =\bra\phi \hat a_j^\dagger \hat a_i\ket\phi
 =\frac{\delta_{ij}}2,
 \label{eq:young-local-reduction-entries}
\end{equation}
which proves \eqref{eq:young-block-reductions}. A uniform choice of
$r$ particles from the $2B$ input particles has block counts
$(j_1,\ldots,j_B)$ with probability
\begin{equation}
 \Pr(j_1,\ldots,j_B)
 =\frac{\prod_{b=1}^B\binom2{j_b}}{\binom{2B}r}.
 \label{eq:young-retained-block-counts}
\end{equation}
Conditioned on these counts, the reduction is
$\bigotimes_b\rho_b^{(j_b)}$, and different count vectors have orthogonal
supports. This proves \eqref{eq:young-global-reduction}.

For the parity statement, let $\mathcal M$ be the family of occupied
sets obtained by choosing either $\{4b-3,4b-2\}$ or
$\{4b-1,4b\}$ in each block. Then
\begin{equation}
 \ket{\Psi_B}=2^{-B/2}\sum_{T\in\mathcal M}\ket T,
 \qquad
 \psi_U(S):=\bra S U_n\ket{\Psi_B}
 =2^{-B/2}\sum_{T\in\mathcal M}\det U_{S,T}.
 \label{eq:young-magic-output-minors}
\end{equation}
For $S=\{s_1<\cdots<s_n\}$, let
$\epsilon(S)=(-1)^{\sum_{i=1}^n(s_i-i)}$. Jacobi's complementary-minor
identity for the unitary matrix $U$ gives
\begin{equation}
 \det U_{S^c,T^c}
 =\epsilon(S)\epsilon(T)\det(U)\,
   \overline{\det U_{S,T}}.
 \label{eq:young-complementary-minors}
\end{equation}
The family $\mathcal M$ is closed under complementation and
$\epsilon(T)=1$ for every $T\in\mathcal M$. Therefore
\begin{equation}
\begin{aligned}
 \psi_U(S^c)
 &=2^{-B/2}\sum_{T\in\mathcal M}\det U_{S^c,T^c}\\
 &=\epsilon(S)\det(U)\,
   \overline{2^{-B/2}\sum_{T\in\mathcal M}\det U_{S,T}}
 =\epsilon(S)\det(U)\,\overline{\psi_U(S)},\\
 p_U(S^c)&=|\psi_U(S^c)|^2=|\psi_U(S)|^2=p_U(S).
\end{aligned}
 \label{eq:young-complement-probabilities}
\end{equation}
The paired-product definition of the Young polynomials gives
$F_{k\alpha}(\boldsymbol1-x)=(-1)^kF_{k\alpha}(x)$. Consequently,
\begin{equation}
\begin{aligned}
 \Tr(\rho_{\rm ideal}F_{k\alpha})
 &=\sum_x p_U(\boldsymbol1-x)F_{k\alpha}(\boldsymbol1-x)\\
 &=(-1)^k\sum_xp_U(x)F_{k\alpha}(x)
 =(-1)^k\Tr(\rho_{\rm ideal}F_{k\alpha}).
\end{aligned}
 \label{eq:young-odd-coefficients}
\end{equation}
Every odd-degree coefficient vanishes, proving
\eqref{eq:young-odd-degrees}.
\end{proof}

\subsubsection{Thermal overlaps and reference proposals}
\label{sec:young-thermal-overlaps}

For $0\le k\le n$ and $|S|=k$, we denote the projected occupation
coefficients by
\begin{equation}
 b_{k,U}(S)
 :=\bra S\Pi_k(U_k\rho_{\rm in}^{(k)}U_k^\dagger)\ket S
 =\bra S U_k\Pi_k(\rho_{\rm in}^{(k)})U_k^\dagger\ket S.
 \label{eq:young-direct-data}
\end{equation}

\begin{lemma}[Thermal degree contributions as reduced overlaps]
\label{lem:young-thermal-overlap}
Let $D\ge0$ be an integer, let $p_L>0$, and set
$t_{n,L,k}=\eta^kr_k(L)/r_0(L)$ and $d=\min(D,n,L,m-L)$.
Write $p^{\rm th}_{k,U}$ for the image of the ideal probability
component $p_{k,U}$ under the thermal channel. For $|y|=L$ and
$1\le k\le\min(n,L,m-L)$, its density relative to
$\mu_{\rm F}(y)=p_Lu_L(y)=r_0(L)$ is
\begin{equation}
 \begin{aligned}
 f^{\rm th}_{k,U}(y)
 &:=\frac{p^{\rm th}_{k,U}(y)}{\mu_{\rm F}(y)}\\
 &=t_{n,L,k}c_{n,k}
 \Tr\!\left[A_{y,k}U_k\Pi_k(\rho_{\rm in}^{(k)})U_k^\dagger\right]\\
 &=t_{n,L,k}c_{n,k}
 \sum_{\substack{S\subseteq\supp(y)\\|S|=k}}b_{k,U}(S).
 \end{aligned}
 \label{eq:young-whole-degree-thermal}
\end{equation}
The contribution vanishes for $k>\min(n,L,m-L)$.
For the magic input, the retained density and signed mass are
\begin{equation}
 f_{D,U}(y)=1+
 \sum_{\substack{1\le k\le d\\k\ {\rm even}}}
 f^{\rm th}_{k,U}(y),\qquad
 q_{D,U}(y)=\mu_{\rm F}(y)f_{D,U}(y).
 \label{eq:young-direct-density}
\end{equation}
Every nonconstant term has zero $u_L$-mean. Hence $q_{D,U}$ has total
mass one and total-count marginal $p_L$.
\end{lemma}
\begin{proof}
On $|x|=n$, substitute
$\rho_x^{(k)}=A_{x,k}/\binom nk$ and
$\beta_{n,k}=\binom nk/c_{n,k}$ into
Eq.~\eqref{eq:common-output-degree-overlap}. With $d_n=\binom mn$,
this gives
\begin{equation}
    d_np_{k,U}(x)=c_{n,k}\Tr\!\left[
 A_{x,k}U_k\Pi_k(\rho_{\rm in}^{(k)})U_k^\dagger\right].
\end{equation}
By Eq.~\eqref{eq:common-polynomial-lift}, the right-hand side is the
unique homogeneous multilinear harmonic extension of this density
to all binary occupations. The thermal action
\eqref{eq:young-output-amplitude} applies by linearity to every such
polynomial: it multiplies this extension by $\eta^kr_k(L)$.
Dividing by $\mu_{\rm F}(y)=r_0(L)$ gives the first overlap formula.
The diagonal test operator selects precisely the coefficients
$b_{k,U}(S)$ with $S\subseteq\supp(y)$, giving the subset sum.
Finally, Lemma~\ref{lem:young-odd-degrees} removes odd degrees for the
magic input, and orthogonality to constants gives the normalization
and count marginal.
\end{proof}

The coefficients $r_k(L)$ in
\eqref{eq:young-forward-coefficient} are computed by multiplying the
$m-2k$ linear factors in $B_1(z)^{L-k}B_0(z)^{m-L-k}$ and extracting
the coefficient of $z^{n-k}$. For a fixed count $L$, successive
multiplication takes $O(m^2(D+1))$ operations across the retained
degrees. The count probabilities $p_L=\binom{m}{L}r_0(L)$ can also
be computed from the binomial law in \eqref{eq:young-reference}.

A proposal from $\mu_{\rm F}$ is obtained by drawing
$K=K_1+K_2$, where $K_1\sim\Bin(n,1-a_-)$ and
$K_2\sim\Bin(m-n,a_+)$ are independent, and then choosing a
uniform $K$-subset of the modes.

\subsection{Deterministic evaluation}
\label{sec:young-enumeration}

\subsubsection{Computing the coefficient arrays}

For each retained even degree $k\le\min(D,n)$, we construct
$\rho_{\rm in}^{(k)}$ using \eqref{eq:young-global-reduction} and the
compound matrix $(U_k)_{S,T}=\det U_{S,T}$. We compute the rotated diagonal
\begin{equation}
\begin{aligned}
 a_{k,U}(S)
 &:=\bra S U_k\rho_{\rm in}^{(k)}U_k^\dagger\ket S\\
 &=\sum_{\substack{|T|=|R|=k}}
   \det U_{S,T}\,
   (\rho_{\rm in}^{(k)})_{T,R}\,
   \overline{\det U_{S,R}}.
\end{aligned}
 \label{eq:young-rotated-diagonal-minors}
\end{equation}
We apply the fermionic projector \eqref{eq:common-projector}
and store the coefficients $b_{k,U}(S)$ from
\eqref{eq:young-direct-data}. As in
Eq.~\eqref{eq:common-output-reduced-coefficients}, the diagonal
is taken after rotation. The particle maps preserve this diagonal
part, so their subsequent action reduces to subset sums:
\begin{equation}
\begin{aligned}
 (\mathcal L a)(T)&=\sum_{i\notin T}a(T\cup\{i\}),\\
 (\mathcal R a)(S)&=\sum_{i\in S}a(S\setminus\{i\}).
\end{aligned}
 \label{eq:young-diagonal-subset-maps}
\end{equation}
Applying these maps in Eq.~\eqref{eq:common-projector}, with the
fermionic coefficients \eqref{eq:common-projector-coefficients},
computes every $b_{k,U}(S)$ from the rotated diagonal.

\subsubsection{Exact rejection sampling}\label{sec:young-exact-rejection}

\begin{lemma}[A uniform envelope for the retained fermionic density]
\label{lem:young-density-envelope}
Let $d=\min(D,n,m-n)$ and let $f^{\rm ideal}_{D,U}$ be the
projection of the ideal density relative to $u_n$ onto Young degrees
at most $D$. On the reference support, let
$f_{D,U}(y)=\bra y\widetilde\rho_D\ket y/\mu_{\rm F}(y)$.
For every input on the $n$-particle sector,
\begin{equation}
 |f^{\rm ideal}_{D,U}(x)|\le\binom md
 \quad(x\in\Omega_{m,n}),\qquad
 |f_{D,U}(y)|\le\binom md
 \quad(\mu_{\rm F}(y)>0).
 \label{eq:young-density-envelope}
\end{equation}
\end{lemma}
\begin{proof}
Let $e_1,\ldots,e_J$ be an orthonormal basis under $u_n$ for the
retained degrees, including the constant function. Their projection
kernel is
\begin{equation}
     K_D(x,z)=\sum_{j=1}^J e_j(x)\overline{e_j(z)},\qquad
 J=\sum_{k=0}^d\left(\binom mk-\binom m{k-1}\right)=\binom md,
\end{equation}

Here $\binom m{-1}=0$, and the dimension sum is the fermionic
case of Eq.~\eqref{eq:common-diagonal-degree-dimension}.
Mode permutations preserve polynomial degree and the harmonic
condition \eqref{eq:common-diagonal-kernel}, so the retained subspace
and its orthogonal projection are permutation invariant.
Thus $K_D(x,x)$ is constant on the transitive slice $\Omega_{m,n}$.
Orthonormality gives
\begin{equation}
 \E_{x\sim u_n}K_D(x,x)=\sum_{j=1}^J\|e_j\|_{L^2(u_n)}^2=J,
 \qquad
 |K_D(x,z)|\le\sqrt{K_D(x,x)K_D(z,z)}=J.
\end{equation}
If $P_U^{\rm ideal}$ denotes the ideal occupation probability, the
projection formula becomes
\begin{equation}
 f^{\rm ideal}_{D,U}(x)
 =\sum_z u_n(z)K_D(x,z)\frac{P_U^{\rm ideal}(z)}{u_n(z)}
 =\sum_z P_U^{\rm ideal}(z)K_D(x,z).
\end{equation}
Its absolute value is at most $J$, since $P_U^{\rm ideal}$ is a
probability distribution.

Let $T(y\mid x)$ be the classical thermal transition probability
on occupations. Under the joint reference law
$\Pr(X=x,Y=y)=u_n(x)T(y\mid x)$, the output marginal is
$\mu_{\rm F}$. The definition \eqref{eq:young-noisy-approximation}
therefore gives, on its support,
\begin{equation}
 f_{D,U}(y)
 =\sum_x\frac{u_n(x)T(y\mid x)}{\mu_{\rm F}(y)}
             f^{\rm ideal}_{D,U}(x)
 =\E[f^{\rm ideal}_{D,U}(X)\mid Y=y].
\end{equation}
A conditional average preserves the bound $J$.
\end{proof}

We evaluate $f_{D,U}(y)$ using
\eqref{eq:young-whole-degree-thermal} and
\eqref{eq:young-direct-density}. Algorithm~\ref{alg:rejection-sampling}
with the exact score $f_{D,U}(y)$ and
$M=\binom m{\min(D,n,m-n)}$ accepts this proposal with probability
$[f_{D,U}(y)]_+/M$. The envelope gives $\tau_M=0$, so
Lemma~\ref{lem:rejection-sampling} bounds the expected number of
proposals by $M$. Corollary~\ref{thm:final-enumeration} gives the full
error guarantee.

\subsubsection{Deterministic arithmetic cost}

Let $Q_D=1+\sum_{1\le k\le\min(D,n),\ k\ {\rm even}}\binom{m}{k}$.
Constructing the input reductions and compound matrices costs
$O(\binom{m}{k}^2\poly(m,k))$ arithmetic operations per retained degree,
while matrix multiplication to obtain the rotated diagonal costs
$O(\binom{m}{k}^3)$. Repeated subset sums in
\eqref{eq:young-diagonal-subset-maps} apply the diagonal projectors in
$\poly(m,D)Q_D$ additional operations. Including the count and
transfer tables, a conservative preprocessing bound is
\begin{equation}
 T_{\rm pre}=\poly(m,n,D)+
 \sum_{\substack{1\le k\le\min(D,n)\\k\ {\rm even}}}
 O\!\left(\binom{m}{k}^3+\binom{m}{k}^2\poly(m,k)\right).
 \label{eq:young-direct-preprocessing}
\end{equation}
Temporary matrix storage is $O(\sum_{k\le\min(D,n)}\binom{m}{k}^2)$, and the
retained coefficient arrays use $O(Q_D)$ words, besides subset labels
and polynomial-sized count tables. A proposal and its density evaluation
cost $O(m+(D+1)Q_D)$ arithmetic operations and subset membership checks.
The expected number of proposals is at most $M$. With the attempt cap,
the worst-case sampling cost after preprocessing is
$O(A[m+(D+1)Q_D])$. These bounds are $m^{O(D)}$ up to the
factor needed to make $e^{-A/M}$ small: polynomial in $m$ for fixed
$D$, and generally quasipolynomial when $D=\Theta(\log m)$.

For rational input entries and noise parameters, the compound
matrices, projectors, and point evaluations use rational arithmetic.
Section~\ref{sec:precision} controls the resulting numerical errors
and bit complexity.

\subsection{Randomized evaluation}
\label{sec:young-sketch}

We estimate the reduced overlaps in Lemma~\ref{lem:young-thermal-overlap}
using a random $m\times m$ matrix $G$. For each retained degree
$0\le k\le n$, the matrix of its $k\times k$ minors, divided by
$\binom nk$, is an unbiased estimate of the input reduction
$\rho_{\rm in}^{(k)}$. A determinant formula computes the projected
overlap without forming this larger matrix. The scalar thermal factors
then give the noisy density, and
Lemma~\ref{lem:young-overlap-second-moment} bounds the Haar-averaged
variance. Each point evaluation requested by
Algorithm~\ref{alg:adaptive-rejection} uses an independent matrix.

\subsubsection{A random-matrix representation of the input}
We denote a four-mode, two-particle magic state by
\(\ket\phi=(\ket{12}+\ket{34})/\sqrt2\). We independently draw a
matrix \(G_b\) by the following rule:
\begin{enumerate}
 \item With probability \(1/2\), we choose \(i\in\{1,2,3,4\}\) uniformly
       and set \(G_b=4\ket i\bra i\).
 \item Otherwise we choose \(\alpha,\gamma\) independently and uniformly
       from \(\{(1,2),(3,4)\}\), choose an independent uniform sign
       \(\sigma\), and set
       \begin{equation}
       G_b=2\sigma\bigl(\ket{\alpha_1}\bra{\gamma_1}
                         +\ket{\alpha_2}\bra{\gamma_2}\bigr).
       \end{equation}
\end{enumerate}
Let \(G=\bigoplus_{b=1}^B G_b\).  These twelve local atoms have
integer entries and dyadic probabilities.  The matrix can be sampled
and stored in linear space before rotation.

\begin{lemma}[Moments of the fermionic input sketch]
\label{lem:young-input-sketch}
For the independent block matrices above and $0\le k\le n$,
\begin{equation}
 \widehat\rho_{\rm in}^{(k)}=\frac{\bigwedge^kG}{\binom nk},\qquad
 \E_G\widehat\rho_{\rm in}^{(k)}=\rho_{\rm in}^{(k)}.
 \label{eq:young-sketch-unbiased}
\end{equation}
\begin{equation}
 \E_G\|\widehat\rho_{\rm in}^{(k)}\|_2^2
 \le\frac{6^k}{\binom nk}.
 \label{eq:young-sketch-input-norm}
\end{equation}
\end{lemma}
\begin{proof}
In the first branch of the above construction, averaging $4\ket i\bra i$
over the four equally likely values of $i$ gives the identity matrix
$I_4$. Each such matrix $4\ket i\bra i$ has rank one, so
the minor formula~\eqref{eq:young-compound-matrix} gives $\bigwedge^jG_b=0$ for $j\ge2$.
In the second branch, averaging the independent sign gives zero for
the first moment. The only nonzero rows of $G_b$ are those in $\alpha$,
and the only nonzero columns are those in $\gamma$. Hence its only
nonzero minor of order two is
\[
 \det(G_b)_{\alpha,\gamma}
 =\det\begin{pmatrix}2\sigma&0\\0&2\sigma\end{pmatrix}
 =4\sigma^2=4.
\]
By \eqref{eq:young-compound-matrix},
$\bigwedge^2G_b=4\ket\alpha\bra\gamma$, independently of the
sign. Each branch has probability $1/2$, and the four ordered choices
of $(\alpha,\gamma)$ in the second branch have conditional probability
$1/4$. Therefore
\begin{equation}
 \begin{aligned}
 \E G_b&=\frac12 I_4+\frac12\,0=\frac{I_4}{2},\\
 \E\bigwedge^2G_b
 &=\frac12\cdot\frac14
   \sum_{\alpha,\gamma\in\{(1,2),(3,4)\}}4\ket\alpha\bra\gamma\\
 &=\frac12(\ket{12}+\ket{34})(\bra{12}+\bra{34})
 =\ket\phi\bra\phi.
 \end{aligned}
\end{equation}
Finally, both branches have rank at most two, so the same minor formula gives
$\bigwedge^jG_b=0$ for every realization when $j\ge3$.
Together with $\bigwedge^0G_b=1$, these moments equal the
unnormalized local reductions in \eqref{eq:young-block-reductions}:
$\E\bigwedge^jG_b=\binom2j\rho_b^{(j)}$ for $j=0,1,2$.
To pass from one block to all $B$ blocks, we decompose the $k$-particle
space according to the numbers $j_b$ of particles in each block.
Since $G=\bigoplus_{b=1}^B G_b$ does not mix the blocks, it preserves
these mutually orthogonal subspaces and acts on them as
\[
 \bigwedge\nolimits^kG
 =\bigoplus_{\substack{j_1+\cdots+j_B=k\\0\le j_b\le4}}
       \bigotimes_{b=1}^B\bigwedge\nolimits^{j_b}G_b.
\]
Independence lets us take the expectation separately in each tensor
factor. Omitting the zero summands with any $j_b\ge3$, we obtain
\[
 \begin{aligned}
 \E_G\bigwedge\nolimits^kG
 &=\bigoplus_{\substack{j_1+\cdots+j_B=k\\j_b\in\{0,1,2\}}}
       \bigotimes_{b=1}^B\E\bigwedge\nolimits^{j_b}G_b\\
 &=\bigoplus_{\substack{j_1+\cdots+j_B=k\\j_b\in\{0,1,2\}}}
       \bigotimes_{b=1}^B\left(\binom2{j_b}\rho_b^{(j_b)}\right)
 =\binom nk\rho_{\rm in}^{(k)}.
 \end{aligned}
\]
The last equality is \eqref{eq:young-global-reduction} with $n=2B$, and
dividing by $\binom nk$ proves \eqref{eq:young-sketch-unbiased}.

For the second moments, the squared Hilbert--Schmidt norm is the
sum of the squared absolute matrix entries. The first branch has
one nonzero entry of magnitude $4$, while the second has two of
magnitude $2$. Its second exterior power has one entry of magnitude
$4$. Weighting the two branches by $1/2$ gives
\[
 \begin{aligned}
 \E\|\bigwedge\nolimits^0G_b\|_2^2&=1,\\
 \E\|G_b\|_2^2&=\frac12\,4^2+\frac12(2^2+2^2)=12,\\
 \E\|\bigwedge\nolimits^2G_b\|_2^2
 &=\frac12\,0+\frac12\,4^2=8.
 \end{aligned}
\]
The squared Hilbert--Schmidt norm adds over orthogonal direct sums
and multiplies over tensor products. Terms with any $j_b\ge3$ vanish,
so
\[
 \left\|\bigwedge\nolimits^kG\right\|_2^2
 =\sum_{\substack{j_1+\cdots+j_B=k\\j_b\in\{0,1,2\}}}
       \prod_{b=1}^B\left\|\bigwedge\nolimits^{j_b}G_b\right\|_2^2.
\]
Independence of the block matrices now gives
\[
 \begin{aligned}
 \E_G\left\|\bigwedge\nolimits^kG\right\|_2^2
 &=\sum_{\substack{j_1+\cdots+j_B=k\\j_b\in\{0,1,2\}}}
       \prod_{b=1}^B\E\left\|\bigwedge\nolimits^{j_b}G_b\right\|_2^2\\
 &=[z^k]\prod_{b=1}^B(1+12z+8z^2).
 \end{aligned}
\]
The last equality follows by expanding the product: choosing the term
of degree $j_b$ from factor $b$ records that block's contribution,
and extracting $z^k$ sums over precisely the allocations with total
particle number $k$. Finally, dividing by
the squared normalization $\binom nk^2$ in
\eqref{eq:young-sketch-unbiased} yields
\begin{equation}
 \E_G\|\widehat\rho_{\rm in}^{(k)}\|_2^2
 =\frac{[z^k](1+12z+8z^2)^B}{\binom nk^2}.
\end{equation}
Since $n=2B$, the coefficientwise bound
$1+12z+8z^2\le(1+6z)^2$ gives
$[z^k](1+12z+8z^2)^B\le\binom nk6^k$.
Substituting this bound proves \eqref{eq:young-sketch-input-norm}.
\end{proof}

\subsubsection{The projected overlap as a determinant}
\begin{lemma}[Determinant formula for a projected fermionic overlap]
\label{lem:young-determinant-overlap}
Let $C$ be any complex $m\times m$ matrix, let $y$ be a binary
occupation with $|y|=L$, and let $P_y=\sum_{i:y_i=1}\ket i\bra i$.
For $0\le k\le\min(L,m-L)$,
\begin{equation}
 \Tr\!\left[A_{y,k}\Pi_k(\bigwedge^kC)\right]
 =\sum_{q=0}^k(-1)^q
   \frac{\rise{L-k+1}q}{\rise{m-2k+2}q}
 \times[z^qw^{k-q}]\det(I_m+zC+wCP_y).
 \label{eq:young-sketch-det}
\end{equation}
\end{lemma}
\begin{proof}
Let $Y=\supp(y)$ and $X=\bigwedge^kC$. We recall from
\eqref{eq:young-test-operator} that
$A_{y,r}=\sum_{S\subseteq Y,\,|S|=r}\ket S\bra S$.
Removing one particle gives, for $1\le r\le k$,
\[
 \begin{aligned}
 \mathcal L(A_{y,r})
 &=\sum_{\substack{J\subseteq Y\\|J|=r}}
       \sum_{i\in J}\ket{J\setminus\{i\}}\bra{J\setminus\{i\}}\\
 &=(L-r+1)\sum_{\substack{S\subseteq Y\\|S|=r-1}}\ket S\bra S
 =(L-r+1)A_{y,r-1}.
 \end{aligned}
\]
The second equality counts the $L-r+1$ ways to extend a fixed
$(r-1)$-subset $S$ to an $r$-subset of $Y$. Iterating gives
\[
 \mathcal L^q(A_{y,k})
 =\left[\prod_{h=0}^{q-1}(L-k+h+1)\right]A_{y,k-q}
 =\rise{L-k+1}q A_{y,k-q},\qquad 0\le q\le k.
\]

We next identify the determinant coefficient. The $j$th column of
$zC+wCP_y$ is $(z+wy_j)$ times the $j$th column of $C$.
By multilinearity, choosing these columns for indices in $J$ and
identity columns for the remaining indices gives
\[
 \det(I_m+zC+wCP_y)
 =\sum_{J\subseteq[m]}\det(C_{J,J})\prod_{j\in J}(z+wy_j).
\]
The identity columns force the rows outside $J$ to match their column
indices, leaving the determinant of $C_{J,J}$ with positive sign.
The term indexed by $J$ has total degree $|J|$, so only $|J|=k$
contributes to $[z^qw^{k-q}]$. For such a $J$, its polynomial factor is
$z^{k-|J\cap Y|}(z+w)^{|J\cap Y|}$. Choosing $k-q$ factors of $w$
therefore gives
\begin{equation}
 [z^qw^{k-q}]\det(I_m+zC+wCP_y)
 =\sum_{|J|=k}\det(C_{J,J})\binom{|J\cap Y|}{k-q}.
 \label{eq:young-sketch-det-coefficient}
\end{equation}

For a set $S$ of size $k-q$, repeated lowering satisfies
\[
 \bra S\mathcal L^q(X)\ket S
 =q!\sum_{\substack{J\supseteq S\\|J|=k}}\bra JX\ket J.
\]
To obtain a fixed $S$ from $J$, the $q$ indices of $J\setminus S$
can be removed in $q!$ orders. In every order the fermionic signs
from the bra and ket cancel, and no off-diagonal entry of $X$
contributes. Summing over the $(k-q)$-subsets of $Y$ now yields
\[
 \begin{aligned}
 \Tr\!\left[A_{y,k-q}\mathcal L^q(X)\right]
 &=q!\sum_{\substack{S\subseteq Y\\|S|=k-q}}
          \sum_{\substack{J\supseteq S\\|J|=k}}\bra JX\ket J\\
 &=q!\sum_{|J|=k}\bra JX\ket J
          \sum_{\substack{S\subseteq J\cap Y\\|S|=k-q}}1\\
 &=q!\sum_{|J|=k}\det(C_{J,J})\binom{|J\cap Y|}{k-q}.
 \end{aligned}
\]
The last equality uses the minor formula
\eqref{eq:young-compound-matrix} and counts the allowed subsets $S$.

Finally, insert the common projector \eqref{eq:common-projector},
with $D_{k,q}=\rise{m-2k+2}{q}$, and use the adjointness of
$\mathcal L$ and $\mathcal R$ from
Proposition~\ref{prop:properties_RL}:
\[
 \begin{aligned}
 \Tr\!\left[A_{y,k}\Pi_k(X)\right]
 &=\sum_{q=0}^k\frac{(-1)^q}{q!\rise{m-2k+2}q}
       \Tr\!\left[A_{y,k}\mathcal R^q\mathcal L^q(X)\right]\\
 &=\sum_{q=0}^k\frac{(-1)^q}{q!\rise{m-2k+2}q}
       \Tr\!\left[\mathcal L^q(A_{y,k})\mathcal L^q(X)\right]\\
 &=\sum_{q=0}^k\frac{(-1)^q\rise{L-k+1}q}{q!\rise{m-2k+2}q}
       q!\sum_{|J|=k}\det(C_{J,J})\binom{|J\cap Y|}{k-q}.
 \end{aligned}
\]
The second equality moves each raising map across the trace. The last
equality substitutes the two lowering calculations above. Cancelling
$q!$ and using \eqref{eq:young-sketch-det-coefficient} gives
\eqref{eq:young-sketch-det}. The calculation includes $q=0$ and $q=k$,
with empty products and the vacuum-sector identity equal to one.
\end{proof}

Let $C=UGU^\dagger$. The exterior-power action
\eqref{eq:particle-mode-action} and covariance of the particle maps
\eqref{eq:common-particle-map-covariance}, applied to the projector
\eqref{eq:common-projector}, give
\[
 \begin{aligned}
 \bigwedge\nolimits^k C
 &=U_k(\bigwedge\nolimits^kG)U_k^\dagger,\\
 \Pi_k(\bigwedge\nolimits^k C)
 &=\binom nk\,U_k\Pi_k(\widehat\rho_{\rm in}^{(k)})U_k^\dagger.
 \end{aligned}
\]
We estimate \eqref{eq:young-whole-degree-thermal} by replacing
$\rho_{\rm in}^{(k)}$ with $\widehat\rho_{\rm in}^{(k)}$ and taking
the real part of the overlap. Applying Lemma~\ref{lem:young-determinant-overlap}
then gives
\begin{equation}
 \begin{aligned}
 \widehat f^{\rm th}_{k,U}(y;G)
 &=\frac{c_{n,k}t_{n,L,k}}{\binom nk}\,\Re
 \sum_{q=0}^k(-1)^q\frac{\rise{L-k+1}q}{\rise{m-2k+2}q}\\
 &\quad\times[z^qw^{k-q}]\det(I_m+zC+wCP_y).
 \end{aligned}
 \label{eq:young-sketch-thermal}
\end{equation}
As in \eqref{eq:young-whole-degree-thermal}, the expression is zero
on slices carrying no degree $k$, and no zero-probability count sector
is evaluated. Linearity and Lemma~\ref{lem:young-input-sketch} give
\begin{equation}
 \begin{aligned}
 \E_G \widehat f^{\rm th}_{k,U}(y;G)
 &=c_{n,k}t_{n,L,k}\,\Re\Tr\!\left[
 A_{y,k}U_k\Pi_k(\E_G\widehat\rho_{\rm in}^{(k)})U_k^\dagger\right]\\
 &=c_{n,k}t_{n,L,k}\Tr\!\left[
 A_{y,k}U_k\Pi_k(\rho_{\rm in}^{(k)})U_k^\dagger\right]
 =f^{\rm th}_{k,U}(y).
 \end{aligned}
 \label{eq:young-sketch-thermal-unbiased}
\end{equation}
In the second equality the mean input is Hermitian, so the overlap
is real. The last equality is Lemma~\ref{lem:young-thermal-overlap}.

To evaluate the determinant, let $H=zC+wCP_y$. With an auxiliary
scalar $s$, the determinant derivative formula and the geometric
series for $(I+sH)^{-1}$ give the formal power-series identity
\[
 \frac{d}{ds}\log\det(I+sH)
 =\Tr\!\left[(I+sH)^{-1}H\right]
 =\sum_{j\ge1}(-1)^{j+1}s^{j-1}\Tr H^j.
\]
Integrating coefficientwise from $s=0$ to $s=1$ gives
$\log\det(I+H)=\sum_{j\ge1}(-1)^{j+1}\Tr H^j/j$.
Each entry of $H$ has total $(z,w)$ degree one, so terms with $j>D$
cannot affect a coefficient of degree at most $D$. Thus
\begin{equation}
 \begin{gathered}
 T_D=\sum_{j=1}^D\frac{(-1)^{j+1}}j\Tr H^j,\\
 \det(I+H)
 =\sum_{\ell=0}^D\frac{T_D^\ell}{\ell!}
 \pmod{\text{total }(z,w)\text{ degree}>D}.
 \end{gathered}
\end{equation}
The exponential truncation is exact to this order because $T_D$ has
no constant term. There are $\binom{D+2}{2}$ monomials $z^aw^b$ with
$a+b\le D$. We store their coefficients in an array, compute the powers
$H^j$ and $T_D^\ell$ successively, and discard higher-degree terms
after each multiplication. This computes all retained determinant
coefficients in $\poly(m,D)$ arithmetic operations.

\subsubsection{Second moments of the point estimator}
\label{sec:young-point-moments}

The Haar bound needed for the point estimator follows from the common
irrep dimensions and norm transfer.

\begin{lemma}[Haar second moment of a Young overlap]
\label{lem:young-overlap-second-moment}
Let $0\le k\le n\le m/2$, let $x$ be a binary occupation with $|x|=n$,
and let $B$ be any complex operator on $\bigwedge^k\mathbb C^m$.
For Haar-distributed $U\in\mathrm U(m)$,
\begin{equation}
 \E_U\left|\Tr\!\left[A_{x,k}U_k\Pi_k(B)U_k^\dagger\right]\right|^2
 =\frac{\binom nk\binom{m-n}k}
 {\binom{m-k+1}k\left(\binom mk^2-\binom m{k-1}^2\right)}
 \|\Pi_k(B)\|_2^2.
 \label{eq:young-overlap-second-moment}
\end{equation}
\end{lemma}
{
\begin{proof}
For $k=0$ the identity is immediate. For $k\ge1$,
Eqs.~\eqref{eq:ker_ident} and~\eqref{eq:common-irrep-dimensions}
identify the range of $\Pi_k$ as the single irrep
$\mathcal V_{k,k}$ of dimension
$v_k=\binom mk^2-\binom m{k-1}^2$.
Its diagonal dimension is
$z_k=\binom mk-\binom m{k-1}$ by
Eq.~\eqref{eq:common-diagonal-degree-dimension}.
It remains to compute the norm of the projected test operator.

Permutation covariance makes $\|\Pi_k(A_{x,k})\|_2$ constant on
the $n$-slice. For a real Hilbert--Schmidt orthonormal basis
$\{B_a\}_{a=1}^{z_k}$ of the diagonal harmonic subspace,
Parseval and the norm transfer \eqref{eq:common-lift-inner-product}
give
\begin{equation}
 \begin{aligned}
 \|\Pi_k(A_{x,k})\|_2^2
 &=\sum_a\E_{X\sim u_n}\Tr(B_aA_{X,k})^2\\
 &=\frac{\binom nk}{c_{n,k}}\left(\binom mk-\binom m{k-1}\right)
 =\frac{\binom nk\binom{m-n}k}{\binom{m-k+1}k}.
 \end{aligned}
 \label{eq:young-test-norm}
\end{equation}
Self-adjointness and covariance of the common projector replace
$A_{x,k}$ by $\Pi_k(A_{x,k})$ in the overlap. Schur orthogonality
(Lemma~\ref{lem:schur-orthogonality}) then divides the product of
the two squared operator norms by $v_k$, proving the stated moment
for any complex $B$.
\end{proof}

For one proposal $y$, we use the real score
\begin{equation}
 Z_{D,U}(y;G)=1+
 \sum_{\substack{1\le k\le\min(D,n)\\k\ {\rm even}}}
                   \widehat f^{\rm th}_{k,U}(y;G),
 \qquad
 S_D^{\rm F}=\sum_{\substack{1\le k\le\min(D,n)\\k\ {\rm even}}}
                   12^k\Gamma_k^{\rm F}.
 \label{eq:young-sketch-budget}
\end{equation}
The same matrix $G$ is used across the retained degrees in this score.

\begin{lemma}[Moments of the fermionic point estimator]
\label{lem:young-point-estimator}
For the magic input \eqref{eq:young-magic-family}, every fixed $U$ and
every $y$ with $\mu_{\rm F}(y)>0$ satisfy
\begin{equation}
 \E_G Z_{D,U}(y;G)=f_{D,U}(y).
 \label{eq:young-point-unbiased}
\end{equation}
For Haar $U$ and the independent block matrix $G$,
\begin{equation}
 \E_{U,y\sim\mu_{\rm F}}\Var_G Z_{D,U}(y;G)\le S_D^{\rm F},
 \qquad
 \E_U\|f_{D,U}\|_{L^2(\mu_{\rm F})}^2\le4.
 \label{eq:young-point-moments}
\end{equation}
\end{lemma}
\begin{proof}
Linearity and \eqref{eq:young-sketch-thermal-unbiased} prove
\eqref{eq:young-point-unbiased}.

For a binary occupation $x$, define the ideal relative-density
estimate below. On $|x|=n$, its mean is $d_np_{k,U}(x)$; on other
slices it is the corresponding harmonic extension.

\begin{equation}
 \widehat f^{\rm ideal}_{k,U}(x;G)
 :=c_{n,k}\Re\Tr\!\left[
 A_{x,k}U_k\Pi_k(\widehat\rho_{\rm in}^{(k)})U_k^\dagger\right].
\end{equation}

The diagonal part of the projected operator belongs to
$(\mathcal V_{k,k})_0$, so the lift
\eqref{eq:common-polynomial-lift} makes this a real homogeneous
harmonic polynomial of degree $k$. On a supported output slice,
\eqref{eq:young-sketch-thermal} and
Lemma~\ref{lem:young-determinant-overlap} give
$\widehat f^{\rm th}_{k,U}(y;G)=t_{n,L,k}\widehat f^{\rm ideal}_{k,U}(y;G)$.

We first average the squared ideal estimate over Haar $U$.
Since $G$ is independent of $U$, conditioning on $G$ fixes
$\widehat\rho_{\rm in}^{(k)}$ without changing the Haar law. For $|x|=n$,
$(\Re z)^2\le |z|^2$ and
Lemma~\ref{lem:young-overlap-second-moment} therefore give
\begin{equation}
 \E_U\!\left[|\widehat f^{\rm ideal}_{k,U}(x;G)|^2\mid G\right]
 \le
 \frac{c_{n,k}^2\binom nk\binom{m-n}k}
 {\binom{m-k+1}k\left(\binom mk^2-\binom m{k-1}^2\right)}
 \|\Pi_k(\widehat\rho_{\rm in}^{(k)})\|_2^2.
\end{equation}
Orthogonal projection does not increase the Hilbert--Schmidt norm,
so Lemma~\ref{lem:young-input-sketch} bounds the expectation of the
last factor by $6^k/\binom nk$. To simplify the resulting expression,
we use $c_{n,k}$ from \eqref{eq:young-overlap-normalization} and
\begin{equation}
 \begin{gathered}
 \binom mk^2-\binom m{k-1}^2
 =\binom mk^2\left(1-\frac{k^2}{(m-k+1)^2}\right)
 =\binom mk^2\frac{(m+1)(m-2k+1)}{(m-k+1)^2},\\
 \binom{m-k+1}k
 =\frac{m-k+1}{m-2k+1}\binom{m-k}k.
 \end{gathered}
\end{equation}
Substitution and cancellation yield
\begin{equation}
\begin{aligned}
 \E_{U,G}|\widehat f^{\rm ideal}_{k,U}(x;G)|^2
 &\le
 \frac{c_{n,k}^2}{\binom mk^2-\binom m{k-1}^2}
 \frac{6^k}{\binom nk}
 \frac{\binom nk\binom{m-n}k}{\binom{m-k+1}k}\\
 &=6^k\frac{m-k+1}{m+1}
              \frac{\binom{m-k}k}{\binom{m-n}k}
 \le 12^k.
\end{aligned}
\end{equation}
The last inequality uses $m=2n$ and
\begin{equation}
 \frac{\binom{m-k}k}{\binom{m-n}k}
 =\prod_{j=0}^{k-1}\frac{2n-k-j}{n-j}
 \le\prod_{j=0}^{k-1}2=2^k,
\end{equation}
because $2n-k-j\le2(n-j)$ for $j<k$. The remaining factor
$(m-k+1)/(m+1)$ is at most one. All denominators above are positive
for $k\le n=m/2$, including $k=n$.

For fixed $G,U$, \eqref{eq:young-output-amplitude} identifies
$\widehat f^{\rm th}_{k,U}$ as the density of
$\mathcal N_{\eta,\beta}^{\otimes m}(u_n\widehat f^{\rm ideal}_{k,U})$ relative to
$\mu_{\rm F}$. Applying \eqref{eq:master-degree-contraction} to this
exact-degree harmonic therefore gives
\begin{equation}
 \|\widehat f^{\rm th}_{k,U}(\cdot;G)\|_{L^2(\mu_{\rm F})}^2
 =\Gamma_k^{\rm F}\|\widehat f^{\rm ideal}_{k,U}(\cdot;G)\|_{L^2(u_n)}^2.
\end{equation}
Averaging the pointwise Haar bound over $x\sim u_n$ gives
\begin{equation}
 \E_{U,G,y\sim\mu_{\rm F}}
        |\widehat f^{\rm th}_{k,U}(y;G)|^2
 \le12^k\Gamma_k^{\rm F},\qquad 1\le k\le n.
 \label{eq:young-sketch-thermal-variance}
\end{equation}

For fixed $U,G$, let
$\Delta_k=\widehat f^{\rm th}_{k,U}(\cdot;G)-f^{\rm th}_{k,U}$.
Distinct Young degrees are orthogonal on every output slice, so
\[
 \langle\Delta_k,\Delta_j\rangle_{L^2(\mu_{\rm F})}
 =\sum_Lp_L\langle\Delta_k,\Delta_j\rangle_{L^2(u_L)}=0
 \qquad(k\ne j).
\]
This remains true when the degrees share the same matrix.
Unbiasedness at fixed $U,y$ therefore gives, with sums over the
retained positive even degrees,
\[
 \begin{aligned}
 \E_{y\sim\mu_{\rm F}}\Var_G Z_{D,U}(y;G)
 &=\E_G\left\|\sum_k\Delta_k\right\|_{L^2(\mu_{\rm F})}^2\\
 &=\sum_k\left(
   \E_G\|\widehat f^{\rm th}_{k,U}(\cdot;G)\|_{L^2(\mu_{\rm F})}^2
       -\|f^{\rm th}_{k,U}\|_{L^2(\mu_{\rm F})}^2\right)\\
 &\le\sum_k\E_G
       \|\widehat f^{\rm th}_{k,U}(\cdot;G)\|_{L^2(\mu_{\rm F})}^2.
 \end{aligned}
\]
Averaging over Haar $U$ and using
\eqref{eq:young-sketch-thermal-variance} gives the variance bound.
Finally, the constant part has norm one and is orthogonal to every
positive degree. Equations~\eqref{eq:young-degree-energy} and
\eqref{eq:young-kappa}, followed by Proposition~\ref{prop:young-flatness},
give
\[
 \begin{aligned}
 \E_U\|f_{D,U}\|_{L^2(\mu_{\rm F})}^2
 &=1+\sum_{k=1}^{\min(D,n)}\Gamma_k^{\rm F}\E_UW_k(U)\\
 &\le1+\E_U\chi^2(\rho_{\rm ideal}\|u_n)\le4.
 \end{aligned}
\]
\end{proof}
Here Haar randomness supplies the overlap second moment in
Lemma~\ref{lem:young-overlap-second-moment}, while flatness supplies the
separate approximation bound.

\subsubsection{Rejection sampling and arithmetic cost}

For each proposal, we draw $y\sim\mu_{\rm F}$ by the two-binomial
count law and uniform-subset procedure above, with $K=|y|\le m$.
Whenever Algorithm~\ref{alg:adaptive-rejection} requests another
estimate at this occupation, we draw an independent block matrix $G$
and evaluate $Z_{D,U}(y;G)$ from \eqref{eq:young-sketch-budget}.
Theorem~\ref{thm:final-sketch} combines
Lemma~\ref{lem:young-point-estimator} and
Corollary~\ref{cor:young-approximation} with
Theorem~\ref{thm:adaptive-rejection}, using $C_{\rm ref}=4$ and
$S_D=S_D^{\rm F}$.

For one matrix and one proposal, the determinant calculation following
\eqref{eq:young-sketch-thermal-unbiased} stores
$\binom{D+2}{2}=O((D+1)^2)$ coefficients. Naive multiplication of two
such arrays costs $O((D+1)^4)$ operations. The first $D$ matrix powers
in the trace--log formula cost $O(m^3(D+1)^5)$ operations, and the
scalar exponential and sums over $k,q$ fit within the same polynomial
bound. Forming $C=UGU^\dagger$ costs $O(m^3)$. A conservative cost
including the scalar factors is therefore
\begin{equation}
 H_{\rm F}(m,D)=O(m^3(D+1)^6)
 \label{eq:young-point-cost}
\end{equation}
per point evaluation. The thermal coefficients in
\eqref{eq:young-forward-coefficient} are computed once per proposal
and reused for its subsequent evaluations.

The algorithm permits at most $A$ proposals and $E_{\max}$ additional
point evaluations beyond the first one at each proposal. Its arithmetic
cost is therefore
\begin{equation}
 O\!\left(A[m+m^2(D+1)+J+1]+(A+E_{\max})H_{\rm F}(m,D)\right)
 \label{eq:young-rejection-cost}
\end{equation}
with $O(m^2(D+1)^2)$ working storage. The matrices are processed one
at a time. Since $\Gamma_k^{\rm F}\le1$,
\[
 S_D^{\rm F}\le\sum_{k=1}^D12^k
 =\frac{12^{D+1}-12}{11}.
\]
The parameter choices in Theorem~\ref{thm:final-sketch} give
$\poly(m,D,\eps^{-1},\delta^{-1})2^{O(D)}$ arithmetic runtime.
Section~\ref{sec:precision} controls numerical evaluation, threshold
comparisons, and proposal draws.

\section{Fock-state boson sampling}
\label{sec:hahn-model}
We specialize the bosonic framework of
Appendix~\ref{sec:particle-representation} to Fock-state inputs and
verify the assumptions in Subsection~\ref{sec:approximation-ingredients}
required for the low-degree approximation theorem.
Hahn polynomials provide coordinates for the diagonal adjoint degrees
identified in Eq.~\eqref{eq:common-diagonal-degree-dimension},
while Eq.~\eqref{eq:common-output-degree-overlap} supplies their
reduced-overlap evaluation formulas. The additional ingredients here
are the action of thermal loss, the required flatness bounds, the
Fock-state reductions, and the algorithms for evaluating these overlaps.

\subsection{Hahn approximation}\label{sec:hahn-approx}
We consider the input $\ket{1^n0^{m-n}}$, with $m\ge2$ and
$1\le n\le m$. We denote the ideal occupation-dephased output by $\rho_U$.
We write $\hat N=\sum_{i=1}^m\hat n_i$ for the total occupation.
For each total occupation $L$, the reference $u_L$ is uniform,
including occupations with collisions:
\begin{equation}
 u_L=\frac1{\binom{m+L-1}{L}}\sum_{|x|=L}\ket x\bra x,
 \qquad L\ge0.
 \label{eq:hahn-reference}
\end{equation}
We use the classical Hahn basis to verify Assumption~2 of
Theorem~\ref{thm:master}, and compute its forward images to prove
orthogonality and suppression after noise. The Haar estimates of
Ref.~\cite{mhiri2026boson} supply Assumption~1 and a sharper bound on the
individual degree contributions.

\subsubsection{Complete orthogonal Hahn observables}
\label{sec:hahn-ingredient-basis}
We fix the Hahn basis recursively in the number of modes. For a coefficient
array $b^{(p)}_{k,\alpha}(r)$, indexed by $r\in\mathbb N_0^p$ with $|r|=k$,
we write
\begin{equation}
 F^{(p)}_{k,\alpha}(x)
 =\sum_{|r|=k}b^{(p)}_{k,\alpha}(r)
                  \prod_{i=1}^p\binom{x_i}{r_i}.
 \label{eq:hahn-coefficient-polynomial}
\end{equation}
In one mode there is only the degree-zero element,
$b^{(1)}_{0,\varnothing}(0)=1$ and $F^{(1)}_{0,\varnothing}=1$.
For $p\ge2$, the degree-$k$ indices are
$\alpha\in\mathbb N_0^{p-1}$ with $|\alpha|=k$.
We write $\alpha=(\alpha',d)$ and $j=|\alpha'|=k-d$. For
$r=(r',\ell)$ with $|r'|+\ell=k$, we set
\begin{equation}
 b^{(p)}_{k,\alpha}(r',\ell)
 =
 \begin{cases}
 \displaystyle
 \frac{(-1)^\ell\binom{p+k+j-2}{\ell}
                  F^{(p-1)}_{j,\alpha'}(r')}
 {\sqrt{\binom{p+k+j-2}{d}\binom{p+2k-2}{d}}},
       &0\le\ell\le d,\\[2ex]
 0,    &\ell>d.
 \end{cases}
 \label{eq:hahn-recursive-coefficients}
\end{equation}
At $p=m$, we omit the superscript and regard $F_{k,\alpha}(\hat n)$
as a polynomial on the full occupation space. We write
\begin{equation}
 h_{L,k}:=\frac{\fall Lk\rise{m+L}k}{\rise m{2k}},
 \qquad 0\le k\le L,
 \qquad F_{0,\mathbf0}=I.
 \label{eq:hahn-basis-definition}
\end{equation}
The recursion is the multivariate Hahn construction of
Ref.~\cite[Eq.~(2.6)]{khare2009rates}, with the mode order reversed and the
normalization of its Eq.~(2.9).

\begin{lemma}[Hahn orthogonality and completeness, adapted from
Proposition~2.1 in Ref.~\cite{khare2009rates}]
\label{lem:hahn-basis}
For every $L\ge0$, the polynomials in \eqref{eq:hahn-coefficient-polynomial}
satisfy, for $0\le k,j\le L$,
\begin{equation}
 \Tr\!\left[u_LF_{k,\alpha}(\hat n)F_{j,\gamma}(\hat n)\right]
 =\delta_{kj}\delta_{\alpha\gamma}h_{L,k}.
 \label{eq:hahn-basis-orthogonality}
\end{equation}
On occupations of total $L$, the elements through degree $k\le L$
span all polynomials of degree at most $k$. In particular, the elements
through degree $L$ form a complete basis. Their coefficient arrays obey
\begin{equation}
 \sum_{|r|=k}b_{k,\alpha}(r)b_{k,\gamma}(r)
 =\delta_{\alpha\gamma}.
 \label{eq:hahn-coefficient-orthogonality}
\end{equation}
\end{lemma}
\begin{proof}
We set all Dirichlet parameters equal to one in Proposition~2.1 of
Ref.~\cite{khare2009rates}. Its weight is then uniform on occupations of total
$L$. With the coefficient normalization in
\eqref{eq:hahn-recursive-coefficients}, its squared norm formula gives
\begin{equation}
 \Tr(u_LF_{k,\alpha}^2)
 =\frac1{\binom{m+k-1}{k}}
   \frac{\fall Lk\rise{m+L}k}{k!\rise{m+k}k}
 =h_{L,k}.
 \label{eq:hahn-classical-norm-adaptation}
\end{equation}
The same proposition gives orthogonality and completeness. At $L=k$,
$F_{k,\alpha}(r)=b_{k,\alpha}(r)$ and $\binom{m+k-1}{k}h_{k,k}=1$, which gives
\eqref{eq:hahn-coefficient-orthogonality}.
\end{proof}

The cancellation needed for the channel calculation follows directly
from this orthogonality. For $|v|<k$,
\begin{equation}
 \begin{aligned}
 \sum_{|r|=k}b_{k,\alpha}(r)\binom rv
 &=\sum_{|r|=k}F_{k,\alpha}(r)\binom rv\\
 &=\binom{m+k-1}{k}\Tr\!\left[u_kF_{k,\alpha}(\hat n)
                         \binom{\hat n}{v}\right]=0.
 \end{aligned}
 \label{eq:hahn-harmonic-coefficients}
\end{equation}
Here $F_{k,\alpha}(r)=b_{k,\alpha}(r)$ when $|r|=k$,
whereas $\binom{\hat n}{v}$ has degree $|v|<k$.
The same definition gives $F_{k,\alpha}(x)=0$ when $|x|<k$.

For the ideal dephased output, Lemma~\ref{lem:hahn-basis} gives
\begin{equation}
 W_k(U)=\frac1{h_{n,k}}\sum_\alpha
   |\Tr[\rho_UF_{k,\alpha}(\hat n)]|^2,\qquad
 \chi^2(\rho_U\|u_n)=\sum_{k=1}^nW_k(U).
 \label{eq:hahn-degree-energy}
\end{equation}

\subsubsection{Orthogonality after thermal loss}
For $0<\eta<1$ and $\beta\ge0$, we put
\begin{equation}
 t=(1-\eta)\beta,\qquad G=1+t,\qquad s=\frac\eta G,
 \qquad \mu_{\rm th}=\mathcal N_{\eta,\beta}^{\otimes m}(u_n).
 \label{eq:hahn-thermal-parameters}
\end{equation}
We define $D(w)=1+t(1-w)$ and $A(w)=1+\eta(w-1)/D(w)$.

\begin{lemma}[Orthogonality after bosonic thermal loss]
\label{lem:hahn-thermal-action}
There are coefficients $\Gamma_k^{\rm th}\ge0$, independent of
$\alpha$, such that, for $0\le k,j\le n$,
\begin{equation}
 \Tr\!\left[
 \mathcal N_{\eta,\beta}^{\otimes m}(u_nF_{k,\alpha})
 \mathcal N_{\eta,\beta}^{\otimes m}(u_nF_{j,\gamma})
 \mu_{\rm th}^{-1}\right]
 =\delta_{kj}\delta_{\alpha\gamma}\Gamma_k^{\rm th}h_{n,k}.
 \label{eq:hahn-image-orthogonality}
\end{equation}
In particular, Assumption~3 of Theorem~\ref{thm:master} holds.
\end{lemma}

\begin{proof}
Applying the coefficient-extraction argument of
\eqref{eq:thermal-coefficient-extraction} to the bosonic sectors gives
\begin{equation}
 \mathcal N_{\eta,\beta}^{\otimes m}(u_nF_{k,\alpha})
 =\frac1{\binom{m+n-1}{n}}[z^n]\,
   \mathcal N_{\eta,\beta}^{\otimes m}(z^{\hat N}F_{k,\alpha}).
 \label{eq:hahn-input-coefficient-extraction}
\end{equation}
We compute this generating expression for $0<z<1$. We write
\begin{equation}
 B(z)=t+(\eta-t)z,\qquad
 C(z)=1+t-(1+t-\eta)z,\qquad B'C-BC'=\eta.
 \label{eq:hahn-forward-generating-functions}
\end{equation}
The one-mode identity \eqref{eq:thermal-pgf} gives
\begin{equation}
 \begin{aligned}
 \sum_{y\ge0}w^y\bra y
       \mathcal N_{\eta,\beta}(z^{\hat n})\ket y
 &=\Tr\!\left[z^{\hat n}
       \mathcal N_{\eta,\beta}^\dagger(w^{\hat n})\right]\\
 &=\frac1{D(w)(1-zA(w))}
 =\frac1{C(z)-wB(z)},\\
 \bra y\mathcal N_{\eta,\beta}(z^{\hat n})\ket y
 &=\frac{B(z)^y}{C(z)^{y+1}}.
 \end{aligned}
 \label{eq:hahn-one-mode-forward-generating}
\end{equation}
Differentiation inserts a factorial polynomial:
\begin{equation}
 \begin{aligned}
 &\bra y\mathcal N_{\eta,\beta}
       \!\left(z^{\hat n}\binom{\hat n}{\ell}\right)\ket y
 =\frac{z^\ell}{\ell!}\partial_z^\ell
       \frac{B(z)^y}{C(z)^{y+1}}\\
 &\quad=\frac{B(z)^y}{C(z)^{y+1}}
    \sum_{j=0}^\ell\binom\ell j\binom yj
    \left(-\frac{zC'(z)}{C(z)}\right)^{\ell-j}
    \left(\frac{\eta z}{B(z)C(z)}\right)^j.
 \end{aligned}
 \label{eq:hahn-forward-factorial-moments}
\end{equation}
The last equality follows by expanding the right-hand side of the
following identity in powers of $w$ and comparing the coefficients of
$w^\ell$:
\begin{equation}
 \frac{B(z+w)^yC(z)^{y+1}}{C(z+w)^{y+1}B(z)^y}
 =\left(1+\frac{C'w}{C}\right)^{-1}
 \left[1+\frac{\eta w}{BC}
       \left(1+\frac{C'w}{C}\right)^{-1}\right]^y.
 \label{eq:hahn-forward-taylor-expansion}
\end{equation}
Tensoring \eqref{eq:hahn-forward-factorial-moments} over the modes and
using \eqref{eq:hahn-harmonic-coefficients}, for $|y|=K\ge k$ we obtain
\begin{equation}
 \begin{aligned}
 &\bra y\mathcal N_{\eta,\beta}^{\otimes m}
             (z^{\hat N}F_{k,\alpha})\ket y\\
 &=\frac{B(z)^K}{C(z)^{m+K}}
   \sum_{|v|\le k}\binom yv
    \left(-\frac{zC'}C\right)^{k-|v|}
    \left(\frac{\eta z}{BC}\right)^{|v|}
    \sum_{|r|=k}b_{k,\alpha}(r)\binom rv\\
 &=\frac{\eta^kz^kB(z)^{K-k}}{C(z)^{m+K+k}}
                F_{k,\alpha}(y).
 \end{aligned}
 \label{eq:hahn-forward-generating}
\end{equation}
For $K<k$, the same sum is zero. We define
\begin{equation}
 r_k(K)=\frac1{\binom{m+n-1}{n}}[z^{n-k}]
            \frac{B(z)^{K-k}}{C(z)^{m+K+k}}\quad(K\ge k),
 \qquad r_k(K)=0\quad(K<k).
 \label{eq:hahn-forward-coefficients}
\end{equation}
Taking coefficients in \eqref{eq:hahn-forward-generating}, including
the case $k=0$, yields
\begin{equation}
 \begin{aligned}
 \bra y\mathcal N_{\eta,\beta}^{\otimes m}
                   (u_nF_{k,\alpha})\ket y
 &=\eta^kr_k(K)F_{k,\alpha}(y),\\
 \mu_{\rm th}(y)&=r_0(K)=\frac{p_K}{\binom{m+K-1}{K}},
 \qquad p_K:=\binom{m+K-1}{K}r_0(K),\qquad |y|=K.
 \end{aligned}
 \label{eq:hahn-thermal-reference}
\end{equation}
Consequently, classical Hahn orthogonality gives
\begin{equation}
 \begin{aligned}
 &\Tr\!\left[
 \mathcal N_{\eta,\beta}^{\otimes m}(u_nF_{k,\alpha})
 \mathcal N_{\eta,\beta}^{\otimes m}(u_nF_{j,\gamma})
 \mu_{\rm th}^{-1}\right]\\
 &=\eta^{k+j}\sum_{\substack{K\ge\max(k,j)\\p_K>0}}
   \frac{\binom{m+K-1}{K}r_k(K)r_j(K)}{r_0(K)}
                  \Tr(u_KF_{k,\alpha}F_{j,\gamma})\\
 &=\delta_{kj}\delta_{\alpha\gamma}h_{n,k}
    \left[\frac{\eta^{2k}}{h_{n,k}}
       \sum_{\substack{K\ge k\\p_K>0}}
          \frac{\binom{m+K-1}{K}h_{K,k}r_k(K)^2}{r_0(K)}\right].
 \end{aligned}
 \label{eq:hahn-forward-gram-calculation}
\end{equation}
The sums converge: positivity gives
$|\mathcal N_{\eta,\beta}^{\otimes m}(u_nF_{k,\alpha})(y)|
\le\|F_{k,\alpha}|_{\mathcal H_n}\|_\infty\mu_{\rm th}(y)$.
The bracket is $\Gamma_k^{\rm th}$, and $\Gamma_0^{\rm th}=1$.
\end{proof}

For the contraction bound and the evaluators, we express the coefficients
in terms of count probabilities. With $D$ and $A$ as above, we define
\begin{equation}
 \sum_{K\ge0}p_{M,N}(K)w^K=D(w)^{-M}A(w)^N,
 \label{eq:hahn-thermal-count-law}
\end{equation}
for $M\in\mathbb N$ and $N\in\mathbb N_0$. Since
$D(w)(1-zA(w))=C(z)-wB(z)$, the two expansions give
\begin{equation}
 \begin{aligned}
 &\sum_{N\ge0}\binom{M+N-1}{N}z^N
          D(w)^{-M}A(w)^N\\
 &\qquad=[C(z)-wB(z)]^{-M}
 =\sum_{L\ge0}\binom{M+L-1}{L}
       w^L\frac{B(z)^L}{C(z)^{M+L}},\\
 \binom{M+N-1}{N}p_{M,N}(L)
 &=\binom{M+L-1}{L}[z^N]
                    \frac{B(z)^L}{C(z)^{M+L}}.
 \end{aligned}
 \label{eq:hahn-count-coefficient-identity}
\end{equation}
Taking $M=m+2k$, $N=n-k$, and $L=K-k$, and using
$\binom{m+L-1}{L}h_{L,k}=\binom{m+L+k-1}{L-k}$, we obtain
\begin{equation}
 \binom{m+K-1}{K}h_{K,k}r_k(K)=h_{n,k}p_{m+2k,n-k}(K-k),
 \qquad p_K=p_{m,n}(K).
 \label{eq:hahn-count-coefficient-transfer}
\end{equation}
Thus
\begin{equation}
 \Gamma_k^{\rm th}
 =\eta^{2k}h_{n,k}
   \sum_{\substack{K\ge k\\p_K>0}}
    \frac{p_{m+2k,n-k}(K-k)^2}{p_Kh_{K,k}}.
 \label{eq:hahn-thermal-orthogonality}
\end{equation}
Expanding a diagonal input operator $\sigma$ on $\mathcal H_n$ in the Hahn basis and
using \eqref{eq:hahn-thermal-reference} gives the moment identity
\begin{equation}
 \sum_{|y|=K}
 \bra y\mathcal N_{\eta,\beta}^{\otimes m}(\sigma)\ket y
          F_{k,\alpha}(y)
 =\eta^kp_{m+2k,n-k}(K-k)\Tr(\sigma F_{k,\alpha}),
 \quad k\le\min(n,K).
 \label{eq:hahn-thermal-transfer}
\end{equation}

\subsubsection{Suppression under noise}
To bound this coefficient, we expand the count law using
$A(z)=1-s+sz/D(z)$:
\begin{equation}
 \begin{aligned}
 D(z)^{-m}A(z)^n
 &=\sum_{r=0}^n\pi_r\frac{z^r}{D(z)^{m+r}},\\
 \pi_r&=\binom nr s^r(1-s)^{n-r},\qquad
 \nu_k(r)=\frac{\fall rk}{\rise{m+r}k}\quad(r\ge k),
 \end{aligned}
 \label{eq:hahn-count-generating-expansion}
\end{equation}
with $\nu_k(r)=0$ for $r<k$. For $0\le r\le\min(n,K)$, the nonnegative coefficients satisfy
\begin{equation}
 \begin{gathered}
 w_{rK}:=\pi_r[z^{K-r}]D(z)^{-m-r}
 =\pi_r\binom{m+K-1}{K-r}\frac{t^{K-r}}{G^{m+K}},\\
 p_K=\sum_{r=0}^{\min(n,K)}w_{rK},\qquad
 \sum_{K\ge r}w_{rK}=\pi_r.
 \end{gathered}
 \label{eq:hahn-positive-count-expansion}
\end{equation}
Setting $r=k+j$ and cancelling factorials gives
\begin{equation}
 \begin{aligned}
 w_{k+j,K}\nu_k(k+j)
 &=\eta^k\frac{\fall nk}{\rise{m+K}k}
   \binom{n-k}{j}s^j(1-s)^{n-k-j}\\
 &\qquad\times\binom{m+K+k-1}{K-k-j}
                  \frac{t^{K-k-j}}{G^{m+K+k}},\\
 \sum_r w_{rK}\nu_k(r)
 &=\eta^k\frac{\fall nk}{\rise{m+K}k}
                         p_{m+2k,n-k}(K-k).
 \end{aligned}
 \label{eq:hahn-count-shift}
\end{equation}
Consequently, for $p_K>0$ and $K\ge k$, we define
\begin{equation}
 \begin{aligned}
 t_{n,K,k}
 &:=\eta^k\sqrt{\frac{h_{n,k}}{h_{K,k}}}
            \frac{p_{m+2k,n-k}(K-k)}{p_K}\\
 &=\sum_{r=k}^{\min(n,K)}\frac{w_{rK}}{p_K}
       \frac{\nu_k(r)}{\sqrt{\nu_k(n)\nu_k(K)}},
 \qquad \Gamma_k^{\rm th}=\sum_{K\ge k}p_Kt_{n,K,k}^2.
 \end{aligned}
 \label{eq:hahn-thermal-multiplier}
\end{equation}
We set $t_{n,K,k}=0$ when $p_K=0$ or $K<k$.

\begin{proposition}[Degree suppression under thermal loss]
\label{prop:hahn-flatness}
For every $Z_k$ in the degree-$k$ Hahn subspace, with $k\le n$,
\begin{equation}
 \Tr\!\left[\mathcal N_{\eta,\beta}^{\otimes m}(u_nZ_k)^2
                    \mu_{\rm th}^{-1}\right]
 =\Gamma_k^{\rm th}\Tr(u_nZ_k^2).
 \label{eq:hahn-exact-noise-contraction}
\end{equation}
Moreover, $1=\Gamma_0^{\rm th}\ge\Gamma_1^{\rm th}\ge\cdots
\ge\Gamma_n^{\rm th}\ge0$, and
\begin{equation}
 \begin{gathered}
 \Gamma_k^{\rm th}\le\Gamma_k(s)\le q_{\rm B}^k,
 \qquad q_{\rm B}=\frac{s(m+n)}{m+sn}<1,\\
 \Gamma_k(s):=\sum_{r=k}^n\pi_r\frac{\nu_k(r)}{\nu_k(n)}.
 \end{gathered}
 \label{eq:hahn-q}
\end{equation}
In particular, Assumption~4 of Theorem~\ref{thm:master} holds.
\end{proposition}

\begin{proof}
The exact contraction follows by expanding $Z_k$ in the Hahn basis
and applying \eqref{eq:hahn-image-orthogonality}.
The ratios in \eqref{eq:hahn-thermal-multiplier} lie in $[0,1]$.
For $k<\min(n,K)$ and $k\le r\le\min(n,K)$, their successive
degree ratios satisfy
\begin{equation}
 \frac{\nu_{k+1}(r)/\sqrt{\nu_{k+1}(n)\nu_{k+1}(K)}}
      {\nu_k(r)/\sqrt{\nu_k(n)\nu_k(K)}}
 =\frac{(r-k)/(m+r+k)}
 {\sqrt{[(n-k)/(m+n+k)][(K-k)/(m+K+k)]}}\le1.
 \label{eq:hahn-thermal-degree-monotonicity}
\end{equation}
The inequality follows from $r\le\min(n,K)$ and the monotonicity of
$x\mapsto(x-k)/(m+x+k)$ for $x\ge k$.
Thus $0\le t_{n,K,k+1}\le t_{n,K,k}\le1$, setting the multiplier
to zero for $K<k$, and hence
$\Gamma_{k+1}^{\rm th}\le\Gamma_k^{\rm th}\le1$.

Jensen's inequality in \eqref{eq:hahn-thermal-multiplier}, followed by
$\nu_k(r)\le\nu_k(K)$ and \eqref{eq:hahn-positive-count-expansion}, gives
\begin{equation}
 \begin{aligned}
 \Gamma_k^{\rm th}
 &\le\sum_{\substack{K\ge k\\k\le r\le\min(n,K)}}
 w_{rK}\frac{\nu_k(r)^2}{\nu_k(n)\nu_k(K)}\\
 &\le\sum_r\pi_r\frac{\nu_k(r)}{\nu_k(n)}=\Gamma_k(s).
 \end{aligned}
 \label{eq:hahn-binomial-bound}
\end{equation}
For pure loss, $t=0$ forces $K=r$, so equality holds.
Shifting the binomial sum by $r=k+B$ yields
\begin{equation}
 \Gamma_k(s)=s^k\rise{m+n}k
    \E_{B\sim\Bin(n-k,s)}\frac1{\rise{m+k+B}k}.
 \label{eq:hahn-shifted-binomial}
\end{equation}
For $k<n$, the ratio of the degree-$(k+1)$ and degree-$k$ summands
at the same $r$ is
$(r-k)(m+n+k)/[(n-k)(m+r+k)]$. Therefore
\begin{equation}
 \frac{\Gamma_{k+1}(s)}{\Gamma_k(s)}
 =\frac{m+n+k}{n-k}\frac{\E[w(B)f(B)]}{\E[w(B)]},
 \qquad w(B)=\frac1{\rise{m+k+B}k},\quad f(B)=\frac B{m+2k+B}.
 \label{eq:hahn-ratio-tilt}
\end{equation}
The function $w$ decreases and $f$ increases. For an independent copy
$B'$,
\begin{equation}
 2\bigl(\E[wf]-\E[w]\E[f]\bigr)
 =\E[(w(B)-w(B'))(f(B)-f(B'))]\le0.
 \label{eq:hahn-negative-covariance}
\end{equation}
Since $f$ is concave, Jensen's inequality gives
\begin{equation}
 \frac{\Gamma_{k+1}(s)}{\Gamma_k(s)}
 \le\frac{m+n+k}{n-k}f(s(n-k))
 =\frac{s(m+n+k)}{m+sn+(2-s)k}\le q_{\rm B}.
 \label{eq:hahn-ratio-bound}
\end{equation}
The final comparison follows from
\begin{equation}
 (m+n)\bigl(m+sn+(2-s)k\bigr)
 -(m+n+k)(m+sn)=k(1-s)(m+2n)\ge0.
 \label{eq:hahn-geometric-ratio-comparison}
\end{equation}
Multiplying the ratio bounds from $\Gamma_0(s)=1$ proves
$\Gamma_k(s)\le q_{\rm B}^k$. Finally,
$m+sn-s(m+n)=m(1-s)>0$, so $q_{\rm B}<1$.
\end{proof}

\subsubsection{Haar flatness and degree contributions}

\begin{proposition}[Haar degree contributions and flatness, adapted from Corollary~1, Theorem~6 and Proposition~11 in Ref.~\cite{mhiri2026boson}]
\label{prop:hahn-haar-flatness}
For Haar-distributed $U$ and $1\le k\le n$,
\begin{equation}
 \begin{aligned}
 \E_UW_k(U)
 &=\frac{\binom{m+k-1}{k}}{\binom{m+k-1}{k}+\binom{m+k-2}{k-1}}\binom nk
   \frac{\rise{m+k}{k}}{\rise{m+n}{k}}H_k,\\
 H_k&:=\binom nk^{-1}\sum_{j=0}^k(-1)^j\binom kj
   \frac{\rise{n-k+1}{j}}{\rise{m+2k-j-1}{j}}.
 \end{aligned}
 \label{eq:hahn-haar-ideal}
\end{equation}
Here $H_k=\|\Pi_k(\rho_{\rm in}^{(k)})\|_2^2$, with the
normalized reduction and orthogonal projector defined in
Eqs.~\eqref{eq:common-normalized-reduction}
and~\eqref{eq:common-projector}. In particular,
\begin{equation}
 0\le H_k\le\binom nk^{-1},\qquad
 \E_U\chi^2(\rho_U\|u_n)\le C_{m,n}-1,\qquad
 C_{m,n}=\min\left\{n+1,\frac{2(m+n-1)}{n+1}\right\}.
 \label{eq:hahn-haar-flatness}
\end{equation}
\end{proposition}
\begin{proof}
The degree formula is Eq.~(434) of Ref.~\cite{mhiri2026boson}, expressed in
our normalization. Corollary~1 there identifies $H_k$ with the squared
projection norm. Since the input reduction is the uniform mixture
over its $\binom nk$ occupied subsets, projection contraction gives
$H_k\le\binom nk^{-1}$. Thus
\begin{equation}
 \E_UW_k(U)
 \le\frac{\binom{m+k-1}{k}}{\binom{m+k-1}{k}+\binom{m+k-2}{k-1}}
       \frac{\rise{m+k}{k}}{\rise{m+n}{k}}\le1,
 \qquad
 1+\E_U\chi^2(\rho_U\|u_n)
 =1+\sum_{k=1}^n\E_UW_k(U)\le n+1.
 \label{eq:hahn-haar-degree-bound}
\end{equation}
For the second bound, Proposition~11, Eq.~(432), of
Ref.~\cite{mhiri2026boson} gives the integral below. With
$f(\theta)=\cos^{m+n-2}\theta$, integration by parts yields
\begin{equation}
 \begin{aligned}
 1+\E_U\chi^2(\rho_U\|u_n)
 &=(m+n-1)\int_0^{\pi/2}f(\theta)
                         \sin((n+1)\theta)\,d\theta\\
 &=\frac{m+n-1}{n+1}
   \left[1+\int_0^{\pi/2}f'(\theta)
                         \cos((n+1)\theta)\,d\theta\right]\\
 &\le\frac{m+n-1}{n+1}
       \left[1+\int_0^{\pi/2}|f'(\theta)|\,d\theta\right]
 =\frac{2(m+n-1)}{n+1}.
 \end{aligned}
 \label{eq:hahn-haar-integral-bound}
\end{equation}
The last equality uses $f(0)=1$, $f(\pi/2)=0$ and $f'\le0$.
\end{proof}
This verifies Assumption~1 of Theorem~\ref{thm:master}.

\subsubsection{Suppression under the interferometer and noise}

\begin{proposition}[Haar degree suppression under thermal loss]
\label{prop:hahn-haar-suppression}
For Haar-distributed $U$, the degree-$k$ contribution to the mean
squared reference error satisfies
\begin{equation}
 E_k^{\rm th}:=\Gamma_k^{\rm th}\E_UW_k(U)
 \le E_k(s):=\Gamma_k(s)\E_UW_k(U)\le s^k,
 \qquad 1\le k\le n.
 \label{eq:hahn-thermal-haar-energy}
\end{equation}
Consequently, the degree-wise bound in
\eqref{eq:ingredient-ensemble-degree-suppression} holds with
$A=1$ and $a=s$.
\end{proposition}
\begin{proof}
The first inequality is \eqref{eq:hahn-binomial-bound}. Substituting
\eqref{eq:hahn-haar-ideal} in the definition of $E_k(s)$ gives
\begin{equation}
 E_k(s)
 =\frac{\binom{m+k-1}{k}H_k}{\binom{m+k-1}{k}+\binom{m+k-2}{k-1}}
   \sum_{r=k}^n\pi_r\binom rk
       \frac{\rise{m+k}k}{\rise{m+r}k}.
 \label{eq:hahn-haar-energy}
\end{equation}
For $r\ge k$ the factorial ratio is at most one. Hence
\begin{equation}
 E_k(s)\le\binom nk^{-1}\sum_{r=k}^n\pi_r\binom rk=s^k,
 \label{eq:hahn-weighted-geometric-bound}
\end{equation}
where the equality is the factorial moment of $\Bin(n,s)$.
\end{proof}

\subsubsection{Low-degree approximation}\label{sec:hahn-truncation}

We define the signed approximation by
\begin{equation}
 \widetilde\rho_{D,U}
 =\mathcal N_{\eta,\beta}^{\otimes m}\!\left[
 u_n\left(I+\sum_{k=1}^{\min(D,n)}\sum_\alpha
     \frac{\Tr[\rho_UF_{k,\alpha}(\hat n)]}{h_{n,k}}
                       F_{k,\alpha}(\hat n)\right)\right].
 \label{eq:hahn-thermal-approximation}
\end{equation}
It has trace one, and its conditional density in each sector has
mass one. By \eqref{eq:master-degree-contraction}, its squared
reference error is
\begin{equation}
 a_D^{\rm th}(U):=\Tr\!\left[
  (\mathcal N_{\eta,\beta}^{\otimes m}(\rho_U)
       -\widetilde\rho_{D,U})^2\mu_{\rm th}^{-1}\right]
 =\sum_{k=D+1}^n\Gamma_k^{\rm th}W_k(U).
 \label{eq:hahn-flatness-tail}
\end{equation}
The tail is zero for $D\ge n$. We set $\Gamma_k^{\rm th}=0$ for $k>n$.

\begin{corollary}[Flatness and Haar approximation bounds]
\label{cor:hahn-approximation}
For any ensemble satisfying $\E_U\chi^2(\rho_U\|u_n)\le C$,
\begin{equation}
 \E_U\|\mathcal N_{\eta,\beta}^{\otimes m}(\rho_U)
        -\widetilde\rho_{D,U}\|_1^2
 \le C\Gamma_{D+1}^{\rm th}\le Cq_{\rm B}^{D+1}.
 \label{eq:hahn-flatness-consequence}
\end{equation}
For Haar $U$, we define
\begin{equation}
 \begin{aligned}
 a_D^{\mathrm H,\mathrm{th}}(\eta,\beta)
 &:=\E_Ua_D^{\rm th}(U)=\sum_{k=D+1}^nE_k^{\rm th},\\
 \E_U\|\mathcal N_{\eta,\beta}^{\otimes m}(\rho_U)
        -\widetilde\rho_{D,U}\|_1^2
 &\le a_D^{\mathrm H,\mathrm{th}}(\eta,\beta)
 \le\min\left\{(C_{m,n}-1)\Gamma_{D+1}^{\rm th},
          \frac{s^{D+1}}{1-s}\right\}.
 \end{aligned}
 \label{eq:hahn-haar-tail}
\end{equation}
\end{corollary}
\begin{proof}
We apply Theorem~\ref{thm:master} with the orthogonality and suppression
bounds established above. Monotonicity of $\Gamma_k^{\rm th}$ gives
\begin{equation}
 \sum_{k>D}\Gamma_k^{\rm th}\E_UW_k(U)
 \le\Gamma_{D+1}^{\rm th}\E_U\chi^2(\rho_U\|u_n).
 \label{eq:hahn-flatness-tail-proof}
\end{equation}
For Haar $U$, we use \eqref{eq:hahn-haar-flatness} for the first bound,
and \eqref{eq:master-degree-suppression} with $A=1$, $a=s$ for the second.
\end{proof}

For a mean squared trace-norm budget $\zeta^2$, we choose $D$ so that
the chosen bound is at most $\zeta^2$. For trace-norm error at most
$\zeta$ with probability at least $1-\delta$ over Haar $U$, we use the
budget $\delta\zeta^2$ and Markov's inequality.

The thermal degree sum also admits a finite upper certificate.
Since $0\le t_{n,K,k}\le1$, for any integer $K_*\ge k$,
\begin{equation}
 0\le\Gamma_k^{\rm th}-\sum_{K=k}^{K_*}p_Kt_{n,K,k}^2
 \le\Prob_{K\sim p}(K>K_*).
 \label{eq:hahn-contraction-count-tail}
\end{equation}
For $K_*\ge n$, adding the count-tail bound in
Lemma~\ref{lem:reference-count-tails} to the finite sum therefore
certifies the thermal coefficient.

Near full input filling, the input reduction to a fixed number of
particles approaches the maximally mixed state. Together with the
suppression of higher degrees by thermal loss, this gives the stronger
concentration bound in Proposition~\ref{prop:main-saturated-concentration}.

\begin{proof}[Proof of Proposition~\ref{prop:main-saturated-concentration}]
Let $h=m-n$. For $1\le k\le n$, the normalized input reduction
$\rho_{\rm in}^{(k)}$ is uniform on the $\binom nk$ subsets of
the occupied input modes, so
$\Tr[(\rho_{\rm in}^{(k)})^2]=\binom nk^{-1}$.
The orthogonal projector $\Pi_k$ annihilates the identity for
$k\ge1$, by Corollary~\ref{cor:common-projectors}. Hence, with
$d_k=\binom{m+k-1}{k}$,
\begin{equation}
 \begin{aligned}
 H_k
 &=\left\|\Pi_k\!\left(\rho_{\rm in}^{(k)}-
                 \frac{I_{\mathcal H_k}}{d_k}\right)\right\|_2^2\\
 &\le\left\|\rho_{\rm in}^{(k)}-
                 \frac{I_{\mathcal H_k}}{d_k}\right\|_2^2
 =\binom nk^{-1}-d_k^{-1}.
 \end{aligned}
 \label{eq:hahn-saturated-harmonic-norm}
\end{equation}
Multiplying by $\binom nk$ gives
\begin{equation}
 \begin{aligned}
 \binom nk H_k
 &\le1-\frac{\binom nk}{\binom{m+k-1}{k}}
 =1-\prod_{j=0}^{k-1}\left(1-\frac{h+2j}{m+j}\right)\\
 &\le\sum_{j=0}^{k-1}\frac{h+2j}{m+j}
 \le\frac{k(h+k-1)}m,
 \end{aligned}
 \label{eq:hahn-saturated-harmonic-bound}
\end{equation}
where we used $1-\prod_j(1-a_j)\le\sum_j a_j$ for $0\le a_j\le1$.

The dimension and factorial ratios in \eqref{eq:hahn-haar-energy}
are at most one. Since $\pi_r$ is the mass function of $\Bin(n,s)$
and $s\le\eta$, the degree contribution therefore satisfies
\begin{equation}
 \begin{aligned}
 E_k^{\rm th}
 &\le H_k\sum_{r=k}^n\pi_r\binom rk
 =s^k\binom nkH_k
 \le\frac{k(h+k-1)}m\eta^k.
 \end{aligned}
 \label{eq:hahn-saturated-degree-energy}
\end{equation}
Summing over the discarded degrees in \eqref{eq:hahn-haar-tail} yields
\begin{equation}
 \begin{aligned}
 a_D^{\mathrm H,\mathrm{th}}(\eta,\beta)
 &\le\frac1m\sum_{k=D+1}^{\infty}k(h+k-1)\eta^k\\
 &=\frac{\eta^{D+1}}m\left[
    \frac{(D+1)(h+D)}{1-\eta}
   +\frac{(h+2D+1)\eta}{(1-\eta)^2}
   +\frac{\eta(1+\eta)}{(1-\eta)^3}
   \right].
 \end{aligned}
 \label{eq:hahn-saturated-finite-tail}
\end{equation}
For bounded $h$ and fixed $0<\eta<1$, the right-hand side is at most
$C\eta^{D/2}/m$, where $C\ge1$ depends only on $\eta$ and an upper
bound on $h$. Thus \eqref{eq:hahn-haar-tail} gives
\begin{equation}
 \E_U\left\|\mathcal N_{\eta,\beta}^{\otimes m}(\rho_U)
              -\widetilde\rho_{D,U}\right\|_1^2
 \le a_D^{\mathrm H,\mathrm{th}}(\eta,\beta)
 \le\frac{C\eta^{D/2}}m,
 \label{eq:hahn-saturated-exponential-tail}
\end{equation}
which proves \eqref{eq:main-saturated-degree-bound}. The bound is
uniform over $\beta\ge0$, and the error is zero for $D\ge n$.

By Markov's inequality, the probability that the trace-norm error
exceeds $\varepsilon$ is at most $C\eta^{D/2}/(m\varepsilon^2)$.
Choosing
\begin{equation}
 D=\max\left\{0,
   \left\lceil
    \frac{2\log[C/(m\delta\varepsilon^2)]}{\log(1/\eta)}
   \right\rceil\right\}
 =O\!\left(\log\!\left[1+\frac1{m\delta\varepsilon^2}\right]\right)
 \label{eq:hahn-saturated-degree-choice}
\end{equation}
makes this probability at most $\delta$, proving
\eqref{eq:main-saturated-cutoff} and \eqref{eq:main-saturated-guarantee}.
In particular, $D=0$ suffices when $m\delta\varepsilon^2\ge C$.
At degree zero, \eqref{eq:hahn-thermal-approximation} gives
$\widetilde\rho_{0,U}=\mathcal N_{\eta,\beta}^{\otimes m}(u_n)$,
which is independent of $U$.
\end{proof}

\begin{corollary}[Fixed-density limit]
\label{cor:hahn-fixed-density}
Suppose $m/n\to C\ge1$, $n\to\infty$, with $0<\eta<1$ and
$\beta\ge0$ fixed. For every fixed $k$,
\begin{equation}
 \binom nkH_k\longrightarrow(1-C^{-1})^k,
 \qquad \E_UW_k(U)\longrightarrow
                  \left(\frac{C-1}{C+1}\right)^k,
 \label{eq:hahn-fixed-density-ideal}
\end{equation}
and
\begin{equation}
 \begin{aligned}
 \Gamma_k^{\rm th}
 &\longrightarrow
 \left[\frac{\eta^2(C+1)}
 {[C(1+t)+\eta](\eta+Ct)}\right]^k,\\
 E_k^{\rm th}&\longrightarrow a_{\rm th}^k,\qquad
 a_{\rm th}=\frac{\eta^2(C-1)}
 {[C(1+t)+\eta](\eta+Ct)}.
 \end{aligned}
 \label{eq:hahn-fixed-density-energy}
\end{equation}
For every fixed cutoff $D$,
\begin{equation}
 a_D^{\mathrm H,\mathrm{th}}(\eta,\beta)
 \longrightarrow\frac{a_{\rm th}^{D+1}}{1-a_{\rm th}},
 \qquad
 a_{\rm th}=\frac{s(C-1)}{C+s}\,
             \frac{\eta}{\eta+C(1-\eta)\beta}.
 \label{eq:hahn-fixed-density-tail}
\end{equation}
\end{corollary}
\begin{proof}
For fixed $k$, the finite sum in \eqref{eq:hahn-haar-ideal} gives
\begin{equation}
 \binom nkH_k
 \longrightarrow\sum_{j=0}^k(-1)^j\binom kjC^{-j}
 =(1-C^{-1})^k,
 \qquad
 \frac{\binom{m+k-1}{k}}{\binom{m+k-1}{k}+\binom{m+k-2}{k-1}}
 \frac{\rise{m+k}k}{\rise{m+n}k}
 \longrightarrow\left(\frac C{C+1}\right)^k.
 \label{eq:hahn-fixed-density-ideal-proof}
\end{equation}
This proves \eqref{eq:hahn-fixed-density-ideal}. We regard $w_{rK}$ in
\eqref{eq:hahn-positive-count-expansion} as the joint law of $R,K$.
Its generating function yields
\begin{equation}
 \begin{gathered}
 R\sim\Bin(n,s),\qquad
 \E[K-R\mid R]=(m+R)t,\qquad
 \operatorname{Var}(K-R\mid R)=(m+R)t(1+t),\\
 \frac Rn\xrightarrow{\Pr}s,\qquad
 \frac Kn\xrightarrow{\Pr}s+(C+s)t=\eta+Ct.
 \end{gathered}
 \label{eq:hahn-fixed-density-counts}
\end{equation}
Consequently,
\begin{equation}
 \frac{\nu_k(R)}{\sqrt{\nu_k(n)\nu_k(K)}}
 \longrightarrow
 \left[\frac{\eta^2(C+1)}
 {[C(1+t)+\eta](\eta+Ct)}\right]^{k/2}
 \quad\text{in squared mean}.
 \label{eq:hahn-fixed-density-multiplier}
\end{equation}
The convergence holds in squared mean because the ratio lies in
$[0,1]$, with value zero when $K<k$. By
\eqref{eq:hahn-thermal-multiplier}, $t_{n,K,k}$ is its conditional
expectation given $K$. Conditional expectation contracts the squared
error from the displayed constant, proving the limit for
$\Gamma_k^{\rm th}=\E[t_{n,K,k}^2]$. Multiplying by the ideal degree
limit proves \eqref{eq:hahn-fixed-density-energy}. Finally,
$0\le E_k^{\rm th}\le s^k$ permits dominated convergence in the
degree sum:
\begin{equation}
 \lim_{n\to\infty}a_D^{\mathrm H,\mathrm{th}}(\eta,\beta)
 =\sum_{k=D+1}^{\infty}a_{\rm th}^k
 =\frac{a_{\rm th}^{D+1}}{1-a_{\rm th}}.
 \label{eq:hahn-fixed-density-tail-proof}
\end{equation}
\end{proof}
For pure loss the second factor in \eqref{eq:hahn-fixed-density-tail}
is one, while a positive bath occupation makes it strictly smaller.
At $C=1$, the tail tends to zero already at $D=0$.
For size-dependent cutoffs, we use the finite-size bound
\eqref{eq:hahn-haar-tail}.

\subsection{Operator overlap formulas}
\label{sec:hahn-evaluation}

\subsubsection{The retained density as operator overlaps}
\label{sec:hahn-degree-overlaps}
We use the bosonic particle spaces and action $U_k$ from
Eq.~\eqref{eq:particle-mode-action}, and the maps $\cL,\cR$ from
Eq.~\eqref{eq:common-lowering-raising}. Throughout, $\Pi_k$ is the
projector in Eq.~\eqref{eq:common-projector} and $v_k$ is the irrep
dimension in Eq.~\eqref{eq:common-irrep-dimensions}. For comparison
with the formulas below, its bosonic denominator satisfies
$D_{k,q}=\fall{m+2k-2}{q}=\rise{m+2k-q-1}{q}$.

The Hahn basis identifies the polynomial degrees, while the common
projector evaluates each whole degree without resolving its basis
coefficients.

For an occupation $x$ with $|x|=L$ and $0\le k\le L$, define
\begin{equation}
 A_{x,k}=\sum_{|r|=k}\binom xr\ket r\bra r
        =\binom Lk\rho_x^{(k)},\qquad
 \binom xr=\prod_i\binom{x_i}{r_i}.
 \label{eq:bosonic-A}
\end{equation}
The equality is Eq.~\eqref{eq:rdm-moments} applied to $\ket x\bra x$;
$\rho_x^{(k)}$ is normalized, with $\rho_x^{(0)}=1$.
To express the common norm factor in the notation used for the
sampling algorithms, set
\begin{equation}
 c_{L,k}=\binom{m+k-1}{k}\frac{\rise{m+k}{k}}{\rise{m+L}{k}}.
 \label{eq:bosonic-lift-c}
\end{equation}

Thus the common inner-product identity
\eqref{eq:common-lift-inner-product} has factor
$\beta_{L,k}=h_{L,k}=\binom Lk/c_{L,k}$.

\begin{proposition}[Ideal and thermal Hahn densities as operator overlaps]
\label{prop:hahn-overlap-representation}
Let $\rho_{\rm in}^{(k)}$ be the normalized $k$-particle reduction of
an input with exactly $n$ particles. For $|x|=L$ and
$0\le k\le\min(n,L)$, we define
\begin{equation}
 f_{L,k,U}(x)=c_{L,k}\Tr\!\left[
 A_{x,k}U_k\Pi_k(\rho_{\rm in}^{(k)})U_k^\dagger\right].
 \label{eq:bosonic-degree-overlap}
\end{equation}

At $L=n$, this is the ideal relative-density component
$f_{n,k,U}(x)=d_np_{k,U}(x)$, where $d_n=\binom{m+n-1}{n}$.
For other $L$, it is the angular profile with the normalization
$c_{L,k}$; thermal noise supplies the additional scalar below.

For a retained output count $K$ with $p_K>0$ and
$1\le k\le\min(n,K)$, we put
\begin{equation}
 \lambda_{n,K,k}:=\frac{\sum_r w_{rK}\nu_k(r)}{p_K\nu_k(K)}
 =\eta^k\frac{\fall nk}{\fall Kk}
     \frac{p_{m+2k,n-k}(K-k)}{p_K},\qquad 0\le\lambda_{n,K,k}\le1.
 \label{eq:hahn-thermal-sketch-weight}
\end{equation}
For an integer $D\ge0$, the thermal approximation to the conditional
density relative to $u_K$ is
\begin{equation}
 f_{K,\le D}(x)
 =1+\sum_{k=1}^{\min(D,n,K)}\lambda_{n,K,k}c_{K,k}
 \Tr\!\left(A_{x,k}U_k\Pi_k(\rho_{\rm in}^{(k)})U_k^\dagger\right).
 \label{eq:hahn-enumerated-density}
\end{equation}
Each positive-degree term has zero mean under $u_K$. Thus
$f_{K,\le D}$ has reference mean one, and its signed distribution has
count marginal $p_K$.
\end{proposition}

\begin{proof}
For $L=n$, Eq.~\eqref{eq:common-output-degree-overlap}, multiplied
by $d_n$, gives Eq.~\eqref{eq:bosonic-degree-overlap}: using
$A_{x,k}=\binom nk\rho_x^{(k)}$, its coefficient is
$d_n(n!/k!)^2/(\alpha_{n,k}\binom nk)=c_{n,k}$.
The common polynomial lift \eqref{eq:common-polynomial-lift}
extends the same harmonic overlap to every count sector.
By linearity, the thermal action \eqref{eq:hahn-thermal-reference}
therefore sends its ideal relative density to
\begin{equation}
    \eta^k\frac{r_k(K)}{r_0(K)}\frac{c_{n,k}}{c_{K,k}}
 f_{K,k,U}(x)=\lambda_{n,K,k}f_{K,k,U}(x).
\end{equation}

Equation~\eqref{eq:hahn-count-coefficient-transfer} and
$h_{L,k}=\binom Lk/c_{L,k}$ identify this prefactor with the second
expression in Eq.~\eqref{eq:hahn-thermal-sketch-weight}.
Equation~\eqref{eq:hahn-count-shift} gives the first, an average of
$0\le\nu_k(r)/\nu_k(K)\le1$ with weights $w_{rK}/p_K$, proving
the bounds. Summing the retained degrees gives
Eq.~\eqref{eq:hahn-enumerated-density}. Orthogonality to constants
then gives reference mean one and count marginal $p_K$.
\end{proof}

\subsubsection{The reference proposal} \label{sec:hahn-reference-proposal}
For $|x|=K$, let $f_{D,U}(x)=f_{K,\le D}(x)$ be the density in
\eqref{eq:hahn-enumerated-density}. The proposal distribution is the
thermal reference
\begin{equation}
 \mu_{\rm th}(x)=p_Ku_K(x),\qquad |x|=K.
 \label{eq:hahn-rejection-reference}
\end{equation}
The positive count expansion \eqref{eq:hahn-positive-count-expansion}
gives an explicit way to sample it. With $t,s$ from
\eqref{eq:hahn-thermal-parameters}, we draw
\begin{equation}
 \begin{gathered}
 J\sim\Bin(n,s),\qquad
 \Pr(V=v\mid J=j)
 =\binom{m+j+v-1}{v}\frac{t^v}{(1+t)^{m+j+v}},\\
 K=J+V,\qquad X\mid K\sim u_K.
 \end{gathered}
 \label{eq:hahn-reference-proposal}
\end{equation}
Thus $V$ is negative binomial with shape $m+J$ and conditional mean
$(m+J)t$. For $t=0$, it is zero. To draw $X\sim u_K$, we choose
$m-1$ bar positions uniformly among $K+m-1$ positions and take the
numbers of photons in the intervening gaps. Each composition corresponds
to one choice of bars. The count-generating function is
\begin{equation}
 \E z^{J+V}
 =\sum_{j=0}^n\pi_jz^jD(z)^{-m-j}
 =D(z)^{-m}A(z)^n,
\end{equation}
which is the required law \eqref{eq:hahn-thermal-count-law}.

For the proposal-abort rule in Section~\ref{sec:precision}, we generate
$V$ as a sum of $m+J$ independent geometric variables of mean $t$,
with the accumulation stopped as soon as the partial count exceeds
$K_*$. The cutoff error is controlled by
\eqref{eq:proposal-count-coupling} and the Hahn count-tail bound
\eqref{eq:reference-count-tail}.

\subsection{Deterministic evaluation}
\label{sec:hahn-enumeration}

\subsubsection{Computing the diagonal entries} \label{sec:hahn-diagonal-evaluation}
Since $A_{x,k}$ is diagonal, only the diagonal of the other operator in
\eqref{eq:hahn-enumerated-density} is needed. We store its entries
\begin{equation}
 a_{k,r}:=\bra rU_k\Pi_k(\rho_{\rm in}^{(k)})U_k^\dagger\ket r,
 \qquad |r|=k.
 \label{eq:hahn-enumeration-harmonic-coefficient}
\end{equation}
The overlap is then $\sum_{|r|=k}a_{k,r}\binom xr$. We compute these
entries directly from occupation moments and the explicit projector.

For $|r|=k\le n$, the input reduction is the uniform mixture over
$k$ occupied input modes. Lemma~\ref{lem:rdm-moments} therefore gives
\begin{equation}
 \mu_r:=\Tr\!\left(\rho_U\prod_i\binom{\hat n_i}{r_i}\right)
 =\sum_{\substack{J\subseteq[n]\\|J|=k}}
 \frac{|\operatorname{Per}U[r,J]|^2}{r!}.
 \label{eq:hahn-enumerated-moment}
\end{equation}
The matrix $U[r,J]$ repeats output row $i$ exactly $r_i$ times,
$r!=\prod_i r_i!$, and $\mu_0=1$. The corresponding diagonal entry of
$U_k\rho_{\rm in}^{(k)}U_k^\dagger$ is $\mu_r/\binom nk$.
Using Ryser’s formula~\cite{ryser1963combinatorial}, $k\times k$ permanent can be computed using
\begin{equation}
 \operatorname{Per}V=(-1)^k\sum_{S\subseteq[k]}(-1)^{|S|}
                  \prod_{i=1}^k\sum_{j\in S}V_{ij}
 \label{eq:hahn-enumeration-permanent}
\end{equation}
in $O(k2^k)$ arithmetic operations by reusing subset sums.

Taking occupation diagonals in the reduced-projector expansion
\eqref{eq:common-reduced-projector-expansion}, and using the raising
coefficients in Eq.~\eqref{eq:common-occupation-projector-lift}
with input sector $k-j$ and output sector $k$, gives

\begin{equation}
 a_{k,r}
 =\sum_{j=0}^k
 \frac{(-1)^j\fall kj}
      {\rise{m+2k-j-1}{j}\binom n{k-j}}
 \sum_{\substack{s\le r\\|s|=k-j}}\binom rs\mu_s.
 \label{eq:hahn-enumeration-diagonal-projector}
\end{equation}
Thus the moment tables through degree $D$ determine the diagonal
entries of every retained exact-degree operator. Substituting them
in \eqref{eq:hahn-enumerated-density} evaluates the approximation
without constructing an orthogonal polynomial basis.

\subsubsection{Deterministic rejection sampling} \label{sec:hahn-deterministic-rejection}
\begin{lemma}[Uniform envelope for the deterministic Hahn density]
\label{lem:hahn-deterministic-envelope}
Let $d=\min(D,n)$. For every interferometer $U$ and every occupation
in the support of $\mu_{\rm th}$,
\begin{equation}
 \begin{aligned}
 |f_{D,U}(x)|
 &\le \mathcal M_D^{\rm H}
 :=1+\sum_{k=1}^{d}c_{n,k}\binom nk\\
 &\le1+\sum_{k=1}^{d}\binom{m+k-1}{k}\binom nk
 \le(d+1)(m+n)^{2d}.
 \end{aligned}
 \label{eq:hahn-deterministic-envelope}
\end{equation}
The same envelope holds for the ideal density before thermal loss.
\end{lemma}
\begin{proof}
For $|x|=n$, the test operator is positive and
\begin{equation}
 \Tr(A_{x,k})=\sum_{|r|=k}\binom xr=\binom nk,
 \qquad \|A_{x,k}\|_2\le\Tr(A_{x,k}).
\end{equation}
The sum is the coefficient of $z^k$ in
$\prod_i(1+z)^{x_i}=(1+z)^n$.
The normalized input reduction has Schatten $2$ norm at most one,
and the harmonic projection is a contraction. The overlap formula
\eqref{eq:bosonic-degree-overlap} therefore gives
\begin{equation}
 \begin{aligned}
 |f_{n,k,U}(x)|
 &\le c_{n,k}\|A_{x,k}\|_2
             \|\Pi_k(\rho_{\rm in}^{(k)})\|_2\\
 &\le c_{n,k}\binom nk.
 \end{aligned}
\end{equation}
Summing the retained degrees proves the ideal bound. Since $k\le n$,
\eqref{eq:bosonic-lift-c} gives
$c_{n,k}\le\binom{m+k-1}{k}$. Each binomial in the last sum in
\eqref{eq:hahn-deterministic-envelope} is at most $(m+n)^k$.

Let $f^{\rm ideal}_{D,U}$ denote the ideal truncated density and
let $P_{\eta,\beta}(y\mid x)$ be the thermal transition kernel on
occupations. Dividing the noisy signed mass by the propagated
reference gives
\begin{equation}
 f_{D,U}(y)
 =\sum_{|x|=n}\frac{u_n(x)P_{\eta,\beta}(y\mid x)}
                        {\mu_{\rm th}(y)}f^{\rm ideal}_{D,U}(x).
\end{equation}
For $\mu_{\rm th}(y)>0$, the coefficients are nonnegative and sum to
one. This conditional average preserves the ideal envelope.
\end{proof}

We use Algorithm~\ref{alg:rejection-sampling} with proposal
\eqref{eq:hahn-reference-proposal}, the exact value $f_{D,U}(X)$,
and a rejection threshold $M\ge\mathcal M_D^{\rm H}$.
Lemma~\ref{lem:rejection-sampling} applies with $\tau_M=0$ and the
approximation certificate \eqref{eq:hahn-haar-tail}. For a fixed $U$,
we use the corresponding deterministic approximation certificate.

\subsubsection{Arithmetic cost}
Let $d_* =\min(D,n)$ and
$Q_D=\sum_{k=0}^{d_*}\binom{m+k-1}{k}=\binom{m+d_*}{d_*}$. There are
$\binom{m+k-1}{k}\binom nk$ permanents for the degree-$k$ moment table.
For $|r|=k$, the number of suboccupations $s\le r$ is
$\prod_i(r_i+1)\le2^k$. Equation
\eqref{eq:hahn-enumeration-diagonal-projector} can therefore be
computed in $O(k2^k)$ additional operations per $r$, after tabulating
its scalar factors. A sufficient preprocessing bound is
\begin{equation}
 O\!\left(\sum_{k=1}^{d_*}k2^k\binom{m+k-1}{k}
                 \left[\binom nk+1\right]\right)
 +\poly(m,n,K_*,D).
 \label{eq:hahn-direct-preprocessing}
\end{equation}
The polynomial term includes the count probabilities and thermal
factors for retained counts. The moment and projected-diagonal tables
use $O(Q_D)$ entries, and the count and transfer tables add
$O((K_*+1)(d_*+1))$ entries, besides the mode matrix.

For a proposed occupation, tabulating the required binomial values
and summing the retained monomials costs
$O(md_*+d_*Q_D+1)$ operations. The proposal itself has polynomial
cost in $m,n,K_*$. A run with at most $A$ attempts therefore uses
\begin{equation}
 A\left[O(md_*+d_*Q_D+1)+\poly(m,n,K_*,D)\right]
 \label{eq:hahn-direct-sample-work}
\end{equation}
operations after preprocessing. With
$A=O(M\log(1/\epsilon_{\rm stop}))$ and
$M=\mathcal M_D^{\rm H}$, this is $m^{O(D)}$ times the displayed
polynomial factors, without an ensemble or flatness assumption for
the evaluation cost. The approximation guarantee still uses the
applicable degree-tail certificate.

\subsection{Randomized evaluation}
\label{sec:hahn-sketch}

We estimate the retained density in
Proposition~\ref{prop:hahn-overlap-representation} at each proposed
occupation. The input and output colors below are generated afresh
for each point estimate.

\subsubsection{Random output operators for bosonic occupations}
\label{sec:bosonic-color-sketch}
We fix a degree $k\ge1$ and an output occupation $x$ with $|x|\ge k$.
Both bosonic samplers use the following random approximation of
$A_{x,k}$ from \eqref{eq:bosonic-A}.

For each mode $i$ we introduce $x_i$ bookkeeping slots, with weights
$\lambda_{i,a}>0$ equal to reciprocals of the roots of the ordinary
Laguerre polynomial $L_{x_i}$. Their defining identity is
\begin{equation}
 \prod_{a=1}^{x_i}(1+\lambda_{i,a}t)
 =L_{x_i}(-t)
 =\sum_{r=0}^{x_i}\frac{\binom{x_i}{r}}{r!}t^r.
 \label{eq:bosonic-laguerre}
\end{equation}
All weights are positive\ \cite[Eqs.~18.5.12 and~18.16.9]{NIST:DLMF}, and their sum in a mode is $x_i$.
For a set $Q$ of slots, let $r_i(Q)$ count its slots in mode $i$ and
let $\lambda_Q$ be their weight product. Coefficient comparison gives
\begin{equation}
 A_{x,k}=\sum_{|Q|=k}r(Q)!\lambda_Q\ket{r(Q)}\bra{r(Q)}.
\end{equation}
Following the color-coding principle~\cite{alon1995color}, we independently color every slot uniformly in $[k]$. The probability that
a fixed $k$-set has all colors is $p_k=k!/k^k$. We define
\begin{equation}
 \widehat A_{x,g}=\frac1{p_k}
 \sum_{\substack{|Q|=k\\g(Q)=[k]}}
 r(Q)!\lambda_Q\ket{r(Q)}\bra{r(Q)}.
 \label{eq:bosonic-output-sketch}
\end{equation}
The retain-and-rescale coloring method and inverse-determinant
evaluation are adapted from Go and Jeong \cite{go2026polynomial}. The weighted
slot decomposition keeps repeated physical modes in the occupation
contraction.

\begin{lemma}[Moments of the bosonic output sketch]
\label{lem:bosonic-output-sketch}
For $k\ge1$ and $|x|\ge k$, the operator in
\eqref{eq:bosonic-output-sketch} satisfies
\begin{equation}
 \E_g\widehat A_{x,g}=A_{x,k},\qquad
 \E_g\|\Pi_k(\widehat A_{x,g})\|_2^2
 \le p_k^{-1}\|A_{x,k}\|_2^2.
 \label{eq:bosonic-output-second}
\end{equation}
\end{lemma}
\begin{proof}
Each fixed $k$-set is retained with probability $p_k$, so the slot
expansion of $A_{x,k}$ proves unbiasedness. Positivity of the weights
gives the pointwise diagonal inequality
$0\preceq\widehat A_{x,g}\preceq p_k^{-1}A_{x,k}$.
Multiplying by $\widehat A_{x,g}$, taking its trace and expectation,
and using unbiasedness and projection contraction proves the second
moment bound. Independence between different retained sets is unnecessary.
\end{proof}

\begin{lemma}[Uniform composition moments]
\label{lem:uniform-composition-moments}
Let $u_L$ be uniform on $\{x\in\mathbb N_0^m:|x|=L\}$, where
$m\ge1$ and $L\ge0$. For $r\in\mathbb N_0^m$,
\begin{equation}
 \E_{X\sim u_L}\prod_i\binom{X_i}{r_i}
 =\frac{\fall L{|r|}}{\rise m{|r|}}.
 \label{eq:hahn-factorial-reference}
\end{equation}
\end{lemma}
\begin{proof}
For $|r|\le L$, multiplying
$\sum_{x_i\ge0}\binom{x_i}{r_i}z^{x_i}
 =z^{r_i}(1-z)^{-r_i-1}$ over the modes gives
\begin{equation}
 \sum_{|x|=L}\prod_i\binom{x_i}{r_i}
 =[z^{L-|r|}](1-z)^{-m-|r|}
 =\binom{m+L-1}{L-|r|}.
\end{equation}
Dividing by the number $\binom{m+L-1}{L}$ of occupations yields
\eqref{eq:hahn-factorial-reference}. Both sides vanish for $|r|>L$.
\end{proof}

\begin{lemma}[Exact uniform-composition frame norm]\label{lem:bosonic-frame}
For $L\ge k$,
\begin{equation}
 \begin{gathered}
 \E_{X\sim u_L}\|A_{X,k}\|_2^2
 =\binom Lk \mathcal B_k(m,L),\\
 \mathcal B_k(m,L)=\sum_{a=0}^k\binom ka\binom{L-k}{a}
                    \frac{a!}{\rise{(m+1)/2}{a}},\\
 \mathcal B_k(m,L)\le\left(1+\frac{2L}{m+1}\right)^k.
 \end{gathered}                                      \label{eq:bosonic-frame}
\end{equation}
In particular $\mathcal B_k(m,L)\le3^k$ when $L\le m$.
\end{lemma}
\begin{proof}
We generate $X$ by drawing $P\sim\operatorname{Dirichlet}(1,\ldots,1)$
and then $L$ independent mode labels conditional on $P$.
The Dirichlet monomial moment is
$\E\prod_iP_i^{x_i}=\prod_i x_i!/\rise mL$ for $|x|=L$.
Multiplication by $L!/\prod_i x_i!$ gives $\Prob(X=x)=\binom{m+L-1}{L}^{-1}$.
The purity of $\rho_X^{(k)}$ is the probability that two
independent uniform $k$-subsets of these labels have equal histograms.
They overlap in $k-a$ positions with probability
$\binom ka\binom{L-k}{a}/\binom Lk$. After removing common positions,
the two disjoint groups of $a$ labels have equal histograms with
averaged probability
\begin{equation}
 \begin{aligned}
 &\frac{(a!)^2}{\rise m{2a}}
 \sum_{|z|=a}\prod_i\binom{2z_i}{z_i}\\
 &\quad=\frac{(a!)^2}{\rise m{2a}}\frac{4^a\rise{m/2}{a}}{a!}\\
 &\quad=\frac{a!}{\rise{(m+1)/2}{a}}.
 \end{aligned}
\end{equation}
The middle equality is the coefficient of $t^a$ in
$(1-4t)^{-m/2}$. We multiply the resulting purity by $\binom Lk^2$ to
obtain the identity. Finally
$\binom{L-k}{a}a!\le L^a$ and
$\rise{(m+1)/2}{a}\ge((m+1)/2)^a$ give the stated binomial bound.
\end{proof}

\subsubsection{Input colors and determinant evaluation}\label{sec:hahn-color-determinant}
We fix a candidate with $|x|=L$ and $1\le k\le\min(L,n)$, where
$n\le m$. Here $L$ is the final thermal photon count and can exceed
$n$. The output sketch is \eqref{eq:bosonic-output-sketch}. We independently
color the occupied input modes $[n]$ uniformly in $[k]$, obtaining $h$,
and write $n_c=|h^{-1}(c)|$. We define
\begin{equation}
 \widehat\rho_h^{(k)}=
 \frac1{p_k\binom nk}
 \sum_{\substack{J\subseteq[n],\ |J|=k\\h(J)=[k]}}\ket J\bra J.
 \label{eq:hahn-input-sketch}
\end{equation}
This operator is stored implicitly through the $n$ colors. Its many
retained subsets must still be contracted without enumeration.
The random scalar to compute is
\begin{equation}
 Z_{L,k}(U,x;h,g)
 =c_{L,k}\Tr\!\left[
 \widehat A_{x,g}U_k\Pi_k(\widehat\rho_h^{(k)})U_k^\dagger\right].
 \label{eq:hahn-Z}
\end{equation}
Every term of the projector uses the same colors, so projection
contraction remains available within each realization.

\begin{lemma}[Input second moment and unbiased Hahn overlap]
\label{lem:hahn-input-sketch}
For $1\le k\le\min(L,n)$, with independent input and output colors,
\begin{equation}
 \E_h\|\widehat\rho_h^{(k)}\|_2^2
 =\frac1{p_k\binom nk}.
 \label{eq:hahn-input-second}
\end{equation}
For each fixed $U$ and $x$ with $|x|=L$,
\begin{equation}
 \E_{h,g}Z_{L,k}(U,x;h,g)=f_{L,k,U}(x).
 \label{eq:hahn-Z-unbiased}
\end{equation}
\end{lemma}
\begin{proof}
A fixed input $k$-set is retained with probability $p_k$, hence
$\E_h\widehat\rho_h^{(k)}=\rho_{\rm in}^{(k)}$.
Orthogonality of the occupation projectors then gives
\eqref{eq:hahn-input-second}. Independence, linearity, and
Lemma~\ref{lem:bosonic-output-sketch} identify the mean of
\eqref{eq:hahn-Z} with the overlap in
Proposition~\ref{prop:hahn-overlap-representation}.
\end{proof}

The full-thermal degree contribution is $\lambda_{n,L,k}Z_{L,k}$,
with the scalar in \eqref{eq:hahn-thermal-sketch-weight}, which is
constant within the count sector.

\begin{proposition}[Finite determinant formula]\label{prop:hahn-det}
We introduce two lists of commuting variables, $y_1,\ldots,y_k$ and
$\xi_1,\ldots,\xi_k$, with $y_c^2=\xi_c^2=0$, and an ordinary variable $t$ retained through
degree $k$. We define
\begin{equation}
 \begin{gathered}
 B_h(y)_{ij}=\sum_{\ell=1}^n y_{h(\ell)}U_{i\ell}\overline{U_{j\ell}},\\
 W_x(\xi)_{ii}=\sum_{a=1}^{x_i}\lambda_{i,a}\xi_{g(i,a)},\\
 P(y,\xi,t)
 =\det\!\left[I_m-W_x(\xi)(B_h(y)+tI_m)\right]^{-1}.
 \end{gathered}
 \label{eq:hahn-det}
\end{equation}
For $C\subseteq[k]$, let $y_C=\prod_{c\in C}y_c$ and $j=k-|C|$.
Then
\begin{equation}
 Z_{L,k}=\frac{c_{L,k}}{\binom nk p_k^2}
 \sum_{C\subseteq[k]}\frac{(-1)^j j!\prod_{c\notin C}n_c}
                              {\rise{m+2k-j-1}{j}}
 \times[\xi_1\cdots\xi_k y_Ct^j]P.
 \label{eq:hahn-det-value}
\end{equation}
It costs $\poly(m,n,L,k)2^{O(k)}$ arithmetic operations.
\end{proposition}
\begin{proof}
Let $\sigma_C$ be the sum of occupation projectors choosing one input
mode of each color in $C$. Lowering the rainbow input gives
\begin{equation}
 \cL^j\widehat\rho_h^{(k)}=
 \frac{j!}{p_k\binom nk}
 \sum_{|C|=k-j}\left(\prod_{c\notin C}n_c\right)\sigma_C.
\end{equation}
The factorial counts deletion orders, and the product counts choices of
the deleted colors. We put $H=B_h(y)+tI_m$. MacMahon's master theorem
\cite[Section~2.3, Eq.~(20)]{chabaud2022quantum} gives
\begin{equation}
 P=\sum_{r\in\mathbb N_0^m}
 \frac{\operatorname{per}(H_{r,r})}{r!}\prod_iW_x(\xi)_{ii}^{r_i},
 \label{eq:hahn-macmahon}
\end{equation}
where $H_{r,r}$ repeats row and column $i$ exactly $r_i$ times.
The identity holds for arbitrary matrices and formal variables, so it
remains valid after imposing the color relations.

Let $z_1,\ldots,z_m,\bar z_1,\ldots,\bar z_m$ be independent
commuting formal variables, with $z^\dagger H z$ abbreviating
$\sum_{i,l}\bar z_iH_{il}z_l$. For an operator $A$ on a fixed
particle sector, we write $A_{rs}=\bra rA\ket s$. Its coherent polynomial
is $\sum_{r,s}A_{rs}\bar z^rz^s/\sqrt{r!s!}$.
Expanding the exponential by input color gives
\begin{equation}
\begin{aligned}
 [y_C]e^{z^\dagger B_h(y)z}
 &=\prod_{c\in C}\sum_{\substack{\ell\in[n]\\h(\ell)=c}}
   \left(\sum_i\bar z_iU_{i\ell}\right)
   \left(\sum_l\overline{U_{l\ell}}z_l\right)\\
 &=\sum_{|r|=|s|=|C|}
   \bra rU_{|C|}\sigma_CU_{|C|}^\dagger\ket s
   \frac{\bar z^rz^s}{\sqrt{r!s!}}.
\end{aligned}
\label{eq:hahn-colored-coherent-polynomial}
\end{equation}
The last equality follows because the product chooses one input mode
of each color. These modes are distinct, so each choice contributes
the coherent polynomial of its rotated occupation projector. The
unrotated projectors sum to $\sigma_C$.

The occupation-basis action
$\hat a_i\ket r=\sqrt{r_i}\ket{r-e_i}$ shows that
$\cR(A)=\sum_i \hat a_i^\dagger A \hat a_i$ multiplies the coherent polynomial
by $\sum_i\bar z_i z_i$. Since $H=B_h(y)+tI_m$,
\begin{equation}
 [y_Ct^j]e^{z^\dagger H z}
 =\frac{(\sum_i\bar z_i z_i)^j}{j!}
   [y_C]e^{z^\dagger B_h(y)z}.
 \label{eq:hahn-coherent-raising}
\end{equation}
Thus this coefficient represents
$U_k\cR^j(\sigma_C)U_k^\dagger/j!$, using the unitary covariance
of $\cR$ and $k=|C|+j$.
The Taylor expansion also gives
$[\bar z^rz^r]e^{z^\dagger H z}
=\operatorname{per}(H_{r,r})/(r!)^2$.
Taking the coefficient of the normalized diagonal monomial
$\bar z^rz^r/r!$ therefore yields, for $|r|=k$,
\begin{equation}
 [y_Ct^j]\frac{\operatorname{per}(H_{r,r})}{r!}
 =\frac1{j!}\bra rU_k\cR^j(\sigma_C)U_k^\dagger\ket r.
\end{equation}
Expanding each mode weight counts all $r_i!$ orders of its selected
Laguerre slots. Therefore
\begin{equation}
 [\xi_1\cdots\xi_k]\prod_iW_x(\xi)_{ii}^{r_i}
 =p_k\bra r\widehat A_{x,g}\ket r.
 \label{eq:hahn-output-color-coefficient}
\end{equation}
Combining these identities gives
\begin{equation}
 [\xi_1\cdots\xi_k y_Ct^j]P
 =\frac{p_k}{j!}\Tr\!\left[
 \widehat A_{x,g}U_k\cR^j(\sigma_C)U_k^\dagger\right].
\end{equation}
We insert this identity and the lowering formula into
\eqref{eq:common-projector} to obtain \eqref{eq:hahn-det-value}.

For the cost, we set $K=W_x(\xi)(B_h(y)+tI)$. Every entry contains a
positive output-color degree, so $K^{k+1}=0$. Thus
\begin{equation}
 P=\sum_{a=0}^k\frac1{a!}
       \left(\sum_{b=1}^k\frac{\Tr K^b}{b}\right)^a.
 \label{eq:hahn-trace-log}
\end{equation}
The coefficient ring has at most $(k+1)4^k$ entries. Direct array
multiplication, truncated after each operation, is single exponential
in $k$. Forming the matrices costs $O(m^2n+L)$, after which polynomially
many ring matrix operations and $2^k$ coefficient reads suffice.
\end{proof}

\subsubsection{Integrated second moments} \label{sec:hahn-point-moments}

\begin{proposition}[Integrated Hahn second moment]\label{prop:hahn-variance}
With independent colors and Haar $U$, for every $L\ge k$,
\begin{equation}
 \begin{aligned}
 S_{L,k}&:=\E_{U,X\sim u_L,h,g}|Z_{L,k}|^2\le V_{L,k},\\
 V_{L,k}&:=\frac{c_{L,k}^2}{v_kp_k^2}
       \frac{\binom Lk}{\binom nk}\mathcal B_k(m,L).
 \end{aligned}
 \label{eq:hahn-variance}
\end{equation}
The full-thermal contributions satisfy
\begin{equation}
 \sum_{L\ge k}p_L\lambda_{n,L,k}^2S_{L,k}
 \le\sum_{r=k}^n \pi_rV_{r,k}
 \le2(3e^2s)^k,
 \qquad \pi_r=\binom nrs^r(1-s)^{n-r}.
 \label{eq:hahn-thermal-sketch-second}
\end{equation}
\end{proposition}
\begin{proof}
By Eqs.~\eqref{eq:ker_ident}
and~\eqref{eq:common-irrep-dimensions}, both projected operators
in Eq.~\eqref{eq:hahn-Z} lie in the same irreducible space of
dimension $v_k$. Conditioning on $X,h,g$, we may therefore apply
Schur orthogonality \eqref{eq:schur-operator-second-moment}.
The independence of the input and output colors separates the
norm factors. Orthogonal projection contracts the input norm
by Corollary~\ref{cor:common-projectors};
Lemmas~\ref{lem:hahn-input-sketch}, \ref{lem:bosonic-output-sketch},
and~\ref{lem:bosonic-frame} then give
\begin{equation}
 \begin{aligned}
 \E_{U,X,h,g}|Z_{L,k}|^2
 &=\frac{c_{L,k}^2}{v_k}
   \E_h\|\Pi_k(\widehat\rho_h^{(k)})\|_2^2
   \E_{X,g}\|\Pi_k(\widehat A_{X,g})\|_2^2\\
 &\le\frac{c_{L,k}^2}{v_kp_k^2\binom nk}
       \E_{X\sim u_L}\|A_{X,k}\|_2^2
 =V_{L,k}.
 \label{eq:hahn-variance-factors}
\end{aligned}
\end{equation}
To control large final counts, we define the normalized frame moment
\begin{equation}
 \mathcal F_{L,k}:=\E_{X\sim u_L}\|\rho_X^{(k)}\|_2^2
 =\frac{\mathcal B_k(m,L)}{\binom Lk}.
 \label{eq:hahn-normalized-frame}
\end{equation}
For $k\le r\le L$, we couple $X_L\sim u_L$ to $X_r$ by uniformly
selecting $r$ particles. For $|y|=r$,
Lemma~\ref{lem:uniform-composition-moments} gives
\begin{equation}
 \Prob(X_r=y)
 =\frac{\E_{X_L\sim u_L}\prod_i\binom{X_{L,i}}{y_i}}{\binom Lr}
 =\frac{r!}{\rise mr}=\frac1{\binom{m+r-1}{r}}.
 \label{eq:hahn-uniform-thinning}
\end{equation}
Thus $X_r\sim u_r$.
Composing this selection with a uniform $k$-particle selection gives
$\E[\rho_{X_r}^{(k)}\mid X_L]=\rho_{X_L}^{(k)}$. Jensen's inequality
for the squared Schatten norm therefore gives
$\mathcal F_{L,k}\le\mathcal F_{r,k}$.
By \eqref{eq:bosonic-lift-c},
$c_{L,k}\binom Lk/\nu_k(L)=\binom{m+k-1}{k}\rise{m+k}k/k!$ is independent of $L$.
Combining these two facts yields
\begin{equation}
 \left(\frac{\nu_k(r)}{\nu_k(L)}\right)^2V_{L,k}
 \le V_{r,k}.
 \label{eq:hahn-frame-attenuation}
\end{equation}

We use the weights $w_{rL}$ of \eqref{eq:hahn-positive-count-expansion},
with $K=L$. That identity gives $\sum_r w_{rL}=p_L$ and
$\sum_{L\ge r}w_{rL}=\pi_r$. For $p_L>0$,
\eqref{eq:hahn-thermal-sketch-weight} gives
$\lambda_{n,L,k}=\sum_r(w_{rL}/p_L)\nu_k(r)/\nu_k(L)$,
with terms $r<k$ set to zero. Zero-probability sectors contribute zero.
Jensen's inequality and
\eqref{eq:hahn-frame-attenuation} therefore yield
\begin{equation}
 \begin{aligned}
 \sum_{L\ge k}p_L\lambda_{n,L,k}^2V_{L,k}
 &\le\sum_{L\ge k}\sum_{r=k}^{\min(n,L)}
 w_{rL}\left(\frac{\nu_k(r)}{\nu_k(L)}\right)^2V_{L,k}\\
 &\le\sum_{r=k}^n \pi_rV_{r,k}.
 \end{aligned}
 \label{eq:hahn-thermal-variance-contraction}
\end{equation}
Finally $r\le n\le m$ in this last sum. Here $c_{r,k}\le \binom{m+k-1}{k}$ and
$\binom{m+k-2}{k-1}/\binom{m+k-1}{k}=k/(m+k-1)\le2/3$, so $c_{r,k}^2/v_k<2$.
Also $p_k^{-1}\le e^k$ and $\mathcal B_k(m,r)\le3^k$.
Retaining the factor $\binom rk/\binom nk$ until averaging gives
$\sum_r \pi_r\binom rk/\binom nk=s^k$, which proves the second bound
in \eqref{eq:hahn-thermal-sketch-second}.
\end{proof}

For a proposed occupation $x$ with $K=|x|$, let
$\xi=(h_k,g_k)_{k=1}^{\min(D,n,K)}$ contain independent input and
output colors for every retained degree. Conditional on $x$, we define
the point estimator
\begin{equation}
 Z_D(U,x;\xi)
 =1+\sum_{k=1}^{\min(D,n,K)}
       \lambda_{n,K,k}Z_{K,k}(U,x;h_k,g_k).
 \label{eq:hahn-point-estimator}
\end{equation}
The variance budget is
\begin{equation}
 \begin{aligned}
 S_D^{\rm H}
 &:=\sum_{k=1}^{\min(D,n)}\sum_{j=k}^n\pi_j
   \frac{c_{j,k}^2}{v_kp_k^2}
   \frac{\binom jk}{\binom nk}\mathcal B_k(m,j)\\
 &\le2\sum_{k=1}^{\min(D,n)}(3e^2s)^k.
 \end{aligned}
 \label{eq:hahn-sketch-budget}
\end{equation}

\begin{lemma}[Hahn point-estimator moments]
\label{lem:hahn-point-estimator}
The point estimator $Z_D$ is real. For every fixed $U,x$ in the
support of $\mu_{\rm th}$,
\begin{equation}
 \E_\xi Z_D(U,x;\xi)=f_{D,U}(x).
 \label{eq:hahn-point-mean}
\end{equation}
For Haar $U$, the integrated variance obeys
\begin{equation}
 \E_{U,X\sim\mu_{\rm th}}\Var_\xi Z_D(U,X;\xi)\le S_D^{\rm H}.
 \label{eq:hahn-point-variance}
\end{equation}
The deterministic density satisfies
\begin{equation}
 \E_U\|f_{D,U}\|_{L^2(\mu_{\rm th})}^2
 \le\min\left\{C_{m,n},\frac1{1-s}\right\}.
 \label{eq:hahn-reference-second-moment}
\end{equation}
\end{lemma}
\begin{proof}
The input and output sketches are Hermitian, and the harmonic
projection and unitary conjugation preserve Hermiticity. Their trace
overlap is therefore real. Lemma~\ref{lem:hahn-input-sketch} makes
each degree contribution unbiased. Their weighted sum is exactly
\eqref{eq:hahn-enumerated-density}, proving \eqref{eq:hahn-point-mean}.
For fixed $U,x$, the independent degree colors give
\begin{equation}
 \begin{aligned}
 \Var_\xi Z_D(U,x;\xi)
 &=\sum_{k=1}^{\min(D,n,K)}\lambda_{n,K,k}^2
       \Var_{h_k,g_k}Z_{K,k}(U,x;h_k,g_k)\\
 &\le\sum_{k=1}^{\min(D,n,K)}\lambda_{n,K,k}^2
       \E_{h_k,g_k}|Z_{K,k}(U,x;h_k,g_k)|^2.
 \end{aligned}
\end{equation}
Averaging over $X\sim\mu_{\rm th}$ and Haar $U$, then applying
\eqref{eq:hahn-thermal-sketch-second}, bounds the last sum by
$S_D^{\rm H}$.

Orthogonality after thermal loss gives
$\|f_{D,U}\|_{L^2(\mu_{\rm th})}^2
 =1+\sum_{k\le\min(D,n)}\Gamma_k^{\rm th}W_k(U)$.
The bound $E_k^{\rm th}\le s^k$ in
\eqref{eq:hahn-thermal-haar-energy} yields the geometric-series bound.
The other bound uses $\Gamma_k^{\rm th}\le1$ and Haar flatness
\eqref{eq:hahn-haar-flatness}.
\end{proof}

\subsubsection{Rejection sampling with fresh point estimates} \label{sec:hahn-fresh-estimates}
For each proposal $X\sim\mu_{\rm th}$ from
\eqref{eq:hahn-reference-proposal}, let $K=|X|$. Each point evaluation
requested by Algorithm~\ref{alg:adaptive-rejection} uses new independent
input and output colors in \eqref{eq:hahn-point-estimator}. The degree
contributions are computed by \eqref{eq:hahn-det-value} and summed with
their thermal weights. Theorem~\ref{thm:final-sketch} combines
Lemma~\ref{lem:hahn-point-estimator} and \eqref{eq:hahn-haar-tail}
with Theorem~\ref{thm:adaptive-rejection}, using
$C_{\rm ref}=\min\{C_{m,n},(1-s)^{-1}\}$ and $S_D=S_D^{\rm H}$.

Only proposals with $K\le K_*$ are evaluated, with the cutoff rule
and its error specified in Section~\ref{sec:precision}.
Proposition~\ref{prop:hahn-det} evaluates each retained degree in
$\poly(m,n,K_*,D)2^{O(D)}$ arithmetic operations. With at most $A$
proposals and $E_{\max}$ additional point evaluations, the total cost is
\begin{equation}
 (A+E_{\max})\poly(m,n,K_*,D,J)2^{O(D)}.
 \label{eq:hahn-rejection-work}
\end{equation}
Here $J$ is the stage limit in \eqref{eq:adaptive-parameters}. The
cost includes the comparisons, proposal generation, and coefficient arithmetic.
The count probabilities and scalar thermal weights can be tabulated in
$\poly(m,n,K_*,D)$ work. Degrees and point evaluations are processed
successively, so the working storage is
$\poly(m,n,K_*,D)2^{O(D)}$. The Laguerre weights are computed once for
each proposed occupation and reused in its subsequent evaluations.
Section~\ref{sec:precision} controls their numerical precision.

\section{Gaussian boson sampling}
\label{sec:gbs-model}

The input is a product of $m\ge2$ squeezed vacua with zero displacement
and common squeezing parameter $r>0$, with mean occupation
$\bar n=\sinh^2r$ per mode. A passive interferometer $U$ is followed
by homogeneous thermal loss with transmissivity $0<\eta<1$ and bath
occupation $\beta\ge0$. We write $\rho_{\mathrm{ideal},U}$ for the
dephased output before loss and $P_U(x)$ for the measured occupation
law after loss.

We first construct the polynomial degrees from the bosonic adjoint
components and their multiplicity across particle sectors. We then
verify the low-degree approximation theorem and derive the overlaps
used by both evaluation algorithms in the common roadmap of
Figure~\ref{fig:dependency-roadmap}.
\subsection{Representation theory and polynomial degrees}
\label{sec:gbs-representation}

On the full bosonic Fock space
$\mathcal F_+=\overline{\bigoplus_{K\ge0}\mathcal H_K}$,
the particle-number-preserving operator space decomposes under passive
transformations as
\cite[Theorem~3, Appendix~B.4]{mhiri2026boson}
\begin{equation}
 \begin{aligned}
 \mathcal W_+:=\bigoplus_{K\ge0}\mathcal W_K
 &=\bigoplus_{K\ge0}\bigoplus_{k=0}^{K}\mathcal V_{K,k}
 \simeq\bigoplus_{k\ge0}\mathcal V_{k,k}\otimes\mathcal M_k,\\
 \mathcal M_k&:=\operatorname{span}\{\mathbf e_K:K\ge k\}.
 \end{aligned}
 \label{eq:gbs-multiplicity-decomposition}
\end{equation}
The multiplicity space $\mathcal M_k$ records the particle sectors
carrying equivalent copies of degree $k$, identified by raising in
Eq.~\eqref{eq:raise_ident}. Passive transformations act on
$\mathcal V_{k,k}$ and trivially on $\mathcal M_k$.
The operator sums are understood algebraically, with completion
in the reference norm when needed.

Occupation dephasing $\mathcal D$ from Eq.~\eqref{eq:occupation-dephasing}
removes every block with $K\ne L$ and selects the zero-weight part
of each remaining irrep. Applying the diagonal decomposition
\eqref{eq:common-diagonal-decomposition} therefore gives
\begin{equation}
 \mathcal D(\mathcal W_+)
 =\bigoplus_{K\ge0}\bigoplus_{k=0}^{K}(\mathcal V_{K,k})_0
 \simeq\bigoplus_{k\ge0}(\mathcal V_{k,k})_0\otimes\mathcal M_k.
 \label{eq:gbs-dephased-multiplicities}
\end{equation}
Thus the measured operator space retains the multiplicity coordinate
while replacing each irrep by its diagonal component.

Choose a Hilbert--Schmidt orthonormal basis of self-adjoint operators
$\{C_{k,\alpha}\}$ in $(\mathcal V_{k,k})_0$.
Its polynomial lifts $h_{C_{k,\alpha}}$ give
angular degree-$k$ bases at fixed total count, by
Eq.~\eqref{eq:common-polynomial-lift}. Orthogonal polynomials of
radial degree $\ell$ in the multiplicity coordinate $K$ then describe
variation across sectors. Their products have total occupation degree
$d=k+\ell$. We next determine their weights and normalization from
the thermal reference.

For $c>0$, let $g_c$ be the geometric law of mean $c$ and let $\tau_c$
be the product thermal reference:
\begin{equation}
 g_c(x)=\frac{c^x}{(1+c)^{x+1}},\qquad
 \mu_c(x)=\prod_{i=1}^m g_c(x_i),\qquad
 \tau_c=\sum_{x\in\mathbb N_0^m}\mu_c(x)\ket x\bra x.
 \label{eq:gbs-thermal-reference}
\end{equation}
We use the input and output reference parameters
\begin{equation}
 a=\bar n+\frac{1+\bar n}{m},\qquad
 b=\eta a+(1-\eta)\beta.
 \label{eq:gbs-parameters}
\end{equation}

For $\nu>0$ and $y\in\mathbb N_0$, we define the negative-binomial weights
\begin{equation}
 w_{\nu,b}(y)=\frac{\rise\nu y}{y!}\frac{b^y}{(1+b)^{\nu+y}}.
 \label{eq:gbs-count-reference}
\end{equation}
We set $w_{\nu,b}(y)=0$ for integer $y<0$. Under $\mu_b$, $K$ has law
$w_{m,b}$. Conditional on $K$, the occupation
has the uniform law $u_K$ on $\{x\in\mathbb N_0^m:|x|=K\}$.

For angular degree $k$, the norm transfer
\eqref{eq:common-lift-inner-product}, with the bosonic factor
\eqref{eq:common-diagonal-beta}, gives the radial reference weight:
with $v=b(1+b)$,
\begin{equation}
 w_{m,b}(K)\beta_{K,k}
 =v^k w_{m+2k,b}(K-k),\qquad K\ge k.
 \label{eq:gbs-radial-tilt}
\end{equation}
Thus the radial coordinate $y=K-k$ carries the negative-binomial
weight $w_{m+2k,b}$, whose orthogonal polynomials are the Meixner
family below.

For $\nu>0$, $y\in\mathbb R$, and $\ell\in\mathbb N_0$, we define
the radial polynomials $H_\ell^{(\nu,b)}$ by their generating series
at $t=0$, and the normalization factors $N_\ell(\nu,b)$ by
\begin{equation}
 \sum_{\ell\ge0}H_\ell^{(\nu,b)}(y)t^\ell
 =\frac{[1+(1+b)t]^y}{[1+bt]^{y+\nu}},\qquad
 N_\ell(\nu,b)=\frac{\rise\nu\ell}{\ell!}v^\ell.
 \label{eq:gbs-radial-polynomials}
\end{equation}
These are the standard Meixner polynomials with normalization
\begin{equation}
 H_\ell^{(\nu,b)}(y)=(-b)^\ell\frac{\rise\nu\ell}{\ell!}
 M_\ell\!\left(y;\nu,\frac{b}{1+b}\right).
 \label{eq:gbs-radial-meixner-convention}
\end{equation}
Their orthogonality under the count reference is
\cite[Eq.~18.23.4 and Table~18.19.1]{NIST:DLMF}
\begin{equation}
 \sum_{y\ge0}w_{\nu,b}(y)
 H_\ell^{(\nu,b)}(y)H_j^{(\nu,b)}(y)
 =\mathbf1_{\ell=j}N_\ell(\nu,b).
 \label{eq:gbs-radial-orthogonality-proof}
\end{equation}
Combining this identity with Eq.~\eqref{eq:gbs-radial-tilt}
and the angular inner product \eqref{eq:common-lift-inner-product}
yields, for diagonal self-adjoint $C,C'\in(\mathcal V_{k,k})_0$,
\begin{equation}
 \begin{aligned}
 &\E_{X\sim\mu_b}\!\left[
 h_C(X)h_{C'}(X)
 H_\ell^{(m+2k,b)}(K-k)H_j^{(m+2k,b)}(K-k)\right]\\
 &=v^k\Tr(CC')\mathbf1_{\ell=j}N_\ell(m+2k,b).
 \end{aligned}
 \label{eq:gbs-radial-hahn-inner-product}
\end{equation}
The integrand is zero for $K<k$. Distinct angular degrees are
orthogonal on each fixed-count sector by
Eq.~\eqref{eq:common-polynomial-filtration}. The diagonal basis
$\{C_{k,\alpha}\}$ therefore gives orthogonal products
$h_{C_{k,\alpha}}(x)H_\ell^{(m+2k,b)}(|x|-k)$ of total degree $k+\ell$.
The dimension $z_k$ in Eq.~\eqref{eq:common-diagonal-degree-dimension}
satisfies
\begin{equation}
 \sum_{k=0}^D(D-k+1)z_k
 =\sum_{k=0}^D(D-k+1)\binom{m+k-2}{k}
 =\binom{m+D}{D}.
 \label{eq:gbs-radial-dimension-count}
\end{equation}
Consequently these products form a complete basis through degree $D$,
with exact degree $d$ spanned by the terms $k+\ell=d$.
This is the radial--Hahn form of the Meixner basis
\cite[Proposition~6.2(i), Eqs.~(6.5)--(6.6)]{griffiths2011multivariate}.
The construction holds for any positive reference mean in place of $b$,
and hence applies to both $\mu_a$ and $\mu_b$.
The next subsection uses product Meixner coordinates for these same
degree spaces to analyze noise. Subsection~\ref{sec:gbs-evaluation}
returns to the radial--angular form to determine the input-dependent
overlaps and radial coefficients.

\subsection{Meixner approximation}
\label{sec:gbs-approx}

We now verify the assumptions of Theorem~\ref{thm:master}. For the noise
analysis, product Meixner coordinates make the action on individual
modes explicit. At each total degree $d$, they span the same orthogonal
space as the radial--angular products with $k+\ell=d$ constructed in
Subsection~\ref{sec:gbs-representation}. We establish their orthogonality
and thermal contraction, then bound the ideal divergence averaged over
$U$ to obtain the low-degree approximation.

\subsubsection{Complete orthogonal Meixner observables}

For any reference mean $c>0$, the one-mode radial family gives the
univariate Meixner polynomials $H_d^{(c)}=H_d^{(1,c)}$, explicitly
\begin{equation}
 H_d^{(c)}(x)=\sum_{j=0}^d\binom dj\binom xj(-c)^{d-j},\qquad
 \sum_{d\ge0}H_d^{(c)}(x)t^d
 =\frac{[1+(1+c)t]^x}{[1+ct]^{x+1}}.
 \label{eq:gbs-product-meixner}
\end{equation}
Here $\binom{x}{j}=0$ for $j>x$. In particular,
$H_0^{(c)}(x)=1$ and $H_1^{(c)}(x)=x-c$.
In the convention of Ref.~\cite[Eq.~18.20.7]{NIST:DLMF},
\begin{equation}
 H_d^{(c)}(x)=(-c)^d M_d\!\left(x;1,\frac{c}{1+c}\right).
 \label{eq:gbs-meixner-convention}
\end{equation}
The generating function in \eqref{eq:gbs-product-meixner} is
\cite[Eq.~18.23.4]{NIST:DLMF} with series variable $-ct$.
For $\alpha\in\mathbb N_0^m$, we use the same product family at any
reference mean $c>0$:
\begin{equation}
 H_\alpha^{(c)}=\prod_{i=1}^m
 H_{\alpha_i}^{(c)}(\widehat n_i),\qquad
 |\alpha|=\sum_i\alpha_i.
 \label{eq:gbs-product-observables}
\end{equation}
The total polynomial degree is $|\alpha|$.

\begin{lemma}[Meixner orthogonality and completeness, with orthogonality from
Table~18.19.1 in Ref.~\cite{NIST:DLMF}]
\label{lem:gbs-meixner-basis}
For $c>0$, the polynomials $H_d^{(c)}$ form a complete orthogonal
basis of $L^2(g_c)$, with
\begin{equation}
 \sum_{x\ge0}g_c(x)H_d^{(c)}(x)H_e^{(c)}(x)
 =\mathbf1_{d=e}[c(1+c)]^d.
 \label{eq:gbs-meixner-norm}
\end{equation}
Consequently, the product polynomials form a complete orthogonal
basis of the diagonal space $L^2(\tau_c)$, with
\begin{equation}
 \Tr\!\left(\tau_c H_\alpha^{(c)}H_\gamma^{(c)}\right)
 =\mathbf1_{\alpha=\gamma}[c(1+c)]^{|\alpha|}.
 \label{eq:gbs-product-orthogonality}
\end{equation}
\end{lemma}
\begin{proof}
We set $p=c/(1+c)$ in the cited orthogonality formula. The normalization
\eqref{eq:gbs-meixner-convention} gives
\begin{equation}
 \sum_{x\ge0}g_c(x)H_d^{(c)}(x)H_e^{(c)}(x)
 =\delta_{de}c^{2d}p^{-d}
 =\delta_{de}[c(1+c)]^d.
 \label{eq:gbs-classical-normalization}
\end{equation}
The defining hypergeometric formula in \cite[Eq.~18.20.7]{NIST:DLMF}
is symmetric in $d,x$, so $M_d(x;1,p)=M_x(d;1,p)$. Applying the
same orthogonality with the degree and occupation interchanged gives
\begin{equation}
 \begin{aligned}
 \sum_{d\ge0}\frac{H_d^{(c)}(x)H_d^{(c)}(y)}{[c(1+c)]^d}
 &=\sum_{d\ge0}p^dM_x(d;1,p)M_y(d;1,p)\\
 &=(1+c)\delta_{xy}p^{-x}
 =\frac{\delta_{xy}}{g_c(x)}.
 \end{aligned}
 \label{eq:gbs-meixner-completeness}
\end{equation}
Thus every point mass belongs to the closed polynomial span.
Tensor products give \eqref{eq:gbs-product-orthogonality} and
completeness in $L^2(\tau_c)$.
\end{proof}

For Theorem~\ref{thm:master}, we take $\tau=\tau_a$ and
\begin{equation}
 X_\alpha=H_\alpha^{(a)},\qquad
 h_\alpha=[a(1+a)]^{|\alpha|},\qquad d_\alpha=|\alpha|.
 \label{eq:gbs-input-basis}
\end{equation}
Lemma~\ref{lem:gbs-meixner-basis} verifies Assumption~2 of that theorem.
There are $\binom{m+D}{D}$ indices with $|\alpha|\le D$.

\subsubsection{Orthogonality after noise}

\begin{lemma}[Orthogonality after bosonic thermal loss]
\label{lem:gbs-thermal-action}
For the reference parameters in \eqref{eq:gbs-parameters},
\begin{equation}
 \mathcal N_{\eta,\beta}^{\otimes m}(\tau_a)=\tau_b,
 \label{eq:gbs-reference-channel}
\end{equation}
and
\begin{equation}
 \begin{aligned}
 &\Tr\!\left[
 \mathcal N_{\eta,\beta}^{\otimes m}(\tau_aH_\alpha^{(a)})
 \mathcal N_{\eta,\beta}^{\otimes m}(\tau_aH_\gamma^{(a)})
 \tau_b^{-1}\right]
 =\delta_{\alpha\gamma}\Gamma_{|\alpha|}[a(1+a)]^{|\alpha|},\\
 &\Gamma_d=\left[\frac{\eta^2a(1+a)}{b(1+b)}\right]^d.
 \end{aligned}
 \label{eq:gbs-forward-gram}
\end{equation}
Thus Assumption~3 of Theorem~\ref{thm:master} holds.
\end{lemma}
\begin{proof}
For every $c>0$, taking the expectation of
\eqref{eq:thermal-pgf} in $g_c(\hat n)$ gives
\begin{equation}
 \begin{aligned}
 \Tr\!\left[z^{\hat n}\mathcal N_{\eta,\beta}(g_c(\hat n))\right]
 &=\frac1{1+(1-\eta)\beta(1-z)}
   \sum_{x\ge0}g_c(x)
   \left(1-\frac{\eta(1-z)}{1+(1-\eta)\beta(1-z)}\right)^x\\
 &=\frac1{1+[\eta c+(1-\eta)\beta](1-z)}.
 \end{aligned}
 \label{eq:gbs-reference-propagation}
\end{equation}
Both outputs are diagonal, so comparison of their probability generating
functions yields
\begin{equation}
 \mathcal N_{\eta,\beta}(g_c(\hat n))
 =g_{\eta c+(1-\eta)\beta}(\hat n).
 \label{eq:gbs-single-mode-reference}
\end{equation}
The Meixner generating function in \eqref{eq:gbs-product-meixner} satisfies
\begin{equation}
 \begin{aligned}
 g_c(x)\sum_{d\ge0}H_d^{(c)}(x)t^d
 &=\frac{[c+c(1+c)t]^x}{[1+c+c(1+c)t]^{x+1}}\\
 &=g_{c+c(1+c)t}(x).
 \end{aligned}
 \label{eq:gbs-thermal-generating-function}
\end{equation}
Applying \eqref{eq:gbs-single-mode-reference} to this identity at $c=a$ gives
\begin{equation}
 \begin{aligned}
 \mathcal N_{\eta,\beta}\!\left(
 g_a(\hat n)\sum_{d\ge0}H_d^{(a)}(\hat n)t^d\right)
 &=g_{b+\eta a(1+a)t}(\hat n)\\
 &=g_b(\hat n)\sum_{d\ge0}H_d^{(b)}(\hat n)
    \left[\frac{\eta a(1+a)}{b(1+b)}t\right]^d.
 \end{aligned}
 \label{eq:gbs-meixner-generating-transfer}
\end{equation}
The series converge in trace norm for $t$ near zero, so coefficients
can be taken through the channel. Extracting $[t^d]$ and tensoring gives
\begin{equation}
 \mathcal N_{\eta,\beta}^{\otimes m}(\tau_aH_\alpha^{(a)})
 =\left[\frac{\eta a(1+a)}{b(1+b)}\right]^{|\alpha|}
       \tau_bH_\alpha^{(b)}.
 \label{eq:gbs-forward-image}
\end{equation}
The case $\alpha=0$ gives \eqref{eq:gbs-reference-channel}. For all
$\alpha,\gamma$, classical orthogonality gives
\begin{equation}
 \begin{aligned}
 &\Tr\!\left[
 \mathcal N_{\eta,\beta}^{\otimes m}(\tau_aH_\alpha^{(a)})
 \mathcal N_{\eta,\beta}^{\otimes m}(\tau_aH_\gamma^{(a)})
 \tau_b^{-1}\right]\\
 &\quad=\delta_{\alpha\gamma}
 \left[\frac{\eta a(1+a)}{b(1+b)}\right]^{2|\alpha|}
 [b(1+b)]^{|\alpha|}\\
 &\quad=\delta_{\alpha\gamma}
 \left[\frac{\eta^2a(1+a)}{b(1+b)}\right]^{|\alpha|}
 [a(1+a)]^{|\alpha|}.
 \end{aligned}
 \label{eq:gbs-forward-gram-proof}
\end{equation}
\end{proof}

\subsubsection{Suppression under noise} \label{sec:gbs-noise-suppression}

\begin{proposition}[Degree suppression under thermal loss]
\label{prop:gbs-degree-suppression}
The contraction factor in Assumption~4 of Theorem~\ref{thm:master}
is exact:
\begin{equation}
 \Gamma_d=q^d,\qquad
 q=\frac{\eta^2a(1+a)}{b(1+b)}
 \le\frac{\eta(1+a)}{1+\eta a}
 =1-\frac{1-\eta}{1+\eta a}<1.
 \label{eq:gbs-suppression-rate}
\end{equation}
In particular, every real polynomial $Z_d$ in the degree-$d$
Meixner subspace satisfies
\begin{equation}
 \Tr\!\left[\mathcal N_{\eta,\beta}^{\otimes m}(\tau_aZ_d)^2\tau_b^{-1}\right]
 =q^d\Tr(\tau_aZ_d^2).
 \label{eq:gbs-exact-suppression}
\end{equation}
\end{proposition}
\begin{proof}
The exact value follows from \eqref{eq:gbs-forward-gram}, and
$b=\eta a+(1-\eta)\beta\ge\eta a$ gives the displayed bound on $q$.
Expanding $Z_d$ in the orthogonal family
$\{H_\alpha^{(a)}:|\alpha|=d\}$ proves
\eqref{eq:gbs-exact-suppression}.
\end{proof}

\subsubsection{Haar flatness} \label{sec:gbs-flatness}

We first bound the collision probability in each photon-number sector.
We denote by $P_{\mathrm{ideal},U}(x\mid2s)$ the normalized ideal occupation
law conditioned on total photon number $2s$.

\begin{lemma}[Sector collision bound, adapted from Proposition~5 and
Eq.~(D114) of Ref.~\cite{kolarovszki2026general}]
\label{lem:gbs-sector-collision}
For the input of $m\ge2$ equally squeezed vacua and Haar-random $U$,
\begin{equation}
 \E_U\sum_{|x|=2s}P_{\mathrm{ideal},U}(x\mid2s)^2
 \le\left[\frac{s!}{\rise{m/2}{s}}\right]^2,
 \qquad s\in \mathbb{N}_0.
 \label{eq:gbs-sector-collision-bound}
\end{equation}
\end{lemma}

\begin{proof}
Let $h=m/2\ge1$. Equation~(59) of \cite{kolarovszki2026general}, with all
$m$ modes squeezed, expresses the collision probability as $s!^2$
times an average of the factors
\begin{equation}
\begin{aligned}
 &\frac1{\rise h s}
 \frac{\rise{h-1/2}j}{\rise h j}
 \frac{\rise{h+s}{s-j}}{\rise{h+1/2}{2s-j}}\\
 &\le\frac1{\rise h s}
 \frac{\rise{h+s}{s-j}}{\rise h{2s-j}}
 =\frac1{(\rise h s)^2},
 \qquad 0\le j\le s.
\end{aligned}
\label{eq:gbs-sector-collision-comparison}
\end{equation}
The averaging weights are nonnegative and sum to one by
Eqs.~(D112)--(D114) of \cite{kolarovszki2026general}, which proves the bound.
\end{proof}

\begin{proposition}[Haar flatness for the squeezed input]
\label{prop:gbs-haar-flatness}
For Haar-random $U$ and the reference parameter $a$ in
\eqref{eq:gbs-parameters},
\begin{equation}
 \E_U\bigl[1+\chi^2(\rho_{\mathrm{ideal},U}\|\tau_a)\bigr]
 \le C_m:=e[1+(m+1)\bar n].
 \label{eq:gbs-total-budget}
\end{equation}
Consequently, Assumption~1 of Theorem~\ref{thm:master} holds with
$C=C_m-1$.
\end{proposition}

\begin{proof}
Let $z=\bar n/(1+\bar n)=\tanh^2r$. The photon-number generating
function of the product of squeezed vacua gives\ \cite{hamilton2017gaussian}
\begin{equation}
 \left(\frac{1-z}{1-zt^2}\right)^{m/2}
 =\sum_{s\ge0}p_{2s}t^{2s},\qquad
 p_{2s}=(1-z)^{m/2}\frac{\rise{m/2}{s}}{s!}z^s,
 \quad p_{2s+1}=0.
 \label{eq:gbs-ideal-count-law}
\end{equation}
The interferometer preserves these sector probabilities. Every occupation
with $|x|=2s$ has reference weight $a^{2s}/(1+a)^{m+2s}$. All summands
are nonnegative, so Tonelli's theorem and
Lemma~\ref{lem:gbs-sector-collision} give
\begin{equation}
\begin{aligned}
 &\E_U\bigl[1+\chi^2(\rho_{\mathrm{ideal},U}\|\tau_a)\bigr]\\
 &=\sum_{s\ge0}\frac{(1+a)^{m+2s}}{a^{2s}}p_{2s}^2
   \E_U\sum_{|x|=2s}P_{\mathrm{ideal},U}(x\mid2s)^2\\
 &\le[(1-z)(1+a)]^m
       \sum_{s\ge0}\left[\frac{z(1+a)}a\right]^{2s}.
\end{aligned}
\label{eq:gbs-sector-budget}
\end{equation}
The choice of $a$ gives
\begin{equation}
 \begin{gathered}
 (1-z)(1+a)=1+\frac1m,\qquad
 \frac{z(1+a)}a
 =\frac{(m+1)\bar n}{1+(m+1)\bar n}<1,\\
 \frac1{1-z(1+a)/a}=1+(m+1)\bar n.
 \end{gathered}
 \label{eq:gbs-reference-slack}
\end{equation}
Thus the series converges, and
\begin{equation}
 \frac{(1+1/m)^m}{1-[z(1+a)/a]^2}
 \le\frac{e}{1-z(1+a)/a}
 =e[1+(m+1)\bar n].
 \label{eq:gbs-geometric-budget}
\end{equation}
The finite bound also shows that the ideal relative density belongs to $L^2(\tau_a)$ for
Haar-almost every $U$. Lemma~\ref{lem:reference-contraction} gives the
corresponding integrability after noise.
\end{proof}

\subsubsection{Low-degree approximation}\label{sec:gbs-truncation}

The contribution of total degree $d$ to the ideal divergence is
\begin{equation}
 \begin{aligned}
 W_d(U)&=\sum_{|\alpha|=d}
       \frac{|\Tr(\rho_{\mathrm{ideal},U}H_\alpha^{(a)})|^2}
            {[a(1+a)]^d},\\
 \sum_{d\ge1}\E_UW_d(U)
 &=\E_U\chi^2(\rho_{\mathrm{ideal},U}\|\tau_a)\le C_m-1
 \quad\text{for Haar }U.
 \end{aligned}
 \label{eq:gbs-ideal-degree-energies}
\end{equation}
For an integer $D\ge0$, we define the signed approximation and its density
relative to the output reference by
\begin{equation}
 \begin{aligned}
 \widetilde\rho_{D,U}
 &=\mathcal N_{\eta,\beta}^{\otimes m}\!\left(
   \tau_a\sum_{|\alpha|\le D}
     \frac{\Tr(\rho_{\mathrm{ideal},U}H_\alpha^{(a)})}
          {[a(1+a)]^{|\alpha|}}H_\alpha^{(a)}\right)\\
 &=\sum_x\mu_b(x)f_{D,U}(x)\ket x\bra x,\\
 f_{D,U}(x)
 &=\sum_{|\alpha|\le D}
     \frac{\eta^{|\alpha|}\Tr(\rho_{\mathrm{ideal},U}H_\alpha^{(a)})}
          {[b(1+b)]^{|\alpha|}}\bra xH_\alpha^{(b)}\ket x.
 \end{aligned}
 \label{eq:gbs-low-degree-state}
\end{equation}
The forward identity \eqref{eq:gbs-forward-image} gives the last line.
Whenever $\rho_{\mathrm{ideal},U}/\tau_a\in L^2(\tau_a)$,
reference contraction allows the complete input expansion to pass
through the channel in the weighted $L^2$ norm. Equation
\eqref{eq:gbs-forward-image} maps each input Meixner polynomial to
the output polynomial of the same index. Hence $f_{D,U}$ is the
orthogonal projection of $P_U/\mu_b$ through total degree $D$.
Since $H_0^{(a)}=H_0^{(b)}=I$, its constant
coefficient is one and $\widetilde\rho_{D,U}$ has trace one.

\begin{corollary}[Meixner approximation under thermal loss]
\label{cor:gbs-approximation}
For any interferometer ensemble satisfying
$\E_U\chi^2(\rho_{\mathrm{ideal},U}\|\tau_a)\le C<\infty$, let
$f_U(x)=P_U(x)/\mu_b(x)$. Then
\begin{equation}
 \begin{aligned}
 \E_U\|\mathcal N_{\eta,\beta}^{\otimes m}(\rho_{\mathrm{ideal},U})-
                         \widetilde\rho_{D,U}\|_1^2
 &\le\E_U\|f_U-f_{D,U}\|_{L^2(\mu_b)}^2\\
 &=\sum_{d>D}q^d\E_UW_d(U)
 \le Cq^{D+1}.
 \end{aligned}
 \label{eq:gbs-flat-bound}
\end{equation}
For Haar $U$, one may take $C=C_m-1$. In particular, trace-norm
error at most $\epsilon>0$ with probability at least $1-\delta$,
where $0<\delta<1$, is ensured by
\begin{equation}
 D\ge\max\left\{0,
 \left\lceil\frac{\log[(C_m-1)/(\delta\epsilon^2)]}{\log(1/q)}\right\rceil-1
 \right\}.
 \label{eq:gbs-flat-cutoff}
\end{equation}
\end{corollary}
\begin{proof}
Lemma~\ref{lem:gbs-meixner-basis}, Lemma~\ref{lem:gbs-thermal-action},
and Proposition~\ref{prop:gbs-degree-suppression} verify Assumptions~2--4
of Theorem~\ref{thm:master} for the input basis
\eqref{eq:gbs-input-basis}. Its degree weights are exactly
\eqref{eq:gbs-ideal-degree-energies}, and $\Gamma_d=q^d$ gives
\eqref{eq:gbs-flat-bound}.
The Haar bound is
\eqref{eq:gbs-total-budget}. Markov's inequality gives
\begin{equation}
 \Prob_U\!\left(
 \|\mathcal N_{\eta,\beta}^{\otimes m}(\rho_{\mathrm{ideal},U})-
                 \widetilde\rho_{D,U}\|_1>\epsilon\right)
 \le\frac{(C_m-1)q^{D+1}}{\epsilon^2}\le\delta
 \label{eq:gbs-cutoff-probability}
\end{equation}
under \eqref{eq:gbs-flat-cutoff}.
\end{proof}

For fixed $\bar n,\eta$ with $\eta<1$, the upper bound on $q$
in \eqref{eq:gbs-suppression-rate} stays below one uniformly in
$m\ge2$, while $C_m=O(m)$. Hence inverse-polynomial accuracy and failure probability
require only $D=O(\log m)$.
The Haar-moment calculation in Subsection~\ref{sec:gbs-degree-moments}
computes the individual contributions after noise and bounds their
remaining tail, which can refine this cutoff.

\subsection{Operator overlap formulas}
\label{sec:gbs-evaluation}

We combine the $k$-particle harmonic projector of
Appendix~\ref{sec:particle-representation} with the radial decomposition
of Subsection~\ref{sec:gbs-representation}. Gaussian occupation moments
determine the scalar radial coefficients, including the noise dependence.
Both algorithms then evaluate reduced angular overlaps and sum the
retained radial degrees, without resolving individual Hahn or
product-Meixner basis elements. The resulting $f_{D,U}$ is a density
relative to $\mu_b$; multiplying by $\mu_b$ gives the approximate
probability mass.

Let $\rho_{\rm out}$ denote the Gaussian state after thermal loss, so
$P_U(x)=\bra x\rho_{\rm out}\ket x$, and write
$\widehat K=\sum_i\hat a_i^\dagger\hat a_i$ for the total-count operator.
Each projection coefficient is its output moment divided by the
reference norm in Eq.~\eqref{eq:gbs-radial-hahn-inner-product}.
These finite Gaussian moments define $f_{D,U}$ for every $U$.

The physical covariance parameters are
\begin{equation}
 n_{\rm out}=\eta\bar n+(1-\eta)\beta,\qquad
 \kappa=\eta^2\bar n(1+\bar n),\qquad \Delta=b-n_{\rm out}.
 \label{eq:gbs-covariance-parameters}
\end{equation}

Thus the output covariances satisfy
\begin{equation}
 \Tr(\rho_{\rm out}\hat a_i^\dagger \hat a_j)=n_{\rm out}\delta_{ij},\qquad
 \Tr(\rho_{\rm out}\hat a_i \hat a_j)=\sqrt\kappa\,(UU^{\mathsf T})_{ij},
 \label{eq:gbs-output-covariances}
\end{equation}
up to a common phase that cancels from number expectations. For the
count generating function, we define the quadratic polynomial
\begin{equation}
 F(z)=[1+n_{\rm out}(1-z)]^2-\kappa(1-z)^2.
 \label{eq:gbs-count-polynomial}
\end{equation}
For even $k=2s\ge0$, we define the normalized pair state, its harmonic
projection, and its angular overlap by
\begin{equation}
\begin{gathered}
 \ket{\Psi_k}=\frac{(\sum_i \hat a_i^{\dagger2})^s\ket0}
 {\sqrt{4^s s!\rise{m/2}{s}}},\\
 T_k=\Pi_k(\ket{\Psi_k}\bra{\Psi_k}),\qquad
 B_k(U,x)=\Tr(U_kT_kU_k^\dagger A_{x,k}).
\end{gathered}
 \label{eq:gbs-pair-angular}
\end{equation}

Here $U_k$ is the bosonic action in Eq.~\eqref{eq:particle-mode-action},
and $\Pi_k$ is the orthogonal projector in
Eq.~\eqref{eq:common-projector}, with the bosonic coefficients of
Eq.~\eqref{eq:common-projector-coefficients}. Its range is
$\mathcal V_{k,k}=\ker\cL$ by Eq.~\eqref{eq:ker_ident}.
We use the test operator $A_{x,k}$ from Eq.~\eqref{eq:bosonic-A}:
$\Tr(CA_{x,k})=h_C(x)$ for diagonal $C$, by
Eq.~\eqref{eq:common-harmonic-polynomial}. Thus $B_k(U,x)$ is the
polynomial lift of the diagonal part of $U_kT_kU_k^\dagger$.
In particular, $B_0(U,x)=1$, and $B_k(U,x)=0$ for $|x|<k$.

We define
\begin{equation}
 \omega_k(U)=\|\operatorname{diag}(U_kT_kU_k^\dagger)\|_2^2,\qquad
 h_k(K)=\fall Kk\rise{m+K}k.
 \label{eq:gbs-angular-definitions}
\end{equation}
The map $\operatorname{diag}$ deletes off-diagonal occupation-basis
entries.

For even $k\ge0$ and $\ell\ge0$, we define
\begin{equation}
 \nu_k=m+2k,\qquad \gamma_k=\frac{\rise{m/2}{k/2}}{(k/2)!},\qquad
 c_{k,\ell}=[t^\ell]\big[(1+\Delta t)^2-\kappa t^2\big]^{-\nu_k/2}.
 \label{eq:gbs-radial-parameters}
\end{equation}
For even $0\le k\le D$, we define the radial coefficient function by
\begin{equation}
 R_{k,D}(K)=\frac{\gamma_k\kappa^{k/2}}{v^k}
       \sum_{\ell=0}^{D-k}
       \frac{c_{k,\ell}H_\ell^{(\nu_k,b)}(K-k)}{N_\ell(\nu_k,b)},\qquad K\ge k.
 \label{eq:gbs-radial}
\end{equation}
Here $[t^\ell]$ denotes the coefficient of $t^\ell$ in the Taylor
expansion at zero. Radial polynomials with $K<k$ need not be evaluated,
since their angular contribution is zero.

\begin{lemma}[Radial--Hahn representation of the Meixner projection, using the basis of Proposition~6.2(i) in Ref.~\cite{griffiths2011multivariate}]
\label{lem:gbs-meixner-representation}
For every integer $D\ge0$, the projection through total degree $D$ is
\begin{equation}
 f_{D,U}(x)=
 \sum_{\substack{0\le k\le\min(D,|x|)\\k\ \mathrm{even}}}
 R_{k,D}(|x|)B_k(U,x).
 \label{eq:gbs-meixner-expansion}
\end{equation}
For each even $k$ and fixed total count $K\ge k$, the angular norm is
\begin{equation}
 \E_{X\sim u_K}B_k(U,X)^2
 =\frac{h_k(K)}{\rise m{2k}}\omega_k(U).
 \label{eq:gbs-angular-norm}
\end{equation}
For each even $k$, the polynomial
$B_k(U,x)H_\ell^{(\nu_k,b)}(|x|-k)$ has total degree $k+\ell$ unless it
vanishes identically. Polynomials with distinct pairs $(k,\ell)$ are
orthogonal in $L^2(\mu_b)$, and their squared norms are
\begin{equation}
 \E_{X\sim\mu_b}\!\left[
 (B_k(U,X)H_\ell^{(\nu_k,b)}(K-k))^2\right]
 =\omega_k(U)v^kN_\ell(\nu_k,b).
 \label{eq:gbs-angular-radial-norm}
\end{equation}
Odd angular degrees vanish for this input, although odd total
Meixner degrees need not vanish.
\end{lemma}

\begin{proof}

The angular norm follows directly from
Eq.~\eqref{eq:common-lift-inner-product}, applied to the diagonal
harmonic operator $\operatorname{diag}(U_kT_kU_k^\dagger)$. The
bosonic factor in Eq.~\eqref{eq:common-diagonal-beta} simplifies to
\begin{equation}
 \beta_{K,k}
 =\frac{\fall Kk\rise{m+K}k}{\rise m{2k}}
 =\frac{h_k(K)}{\rise m{2k}},
 \label{eq:gbs-angular-normalization-proof}
\end{equation}
which gives Eq.~\eqref{eq:gbs-angular-norm}. The radial--angular
orthogonality, norms, and cutoff $k+\ell\le D$ are established in
Eqs.~\eqref{eq:gbs-radial-hahn-inner-product}
and~\eqref{eq:gbs-radial-dimension-count}. It remains to evaluate
the Gaussian moments selecting the angular direction and its radial
coefficients.

To compute the projection coefficients, we first observe that the total-count
generating function is
\begin{equation}
 \E_{P_U}z^K=F(z)^{-m/2}.
 \label{eq:gbs-pgf}
\end{equation}
Homogeneous thermal loss commutes with passive interferometers, so
we may apply it before $U$. At that point the modes are independent,
with normal covariance $n_{\rm out}$ and anomalous covariance
$\sqrt\kappa$. The interferometer preserves $K$, so its count law
is unchanged and is independent of $U$. The one-mode number generating function is $F(z)^{-1/2}$,
so multiplication over the $m$ modes proves \eqref{eq:gbs-pgf}.
This identity follows either from the Gaussian determinant formula
of \cite{cardin2024photon} or by summing the Wick expansion of
$z^{\widehat n}={:}\exp((z-1)\widehat n){:}$.

To identify the angular coefficients, we consider real $z>0$ in a
neighborhood of $1$ where the generating functions converge. We weight
the Gaussian state by $z^{\widehat K/2}$ on both sides and normalize
by $F(z)^{-m/2}$. Its normal covariance remains a scalar
matrix, and its anomalous covariance becomes
\begin{equation}
\begin{gathered}
 n_z I,\qquad m_zUU^{\mathsf T},\\
 n_z=\frac{z[n_{\rm out}+(1-z)(n_{\rm out}^2-\kappa)]}{F(z)},\qquad
 m_z=\frac{z\sqrt\kappa}{F(z)}.
 \end{gathered}
 \label{eq:gbs-tilted-covariances}
\end{equation}
To verify the transformation, the normal-ordering identity
for a balanced moment with $k$ creation and $k$ annihilation operators is
\begin{equation}
 z^{\widehat K}(\hat a^\dagger)^s \hat a^r
 =z^k{:}\exp((z-1)\widehat K)(\hat a^\dagger)^s \hat a^r{:},\qquad
 |r|=|s|=k.
 \label{eq:gbs-normal-ordered-tilt}
\end{equation}
For diagonal moments this reduces to
$\sum_{j\ge0}(z-1)^j\fall{x}{r+j}/j!=\fall{x}r z^{x-r}$.
The same expansion on occupation matrix elements proves the balanced
operator identity. Wick's rule for the quadratic exponential changes
the one-mode contraction matrix by
\begin{equation}
\begin{aligned}
 z\begin{pmatrix}n_{\rm out}&\sqrt\kappa\\
                   \sqrt\kappa&n_{\rm out}\end{pmatrix}
 \left[I+(1-z)
   \begin{pmatrix}n_{\rm out}&\sqrt\kappa\\
                   \sqrt\kappa&n_{\rm out}\end{pmatrix}\right]^{-1}
 =\begin{pmatrix}n_z&m_z\\m_z&n_z\end{pmatrix}.
 \end{aligned}
 \label{eq:gbs-tilted-covariance-inverse}
\end{equation}
The determinant of the inverted matrix is $F(z)$. Multiplying its
$2\times2$ inverse verifies the entries in
\eqref{eq:gbs-tilted-covariances}. The quadratic Wick identity itself
can be checked by expanding the inverse as a geometric series: a chain
of $j$ internal contractions contributes
$z(z-1)^j C^{j+1}$, where $C$ is the displayed contraction matrix.
Closed chains give the normalizing determinant $F(z)^{-1/2}$.

Let $J_k(z)$ be the unnormalized $k$-particle moment operator, defined by
\begin{equation}
\begin{gathered}
 \bra rJ_k(z)\ket s
 =\frac{\Tr[\rho_{\rm out}z^{\widehat K}(\hat a^\dagger)^s \hat a^r]}
        {\sqrt{r!s!}},\quad
 |r|=|s|=k.
 \end{gathered}
 \label{eq:gbs-moment-operator}
\end{equation}
Its diagonal entry is
$\E_{P_U}[z^K\prod_i\binom{X_i}{r_i}]$. Consequently, for any diagonal
$C$ with $\cL(C)=0$,
\begin{equation}
 \E_{P_U}\!\left[z^K\Tr(CA_{X,k})\right]
 =\Tr(CJ_k(z))
 =\Tr(C\Pi_k(J_k(z))).
 \label{eq:gbs-harmonic-moment-test}
\end{equation}
To compute this projection, we first set $U=I$ and introduce the homogeneous
coherent vector $\ket{u;k}=\sum_{|r|=k}u^r\ket r/\sqrt{r!}$.
Wick contractions~\cite{hackl2021bosonic} in $\bra{u;k}J_k(z)\ket{u;k}$ either join one creation
operator to one annihilation operator, contributing
$n_z\sum_i|u_i|^2$, or join two operators of the same type, contributing
$m_z\sum_i u_i^2$ or its conjugate.

Multiplication of an operator's coherent symbol by
$\sum_i|u_i|^2$ realizes the raising map in
Eq.~\eqref{eq:common-lowering-raising}. By adjointness and
Eq.~\eqref{eq:ker_ident}, $\operatorname{range}\cR$ is orthogonal
to the range of $\Pi_k$. Thus the projector removes every term
containing a normal contraction.

If $k$ is odd, every Wick pairing has such a normal contraction,
so the projection is zero. If $k=2s$, the remaining contractions pair all
creation operators among themselves and all annihilation operators among
themselves. There are $[(2s)!/(2^s s!)]^2$ such pairings. Including the
factor $1/(k!)^2$ from the coherent symbol gives
\begin{equation}
 F(z)^{-m/2}\frac{m_z^{2s}}{4^s(s!)^2}
          \left|\sum_i u_i^2\right|^{2s}.
 \label{eq:gbs-anomalous-pairings}
\end{equation}
The coherent symbol of $\ket{\Psi_{2s}}\bra{\Psi_{2s}}$ is
$|\sum_i u_i^2|^{2s}/[4^s s!\rise{m/2}s]$.
Their ratio is $\gamma_k m_z^k$, with $\gamma_k$ from
\eqref{eq:gbs-radial-parameters}. Applying $\Pi_k$ and restoring $U$ by covariance
therefore gives
\begin{equation}
 \Pi_k(J_k(z))
 =\gamma_k\kappa^{k/2}z^kF(z)^{-(m+2k)/2}
                  U_kT_kU_k^\dagger,
 \qquad k\text{ even}.
 \label{eq:gbs-projected-moment-operator}
\end{equation}
Taking $C=\operatorname{diag}(U_kT_kU_k^\dagger)$ in
\eqref{eq:gbs-harmonic-moment-test} yields
\begin{equation}
 \E_{P_U}[B_k(U,X)z^K]
 =\gamma_k\kappa^{k/2}\omega_k(U)
             z^kF(z)^{-(m+2k)/2}.
 \label{eq:gbs-angular-count-moment}
\end{equation}
Equation~\eqref{eq:gbs-projected-moment-operator} also identifies the
moments against every other angular test. Thus the angular coefficient
vector is in the direction specified by $T_k$.

It remains to calculate its radial coefficients. We set
$z=[1+(1+b)t]/[1+bt]$. The radial generating function is
$z^{K-k}(1+bt)^{-\nu_k}$, and direct substitution gives
\begin{equation}
 (1+bt)^2F(z)=(1+\Delta t)^2-\kappa t^2.
 \label{eq:gbs-radial-substitution}
\end{equation}
Multiplying \eqref{eq:gbs-angular-count-moment} by
$z^{-k}(1+bt)^{-\nu_k}$ and comparing powers of $t$ now gives
\begin{equation}
 \E_{P_U}[B_k(U,X)H_\ell^{(\nu_k,b)}(K-k)]
 =\gamma_k\kappa^{k/2}\omega_k(U)c_{k,\ell}.
 \label{eq:gbs-angular-radial-coefficient}
\end{equation}

For the reference norm, take
$C=C'=\operatorname{diag}(U_kT_kU_k^\dagger)$ and $j=\ell$ in
Eq.~\eqref{eq:gbs-radial-hahn-inner-product}. This operator lies
in $(\mathcal V_{k,k})_0$ by the covariance and diagonal decomposition
in Eqs.~\eqref{eq:common-particle-map-covariance}
and~\eqref{eq:common-diagonal-decomposition}. Since
$h_C(x)=B_k(U,x)$ and $\Tr(C^2)=\omega_k(U)$, this gives
Eq.~\eqref{eq:gbs-angular-radial-norm}.

Dividing the coefficient in
\eqref{eq:gbs-angular-radial-coefficient} by this squared norm gives
$\gamma_k\kappa^{k/2}c_{k,\ell}/[v^kN_\ell(\nu_k,b)]$.
Summing over $k+\ell\le D$ proves
\eqref{eq:gbs-meixner-expansion}. If $\omega_k(U)=0$, the corresponding
angular polynomial is zero and contributes nothing.

\end{proof}

\subsection{Deterministic evaluation}
\label{sec:gbs-enumeration}

\subsubsection{Exact pointwise density}\label{sec:gbs-exact-density}
\begin{lemma}[Exact deterministic GBS density]
\label{lem:gbs-deterministic-evaluation}
For an integer $D\ge0$, let
$b_{k,r}(U)=\bra rU_kT_kU_k^\dagger\ket r$ for even $k\le D$ and
$|r|=k$, with $b_{0,0}=1$. The retained density is
\begin{equation}
 f_{D,U}(x)=
 \sum_{\substack{0\le k\le\min(D,|x|)\\k\ \mathrm{even}}}
 R_{k,D}(|x|)\sum_{|r|=k}b_{k,r}(U)
                                \prod_i\binom{x_i}{r_i}.
 \label{eq:gbs-enumerated-density}
\end{equation}
The value at any occupation $x$ admits exact evaluation using finite
arrays of occupation coefficients through degree $D$.
\end{lemma}
\begin{proof}
Let $B=UU^{\mathsf T}$ and $Q(z)=\sum_{i,j=1}^mB_{ij}z_iz_j$ for
commuting variables $z_1,\ldots,z_m$. For $k=2s$, rotation of the
creation operators in \eqref{eq:gbs-pair-angular} gives
\begin{equation}
 U_k\ket{\Psi_k}
 =\frac{\bigl(\sum_{i,j}B_{ij}\hat a_i^\dagger \hat a_j^\dagger\bigr)^s\ket0}
        {\sqrt{4^ss!\rise{m/2}s}}
 =\sum_{|r|=k}\frac{\sqrt{r!}\,[z^r]Q(z)^s}
                        {\sqrt{4^ss!\rise{m/2}s}}\ket r.
 \label{eq:gbs-rotated-pair-coefficients}
\end{equation}
The factor $\sqrt{r!}$ follows from
$\prod_i(\hat a_i^\dagger)^{r_i}\ket0=\sqrt{r!}\ket r$.
Consequently,
\begin{equation}
 \operatorname{diag}\!\left(U_k\ket{\Psi_k}\bra{\Psi_k}U_k^\dagger\right)
 =\sum_{|r|=k}\frac{r!\,|[z^r]Q(z)^s|^2}{4^ss!\rise{m/2}s}
                       \ket r\bra r.
 \label{eq:gbs-rotated-pair-diagonal}
\end{equation}
The squared amplitudes involve only rational factorial factors and
entries of $U$ and their conjugates.

We apply the bosonic projector of Eq.~\eqref{eq:common-projector}
to this diagonal operator. On coefficient arrays
$C=\sum_{|r|=j}c_r\ket r\bra r$, the maps from
Eq.~\eqref{eq:common-lowering-raising} act as follows; their entries
also follow from Eq.~\eqref{eq:common-occupation-projector-lift}
and adjointness:

\begin{equation}
 \begin{aligned}
 \bra r\cL(C)\ket r&=\sum_{i=1}^m(r_i+1)c_{r+e_i},
       &&|r|=j-1,\\
 \bra r\cR(C)\ket r&=\sum_{i:r_i>0}r_i c_{r-e_i},
       &&|r|=j+1.
 \end{aligned}
 \label{eq:gbs-diagonal-particle-maps}
\end{equation}
Here $e_i$ has one in coordinate $i$ and zeros elsewhere. Each map
therefore acts on a finite array of occupation coefficients.

Covariance and the diagonal-degree decomposition,
Eqs.~\eqref{eq:common-particle-map-covariance}
and~\eqref{eq:common-diagonal-decomposition}, allow the projector
to be applied after rotation and dephasing. Hence the required
coefficients are

\begin{equation}
 b_{k,r}(U)=\bra r\Pi_k\!\left(
   \operatorname{diag}\!\left(U_k\ket{\Psi_k}\bra{\Psi_k}U_k^\dagger\right)
                    \right)\ket r,
 \qquad |r|=k.
 \label{eq:gbs-enumerated-angular-coefficients}
\end{equation}

Equations~\eqref{eq:common-projector}
and~\eqref{eq:gbs-diagonal-particle-maps} compute them using only
the arrays in Eq.~\eqref{eq:gbs-rotated-pair-diagonal}.

The coefficients are real, since the projected operator is Hermitian.

Substituting the diagonal entries into
\eqref{eq:gbs-meixner-expansion} proves \eqref{eq:gbs-enumerated-density}.
\end{proof}

\subsubsection{A deterministic norm bound for rejection sampling} \label{sec:gbs-deterministic-rejection}

\begin{lemma}[Uniform norm bound for the retained GBS density]
\label{lem:gbs-deterministic-norm}
For every mode unitary $U$ and integer $D\ge0$,
\begin{equation}
 \begin{aligned}
 \sum_x\mu_b(x)f_{D,U}(x)^2&\le B_{m,D},\\
 B_{m,D}&:=\sum_{d=0}^D\binom{m+d-1}{d}[1+2(m+d)]^{2d}
          =(m+D)^{O(D)}.
 \end{aligned}
 \label{eq:gbs-deterministic-norm}
\end{equation}
\end{lemma}
\begin{proof}
Let $X$ have the ideal occupation law for the fixed $U$, and let
$K=|X|$. The interferometer preserves $K$. Let
$c_\pm=\bar n\pm\sqrt{\bar n(1+\bar n)}$. Equation
\eqref{eq:gbs-ideal-count-law} gives the factorial-moment generating function
\begin{equation}
 \E(1+t)^K=[1-\bar n(2t+t^2)]^{-m/2}
 =(1-c_+t)^{-m/2}(1-c_-t)^{-m/2}.
\end{equation}
Since $|c_-|\le c_+$, the absolute value of each coefficient in the
product is at most the corresponding coefficient of
$(1-c_+t)^{-m}$. Comparing coefficients of $t^j$ gives
\begin{equation}
 \E\binom Kj\le\binom{m+j-1}{j}c_+^j.
\end{equation}
For a multi-index $\alpha$ with $|\alpha|=d$, the product expansion
\eqref{eq:gbs-product-meixner} and positivity of binomial moments give
\begin{equation}
 \begin{aligned}
 \bigl|\E H_\alpha^{(a)}(X)\bigr|
 &\le\sum_{\mathbf j\le\alpha}
   \left(\prod_i\binom{\alpha_i}{j_i}\right)
   a^{d-|\mathbf j|}\E\prod_i\binom{X_i}{j_i}\\
 &\le\sum_{j=0}^d\binom dj a^{d-j}\E\binom Kj\\
 &\le\sum_{j=0}^d\binom dj a^{d-j}
                         \binom{m+j-1}{j}c_+^j.
 \end{aligned}
\end{equation}
The second line uses
$\prod_i\binom{X_i}{j_i}\le\binom K{|\mathbf j|}$ and
$\sum_{|\mathbf j|=j,\,\mathbf j\le\alpha}
 \prod_i\binom{\alpha_i}{j_i}=\binom dj$.
Because $a\ge\bar n>0$,
\begin{equation}
 \frac{a}{\sqrt{a(1+a)}}\le1,\qquad
 \frac{c_+}{\sqrt{a(1+a)}}
 \le\frac{\bar n+\sqrt{\bar n(1+\bar n)}}
          {\sqrt{\bar n(1+\bar n)}}\le2.
\end{equation}
After division by $[a(1+a)]^{d/2}$ and using
$\binom{m+j-1}{j}\le(m+d)^j$, the preceding sum is at most
\begin{equation}
 \frac{\bigl|\E H_\alpha^{(a)}(X)\bigr|}{[a(1+a)]^{d/2}}
 \le\sum_{j=0}^d\binom dj[2(m+d)]^j
 =[1+2(m+d)]^d.
\end{equation}
Finally, \eqref{eq:gbs-low-degree-state} and product orthogonality give
\begin{equation}
 \|f_{D,U}\|_{L^2(\mu_b)}^2=\sum_{d=0}^D q^d\sum_{|\alpha|=d}
  \frac{|\Tr(\rho_{\mathrm{ideal},U}H_\alpha^{(a)})|^2}{[a(1+a)]^d}.
\end{equation}
The factor $q$ in \eqref{eq:gbs-suppression-rate} is at most one, and
there are $\binom{m+d-1}{d}$ indices of degree $d$. Substitution proves
the bound, with the degree-zero term equal to one.
\end{proof}

The proposal law is $\mu_b$: its $m$ occupations are independent
geometric variables of mean $b$. At each proposal $x$, we evaluate
$f_{D,U}(x)$ by Lemma~\ref{lem:gbs-deterministic-evaluation} and use
it as the score in Algorithm~\ref{alg:rejection-sampling}.
It has reference mean one by \eqref{eq:gbs-low-degree-state}.
Corollary~\ref{thm:final-enumeration} chooses the threshold and proposal
limit using Lemma~\ref{lem:gbs-deterministic-norm}.
Corollary~\ref{cor:gbs-approximation} supplies the degree-truncation bound.

\subsubsection{Arithmetic cost}
There are at most
$Q_D=\sum_{k=0}^D\binom{m+k-1}{k}=\binom{m+D}{D}$ stored angular
coefficients. Forming $B=UU^{\mathsf T}$ costs $O(m^3)$ operations.
Computing the powers $Q(z)^s$ through degree $D$ by ordinary polynomial
arithmetic costs $\poly(m,D)Q_D^2$ operations with naive multiplication.
Each application of \eqref{eq:gbs-diagonal-particle-maps} takes at
most $O(m\binom{m+k-1}{k})$ operations on a sector of dimension
$\binom{m+k-1}{k}$, so the finite projector costs at most
$O(mk^2\binom{m+k-1}{k})$ for degree $k$.

Let $K_*$ be the proposal-count cap from Section~\ref{sec:precision}.
After summing the retained degrees and preparing the radial tables,
a safe preprocessing and storage bound is
\begin{equation}
 \poly(m,D)Q_D^2+\poly(m,D,K_*),\qquad
 \text{storage }\poly(m,D)Q_D+\poly(m,D,K_*).
 \label{eq:gbs-enumeration-preprocessing}
\end{equation}
Each point evaluation costs $Q_D\poly(m,D,K_*)$, and at most $A$
proposals are evaluated. The subsequent work per sample is therefore
\begin{equation}
 A Q_D\poly(m,D,K_*).
 \label{eq:gbs-enumeration-sample-cost}
\end{equation}
Taking $M,A$ at the lower bounds in
\eqref{eq:deterministic-rejection-threshold} and
\eqref{eq:deterministic-rejection-attempts}, with error budgets at most
$1/2$, gives $(m+D)^{O(D)}\poly(K_*,\eps_{\rm clip}^{-1},
\log(1/\eps_{\rm stop}))$. Section~\ref{sec:precision} controls proposals
exceeding the count cap and the arithmetic errors in the point evaluations.

\subsection{Randomized evaluation}
\label{sec:gbs-sketch}
Lemma~\ref{lem:gbs-meixner-representation} expresses the retained
density through deterministic radial coefficients \eqref{eq:gbs-radial}
and angular overlaps $B_k(U,x)$. We evaluate those overlaps
without enumerating $\binom{m+k-1}{k}$ occupation monomials. The input
is the explicit pair state~\eqref{eq:gbs-pair-angular}. Randomness enters
through the output test operator. Its first and second moments are
supplied by Lemma~\ref{lem:bosonic-output-sketch}.

For even $k\ge2$, let $\widehat A_{x,g}$ be the Laguerre-slot rainbow
operator of~\eqref{eq:bosonic-output-sketch}, using $k$ colors. We set
\begin{equation}
 \widehat B_k(U,x,g)=\Tr[U_kT_kU_k^\dagger\widehat A_{x,g}],\qquad
 p_k=k!/k^k.
\end{equation}
All harmonic subtraction terms use the same coloring. As in the Fock
branch, the slots encode repeated occupations in the polynomial contraction.

\subsubsection{The projected pair state}

\begin{lemma}[Exact-degree projection of the pair state]
\label{lem:gbs-pair-projection}
Let $k=2s$ with $s\in\mathbb N_0$, let $\hat A=\sum_{i=1}^m \hat a_i^2$, and let
$I_j$ denote the identity on $\Sym^j\mathbb C^m$. We define
\begin{equation}
 G_{k,a}=\frac{(\hat A^\dagger)^a I_{k-2a}\hat A^a}{4^a(a!)^2},\qquad 0\le a\le s,
 \label{eq:gbs-pair-raising-operators}
\end{equation}
and define the real coefficients $\ell_{k,a}$ recursively by
\begin{equation}
 \ell_{k,s}=\frac{s!}{\rise{m/2}s},\qquad
 \ell_{k,a}=-\frac{k-2a-1}{m+k+2a-1}\ell_{k,a+1}\quad(0\le a<s).
 \label{eq:gbs-harmonic-coefficients}
\end{equation}
Then the orthogonal projection $T_k=\Pi_k(\ket{\Psi_k}\bra{\Psi_k})$
has the expansion and squared Schatten norm
\begin{equation}
 T_k=\sum_{a=0}^{s}\ell_{k,a}G_{k,a},
 \label{eq:gbs-pair-projector-expansion}
\end{equation}
\begin{equation}
 \|T_k\|_2^2=\sum_{a=0}^{s}\ell_{k,a}
       \frac{\fall sa\rise{m/2+s-a}a}{(a!)^2}.
 \label{eq:gbs-pair-norm}
\end{equation}
In particular, $\|T_0\|_2^2=1$.
\end{lemma}

\begin{proof}
We first show that the combination in
\eqref{eq:gbs-pair-projector-expansion} belongs to $\ker\cL$.
In the coherent vectors used above, we set
$t=\sum_i|u_i|^2$ and $Q=|\sum_i u_i^2|^2$. The symbols of the
operators in \eqref{eq:gbs-pair-raising-operators} are
\begin{equation}
 \bra{u;k}G_{k,a}\ket{u;k}
 =\frac{t^{k-2a}Q^a}{(k-2a)!4^a(a!)^2}.
 \label{eq:gbs-pair-operator-symbol}
\end{equation}
Lowering corresponds to applying the complex Laplacian to the symbol.
Differentiating its two factors gives
\begin{equation}
 \sum_i\partial_{u_i}\partial_{\bar u_i}(t^pQ^a)
 =p(m+p-1+4a)t^{p-1}Q^a+4a^2t^{p+1}Q^{a-1}.
 \label{eq:gbs-pair-laplacian}
\end{equation}
For $0\le a<s$, the coefficient of $t^{k-2a-1}Q^a$ in the symbol of
$\cL(\sum_a\ell_{k,a}G_{k,a})$ is therefore
\begin{equation}
 \frac{\ell_{k,a}(m+k+2a-1)+(k-2a-1)\ell_{k,a+1}}
      {(k-2a-1)!4^a(a!)^2}.
 \label{eq:gbs-pair-laplacian-cancellation}
\end{equation}
The recurrence in \eqref{eq:gbs-harmonic-coefficients} makes every such
coefficient zero. The chosen coefficient at $a=s$ satisfies
\begin{equation}
 \ell_{k,s}G_{k,s}=\ket{\Psi_k}\bra{\Psi_k}.
 \label{eq:gbs-pair-leading-term}
\end{equation}
For $a<s$, the symbol in \eqref{eq:gbs-pair-operator-symbol} contains a
factor $t$, so $G_{k,a}$ belongs to $\operatorname{range}\cR$.

The combination is therefore harmonic and differs from the pair
projector by an element of $\operatorname{range}\cR$.
Adjointness and Eq.~\eqref{eq:ker_ident} identify it with the
orthogonal projection in Eq.~\eqref{eq:common-projector}, proving
Eq.~\eqref{eq:gbs-pair-projector-expansion}.

The commutator
$[\hat A,\hat A^\dagger]=4\widehat K+2m$ gives the ladder identity
\begin{equation}
 \hat A\ket{\Psi_{2s}}
 =2\sqrt{s(m/2+s-1)}\ket{\Psi_{2s-2}}\qquad(s\ge1).
 \label{eq:gbs-pair-ladder}
\end{equation}
Applying it $a$ times and taking the squared norm yields
\begin{equation}
 \bra{\Psi_k}G_{k,a}\ket{\Psi_k}
 =\frac{\|\hat A^a\ket{\Psi_k}\|_2^2}{4^a(a!)^2}
 =\frac{\fall sa\rise{m/2+s-a}a}{(a!)^2}.
 \label{eq:gbs-pair-reduction-moment}
\end{equation}
Since $\Pi_k$ is an orthogonal projector, its self-adjointness and
idempotence imply $\Tr(T_k^2)=\Tr(T_k\ket{\Psi_k}\bra{\Psi_k})$.
Substituting \eqref{eq:gbs-pair-projector-expansion} and
\eqref{eq:gbs-pair-reduction-moment} proves \eqref{eq:gbs-pair-norm}.
\end{proof}

\subsubsection{A determinant evaluates the colored overlap} \label{sec:gbs-color-determinant}
Let $w_i=\sum_{a=1}^{x_i}\lambda_{i,a}\xi_{g(i,a)}$ and
$W=\operatorname{diag}(w_1,\ldots,w_m)$, where $\xi_c^2=0$.
We set $B=UU^{\mathsf T}$ and form the formal series
\begin{equation}
\mathcal G_k(u,z,\xi)
 =\det\!\left[I_{2m}-
 \begin{pmatrix}uI&zB\\z\overline B&uI\end{pmatrix}
 \begin{pmatrix}W&0\\0&W\end{pmatrix}\right]^{-1/2}.
 \label{eq:gbs-mode-determinant}
\end{equation}
The branch has constant coefficient one.

\begin{lemma}[Colored determinant estimator for an angular overlap]
\label{lem:gbs-colored-overlap}
For even $2\le k\le|x|$, the estimator $\widehat B_k(U,x,g)$ is real and
\begin{equation}
 \E_g\widehat B_k(U,x,g)=B_k(U,x).
 \label{eq:gbs-angular-sketch-unbiased}
\end{equation}
For every fixed coloring, it admits the finite determinant formula
\begin{equation}
\widehat B_k(U,x,g)
 =\frac1{p_k}\sum_{a=0}^{k/2}\ell_{k,a}
 [\xi_1\cdots\xi_k u^{k-2a}z^{2a}]\mathcal G_k(u,z,\xi).
 \label{eq:gbs-colored-contraction}
\end{equation}
It can be evaluated in $\poly(m,|x|,k)2^{O(k)}$ arithmetic operations.
\end{lemma}
\begin{proof}
The overlap is real because its two operator factors are Hermitian.
Lemma~\ref{lem:bosonic-output-sketch} and linearity give
\eqref{eq:gbs-angular-sketch-unbiased}.
For the determinant identity, we temporarily keep separate slot variables.
Gaussian Wick contraction shows that the coefficient of $u^{k-2a}z^{2a}$ is the
occupation generating function of $U_kG_{k,a}U_k^\dagger$, with $G_{k,a}$
defined in Lemma~\ref{lem:gbs-pair-projection}. Taking one variable of
each color leaves exactly the rainbow subsets, with the repeated-mode
factorials present in $\widehat A_{x,g}$. That coefficient is therefore
$p_k\Tr[U_kG_{k,a}U_k^\dagger\widehat A_{x,g}]$. The harmonic linear
combination gives~\eqref{eq:gbs-colored-contraction}.
Every matrix factor contains positive color degree. Hence, in this
quotient ring,
\begin{equation}
 \det(I-H)^{-1/2}
 =\exp\left(\frac12\sum_{j=1}^{k}\frac{\Tr H^j}{j}\right)
\end{equation}
terminates after total color degree $k$. There are
$2^k\poly(k)$ coefficients, counting color subsets and powers of $u,z$.
Matrix multiplication and truncated series arithmetic therefore take
$\poly(m,|x|,k)2^{O(k)}$ operations.
\end{proof}

When $m<k$, we use the deterministic evaluation of
Lemma~\ref{lem:gbs-deterministic-evaluation}. The total number of occupation
coefficients through degree $k$ is $\binom{m+k}{k}\le4^k$, so its
preprocessing and point evaluation also cost
$\poly(m,|x|,k)2^{O(k)}$. This branch has zero sketch variance.

\subsubsection{Haar moments and the error of the degree truncation}
\label{sec:gbs-degree-moments}
For $d\in\mathbb N_0$, we define the contribution of exact total degree $d$
to the mean squared reference norm by
\begin{equation}
 \mathcal E_d:=\E_U\sum_x\mu_b(x)
       |f_{d,U}(x)-f_{d-1,U}(x)|^2,\qquad f_{-1,U}:=0.
 \label{eq:gbs-degree-energy-definition}
\end{equation}
Here and below $U$ is Haar-distributed, and the functions $f_{d,U}$
are the Meixner projections defined above. In particular,
$f_{0,U}=1$ and $\mathcal E_0=1$.

\begin{lemma}[Haar-averaged angular moments and degree contributions]
\label{lem:gbs-degree-energies}
For even $k\ge0$, let $U$ be Haar-distributed and let $X\sim u_K$
be independent of $U$, where $u_K$ is the uniform law on occupations
with total count $K$. With $h_k$ as in
\eqref{eq:gbs-angular-definitions}, one has
\begin{equation}
 \E_{U,X}B_k(U,X)^2
 =\frac{h_k(K)}{\rise m{2k}}
       \frac{\|T_k\|_2^2}{\binom{m+k-1}{k}+\binom{m+k-2}{k-1}}.
 \label{eq:gbs-angular-haar-second-moment}
\end{equation}
For every $d\ge0$, the exact-degree contribution is
\begin{equation}
 \mathcal E_d=\sum_{\substack{0\le k\le d\\k\ \mathrm{even}}}
 \frac{\gamma_k^2\kappa^k}{v^k}
 \frac{c_{k,d-k}^2}{N_{d-k}(m+2k,b)}
 \frac{\|T_k\|_2^2}{\binom{m+k-1}{k}+\binom{m+k-2}{k-1}}.
 \label{eq:gbs-exact-energy}
\end{equation}
Consequently, the averaged squared truncation error at degree $D$ is
\begin{equation}
 a_D:=\E_U\sum_x\mu_b(x)|f_U(x)-f_{D,U}(x)|^2
 =\sum_{d>D}\mathcal E_d.
 \label{eq:gbs-exact-weighted-tail}
\end{equation}
\end{lemma}

\begin{proof}

By Eq.~\eqref{eq:ker_ident}, $T_k$ lies in the irrep
$\mathcal V_{k,k}$. Its full and diagonal dimensions are
$v_k=d_k^2-d_{k-1}^2$ and $z_k=d_k-d_{k-1}$, respectively, by
Eqs.~\eqref{eq:common-irrep-dimensions}
and~\eqref{eq:common-diagonal-degree-dimension}.
Applying Lemma~\ref{lem:schur-orthogonality} to an orthonormal
basis of $(\mathcal V_{k,k})_0$ and summing by Parseval gives

\begin{equation}
 \E_U\omega_k(U)
 =\frac{z_k}{v_k}\|T_k\|_2^2
 =\frac{\|T_k\|_2^2}{\binom{m+k-1}{k}+\binom{m+k-2}{k-1}}.
 \label{eq:gbs-angular-haar-energy}
\end{equation}

Averaging \eqref{eq:gbs-angular-norm} over $U$ proves
\eqref{eq:gbs-angular-haar-second-moment}, including $K<k$, when both
sides vanish.

For each $k+\ell=d$, we multiply the squared coefficient in
\eqref{eq:gbs-radial} by the norm in \eqref{eq:gbs-angular-radial-norm}
and average using \eqref{eq:gbs-angular-haar-energy}. Summing these
orthogonal contributions proves \eqref{eq:gbs-exact-energy}.
Parseval gives the tail as the sum of the squared exact-degree norms
for Haar-almost every $U$. Tonelli's theorem passes the expectation
through this nonnegative sum, proving \eqref{eq:gbs-exact-weighted-tail}.
\end{proof}

With $q$ and $C_m$ from \eqref{eq:gbs-suppression-rate} and
\eqref{eq:gbs-total-budget}, the nonnegative quantities $W_d(U)$ in
\eqref{eq:gbs-ideal-degree-energies} obey
$\mathcal E_d=q^d\E_UW_d(U)$ and
$\sum_{d\ge1}\E_UW_d(U)\le C_m-1$.

\begin{corollary}[A finite bound on the degree tail]
\label{cor:gbs-finite-tail}
For integers $J\ge D\ge0$, the truncation error in
Lemma~\ref{lem:gbs-degree-energies} satisfies
\begin{equation}
 a_D\le\sum_{d=D+1}^J\mathcal E_d+
 q^{J+1}\left(C_m-1-\sum_{d=1}^J\frac{\mathcal E_d}{q^d}\right)
 \le(C_m-1)q^{D+1}.
 \label{eq:gbs-finite-tail-certificate}
\end{equation}
The bracket is nonnegative.
\end{corollary}

\begin{proof}
We separate the known degrees from the remaining ones:
\begin{equation}
 \begin{aligned}
 a_D&=\sum_{d=D+1}^J\mathcal E_d+\sum_{d>J}q^d\E_UW_d(U)\\
 &\le\sum_{d=D+1}^J\mathcal E_d+
       q^{J+1}\sum_{d>J}\E_UW_d(U)\\
 &\le\sum_{d=D+1}^J\mathcal E_d+
       q^{J+1}\left(C_m-1-\sum_{d=1}^J\frac{\mathcal E_d}{q^d}\right).
 \end{aligned}
 \label{eq:gbs-finite-tail-proof}
\end{equation}
The flatness bound proves nonnegativity of the bracket. In the finite
sum, we write $\mathcal E_d=q^d\E_UW_d(U)$ and use $q^d\le q^{D+1}$.
We also have $q^{J+1}\le q^{D+1}$. The resulting upper bound is
$q^{D+1}(C_m-1-\sum_{d=1}^D\E_UW_d(U))$, which is at most the last
quantity in \eqref{eq:gbs-finite-tail-certificate}.
\end{proof}

Computing finitely many degree contributions can thus sharpen the cutoff
\eqref{eq:gbs-flat-cutoff}. By \eqref{eq:gbs-flat-bound} and Markov's
inequality, making the first upper bound in
\eqref{eq:gbs-finite-tail-certificate} at most $\delta\epsilon^2$
gives trace-norm error at most $\epsilon$ with Haar probability at least
$1-\delta$.

\subsubsection{The variance bound includes the radial factors} \label{sec:gbs-point-variance}
For a proposed occupation $x$, we draw independent colors for different
retained angular degrees, with the exact branch when $m<k$, and define
\begin{equation}
 Z_D(U,x,g)=R_{0,D}(|x|)
 +\sum_{\substack{2\le k\le\min(D,|x|)\\k\ {\rm even}}}
 R_{k,D}(|x|)\widehat B_k(U,x,g_k).
 \label{eq:gbs-point-sketch}
\end{equation}

\begin{lemma}[Unbiased GBS sketch with bounded integrated variance]
\label{lem:gbs-sketch-variance}
For every integer $D\ge0$ and fixed $U,x$,
$\E_{g\mid x}Z_D(U,x,g)=f_{D,U}(x)$. Let $U$ be Haar-distributed and
$X\sim\mu_b$ be independent of $U$, with colors drawn conditionally
on $X$ as above. Then the integrated variance satisfies
\begin{equation}
 \E_{U,X\sim\mu_b,g}|Z_D-f_{D,U}|^2
 \le S_D^{\mathrm G}:=3(6e)^D C_m.
 \label{eq:gbs-variance-budget}
\end{equation}
The retained density satisfies
\begin{equation}
 \E_U\|f_{D,U}\|_{L^2(\mu_b)}^2\le C_m.
 \label{eq:gbs-reference-second-moment}
\end{equation}
For $D\le1$, the sketch is exact and its variance is zero.
\end{lemma}
\begin{proof}
Unbiasedness follows by substituting
Lemma~\ref{lem:gbs-colored-overlap} into the representation of
Lemma~\ref{lem:gbs-meixner-representation}. The branch $m<k$ is exact.
We first prove a bound valid for every radial multiplier. For
$m\ge k\ge2$, we condition on the output colors, which are independent
of $U$, and apply Lemma~\ref{lem:schur-orthogonality} to
$\Pi_k(\widehat A_{x,g})$ and $T_k$.

The irrep dimension in Eq.~\eqref{eq:common-irrep-dimensions},
contractivity of the orthogonal projector in
Eq.~\eqref{eq:common-projector}, and the moment bound of
Lemma~\ref{lem:bosonic-output-sketch} then give

\begin{equation}
 \E_{U,g}\widehat B_k(U,x,g)^2
 \le\frac{\|T_k\|_2^2}
 {\bigl(\binom{m+k-1}{k}^2-\binom{m+k-2}{k-1}^2\bigr)p_k}\|A_{x,k}\|_2^2.
\end{equation}
The unrestricted bosonic frame estimate of Lemma~\ref{lem:bosonic-frame}
holds even for $K>m$ and implies
\begin{equation}
 \frac{\E_{u_K}\|A_{X,k}\|_2^2}{h_k(K)}
 \le\frac{(1+2K/(m+1))^k}{k!\rise{m+K}k}
 \le\frac{2^k}{k!(m+1)^k}.
\end{equation}
Comparing with the Haar-averaged angular moment in
Lemma~\ref{lem:gbs-degree-energies}, using
$\binom{m+k-2}{k}=\rise{m-1}k/k!$, and then averaging in $K$, yields for every
real $R(K)$ with finite squared integrals
\begin{equation}
 \begin{aligned}
 &\E_{U,X\sim\mu_b,g}[R(K)\widehat B_k(U,X,g)]^2\\
 &\le \frac{2^k}{p_k}
 \frac{\rise m{2k}}{\rise{m-1}k(m+1)^k}
 \E_{U,X\sim\mu_b}[R(K)B_k(U,X)]^2\\
 &\le\frac{3\,6^k}{p_k}
 \E_{U,X\sim\mu_b}[R(K)B_k(U,X)]^2.
 \end{aligned}
 \label{eq:gbs-relative-sketch-variance}
\end{equation}
For the last inequality, the factorial ratio factors as
\begin{equation}
 \frac{m+k-1}{m-1}
 \prod_{a=0}^{k-1}\frac{m+k+a}{m+1}\le 3\,3^k
\end{equation}
when $m\ge k$ and $m\ge2$. For $m<k$ the exact branch has relative factor
one. We apply this estimate with $R(K)=R_{k,D}(K)$.

For fixed $U,x$, the color families are independent and the radial
term $R_{0,D}(|x|)$ is deterministic. Thus their conditional variances
add. With zero contributions when $|x|<k$, the preceding comparison gives
\begin{equation}
 \begin{aligned}
 \E_{U,X,g}|Z_D-f_{D,U}|^2
 &=\sum_{\substack{2\le k\le D\\k\ {\rm even}}}
   \E_{U,X}\operatorname{Var}_{g_k\mid X}
       \!\left(R_{k,D}(K)\widehat B_k(U,X,g_k)\right)\\
 &\le\sum_{\substack{2\le k\le D\\k\ {\rm even}}}
       3(6e)^k\E_{U,X}[R_{k,D}(K)B_k(U,X)]^2\\
 &\le3(6e)^D\E_U\|f_{D,U}\|_{L^2(\mu_b)}^2
 \le3(6e)^D C_m.
 \end{aligned}
\end{equation}
Here $p_k^{-1}\le e^k$. The second inequality uses orthogonality of
angular degrees even after multiplication by functions of $K$, as in
Lemma~\ref{lem:gbs-meixner-representation}. Orthogonal projection, reference contraction, and
\eqref{eq:gbs-total-budget} give \eqref{eq:gbs-reference-second-moment},
which proves the last inequality. The degree-zero contribution is exact.
For $D\le1$ it is the only term and the variance is zero.
\end{proof}

\subsubsection{Rejection sampling from the point estimator} \label{sec:gbs-fresh-estimates}
For each proposal, the occupations $X_1,\ldots,X_m$ are independent
geometric variables of mean $b$, so $X\sim\mu_b$. Let $K=|X|$.
Algorithm~\ref{alg:adaptive-rejection} obtains each value of
\eqref{eq:gbs-point-sketch} using fresh independent colors for the
retained angular degrees. The exact branches are reused.
Theorem~\ref{thm:final-sketch} combines
Lemma~\ref{lem:gbs-sketch-variance} and
Corollary~\ref{cor:gbs-approximation} with
Theorem~\ref{thm:adaptive-rejection}, using $C_{\rm ref}=C_m$ and
$S_D=S_D^{\rm G}$. The finite truncation certificate
\eqref{eq:gbs-finite-tail-certificate} may sharpen the choice of $D$.
For $D\le1$, we use the zero variance bound and the exact score.
Only proposals with $K\le K_*$ are evaluated. The cutoff rule and
its error are specified in Section~\ref{sec:precision}.

The radial factors can be tabulated in $\poly(D,K_*)$ operations for
proposals of total count at most $K_*$. With negative-index terms zero,
their recurrences are
\begin{equation}
\begin{aligned}
 \ell c_{k,\ell}
 &=-\Delta(2\ell-2+\nu_k)c_{k,\ell-1}
 -(\Delta^2-\kappa)(\ell-2+\nu_k)c_{k,\ell-2},\\
 \ell H_\ell^{(\nu,b)}(y)
 &=[y-\nu b-(1+2b)(\ell-1)]H_{\ell-1}^{(\nu,b)}(y)
 -v(\nu+\ell-2)H_{\ell-2}^{(\nu,b)}(y).
\end{aligned}
\end{equation}
Both start with constant term one. These recurrences and
\eqref{eq:gbs-harmonic-coefficients} supply the scalar coefficients.
The point determinant for degree $k$ uses
$\poly(m,K_*,k)2^{O(k)}$ arithmetic operations. Summing the retained
degrees gives one point evaluation in
$\poly(m,K_*,D)2^{O(D)}$ work. At most $A$ proposals and
$E_{\max}$ additional point evaluations therefore cost
\begin{equation}
 (A+E_{\max})\poly(m,K_*,D,J)2^{O(D)}.
 \label{eq:gbs-sketch-work}
\end{equation}
Here $J$ is the stage limit in \eqref{eq:adaptive-parameters}.
The mode matrix $B=UU^{\mathsf T}$ and radial coefficient tables are
prepared once. The Laguerre weights are computed once per proposed
occupation, and each new point evaluation forms new colored matrices.
Preprocessing and working storage are
$\poly(m,K_*,D)2^{O(D)}$, with point evaluations processed successively.
Section~\ref{sec:precision} bounds the reference proposal count tail
and the arithmetic errors.

\section{Finite precision and end-to-end guarantees}\label{sec:precision}
We now implement the point evaluations with finite precision and bounded
reference counts, then combine the model bounds with Section~\ref{sec:sampling}.

\subsection{Finite precision}
For the bit-complexity statements, the physical parameters $\eta,\beta$
and, for GBS, $\bar n$ are nonnegative rationals with at most $B$ bits,
in the nondegenerate ranges stated earlier. Each real and imaginary
part of the exactly unitary matrix $U$ can be computed to absolute
error $2^{-b}$ in time polynomial in $m,B,b$. Ensemble statements refer
to this underlying Haar unitary.

By \eqref{eq:rejection-precision}, a numerical error budget
$\eps_{\rm num}$ is met if the value and threshold errors are at most
$0<\zeta<1$, with
\begin{equation}
 \zeta\le\frac{M\eps_{\rm num}}{8A(J+1)}.
 \label{eq:point-precision-budget}
\end{equation}
We allocate $\zeta/2$ to each point evaluation and $\zeta/2$ to
forming its sample mean. We implement the acceptance probability in
Algorithm~\ref{alg:rejection-sampling} by comparison with a uniform
threshold on $[0,M]$, as in the coupling proof above. For both
algorithms, the implemented threshold is $M2^{-b_T}I$, where $I$ is
uniform on $\{0,\ldots,2^{b_T}-1\}$ and $M2^{-b_T}\le\zeta$.
Coupling it to $T=M\theta$ through
$I=\lfloor2^{b_T}\theta\rfloor$, with
$\theta\sim\mathrm{Unif}[0,1]$, gives threshold error at most $\zeta$.
We replace $e$ by $3$ in upper bounds used to choose budgets, so the stage
radii and integer work limits can then be computed exactly from
rational parameters.

\begin{lemma}[Finite-precision evaluation]
\label{lem:finite-precision-evaluation}
Under these input assumptions and the count bound $K_*$, both the
deterministic and randomized point evaluations admit polynomial bit
overhead. If $T_{\rm ar}$ is the number of arithmetic operations per
evaluation and $n_J$ bounds the number of evaluations in one cumulative
average, then point errors at most $\zeta/2$ and average errors at most
$\zeta$ are obtained using
\begin{equation}
 b_{\rm ar}=\poly(m,n,K_*,D,B)
 +O\!\left(\log(T_{\rm ar}+1)+\log(1/\zeta)+\log(n_J+1)\right)
 \label{eq:sufficient-bits}
\end{equation}
fractional bits and $O(b_{\rm ar})$ integer bits.
For deterministic evaluation, $n_J=1$. The parameter $n$ is omitted
for GBS.
\end{lemma}
\begin{proof}
For $s\ge1$, the Laguerre weights are the
simple positive roots of the integer polynomial
$Q_s(u)=s!u^sL_s(1/u)$, whose coefficient bit length is
$O(s\log(s+1))$. Theorem~3 of Ref.~\cite{sagraloff2016computing}
therefore computes them to absolute error $2^{-b}$ in
$\poly(s,b)$ bit operations. Their sum is $s$ by
\eqref{eq:bosonic-laguerre}, so each weight is at most $s\le K_*$.
There are no weights for $s=0$.

The randomized formulas use trace-log and exponential expansions
truncated to degree $O(D)$, with
$2^{O(D)}\poly(m,n,K_*,D)$ coefficients. The deterministic formulas in
Subsections~\ref{sec:young-enumeration}, \ref{sec:hahn-enumeration},
and \ref{sec:gbs-enumeration} use minors, permanents, or pair-polynomial
coefficients of order at most $D$, followed by the displayed
projections. Their expansions have $m^{O(D)}$ terms of degree $O(D)$
in the matrix entries and their conjugates. This also covers the GBS
fallback to deterministic evaluation.

All scalar normalizers can be computed exactly as rationals of
polynomial bit length in $m,n,K_*,D,B$. They consist of factorials,
binomial and rising-factorial factors, radial coefficients, powers
of $b(1+b)$ for GBS, and positive thermal sector weights. In particular,
the Hahn quantities in \eqref{eq:hahn-positive-count-expansion} and
\eqref{eq:hahn-thermal-sketch-weight} are finite rational sums and
ratios, and only sectors with $p_K>0$ are queried. Every division is
therefore by a known nonzero scalar whose reciprocal has polynomial
bit length.

We bound error amplification by replacing coefficients and input
entries by their absolute values. A trace of $j\le O(D)$ matrices
has at most $m^{O(j)}$ terms. The truncated exponential and projections
add at most $2^{\poly(D)}$ terms with polynomial-bit coefficients.
The bounds $|U_{ij}|\le1$, $|G_{ij}|\le4$, and
$\lambda_{i,a}\le K_*$ then give a common bound $H\ge1$ on the
values and their first derivatives, with
\begin{equation}
 \log H\le\poly(m,n,K_*,D,B).
 \label{eq:precision-majorant}
\end{equation}
The reciprocal normalizers satisfy the same logarithmic bound.
Enlarging each input bound by one covers the segment between its exact
and rounded values.

The same estimate controls intermediate rounding. Constant and
structural zero coefficients are kept exact. Every other formal
factor has positive degree, so a retained product contains at most
$O(D)$ such factors. Applying the absolute-coefficient bound to the
remaining computation after any rounded operation still gives
\eqref{eq:precision-majorant}, after increasing $H$. The mean-value
bound and a sum over the $T_{\rm ar}$ local errors bound the point
error by $H(T_{\rm ar}+1)2^{-b_{\rm ar}}$, with primitive input errors
included by increasing $H$. Equation~\eqref{eq:sufficient-bits} makes
this at most $\zeta/2$. The additional $\log(n_J+1)$ bits bound the
rounding error of the cumulative sum and division by $\zeta/2$.
\end{proof}

For discrete random choices, a uniform integer in a finite interval
is obtained by binary rejection, which succeeds with probability at
least $1/2$ per attempt. Capping each draw at $L_{\rm rng}$ attempts
changes its law by at most $2^{-L_{\rm rng}}$. If $T_{\rm draw}$
bounds the total number of these draws, the choice
\begin{equation}
 L_{\rm rng}\ge
 \left\lceil\log_2\frac{\max\{1,T_{\rm draw}\}}{\eps_{\rm rng}}\right\rceil
 \label{eq:random-choice-cap}
\end{equation}
limits their combined discrepancy to $\eps_{\rm rng}$. No draws are
needed when $T_{\rm draw}=0$. All integer ranges have polynomial bit
length. There are at most $A$ reference draws and $A+E_{\max}$
randomized point evaluations, each using polynomially many such
choices. Deterministic sampling uses $A$ evaluations. The dyadic
thresholds and Young matrix atoms use a fixed number of unbiased
bits. Rational Bernoulli trials generate the bosonic geometric counts
and stop as soon as the accumulated count exceeds $K_*$, so these
draws also have a bounded number of trials.

\subsection{Reference proposal tails} \label{sec:reference-proposal-tails}
If a proposed count exceeds $K_*$, the finite implementation returns
the fallback occupation. We choose $K_*$ so that
\begin{equation}
 A\Prob_\mu(K>K_*)\le\eps_{\rm prop}.
 \label{eq:proposal-count-coupling}
\end{equation}
Coupling at most $A$ proposals to the uncapped procedure then bounds
the additional total variation by $\eps_{\rm prop}$.
The following bound determines a sufficient cutoff for both bosonic
reference distributions.

\begin{lemma}[Reference photon-count tails]
\label{lem:reference-count-tails}
Let $K$ be the total count of a reference proposal. For integer $K_*\ge n$
in Hahn, with $t=(1-\eta)\beta$, and $K_*\ge0$ in GBS,
\begin{equation}
 \Prob_\mu(K>K_*)\le
 \begin{cases}
  2^{m+n}\left(\dfrac{2t}{1+2t}\right)^{K_*-n+1},&\text{Hahn},\\[4pt]
  2^m\left(\dfrac{2b}{1+2b}\right)^{K_*+1},&\text{GBS}.
 \end{cases}
 \label{eq:reference-count-tail}
\end{equation}
\end{lemma}
\begin{proof}
For $r$ independent geometric variables $G_i$ of mean $c>0$,
$\E z^{\sum_iG_i}=[1+c(1-z)]^{-r}$. Markov's inequality at
$z=1+1/(2c)$ gives, for integer $u\ge0$,
\begin{equation}
 \Prob\!\left(\sum_{i=1}^rG_i>u\right)
 \le 2^r\left(\frac{2c}{1+2c}\right)^{u+1}.
\end{equation}
The Hahn construction \eqref{eq:hahn-reference-proposal} has
$J\le n$ and $K=J+\sum_{i=1}^{m+J}G_i$ with mean-$t$ geometric
variables independent of $J$. Using the same variables couples $K$
below $n+\sum_{i=1}^{m+n}G_i$. For GBS, $K$ is already the sum of
$m$ mean-$b$ geometric variables. At zero geometric mean, all $G_i$
vanish, so $K\le n$ for Hahn and $K=0$ for GBS.
\end{proof}

Thus \eqref{eq:proposal-count-coupling} holds with
$K_*=n+O((1+t)[m+n+\log(A/\eps_{\rm prop})])$ for Hahn and
$K_*=O((1+b)[m+\log(A/\eps_{\rm prop})])$ for GBS.
For pure loss in Hahn, $K_*=n$ gives zero tail. Young requires no
truncation since $K\le m$.

\subsection{Putting the guarantees together}\label{sec:combined-guarantees}
Let $a_D$ bound the Haar mean squared weighted $L^2$ truncation error,
$S_D$ the Haar mean integrated point-estimator variance, and
$C_{\rm ref}$ the Haar mean of $\|f_{D,U}\|_{L^2(\mu)}^2$.
Table~\ref{tab:haar-guarantees} gives sufficient choices.
\begin{table}[t]
\caption{Bounds used to choose the degree and rejection-sampling parameters.}
\label{tab:haar-guarantees}
\centering\small
\begin{tabular}{p{22mm}p{38mm}p{56mm}p{21mm}}
\toprule
Model & Truncation $a_D$ & Point variance $S_D$ & $C_{\rm ref}$\\
\midrule
Young & $3\kappa_{\rm F}^{D+1}$
 & $\displaystyle\sum_{\substack{1\le k\le\min(D,n)\\k\ \mathrm{even}}}12^k\Gamma_k^{\rm F}$
 & $4$\\[5pt]
Hahn & $s^{D+1}/(1-s)$
 & $\displaystyle2\sum_{k=1}^{\min(D,n)}(3e^2s)^k$
 & $(1-s)^{-1}$\\[5pt]
GBS & $(C_m-1)q^{D+1}$
 & $3(6e)^D C_m$ & $C_m$\\
\bottomrule
\end{tabular}
\end{table}
Here $s=\eta/[1+(1-\eta)\beta]$, while $q$ and $C_m$ are the GBS
contraction and total Haar budget. Young and Hahn have zero truncation
error at their finite exact degree cutoff. Exact thermal degree sums
and the GBS finite spectrum can sharpen the tabulated truncation bounds.
The reference-norm bounds follow from Young flatness, the Hahn
energies $E_k^{\rm th}\le s^k$, and the GBS total reference norm.

\begin{corollary}[Rejection sampling with deterministic evaluation]
\label{thm:final-enumeration}
For a fixed interferometer $U$ with the model's specified input access,
let $A_D\ge\|f_{D,U}-P_U/\mu\|_{L^2(\mu)}^2$ be a deterministic bound.
We use Algorithm~\ref{alg:rejection-sampling} with the model's exact
point evaluator and reference sampler. For Young and Hahn, let $M$
be the envelope in Lemma~\ref{lem:young-density-envelope} or
Lemma~\ref{lem:hahn-deterministic-envelope} and set $\eps_{\rm clip}=0$.
For GBS, let $0<\eps_{\rm clip}\le1/2$ and choose
\begin{equation}
 M\ge\max\{1,B_{m,D}/(4\eps_{\rm clip})\},
 \label{eq:deterministic-rejection-threshold}
\end{equation}
with $B_{m,D}$ from Lemma~\ref{lem:gbs-deterministic-norm}.
For $0<\eps_{\rm stop}<1$ and
\begin{equation}
 A\ge\left\lceil\frac{M}{1-\eps_{\rm clip}}
                         \log\frac1{\eps_{\rm stop}}\right\rceil,
 \label{eq:deterministic-rejection-attempts}
\end{equation}
the finite implementation satisfies
\begin{equation}
 \TV(Q,P_U)\le\frac{\sqrt{A_D}}2+
 \eps_{\rm clip}+\eps_{\rm stop}+\eps_{\rm num}
 +\eps_{\rm rng}+\eps_{\rm prop}.
 \label{eq:deterministic-rejection-error}
\end{equation}
It uses deterministic preprocessing and at most $A$ point evaluations
and reference draws, with the polynomial bit overhead established above.
\end{corollary}
\begin{proof}
Weighted Cauchy--Schwarz gives $\|q_{D,U}-P_U\|_1\le\sqrt{A_D}$.
The envelope or \eqref{eq:deterministic-rejection-threshold} gives
$\tau_M\le\eps_{\rm clip}$. Lemma~\ref{lem:rejection-sampling}
then gives the first three error terms. Equations
\eqref{eq:rejection-precision}, \eqref{eq:random-choice-cap}, and
\eqref{eq:proposal-count-coupling} add the implementation budgets.
\end{proof}

At fixed transmissivity and fixed squeezing in GBS, the deterministic
evaluators and their
envelope or norm bounds have size $(m+n+D)^{O(D)}$, with $n$ omitted
for GBS. Choosing $M,A$ within constant factors of the stated bounds
therefore preserves this cost up to polynomial factors in count limits
and inverse error budgets. For a fixed interferometer with flatness
$\chi^2\le C$, let $\mathcal S=m+n$ for Young and Hahn and $\mathcal S=m$
for GBS. At fixed transmissivity $\eta<1$ and fixed squeezing in GBS,
a sufficient degree and bit runtime are
\begin{equation}
 T_{\rm det}\le(\mathcal S+D)^{O(D)}
                  \poly(1+\beta,B,\eps^{-1}),
 \qquad D=O(\log[(1+C)/\eps^2]).
 \label{eq:flatness-runtime}
\end{equation}
For $0\le\beta\le\mathcal S^{\mathcal{O}(1)}$, this is quasipolynomial for
polynomial $C$ and inverse-polynomial accuracy, and polynomial
for bounded $C$ and fixed accuracy. Under an
ensemble bound $\E\chi^2\le C_0$, Markov's inequality permits
$C=C_0/\delta$ outside a set of probability at most $\delta$.

Homogeneous thermal loss can be moved through the passive layers and
combined into a single channel. At logarithmic depth, the resulting
transmissivity is polynomially small in $m$, so a much smaller degree
cutoff suffices.

\begin{proof}[Proof of Corollary~\ref{cor:main-logarithmic-depth}]
A thermal environment with the same occupation $\beta$ in every mode
is invariant under passive transformations. The coupling
\eqref{eq:thermal-mode-coupling} therefore shows that homogeneous
thermal loss commutes with passive interferometers. It also gives
\begin{equation}
 \mathcal N_{\eta_2,\beta}\circ\mathcal N_{\eta_1,\beta}
 =\mathcal N_{\eta_1\eta_2,\beta}.
 \label{eq:layered-thermal-composition}
\end{equation}
Indeed, the two environment contributions have total squared
coefficient $1-\eta_1\eta_2$, and their normalized combination is
again a thermal mode of occupation $\beta$. This holds for both
particle statistics. We can thus combine all losses into one channel
with transmissivity $\eta=\eta_0^L$.

By \eqref{eq:young-kappa}, \eqref{eq:hahn-q}, and
\eqref{eq:gbs-suppression-rate},
\begin{equation}
 \kappa_{\rm F}\le\eta,\qquad
 q_{\rm B}\le2\eta,\qquad
 q\le(1+a)\eta.
 \label{eq:layered-degree-contractions}
\end{equation}
Here we used $n\le m$ and $s\le\eta$ for Hahn; for GBS,
$a=\bar n+(1+\bar n)/m$ is bounded because the squeezing is fixed.

Beyond a depth $L_*=O(\log m+\log(1/\eps))$, the output distribution
is $\eps/2$-close in total variation to the occupation distribution
of the product thermal state $\vartheta_\beta^{\otimes m}$, which
can be sampled efficiently; we therefore assume $L<L_* = \mathcal{O}(\log m)$ below.

Since $L=\Theta(\log m)$, all three contraction factors are at most
$m^{-c}$ for some constant $c>0$ and sufficiently large $m$.
Choose a constant $p\ge0$ such that
$\chi^2(\rho_U\|\tau)\le m^p$.
The weighted estimate \eqref{eq:master-tail-proof} gives
\begin{equation}
 \|f_{D,U}-P_U/\mu\|_{L^2(\mu)}^2
 \le m^{p-c(D+1)}.
 \label{eq:layered-flatness-tail}
\end{equation}
We choose $D$ so that the right-hand side is at most $\eps^2/4$.
This requires only
\begin{equation}
 D=O\!\left(1+\frac{\log(1/\eps)}{\log m}\right).
\end{equation}
For Young and Hahn, we may stop at the finite exact degree cutoff,
where the truncation error vanishes.

The retained density has reference mean one. Orthogonal projection
and Lemma~\ref{lem:reference-contraction} also give
\begin{equation}
 \|f_{D,U}\|_{L^2(\mu)}^2
 \le 1+\chi^2(\rho_U\|\tau)\le1+m^p.
\end{equation}
Use Algorithm~\ref{alg:rejection-sampling} with deterministic point
evaluations and
\begin{equation}
 M=\frac{1+m^p}{\eps},\qquad
 A=\left\lceil2M\log\frac4\eps\right\rceil.
\end{equation}
Weighted Cauchy--Schwarz bounds the approximation contribution to
total variation by $\eps/4$. Lemma~\ref{lem:rejection-sampling}
bounds each of the clipping and stopping contributions by $\eps/4$,
leaving $\eps/4$ for the finite implementation.

It remains to check the work. The Young and Hahn evaluators in
Subsections~\ref{sec:young-enumeration} and
\ref{sec:hahn-enumeration} cost $m^{O(D)}$ times polynomial factors,
since $n\le m$ and the retained degrees are at most $n$.
For GBS, \eqref{eq:gbs-enumeration-preprocessing} and
\eqref{eq:gbs-enumeration-sample-cost} are polynomial in
$m,D,K_*,A$ and
\begin{equation}
 Q_D=\binom{m+D}{D}\le(m+1)^D
     =\poly(m,\eps^{-1}).
\end{equation}
The same degree choice makes $m^{O(D)}$ polynomial in $m$ and
$\eps^{-1}$. Under the assumed polynomial bound on $\beta$,
Lemma~\ref{lem:reference-count-tails} permits a total-count cutoff
$K_*$ polynomial in $m$ and $\log(1/\eps)$.
Finally, forming $\eta_0^L$ requires $O(B\log m)$ bits, and
Lemma~\ref{lem:finite-precision-evaluation} adds only polynomial bit
overhead in $m,B,\eps^{-1}$ under the stated input-access assumptions.
Allocating the remaining $\eps/4$ to the implementation errors in
Section~\ref{sec:precision} proves the claimed guarantee.
\end{proof}

For randomized evaluation and $0<\eps,\delta<1$, we use the schedule
\eqref{eq:adaptive-parameters} and limits \eqref{eq:adaptive-work-limits}
with
\begin{equation}
 \begin{gathered}
 a_D\le\frac{\delta\eps^2}{64},\qquad
 \overline C=\frac{4C_{\rm ref}}\delta,\qquad
 \overline V=\frac{4S_D}\delta,\\
 M=\frac{2\overline C}\eps,\qquad
 \gamma=\frac\eps8,\qquad
 \beta_{\rm p}=\beta_{\rm e}=\frac\eps{16}.
 \end{gathered}
 \label{eq:final-budget}
\end{equation}
The numerical, random-choice, and reference-count budgets satisfy
\begin{equation}
 \eps_{\rm num}+\eps_{\rm rng}+\eps_{\rm prop}\le\eps/2.
 \label{eq:final-implementation-budget}
\end{equation}

\begin{theorem}[Rejection sampling with adaptive evaluations]
\label{thm:final-sketch}
For each family in Table~\ref{tab:haar-guarantees},
Algorithm~\ref{alg:adaptive-rejection} with
\eqref{eq:final-budget}--\eqref{eq:final-implementation-budget}
has output law within total variation $\eps$ of $P_U$ outside a set
of Haar interferometers of probability at most $\delta$.
Its bounded arithmetic cost per output is
\begin{equation}
 (A+E_{\max})\poly(m,n,K_*,D,J)2^{O(D)},
 \label{eq:final-sketch-cost}
\end{equation}
with $n$ omitted for GBS, where
\begin{equation}
 A+E_{\max}=O\!\left(
 \frac{C_{\rm ref}}{\delta\eps}\log\frac{16}\eps
 +\frac{S_D}{\delta\eps^2}
       \log^2\frac{16C_{\rm ref}}{\delta\eps^2}\right).
 \label{eq:adaptive-final-evaluations}
\end{equation}
Young point evaluation is polynomial in $D$. The factor $2^{O(D)}$
appears only in the bosonic cost bounds. Bit complexity has the
polynomial overhead established above. The output law averages all
sampler randomness, while the exceptional probability concerns only $U$.
\end{theorem}
\begin{proof}
For a fixed interferometer, let
\begin{equation}
 C(U)=\|f_{D,U}\|_{L^2(\mu)}^2,\qquad
 V(U)=\E_{X\sim\mu}\Var_{\xi\mid X}Z_U(X,\xi).
\end{equation}
The model lemmas give conditional unbiasedness,
$\E_U V(U)\le S_D$, and $\E_U C(U)\le C_{\rm ref}$.
Markov's inequality gives
\begin{equation}
 \begin{aligned}
 \Prob_U\{\|q_{D,U}-P_U\|_1>\eps/4\}
 &\le\frac{16a_D}{\eps^2}\le\delta/4,\\
 \Prob_U\{C(U)>\overline C\}
 &\le\frac{C_{\rm ref}}{\overline C}=\delta/4,\\
 \Prob_U\{V(U)>\overline V\}&\le\delta/4.
 \end{aligned}
 \label{eq:final-rejection-events}
\end{equation}
The last bound uses $S_D/\overline V=\delta/4$ for $S_D>0$.
If $S_D=0$, then $V(U)=0$ almost surely. Outside these events,
Theorem~\ref{thm:adaptive-rejection} gives the ideal error bound
\begin{equation}
 \TV(Q,P_U)\le\frac\eps8+\frac{\overline C}{4M}
 +\gamma+\beta_{\rm p}+\beta_{\rm e}=\frac\eps2.
 \label{eq:final-rejection-ideal-error}
\end{equation}
The implementation adds at most $\eps/2$, and the exceptional Haar
probability is at most $3\delta/4\le\delta$.
Since $J+1=O(\log[16C_{\rm ref}/(\delta\eps^2)])$ and
$z_*=1-\eps/4$, substitution in \eqref{eq:adaptive-work-limits}
gives \eqref{eq:adaptive-final-evaluations}. Multiplying by the
model point-evaluation costs gives \eqref{eq:final-sketch-cost}.
\end{proof}

At fixed nonzero loss and fixed squeezing in GBS, Young and Hahn
permit $D=O(\log(1/(\delta\eps^2)))$, capped at their exact degree.
GBS permits $D=O(\log(m/(\delta\eps^2)))$, since $C_m=O(m)$ and
its contraction stays below one. These bounds hold uniformly over
the allowed bath occupations. In all three models, $S_D$ is at most
a polynomial system-size factor times $2^{O(D)}$.
For $0\le\beta\le m^{\mathcal{O}(1)}$, the reference count cutoffs are polynomial
in system size and the logarithms of the inverse error budgets.
The bounded bit runtime is therefore polynomial in system size,
$\eps^{-1}$, and $\delta^{-1}$, with exponents depending on the
fixed loss and squeezing parameters and the polynomial bound on
the bath occupation.
Processing proposals and point estimates successively uses
$(m+n+D)^{O(D)}$ deterministic storage, up to count and precision
factors, or $\poly(m,n,K_*,D)2^{O(D)}$ randomized storage, with $n$
omitted for GBS.

\end{document}